\documentclass[preprint]{aastex}

\pdfoutput=1

\usepackage{amsmath}
\usepackage{lineno}
\usepackage{lscape}
\usepackage{hyperref}

\newcommand{\fermi}{\emph{Fermi}}
\newcommand{\DG}{$^{\circ}$}
\newcommand{\al}{$\alpha$}
\newcommand{\ze}{$\zeta$}
\newcommand{\Per}{$P$}
\newcommand{\Pd}{$\dot{P}$}
\newcommand{\Blc}{$B_{\rm LC}$}
\newcommand{\Rlc}{$R_{\rm LC}$}
\newcommand{\Rns}{$R_{\rm NS}$}
\newcommand{\Ed}{$\dot{E}$}
\newcommand{\Tpc}{$\Theta_{\rm PC}$}
\newcommand{\fO}{$f_{\Omega}$}
\newcommand{\lnl}{$-\ln(\mathcal{L})$}

\begin{document}

\title{Constraints on the Emission Geometries and Spin Evolution of Gamma-ray Millisecond Pulsars}

\author{
T.~J.~Johnson\altaffilmark{1,2}, 
C.~Venter\altaffilmark{3,4}, 
A.~K.~Harding\altaffilmark{5,6}, 
L.~Guillemot\altaffilmark{7}, 
D.~A.~Smith\altaffilmark{8}, 
M.~Kramer\altaffilmark{9,10}, 
\"O.~\c{C}elik\altaffilmark{5,11,12},
P.~R.~den~Hartog\altaffilmark{13}, 
E.~C.~Ferrara\altaffilmark{5}, 
X.~Hou\altaffilmark{8}, 
J.~Lande\altaffilmark{14}, 
P.~S.~Ray\altaffilmark{15}, 
}
\altaffiltext{1}{National Research Council Research Associate, National Academy of Sciences, Washington, DC 20001, resident at Naval Research Laboratory, Washington, DC 20375, USA}
\altaffiltext{2}{email: tyrel.j.johnson@gmail.com}
\altaffiltext{3}{Centre for Space Research, North-West University, Potchefstroom Campus, Private Bag X6001, 2520 Potchefstroom, South Africa}
\altaffiltext{4}{email: Christo.Venter@nwu.ac.za}
\altaffiltext{5}{NASA Goddard Space Flight Center, Greenbelt, MD 20771, USA}
\altaffiltext{6}{email: ahardingx@yahoo.com}
\altaffiltext{7}{Laboratoire de Physique et Chimie de l'Environnement, LPCE UMR 6115 CNRS, F-45071 Orl\'eans Cedex 02, and Station de radioastronomie de Nan\c{c}ay, Observatoire de Paris, CNRS/INSU, F-18330 Nan\c{c}ay, France}
\altaffiltext{8}{Centre d'\'Etudes Nucl\'eaires de Bordeaux Gradignan, IN2P3/CNRS, Universit\'e Bordeaux 1, BP120, F-33175 Gradignan Cedex, France}
\altaffiltext{9}{Max-Planck-Institut f\"ur Radioastronomie, Auf dem H\"ugel 69, 53121 Bonn, Germany}
\altaffiltext{10}{Jodrell Bank Centre for Astrophysics, School of Physics and Astronomy, The University of Manchester, M13 9PL, UK}
\altaffiltext{11}{Center for Research and Exploration in Space Science and Technology (CRESST) and NASA Goddard Space Flight Center, Greenbelt, MD 20771, USA}
\altaffiltext{12}{Department of Physics and Center for Space Sciences and Technology, University of Maryland Baltimore County, Baltimore, MD 21250, USA}
\altaffiltext{13}{W. W. Hansen Experimental Physics Laboratory, Kavli Institute for Particle Astrophysics and Cosmology, Department of Physics and SLAC National Accelerator Laboratory, Stanford University, Stanford, CA 94305, USA}
\altaffiltext{14}{Twitter Inc., 1355 Market St \#900, San Francisco, CA 94103, USA}
\altaffiltext{15}{Space Science Division, Naval Research Laboratory, Washington, DC 20375-5352, USA}

\begin{abstract}
Millisecond pulsars (MSPs) are a growing class of gamma-ray emitters.  Pulsed gamma-ray signals have been detected from more than 40 MSPs with the \fermi\ Large Area Telescope (LAT).  The wider radio beams and more compact magnetospheres of MSPs enable studies of emission geometries over a broader range of phase space than non-recycled radio-loud gamma-ray pulsars.  We have modeled the gamma-ray light curves of 40 LAT-detected MSPs using geometric emission models assuming a vacuum retarded-dipole magnetic field.  We modeled the radio profiles using a single-altitude hollow-cone beam, with a core component when indicated by polarimetry; however, for MSPs with gamma-ray and radio light curve peaks occurring at nearly the same rotational phase we assume that the radio emission is co-located with the gamma rays and caustic in nature.  The best-fit parameters and confidence intervals are determined using a maximum likelihood technique.  We divide the light curves into three model classes, with gamma-ray peaks trailing (Class I), aligned (Class II) or leading (Class III) the radio peaks.  Outer gap and slot gap (two-pole caustic) models best fit roughly equal numbers of Class I and II, while Class III are exclusively fit with pair-starved polar cap models.  Distinguishing between the model classes based on typical derived parameters is difficult.  We explore the evolution of magnetic inclination angle with period and spin-down power, finding possible correlations.  While the presence of significant off-peak emission can often be used as a discriminator between outer gap and slot gap models, a hybrid model may be needed.
\end{abstract}


\section{INTRODUCTION}\label{intro}
Millisecond pulsars \citep[MSPs, first discovered by][]{Backer82} are thought to be old pulsars, spun up to extremely short periods ($P \lesssim 25$ ms) via accretion from a companion \citep[e.g.,][]{Alpar82}, and are often dubbed ``recycled'' pulsars.  The recycled pulsar scenario is supported by the fact that $\sim80$\% of MSPs are in binary systems and the detection of millisecond X-ray pulsations from neutron stars in low-mass X-ray binaries \citep[LMXBs, e.g.,][]{WK98,Chak05}, presumed to be the progenitors of radio MSPs.
Further evidence supporting this model was provided by the discovery of a radio MSP that had shown LMXB behavior, and no pulsations, in the past \citep{Archibald09}, and is thought to be a missing link in the LMXB-to-MSP evolutionary chain.  More recent observations of PSR J1824$-$2452I in the globular cluster M28 (NGC 6626) transitioning from rotation-powered radio pulsar to accretion-powered X-ray pulsar, and back, have made this interpretation even more certain \citep{Papitto13,ATEL5069}.

The rotation periods of MSPs are observed to be increasing at a much slower rate than non-recycled pulsars \citep[\Pd\ typically $\sim 10^{-20}$ s s$^{-1}$ for MSPs versus $\sim 10^{-15}$ s s$^{-1}$ for non-recycled pulsars,][]{handbook}.  This leads to weaker inferred surface magnetic fields ($B_{\rm surf}= (1.5 I c^{3} \dot{P} P)^{1/2}/(2\pi R_{\rm NS}^{3}) \lesssim 10^{9}$ G, for an orthogonal rotator assuming dipole spin down and with \Rns\ the neutron star radius, $c$ the speed of light in vacuum, and $I$ the neutron star moment of inertia).  \citet{Lee12} have empirically defined MSPs as those pulsars satisfying:

\begin{equation}\label{eqn-LeeMSPs}
\frac{\dot{P}}{10^{-17}}\ \leq\ 3.23\Big(\frac{P}{100\ \rm{ms}}\Big)^{-2.34}.
\end{equation}

The light-cylinder radii of MSPs (cylindrical radius where co-rotation with the neutron star requires moving at the speed of light, $R_{\rm LC} = c/\Omega$ with $\Omega = 2\pi/P$) are on the order of tens to hundreds of kilometers as opposed to many thousands of kilometers in non-recycled pulsars.  The radio beams and polar cap sizes of MSPs are also very broad, making them detectable over a larger range of viewing geometries than non-recycled pulsars.  Therefore, MSPs provide excellent opportunities to study the global pulsar magnetosphere in detail through analysis of pulse profiles at different wavelengths.

Using timing solutions from radio observatories around the world \citep{Smith08}, MSPs have been established as a class of high-energy (HE, $\geq$0.1 GeV) emitters \citep[e.g.,][]{MSPpop,1PC,J0034,Guillemot12} via observations with the Large Area Telescope \citep[LAT,][]{Atwood09}, the main instrument on the \emph{Fermi Gamma-ray Space Telescope}.  Additionally, steady point-source emission has been detected from the vicinity of more than a dozen globular clusters \citep{47Tuc,Kong10,GCpop,Tam11,2FGL} consistent with emission from the combination of many MSPs.  HE pulsations have been detected from two extremely luminous MSPs in globular clusters \citep{Freire11,wuM28A,JohnsonB1821}.  The population of known radio MSPs in the Galactic field\footnote{Those not in globular clusters.} has been increased by $\sim$50\% through follow-up searches of unassociated LAT sources with pulsar-like characteristics \citep[e.g.,][]{Ransom11,Cognard11,Keith11,Ray12,Barr13,Bhatt13}, suggesting that MSPs are generally gamma-ray emitters.  Of these new radio MSPs, over 75\% are in binaries and 11 are ``black-widow'' systems \citep[with extremely low-mass companions thought to have been ablated by the pulsar wind, see][for a review]{Roberts11}, further supporting the recycling scenario.

The origin of HE pulsed emission remains an important question in gamma-ray pulsar physics.  It is commonly accepted that the observed HE gamma rays are primarily the result of curvature radiation from electrons/positrons accelerated along the magnetic field lines by the rotationally-induced electric field, but the recent detections of pulsations from the Crab pulsar at energies up to $\sim$400 GeV by VERITAS \citep{Aliu11} and MAGIC \citep{Aleksic11,Aleksic12} suggest that either an additional component is necessary or a different process is at work \citep[e.g.,][]{Lyutikov12}.  The exact location in the magnetosphere where the acceleration occurs is still uncertain.  The light curves of gamma-ray pulsars detected with the LAT strongly suggest that emission occurs in the outer magnetosphere rather than near the polar caps, in narrow gaps bordering the closed-field-line boundary, but it is not yet clear how the acceleration or emission is distributed.  Several authors have attempted to address these questions by generating simulated light curves using either geometric models \citep[e.g.,][]{Cheng86b,Venter09,BS10b,WR11} or full radiation models \citep[e.g.,][]{Harding08,Wang11}.

We have simulated gamma-ray and radio MSP light curves assuming the vacuum retarded-dipole (VRD) magnetic field geometry of \citet{Deutsch55}.  These simulations have been used to fit the observed light curves of the 40 gamma-ray MSPs from which significant pulsed signals have been detected with the LAT in three years of sky-survey operations as reported in the second LAT catalog of gamma-ray pulsars \citep[2PC hereafter,][]{2PC}, all of which satisfy the inequality in Equation \ref{eqn-LeeMSPs}.

We fit the simulated light curves to the observed profiles and estimate uncertainties on the best-fit model parameters, using either one or two-dimensional likelihood profiles, and discuss the implications of trends in the best-fit parameters.
Similar to \citet{Venter12}, we define three model classes as follows: MSPs with gamma-ray peaks trailing the radio peaks (by $\lesssim0.5$ in phase) are Class I, those with gamma-ray and radio peaks nearly aligned in phase (to within 0.1) are Class II, and those with gamma-ray peaks leading the radio peaks (by between 0.3 and 0.1 in phase, corresponding to radio lags between 0.7 and 0.9 in phase) are Class III.  In Appendix \ref{app-fits} we provide, for each MSP, the observed and best-fit light curves, summarize the observational characteristics, discuss how the models match the data, and compare our results to those from other methods when possible.  Appendices \ref{app-contours} and \ref{app-skymaps} provide confidence contours and maps of simulated emission on the sky for selected MSPs, respectively.

Due to the choice of radio phase zero in 2PC, PSRs J0034$-$0534 and J1810$-$1744 have radio lags of 0.866 and 0.849, respectively, but have wide double-peaked radio and gamma-ray light curves with similar morphology and thus are considered Class II for our purposes.  Additionally, PSRs J1744$-$1134 and J2214+3000 have radio lags of 0.2 to 0.3 in phase but are considered Class III in our studies.  For PSR J1744$-$1134 this is due to the use of the `h' method for shifting the profile, which puts phase zero at the weaker radio interpulse.  For PSR J2214+3000, phase zero is placed at the peak radio intensity, the `p' method, and the radio lag is referenced to the first gamma-ray peak appearing later in phase, but we note that the highest gamma-ray peak occurs just before the highest radio peak.

\citet{Espinoza13} separated the gamma-ray MSPs known at the time into three types based on the characteristics of their gamma-ray and radio light curves.  Their A-type MSPs, those that have the main gamma-ray peak aligned with the main radio pulse, overlap exactly with our Class II MSPs with the exception of PSR J1810+1744, which was not known to be a gamma-ray emitter prior to 2PC.  Their N-type and W-type MSPs are a mix of our Class I (3) and III (1) MSPs.  Both N-type and W-type MSPs are defined as those without the main gamma-ray and radio peaks occurring at or near the same phase, with the radio profile of the former dominated by a single pulse and the latter consisting of wide peaks covering most of the pulse phase.  \citet{Espinoza13} hypothesized that the W-type MSPs were possibly aligned rotators, explaining the wide radio profiles and high duty cycles.

\section{PULSAR GEOMETRY}\label{psrgeom}
The viewing geometry of a pulsar is defined by two angles, that between the spin axis and the magnetic dipole (\al) and that between the spin axis and the observer's line of sight (\ze).  In studies of radio pulsars, the geometry is often characterized using \al\ and $\beta\equiv\zeta-\alpha$, with $\beta$, the impact parameter, describing the closest approach between the line of sight and the magnetic axis.

Using geometric simulations, it is generally possible to reproduce almost any light curve shape for some combination of model parameters and viewing geometry, especially when the number of free parameters is large.  As such, it is important to use all the multi-wavelength information available to restrict the parameter space and more strictly test the models.  For reasons discussed below, strong constraints on viewing geometry exist for only a few MSPs and other considerations guide our modeling.

Pulsar viewing geometries can be constrained from radio polarimetry by fitting the observed position-angle swings with predictions from the rotating-vector model \citep[RVM,][]{RVM}.  However, the position-angle swings of MSPs are typically not well fit by simple RVM predictions resulting in weak or no geometric constraints \citep[e.g.,][]{Xil98,Stairs99,Ord2004,Yan11}.  Relativistic and higher-altitude corrections can be applied to the RVM and lead to better constraints in some cases \citep[e.g.,][]{Blaskiewicz91,CR12}.

Polarization measurements can be used to guide the type of radio model used \citep[i.e., a cone and/or core beam following][]{Rankin83}.  The characteristic S-shaped or nearly flat linear polarization position-angle swings are indicative of a hollow-cone beam viewed near or far from the magnetic dipole axis, respectively.  Sense-reversing circular polarization has been interpreted as evidence for a core beam, and we take it as such here.  \citet{Han98} note that this is not restricted to core beams, but when observed in conal components the sense of circular polarization is correlated with that of the linear polarization, which is not true for core components.  Additionally, \citet{BP12} have shown that the sense reversal can be due to propagation effects as the radiation travels through the pulsar magnetosphere.  Most Class II MSPs are observed to have little or no polarized emission, which may be indicative of caustic emission \citep[see Section \ref{models} and][]{Dyks04a,Venter12}.

In binary systems the inclination angle of the orbit can be constrained if significant Shapiro delay is measured \citep[e.g,][]{Demorest10}.  This angle should approximate \ze\ well, assuming that the spin and orbital axes are at least nearly aligned, such that $i \sim \zeta$, providing a constraint against which to test our results.  For ``black-widow'' and ``redback'' systems \citep{Roberts11,Ray12}, observing radio eclipses indicates that our line of sight is near the orbital plane, and therefore the pulsar spin equator, suggesting $\zeta\gtrsim60^{\circ}$.

For non-recycled pulsars with observed X-ray pulsar wind nebulae, it is possible to constrain \ze\ by modeling the shape of the torus \citep[e.g.,][]{NR08}.  However, MSPs do not typically power bright pulsar wind nebulae as, given their advanced ages, there is little material for the particle wind to interact with.  If thermal X-ray pulsations are detected, models of the hot spot on the polar cap can be used to fit the light curve and estimate \al, \ze, and other parameters \citep[e.g.,][]{Bogdanov07}.  However, the resulting fits are typically unconstrained.

For MSPs with helium white dwarf companions, \citet{TS99} have derived a relationship between the mass of the white dwarf ($M_{\rm WD}$) and the period of the binary orbit.  With this estimate of $M_{\rm WD}$ and the binary mass function of a pulsar, it is possible to estimate the inclination angle of the orbital plane with respect to our line sight if the mass of the pulsar is known \citep[e.g.,][]{GT13}.  Even if the mass of the pulsar is unknown, knowing the orbital period and assuming reasonable values for the pulsar mass provides constraints on the orbital inclination angle.  Then, assuming that the orbit and spin axes have aligned over time, this provides a constraint on \ze.  \citet{GT13} have shown that \ze\ constraints derived in this manner tend to agree well with those from fitting the gamma-ray and radio light curves of MSPs as described in this paper (see Section \ref{lcfit}) and others \citep[e.g.,][]{thesis}.

While full radiation models have been used to model particular LAT pulsar light curves \citep[e.g.,][]{Hirotani11,Du11,Du13}, such analyses are time consuming and require tailoring of parameters such as \Per\ and \Pd\ for each pulsar.  For fitting large numbers of pulsar light curves, geometric models prove to be much more practical \citep[e.g.,][]{Venter09,RW10,thesis,Kala12b,Pierbattista13}.

\section{EMISSION MODELS}\label{models}
Polar cap models \citep[e.g.,][]{DH96} assume that gamma rays come from near the stellar surface above the magnetic polar cap.  In non-recycled pulsars, such models predict that the gamma-ray flux above a few GeV should be strongly attenuated due to magnetic pair creation.  However, observations with the LAT strongly disfavor such models \citep{1PC,2PC} as the primary source of HE photons.  The inferred dipolar magnetic field strengths of MSPs are not strong enough to lead to a significant attenuation of the gamma-ray flux.  However, comparisons of the observed gamma-ray and radio light curves of MSPs do not agree with standard polar cap models, which predict broad peaks from wide emission regions.

Outer-magnetospheric emission models assume that the HE emission is concentrated at high altitudes out to \Rlc.  The most commonly used models, for MSPs, are the outer gap \citep[OG, e.g.,][]{Cheng86a} model, slot gap \citep[SG,][]{MH03,MH04} or two-pole caustic \citep[TPC,][]{DR03} model, and the pair-starved polar cap model \citep[PSPC,][]{Harding05}.

In the OG and SG models, the magnetosphere is assumed to be filled with the charge density of \citet{GJ69} except in narrow vacuum gaps bordering the surface of last-closed field lines (those that close at the light cylinder) where particle acceleration is possible.  In the OG model, the gap is bounded below by the null-charge surface (NCS, the geometric surface across which the charge density changes sign, defined by the condition $\vec{\Omega}\cdot\vec{B}\ =\ 0$).  In the SG model, the gap extends from the stellar surface out to the light cylinder.  In this paper we shall consider the TPC model to be a geometric representation of the SG model.  These models all predict relatively narrow gamma-ray light curve peaks that appear at later phase than features in the radio profiles, assuming low-altitude radio emission.

In the TPC and OG models the narrow bright peaks observed in gamma-ray pulsar light curves are the result of caustic emission \citep{Morini83}.  This occurs when emission from different altitudes on the trailing field lines arrives closely spaced in phase due to relativistic aberration and the finite speed of light.

In the PSPC model, the charged particle density is not sufficient to screen the accelerating electric field over the entire open-field-line region making it available for particle acceleration out to high altitudes.  This model is only viable for MSPs and non-recycled pulsars with relatively low values of \Ed\ \citep[i.e., those pulsars below the ``pair-creation death line'',][]{HMZ02,Harding05}.  PSPC models predict broad gamma-ray peaks that may appear at earlier phase than those in the radio.

With the first gamma-ray MSP detections it was clear that acceleration was occurring in narrow gaps for most sources \citep{Venter09}.  \citet{HM11a} showed that an offset of the magnetic dipole  axis from the center of the polar cap could lower the death line and lead to narrow peaks in MSP light curves, depending on the amount of offset.

A subclass of gamma-ray MSPs has emerged in which the gamma-ray and radio peaks are (nearly) aligned in phase \citep[e.g.,][]{J0034,Guillemot12}, a phenomenon only observed in one non-recycled gamma-ray pulsar \citep[the Crab, e.g.,][]{FermiCrab}.  For this subclass of MSPs, \citet{J0034} introduced the altitude-limited TPC and OG models (alTPC and alOG, respectively) in which the radio emission is assumed to originate in regions that are significantly extended in altitude and co-located with the gamma-ray emission regions.  These models were refined by \citet{Venter12} who noted that the caustic nature of the radio emission has interesting implications for the expected polarization properties.  In particular, for most geometries the caustic mixing of emission from different altitudes leads to the observed emission being largely depolarized \citep{Dyks04a}.

In addition to the altitude-limited models, \citet{Venter12} explored a low-altitude slot gap (laSG) model in which emission occurs within a few \Rns\ of the polar cap, from pair cascades along the SG inner edge.  The polarization pattern expected for the laSG model is basically that of the RVM \citep{Dyks04a}, producing high levels of linear polarization with the classic S-shaped position-angle swing.

\citet{Qiao04} and \citet{Qiao07} proposed the annular gap model \citep[later expounded on by][]{Du10,Du11,Du12,Du13} to explain gamma-ray and radio pulsar emission.  In this model the open volume of a pulsar magnetosphere is separated into two regions by the critical field lines, those which cross the light cylinder at the point of intersection with the NCS.  The region between the magnetic axis and the critical field lines is called the core gap while the region between the critical and last-closed field lines is called the annular gap.  If the binding energy on the surface of the neutron star is strong enough, only one or the other acceleration regions can form (only the core gap if $\vec{\Omega}\cdot\vec{\mu}>0$ and only the annular gap if $\vec{\Omega}\cdot\vec{\mu}<0$, where $\vec{\Omega}$ is the spin angular momentum vector and $\vec{\mu}$ is the magnetic axis vector).  For low surface binding energy both acceleration regions can form.   The size of the annular gap grows with decreasing spin period and is thus thought to be most important for millisecond and young, non-recycled pulsars known to emit gamma rays.  If the annular gap is large enough, pair-production is possible, leading to the production of secondaries that are accelerated from the stellar surface to the NCS (or even beyond, possibly out to the light cylinder) where gamma rays are produced.  While we do not simulate MSP light curves using the annular gap model we do discuss comparisons with results of light curve fitting using this model by other authors (see Section \ref{res}).

\citet{Petri09} proposed the striped wind model to explain HE pulsar emission.  In this model, the observed gamma rays are produced via inverse Compton scattering of the cosmic microwave background off an electron-positron pair wind outside the light cylinder.  This model was able to successfully reproduce the phase-resolved spectra of the Geminga pulsar as observed by EGRET.  When comparing the simulated gamma-ray and radio light curves from this model, assuming a polar cap model for the radio beam, \citet{Petri11} found that the basic profile characteristics could be reproduced but the predicted phase lags were usually too large.

\section{LIGHT CURVE SIMULATIONS}\label{sims}
We assume that the VRD magnetic field geometry of \citet{Deutsch55} is valid in the rest frame of an inertial observer, to first order in $r/R_{\rm LC}$.   For a given spin period and \al, we begin by finding the rim of the polar cap, defined to be the contour on the surface of the neutron star from which the last-closed field lines emerge.  This is done iteratively by selecting a magnetic polar angle ($\theta^{\prime}$) which is close to the polar cap opening angle \Tpc$\approx (\Omega R_{\rm NS}/c)^{1/2}$, as well as a particular magnetic azimuthal angle ($\phi^{\prime}$) and then integrating along the magnetic field line originating at $(\theta^\prime,\phi^\prime)$ using a fourth-order Runge-Kutta routine.  If the line closes inside the light cylinder, the code chooses a smaller value of $\theta^\prime$, and vice versa. This is repeated until a field line is found which closes near the light cylinder, and this footpoint position is stored. The code then moves to the next $\phi^\prime$, looking for the $\theta^\prime$ of the next last-closed field line, and continues this process until the rim of the polar cap has been fully defined.

Since the VRD polar cap is not symmetric about the magnetic axis, a new coordinate definition is used to label positions on the stellar surface. A newer implementation of the so-called open-volume coordinates $(r_{\rm ovc}, l_{\rm ovc})$ was introduced by \citet{Dyks04a}, following initial work by \citet{Y97,Cheng00}. The ``radial'' coordinate is formally defined as $r_{\rm ovc} = 1 \pm d_{\rm ovc}$, where $d_{\rm ovc}$ is the minimum distance of a point from the polar cap rim, normalized by the polar cap radius $R_{\rm PC} = R_{\rm NS}\Theta_{\rm PC}\approx (\Omega R_{\rm NS}^3/c)^{1/2}$. This is similar to a scaled co-latitude which ranges from 0 to 1, labeling points (or cones) from the magnetic axis to the rim. However, given the asymmetric polar cap shape, contours of constant $r_{\rm ovc}$ now label concentric, self-similar, deformed rings interior or exterior to the polar cap instead of circles of constant co-latitude on the stellar surface \citep[see Figure 2 of][]{Dyks04a}. On the rim, $r_{\rm ovc}\equiv1$. The second coordinate $l_{\rm ovc}$ is analogous to the azimuthal angle, and measures arc length along a deformed ring of fixed $r_{\rm ovc}$ in the direction of increasing azimuthal angle, with $l_{\rm ovc} = 0$ coinciding with zero azimuth.

Rings of constant $r_{\rm ovc}$ are defined between specified values of $r_{\rm ovc}^{\rm min}$ and $r_{\rm ovc}^{\rm max}$ depending on the model and waveband to be simulated, as described in Sections \ref{gmod} and \ref{rmod}.  Magnetic field line footpoints are positioned equidistantly along the arc length of each ring (non-uniformly in azimuth).  Therefore, there will automatically be fewer magnetic field lines positioned on the inner rings, since their circumferences are smaller, leading to a uniform placement of field lines over the PC such that one would not overestimate the emission originating at small colatitudes.

We next follow a hypothetical electron/positron as it is accelerated along each field line between $r_{\rm ovc}^{\rm min}$ and $r_{\rm ovc}^{\rm max}$ and collect the resulting emission in bins of pulse phase and \ze.  The emission is assumed to be tangent to the local magnetic field line direction at the point of emission in a frame that co-rotates with the star.  To calculate the phase and \ze\ at which an emitted photon would be observed, this local direction is first transformed from the inertial observer's frame to the co-rotating frame \citep[for details of this calculation see][]{thesis} and used to calculate the emission direction as advocated by \citet{BS10a}.  This direction is then transformed back to the lab frame correcting for relativistic aberration and time-of-flight delays.

The corresponding phase and \ze\ bins are incremented by a number of photons for gamma-ray models, either proportional to physically-motivated emissivity profiles \citep[see e.g.,][]{Venter09,Venter12} or to the step length along the magnetic field line for uniform emissivity, and flux level for radio models (and divided by the solid angle spanned by each bin).  Our models assume uniform emissivity along the field lines in the co-rotating frame except for the PSPC and laSG models.

Simulated light curves are constructed by plotting the bin contents for a given \ze\ from the skymap generated for a given \al.  The \ze\ bins are defined with integer boundaries (e.g., [0\DG,1\DG), [1\DG,2\DG), etc.) but only the lower boundary is reported as the best-fit \ze\ in Section \ref{res}.  \citet{Venter09} have shown how the emission skymaps of models used to fit Class I and III MSPs change for different parameters, and \citet{Venter12} have done the same for models used to fit Class II MSPs.  Skymaps corresponding to the best-fit geometries of select MSPs can be found in Appendix \ref{app-skymaps}.

We assume a neutron star mass and radius of 1.4 M$_{\odot}$ and \Rns=10 km, respectively, which lead to $I = 10^{45}$ g cm$^{2}$ assuming a spherical star with uniform mass density.  These values only affect the PSPC model where we use an analytic form of the accelerating electric field to calculate the number of emitted photons at each spatial step \citep{Venter09}.  Different assumptions for the neutron star equation of state will mainly influence the stellar compactness \citep[$\propto I R_{\rm NS}^{-3}$,][]{MH03}, which in turn would change the electric field, the latter being roughly linearly dependent on the compactness.

\subsection{GAMMA-RAY SIMULATIONS}\label{gmod}
For OG and TPC models we assume uniform emissivity along the field lines in the co-rotating frame such that the number of emitted photons is proportional to the step length along the field line.  For all gamma-ray emission models, particles are followed out to a radial distance of 1.2 \Rlc\ (except for the altitude-limited models where this becomes a free parameter) but not beyond a cylindrical distance of 0.95 \Rlc\ as the structure of the magnetic field near the light cylinder is not well known.

We use a two-layer OG model in which particles are accelerated in a vacuum gap between the surface of last-closed field lines (at the rim with $r_{\rm ovc} \equiv 1$) and a specified $r_{\rm ovc}^{\rm max} \leq 1$.  The particles are then assumed to emit HE photons via curvature radiation above the NCS on those field lines emerging from the stellar surface between $r_{\rm ovc}^{\rm max}$ and a specified $r_{\rm ovc}^{\rm min} \leq r_{\rm ovc}^{\rm max}$, in the standard OG model $r_{\rm ovc}^{\rm min}$ = $r_{\rm ovc}^{\rm max}$.  Following \citet{Wang10}, this emission layer should be small compared to the vacuum gap width ($w_{\rm acc} \equiv 1-r_{\rm ovc}^{\rm max}$) and thus we constrain the size of this region ($w_{\rm em}$) to be no more than half the size of the vacuum gap (i.e., $r_{\rm ovc}^{\rm max}-r_{\rm ovc}^{\rm min} \leq 0.5(1-r_{\rm ovc}^{\rm max})$).

In the TPC model the emission layer and accelerating gap are the same.  Particles are followed along field lines that originate on the stellar surface between $r_{\rm ovc}^{\rm max} = 1$ and some specified $r_{\rm ovc}^{\rm min} \leq 1$.  For TPC models the gap width is defined as $w \equiv 1 -r_{\rm ovc}^{\rm min}$.

In PSPC models the entire open volume is available for particle acceleration and emission is collected from all field lines with $0 \leq r_{\rm ovc} \leq 1$.  The accelerating field is expected to change drastically over this range of field lines and thus the uniform emissivity assumption is no longer reasonable.  Therefore, we use the solution of the accelerating field and emission prescription outlined in \citet{Venter09} to calculate the number of photons emitted per step length along a field line.

Class II MSP light curves are simulated using alTPC, alOG, and laSG models \citep{Venter12}.  The altitude-limited simulations are carried out similar to the standard OG and TPC models, with the exception that the maximum radial distance ($R_{\rm max}^{\gamma}$) to which the emission is followed is a specified parameter, taking on values from 0.7 to 1.2 \Rlc.

In the laSG model, the emission peaks at a distance $s_{\rm f}$ above the stellar surface \citep[typically 1 to 2 \Rns,][]{MH03,Venter12} and falls off exponentially above and below $s_{\rm f}$ with a characteristic length $\sigma_{\rm in}$ towards and $\sigma_{\rm out}$ away from the star, both in units of \Rns.  We use this emission profile to modify the number of photons that would have been added to the skymap under the assumption of uniform emissivity.  This emission profile is motivated by physical models involving particle acceleration, curvature emission, and pair cascade formation.  The acceleration and emission is taken to occur between $r_{\rm ovc}^{\rm max}$ = 0.95 \Tpc\ and a specified $r_{\rm ovc}^{\rm min}$.  

\subsection{RADIO SIMULATIONS}\label{rmod}
A single-altitude hollow-cone and/or core beam is used to model the radio profiles of Class I and III MSPs.  We follow the prescription of \citet{Story07} \citep[who built upon the work of][]{Arz02,Gonthier04} but do not assume the same core-to-cone flux ratio.  In particular, the two components are generated separately and, if both are used, the normalizations are fit separately.

The polar cap is divided into rings, as described previously, between $r_{\rm ovc}^{\rm \rm max} =$1.2 \Tpc\ and $r_{\rm ovc}^{\rm min} =$0.1 \Tpc\ (cone) or 0.0 (core).  These values are chosen to allow the Gaussian tails of the beam components to be explored.  The field line emerging from each ring segment is followed out to the emission height calculated by \citet{KG03}:

\begin{equation}\label{eqn-rkg}
r_{\rm KG}\ =\ 40\Big(\frac{\dot{P}}{10^{-15}\ \rm s\ s^{-1}}\Big)^{0.07}\Big(\frac{P}{1\ \rm{s}}\Big)^{0.3}\Big(\frac{\nu}{1\ \rm{GHz}}\Big)^{-0.26},
\end{equation}
\noindent{}in units of \Rns\, where $\nu$ is the emitted frequency.  Once this height is reached along a field line the observed phase and \ze\ values are calculated and that bin is incremented by the appropriate flux level \citep[given in][]{Harding08}.

Equation \ref{eqn-rkg} depends most strongly on \Per\ and $\nu$, but for typical values the expected emission heights are $\lesssim$0.3 \Rlc.  Such models do not produce radio peaks at the same phase as those from outer-magnetospheric emission models and, thus, cannot explain the Class II MSPs.  The flux of the core component is maximum along the magnetic axis, dropping off as a Gaussian with characteristic width $\sigma_{\rm core} = 1\fdg5/(P/1 \rm{s})^{1/2}$.  The characteristic size of the cone beam is given by $\rho_{\rm cone} = 1\fdg24 (r_{\rm KG})^{1/2} (P/1 \rm{s})^{-1/2}$ \citep[the width at 0.1\% of the peak intensity,][]{Story07}, which has overall $P^{-0.35}$ and $\nu^{-0.13}$ dependencies, consistent with \citet{Kramer98} and \citet{Kramer99}, respectively.  The flux of the cone component is a maximum at magnetic polar angle $\theta_{\mu} = 0.52 \rho_{\rm cone}$ and falls off as a Gaussian on either side with characteristic width $\sigma_{\rm cone} = 0.18 \rho_{\rm cone}$ \citep{Harding08}. 

The radio profiles of Class II MSPs are simulated with alTPC, alOG, and laSG models.  For the alTPC and alOG models, this implies that the radio emission can extend to substantially larger altitudes than given by Equation \ref{eqn-rkg}.  In these models the minimum radius of emission is also a parameter of the simulations.  In particular, for alTPC models emission is only collected between a specified $R_{\rm min}^{\rm R}$ and $R_{\rm max}^{\rm R}$.  For alOG models the minimum radius of emission is taken to be $\max\lbrace R_{\rm min}^{\rm R},R_{\rm NCS}\rbrace$, where $R_{\rm NCS}$ is the radius of the NCS and is a function of magnetic azimuth and \al.

\subsection{SIMULATION PARAMETERS}\label{spar}
We have generated simulations using the models described previously with parameter resolutions and ranges given in Table \ref{tbl-res1} for the TPC and OG models; Table \ref{tbl-res2} for the alTPC, alOG, and laSG models; and Table \ref{tbl-res3} for the PSPC model.  When fitting light curves of Class II MSPs, the radio and gamma-ray emitting regions are allowed to have different gap widths.  When using laSG models to fit the Class II MSP light curves, we require that $\sigma_{\rm out} > \sigma_{\rm in}$ and require that all parameters, except the gap widths, be the same for the radio and gamma-ray emitting regions.


\begin{deluxetable}{lccccc}
\tabletypesize{\scriptsize}
\tablecaption{Simulation Parameter Resolutions and Ranges for Class I MSP Models\label{tbl-res1}}
\tablewidth{0pt}
\tablehead{\colhead{Model} & \colhead{\Per} & \colhead{\al}  & \colhead{\ze}  & \colhead{$w_{\rm acc}$} & \colhead{$w_{\rm em}$}\\
\colhead{} & \colhead{(ms)} & \colhead{($\circ$)} & \colhead{($\circ$)} & \colhead{(\%\Tpc)}  & \colhead{(\%\Tpc)}}
\startdata
\underline{Resolutions:}\\
TPC & 1.0 & 1 & 1 & 2.5 & \nodata \\
OG & 1.0 & 1 & 1 & 2.5 & 2.5 \\
\underline{Ranges:}\\
TPC & $[1.5,5.5]$ & $[1,90]$ & $[0,180]$ & $[0.0,10.0]$ & \nodata\\
OG & $[1.5,5.5]$ &  $[1,90]$ & $[0,180]$ & $[0.0,10.0]$ & $[0.0,5.0]$\\
\enddata
\end{deluxetable}
\normalsize


\begin{deluxetable}{lcccccccccc}
\tabletypesize{\scriptsize}
\tablecaption{Simulation Parameter Resolutions and Ranges for Class II MSP Models\label{tbl-res2}}
\tablewidth{0pt}
\tablehead{
\colhead{Model} & \colhead{\al}  & \colhead{\ze}  & \colhead{$w_{\rm acc}$} & \colhead{$w_{\rm em}$} & \colhead{$R_{\rm max}^{\gamma}$} & \colhead{$R_{\rm min}^{\rm R}$} & \colhead{$R_{\rm max}^{\rm R}$} & \colhead{$\sigma_{\rm in}$} & \colhead{$\sigma_{\rm out}$} & \colhead{$s_{\rm f}$} \\
\colhead{}    & \colhead{($\circ$)} & \colhead{($\circ$)} & \colhead{(\%\Tpc)}  & \colhead{(\%\Tpc)} & \colhead{(\Rlc)} & \colhead{(\Rlc)} & \colhead{(\Rlc)} & \colhead{(\Rns)}  & \colhead{(\Rns)}  & \colhead{(\Rns)}
}
\startdata
\underline{Resolutions:}\\
alTPC & 1 & 1 & 2.5 & \nodata & 0.05 & 0.05 & 0.05 & \nodata & \nodata & \nodata \\
alOG & 1 & 1 & 2.5 & 2.5 &  0.05 & 0.05 & 0.05 & \nodata & \nodata & \nodata \\
laSG & 1 & 1 & 2.5 & \nodata & \nodata & \nodata & \nodata & 0.1 & 0.3 & 0.2 \\
\underline{Ranges:}\\
alTPC & $[1,90]$ & $[0,180]$ & $[0.0,10.0]$ & \nodata & $[0.7,1.2]$ & $[0.14,1.15]$ & $[0.2,1.2]$ & \nodata & \nodata & \nodata \\
alOG &  $[1,90]$ & $[0,180]$ & $[0.0,10.0]$ & $[0.0,5.0]$ & $[0.7,1.2]$ & $[0.14,1.15]$ & $[0.2,1.2]$ & \nodata & \nodata & \nodata \\
laSG & $[1,90]$ & $[0,180]$ & $[0.0,10.0]$ & \nodata & \nodata & \nodata & \nodata & $[0.1,1.0]$ & $[0.3,2.1]$ & $[1.2,2.0]$ \\
\enddata
\tablecomments{All alTPC, alOG, and laSG simulations are done using a 1.5 ms spin period. For the $w_{\rm acc}$ and $w_{\rm em}$ parameters the radio and gamma-ray simulations have the same resolutions and ranges.  The lower limit for $R_{\rm min}^{\rm R}$ of 0.14 \Rlc\ corresponds to \Rns\ for a 1.5 ms spin period.}
\end{deluxetable}
\normalsize


\begin{deluxetable}{lccc}
\tabletypesize{\scriptsize}
\tablecaption{Simulation Parameter Resolutions and Ranges for Class III MSP Models\label{tbl-res3}}
\tablewidth{0pt}
\tablehead{
\colhead{Model} & \colhead{\Per} & \colhead{\al}  & \colhead{\ze}\\
\colhead{} & \colhead{(ms)} & \colhead{($\circ$)} & \colhead{($\circ$)}
}
\startdata
\underline{Resolutions:}\\
PSPC & 1.0 & 1 & 1 \\
\underline{Ranges:}\\
PSPC & $[1.5,5.5]$ & $[1,90]$ & $[0,180]$ \\
\enddata
\end{deluxetable}
\normalsize
\clearpage

\section{LIGHT CURVE FITTING}\label{lcfit}
We fit the gamma-ray and radio light curves by scanning over our model-parameter grid.  We define phase zero in the simulated light curves to be the closest approach of the magnetic axis.  When scanning over the parameter space, we introduce a phase shift ($\Phi$, in units of light curve bins) of the simulated light curves as an additional fit parameter.  For a given set of model parameters, we scan through all possible values of $\Phi$ (from 0 to $n_{\gamma}-1$, with $n_{\gamma}$ the number of bins used in a given gamma-ray light curve) optimizing the normalizations of the simulated radio and gamma-ray light curves, on top of estimated background levels, separately.  The gamma-ray light curves are fit with Poisson likelihood while the radio profiles are fit with a $\chi^{2}$ statistic and then the two statistics are combined as follows.

In maximum likelihood techniques, problems are typically recast as minimizations of \lnl, where $\mathcal{L}$ is the likelihood.  For the gamma-ray light curves we minimize:
\begin{equation}\label{grayFunc}
-\ln(\mathcal{L}_{\gamma})\ =\ -\ln\bigg[\prod_{i=0}^{n_{\gamma}-1}\frac{(c_{\gamma,i}\lambda_{\psi}+b_{\gamma})^{d_{\gamma,i}}\exp\Big\lbrace-(c_{\gamma,i}\lambda_{\psi}+b_{\gamma})\Big\rbrace}{d_{\gamma,i}!}\bigg].
\end{equation}
\noindent{}In Equation \ref{grayFunc}, $d_{\gamma,i}$ is the value of the $i^{th}$ gamma-ray light curve bin, $b_{\gamma}$ is the background estimate, and $c_{\gamma,i}\lambda_{\psi}$ gives the value of the $i^{th}$ model light curve bin $\lambda_{i}$.  The $c_{\gamma,i}$ values are defined such that, for some reference bin $\lambda_{\psi}\neq 0$, $c_{\gamma,i}\equiv\lambda_{i}/\lambda_{\psi}$.  When minimizing Equation \ref{grayFunc}, the only parameter that is optimized is $\lambda_{\psi}$.

For the radio light curves we minimize:
\begin{equation}\label{radioFunc}
-\ln(\mathcal{L}_{\rm R})\ =\ \frac{0.5}{\sigma_{\rm{R}}^{2}} \sum_{i=0}^{n_{\rm{R}}}\Big((c_{\rm{R},i}R_{\Psi}+b_{\rm{R}}-d_{\rm{R},i}\Big)^{2}.
\end{equation}
\noindent{}In Equation \ref{radioFunc}, $d_{\rm{R},i}$ is the value of the $i^{th}$ radio light curve bin, $b_{\rm{R}}$ is the background estimate, $\sigma_{\rm{R}}$ is the error used for each radio bin (see the following discussion), and $c_{\rm{R},i}R_{\Psi}$ gives the value of the $i^{th}$ model light curve bin $R_{i}$.  The $c_{\rm{R},i}$ values are defined such that, for some reference bin $R_{\Psi}\neq 0$, $c_{\rm{R},i}\equiv R_{i}/R_{\Psi}$.  When minimizing Equation \ref{radioFunc}, the only parameter that is optimized is $R_{\Psi}$.  We combine the gamma-ray and radio log-likelihood functions by adding Equations \ref{grayFunc} and \ref{radioFunc}.

When attempting a joint fit of the radio and gamma-ray light curves one difficulty arises due to the fact that the statistical uncertainty of the radio data is much less than that of the gamma rays.  Without adjusting the radio uncertainty, this results in the likelihood essentially ignoring the information in the gamma-ray light curve when choosing the best-fit geometry.  Given the simplistic nature of our radio model for Class I and III MSPs, this leads to inadequate solutions.

To balance the relative contributions of the radio and gamma-ray profiles, we define an on-peak interval for the gamma-ray light curve and calculate the average relative uncertainty in this interval ($\sigma_{\gamma,\rm ave}$).  We then calculate $\sigma_{\rm R} = (r_{\rm max}\times\sigma_{\gamma,\rm ave}$), where $r_{\rm max}$ is the value of the highest radio light curve bin, and use this as the absolute uncertainty for each radio bin.  For those MSPs where we use more radio bins than gamma-ray bins, this uncertainty is decreased by the ratio of gamma-ray to radio bin numbers.

The bins in the on-peak interval of the gamma-ray light curve are the dominant contribution to that part of the likelihood, and using $\sigma_{\rm R}$ as defined leads to a comparable contribution from the radio profile.  However, this $\sigma_{\rm R}$ is still somewhat arbitrary so we have attempted to estimate systematic biases in our fits due to this choice, see Section \ref{systs}.

We construct likelihood profiles to estimate uncertainties on the best-fit model parameters.  For \al\ and \ze\ we use a two-dimensional likelihood profile to produce confidence contours that can be compared to those from polarization fits, otherwise we use one-dimensional profiles to estimate the uncertainties.  In order to be conservative, we have chosen to report 95\% confidence-level uncertainties.

Our geometric models are simple when compared to the complex issue we are trying to address and the likelihood surfaces can be steep near the best-fit values leading to unreasonably small uncertainties.  Thus, when estimating the uncertainties we multiply the log-likelihood differences by $(n_{\rm dof}/2)/(-\ln(\mathcal{L}_{max}))$, where $\mathcal{L}_{max}$ is the maximum likelihood value and $n_{\rm dof}$ is the degrees of freedom in a particular fit.  Assuming the log-likelihood differences follow a $\chi^{2}$ distribution ($-\Delta\ln(\mathcal{L})=\Delta\chi^{2}/2$), this results in the best-fit corresponding to a reduced $\chi^{2}$ of 1.  We use these rescaled log-likelihood differences to estimate uncertainties on the best-fit model parameters.

The \al-\ze\ confidence contours are often not simple shapes with confidence regions that are not simply connected.  Thus, while we do report $\pm$ error bars for these angles in Section \ref{res}, when comparing to other constraints (such as RVM fits) it is best to use the confidence contours directly, provided in the auxiliary online material\footnote{Available from the journal web page or from the entry for this paper at \url{http://www-glast.stanford.edu/cgi-bin/pubpub.}}.

\subsection{EVALUATION OF SYSTEMATIC BIASES}\label{systs}
We have investigated multiple sources of systematic biases in our fitting procedure.  The first is related to our choice of $\sigma_{\rm R}$.  Choosing a smaller (larger) value makes matching the radio profile more (less) important to the overall likelihood value and can thus affect the best-fit geometry.  To assess the importance of this systematic, we refit a sub-sample of MSPs, one-third of each model class, varying the radio uncertainty by a factor of 2.

For Class I MSPs, changes in \al\ and \ze, due to varying $\sigma_{\rm R}$ by a factor of 2, were generally $\lesssim30^{\circ}$ and typically in different directions (i.e., if \al\ increases \ze\ usually decreases and vice versa).  There were a few extreme cases where the change in one parameter was as much as $\sim60^{\circ}$.  In such cases the best-fit was jumping from one local maximum to another, but such changes are encompassed by the confidence contours (see Appendix \ref{app-cls1cont} for examples).  Depending on the gamma-ray light curve, only fits with one of the models (TPC or OG) may be affected, but the general behavior of the two models was similar.  The gap width parameters were generally unaffected, with any changes within the estimated statistical uncertainties.

For Class II MSPs, varying $\sigma_{\rm R}$ by a factor of 2 led to changes in \al\ or \ze\ $\leq2^{\circ}$.  The additional model parameters were either unchanged or changed by less than the estimated statistical uncertainties.

For Class III MSPs, \al\ and \ze\ typically changed by only a few degrees when varying $\sigma_{\rm R}$ by a factor of 2.  However, for fainter MSPs in this class the geometry was observed to change by as much as $\sim50^{\circ}$.

The estimated background level in the gamma-ray light curves can also strongly affect the best-fit parameters.  These values come from the weights estimated using spectral fits that can have sizable statistical uncertainty.  From investigations in 2PC, we estimate that these background levels could be wrong by $\sim$5\%.  This could change which model is preferred as it alters the off-peak emission level for which OG and TPC models give different predictions.  To assess the importance of this systematic, we refit a selection of MSPs varying the gamma-ray background level by $\pm$5\%.

For Class I and II MSPs, a modified background level led to changes in \al\ and \ze\ typically less than a few degrees and $\sim10^{\circ}$ at most.  The most significant effect was in the \lnl\ values, which were seen to change by $\lesssim7$ in most cases.  In one extreme case, for the Class II MSP PSR J1939+2134, which has a very high gamma-ray background level; the changes in the \lnl\ values were on the order of $\sim20$.  To be conservative we only say one model is significantly preferred over another for a given MSP if the log-likelihood difference is $\geq15$.  Our models are not nested, so assigning a significance to any log-likelihood difference is not straightforward; however, we note that a log-likelihood difference of 15 amounts to the preferred model being more likely by a factor of e$^{15}$ (compare this to a 5$\sigma$ result for nested models with one degree of freedom corresponding to the preferred model more likely by a factor of $\sim$e$^{12.5}$).  As with changes in $\sigma_{\rm R}$, the gamma-ray gap width parameters were not significantly affected by this systematic.  The emission altitude parameters in the fits of Class II MSPs were also not strongly affected, with any differences within the estimated statistical uncertainties.

The best-fit geometries of Class III MSPs are insensitive to this systematic.  If any change was observed it was $\leq1^{\circ}$.

The VRD magnetic field geometry is only an approximation.  Charges will be pulled from the surface of the neutron star, populate the magnetosphere, and modify the field structure.  Attempts to construct an analytic model of a pulsar magnetosphere, accounting for the presence of charges, have only been successful for aligned rotators \citep[e.g.,][]{Michel74,MP92}.  However, using magneto-hydrodynamic simulations several authors have successfully produced numeric models assuming force-free conditions \citep[i.e., no acceleration,][]{Spitkovsky06} and for finite-conductivity spanning the range between vacuum and force-free \citep{Kala12a,Li12}.  One result of including charges is an increase in the size of the polar cap.  This leads to larger predicted phase lags between the gamma-ray and radio pulse profiles \citep[by as much as 0.1 in phase,][]{Harding11,Kala12b}.

In order to evaluate how the use of the VRD approximation may skew the best-fit geometries, we repeated the fits for a subset of Class I MSPs allowing phase zero of the gamma-ray models to be different than the radio models by $\leq0.1$.  Additionally, we refit some of the Class III MSPs with OG and TPC models with this same strategy in an attempt to address if an increased phase lag allows the Class I models to explain the Class III light curves or if there really is evidence for pair-starved magnetospheres in these MSPs.

For Class I MSPs, allowing the gamma-to-radio phase lag to increase beyond the VRD prediction led to changes in \al\ and \ze\ $\leq10^{\circ}$, almost always in the same direction (i.e., both increased or decreased).  In the case of one faint MSP, \al\ was observed to change by $\sim$30\DG.

For Class III MSPs, allowing the gamma-to-radio phase lag to increase beyond the VRD prediction and fitting with TPC and OG models did not result in acceptable fits.  The maximum allowed change of 0.1 was not sufficient to match the observations, suggesting that the light curves of these MSPs cannot be explained only by deficiencies in the VRD approximation.

\section{SOURCES AND DATA PREPARATION}\label{srcs}
For each of the MSPs in Table \ref{tbl-vitals} we have used the same LAT data set and radio profiles as 2PC.  The \Pd\ and spin-down power ($\dot{E} = 4\pi^2 I \dot{P}/P^{3}$) values reported in columns 3 and 6 are corrected for the Shklovskii effect \citep{Shklovskii70} and Galactic acceleration following 2PC.  The observed frequency and observatory responsible for generating the radio profile are given in columns 4 and 5, respectively.  The corrected \Pd\ values are used to calculate the magnetic field strength at the light cylinder ($B_{\rm LC} = 4\pi^{2}((1.5 I \dot{P})/(c^{3}P^{5}))$, for an orthogonal rotator) given in column 7.  The gamma-ray luminosities ($L_{\gamma} = 4\pi f_{\Omega} G_{100} d^{2}$, with $G_{100}$ the 2PC gamma-ray energy flux from 0.1 to 100 GeV, $d$ the pulsar distance, and \fO\ a beaming factor defined in Section \ref{eff}) in column 8 are derived using the distance estimates adopted in 2PC.  Column 9 gives the efficiency with which rotational energy is turned into gamma rays ($\eta_{\gamma} = L_{\gamma}/\dot{E}$) .


\begin{landscape}
\begin{deluxetable}{lccccccccc}
\tabletypesize{\scriptsize}
\tablecaption{Parameters of 2PC MSPs\label{tbl-vitals}}
\tablewidth{0pt}

\tablehead{
\colhead{PSR} & \colhead{$P_{\rm obs}$}  & \colhead{$\dot{P}_{\rm obs}$}  & \colhead{$\nu_{\rm obs}$} & \colhead{Observatory\tablenotemark{b}} & \colhead{\Ed} & \colhead{\Blc} & \colhead{$d$} & \colhead{$L_{\gamma}/f_{\Omega}$} & \colhead{$\eta_{\gamma}/f_{\Omega}$}\\
\colhead{}    & \colhead{(ms)} & \colhead{($10^{-20}$ s s$^{-1}$)} & \colhead{(MHz)}  & \colhead{} & \colhead{($10^{32}$ erg s$^{-1}$)} & \colhead{($10^{4}$ G)} & \colhead{(pc)} & \colhead{($10^{32}$ erg s$^{-1}$)} & \colhead{(\%)}
}

\startdata
J0023+0923 & 3.05 & 1.08 & 2000 & GBT & 151.00 & 5.99 & 690$_{-110}^{+210}$ & $4.56\pm0.70$ & $3.00\pm0.50$\\
J0030+0451\tablenotemark{a} & 4.87 & 0.11 & 1400 & NRT & 36.40 & 5.82 & 280$_{-60}^{+100}$ & $5.75\pm0.17$ & $15.8\pm0.50$\\
J0034$-$0534\tablenotemark{a} & 1.88 & 0.29 & 1400 & NRT & 173.00 & 10.35 & 540$_{-100}^{+110}$ & $5.67\pm0.41$ & $3.30\pm0.20$\\
J0101$-$6422\tablenotemark{a} & 2.57 & 0.44 & 1400 & PKS & 101.00 & 5.83 & 550$_{-80}^{+90}$ & $3.79\pm0.33$ & $3.80\pm0.30$\\
J0102+4839 & 2.96 & 1.17 & 1500 & GBT & 175.00 & 6.68 & 2319$_{-430}^{+500}$ & $85.1\pm10.2$ & $48.6\pm5.8$\\
J0218+4232\tablenotemark{a} & 2.32 & 7.69 & 1400 & NRT & 2432.00 & 31.51 & 2640$_{-640}^{+1080}$ & $380\pm20$ & $15.6\pm0.8$\\
J0340+4130 & 3.30 & 0.59 & 1500 & GBT & 78.70 & 3.62 & 1730$_{-300}^{+290}$ & $72.9\pm5.5$ & $92.6\pm7.0$\\
J0437$-$4715\tablenotemark{a} & 5.76 & 1.41 & 1400 & PKS & 29.05 & 1.39 & 156$\pm1$ & $0.49\pm0.03$ & $1.70\pm0.10$\\
J0610$-$2100\tablenotemark{a} & 3.86 & 0.12 & 1400 & NRT & 8.00 & 1.10 & 3540$_{-1000}^{+5460}$ & $98.5\pm16.0$ & $1230\pm201$\\
J0613$-$0200\tablenotemark{a} & 3.06 & 0.87 & 1400 & NRT & 120.00 & 5.30 & 900$_{-200}^{+400}$ & $29.0\pm1.8$ & $24.1\pm1.5$\\
J0614$-$3329 & 3.15 & 1.78 & 1400 & NRT & 220.00 & 7.06 & 1900$_{-350}^{+440}$ & $472\pm12$ & $215\pm5$\\
J0751+1807\tablenotemark{a} & 3.48 & 0.77 & 1400 & NRT & 72.14 & 3.62 & 400$_{-100}^{+200}$ & $2.54\pm0.23$ & $3.50\pm0.30$\\
J1024$-$0719\tablenotemark{a} & 5.16 & 0.16 & 1400 & NRT & 4.00 & 0.62 & 386$_{-38}^{+39}$ & $0.57\pm0.18$ & $14.2\pm4.5$\\
J1124$-$3653 & 2.41 & 0.58 & 820 & GBT & 171.00 & 7.83 & 1720$_{-360}^{+430}$ & $42.9\pm4.7$ & $25.1\pm2.7$\\
J1125$-$5825 & 3.10 & 6.09 & 1400 & NRT & 805.00 & 13.59 & 2619$_{-370}^{+400}$ & $73.1\pm16.4$ & $9.10\pm2.00$\\
J1231$-$1411\tablenotemark{a} & 3.68 & 0.65 & 1400 & NRT & 51.46 & 2.89 & 438$_{-50}^{+52}$ & $23.6\pm0.6$ & $45.9\pm1.2$\\
J1446$-$4701 & 2.19 & 0.99 & 1400 & PKS & 368.00 & 13.03 & 1460$_{-220}^{+230}$ & $19.0\pm3.7$ & $5.10\pm1.00$\\
J1514$-$4946 & 3.59 & 1.87 & 1400 & PKS & 160.00 & 5.22 & 940$_{-120}^{+110}$ & $47.6\pm3.0$ & $29.7\pm1.8$\\
J1600$-$3053\tablenotemark{a} & 3.60 & 0.86 & 1400 & NRT & 72.98 & 3.51 & 1630$_{-270}^{+310}$ & $16.9\pm8.9$ & $23.1\pm12.2$\\
J1614$-$2230\tablenotemark{a} & 3.15 & 0.30 & 1400 & NRT & 37.79 & 2.90 & 650$\pm50$ & $12.3\pm1.0$ & $32.6\pm2.6$\\
J1658$-$5324 & 2.43 & 1.10 & 1400 & PKS & 302.00 & 10.61 & 930$_{-130}^{+110}$ & $29.9\pm2.4$ & $9.90\pm0.80$\\
J1713+0747\tablenotemark{a} & 4.57 & 0.83 & 1400 & NRT & 34.20 & 1.90 & 1050$_{-50}^{+60}$ & $13.4\pm1.9$ & $39.2\pm5.6$\\
J1741+1351\tablenotemark{a} & 3.75 & 2.91 & 1400 & AO & 217.60 & 5.83 & 1080$_{-50}^{+40}$ & $3.36\pm1.13$ & $1.50\pm0.50$\\
J1744$-$1134\tablenotemark{a} & 4.07 & 0.70 & 1400 & NRT & 41.08 & 2.33 & 417$_{-17}^{+18}$ & $6.76\pm0.52$ & $16.5\pm1.3$\\
J1747$-$4036 & 1.64 & 1.33 & 1400 & PKS & 1160.00 & 31.19 & 3390$_{-760}^{+820}$ & $136\pm31$ & $11.7\pm2.7$\\
J1810+1744 & 1.66 & 0.46 & 2000 & GBT & 397.00 & 17.85 & 2000$_{-280}^{+310}$ & $112\pm8$ & $28.2\pm2.1$\\
J1823$-$3021A & 5.44 & 338.00 & 1400 & NRT & 8280.00 & 24.81 & 7599$\pm400$ & $740\pm103$ & $8.90\pm1.30$\\
J1858$-$2216 & 2.38 & 0.39 & 820 & GBT & 113.00 & 6.63 & 940$_{-130}^{+200}$ & $7.64\pm1.57$ & $6.80\pm1.40$\\
J1902$-$5105 & 1.74 & 0.90 & 1400 & PKS & 686.00 & 22.13 & 1179$_{-210}^{+220}$ & $35.9\pm2.5$ & $5.20\pm0.40$\\
J1939+2134\tablenotemark{a} & 1.56 & 10.55 & 1400 & NRT & 10970.00 & 99.54 & 3559$\pm350$ & $139\pm50$ & $1.30\pm0.50$\\
J1959+2048\tablenotemark{a} & 1.61 & 0.81 & 1400 & NRT & 763.30 & 25.49 & 2490$_{-490}^{+160}$ & $126\pm14$ & $16.5\pm1.8$\\
J2017+0603 & 2.90 & 0.83 & 1400 & NRT & 130.00 & 5.93 & 1570$_{-150}^{+160}$ & $98.2\pm6.3$ & $75.6\pm4.8$\\
J2043+1711\tablenotemark{a} & 2.38 & 0.43 & 1400 & AO & 126.50 & 6.99 & 1760$_{-320}^{+150}$ & $100\pm6$ & $79.2\pm4.7$\\
J2047+1053 & 4.29 & 2.10 & 820 & GBT & 105.00 & 3.54 & 2050$_{-290}^{+320}$ & $31.0\pm6.9$ & $29.5\pm6.6$\\
J2051$-$0827\tablenotemark{a} & 4.51 & 1.26 & 1380 & WSRT & 54.30 & 2.42 & 1040$_{-150}^{+160}$ & $4.37\pm1.33$ & $8.10\pm2.50$\\
J2124$-$3358\tablenotemark{a} & 4.93 & 1.12 & 1400 & NRT & 37.00 & 1.83 & 300$_{-50}^{+70}$ & $3.96\pm0.17$ & $10.7\pm0.5$\\
J2214+3000 & 3.12 & 1.50 & 1400 & NRT & 192.00 & 6.63 & 1540$_{-180}^{+190}$ & $92.9\pm4.3$ & $48.4\pm2.2$\\
J2215+5135 & 2.61 & 2.34 & 2000 & GBT & 519.00 & 12.95 & 3010$_{-370}^{+330}$ & $128\pm16$ & $24.6\pm3.0$\\
J2241$-$5236 & 2.19 & 0.87 & 1400 & NRT & 260.00 & 12.24 & 513$_{-76}^{+80}$ & $10.5\pm0.5$ & $4.00\pm0.20$\\
J2302+4442 & 5.20 & 1.33 & 1400 & NRT & 38.20 & 1.74 & 1190$_{-230}^{+90}$ & $62.2\pm2.9$ & $163\pm8$\\
\enddata

\tablenotetext{a}{\Pd, \Ed, and \Blc\ have been corrected for the Shklovskii effect and Galactic acceleration as in 2PC.}
\tablenotetext{b}{Observatory responsible for generating the radio profile: AO = Arecibo Observatory, GBT = Green Bank Telescope, NRT = Nan\c cay Radio Telescope, PKS = Parkes Radio Telescope, and WSRT = Westerbork Synthesis Radio Telescope.}
\tablenotetext{c}{All distance and distance uncertainty estimates are taken from \citet{2PC}.}
\end{deluxetable}
\end{landscape}
\normalsize

\begin{figure}[h]
\begin{center}
\includegraphics[height=0.5\textheight]{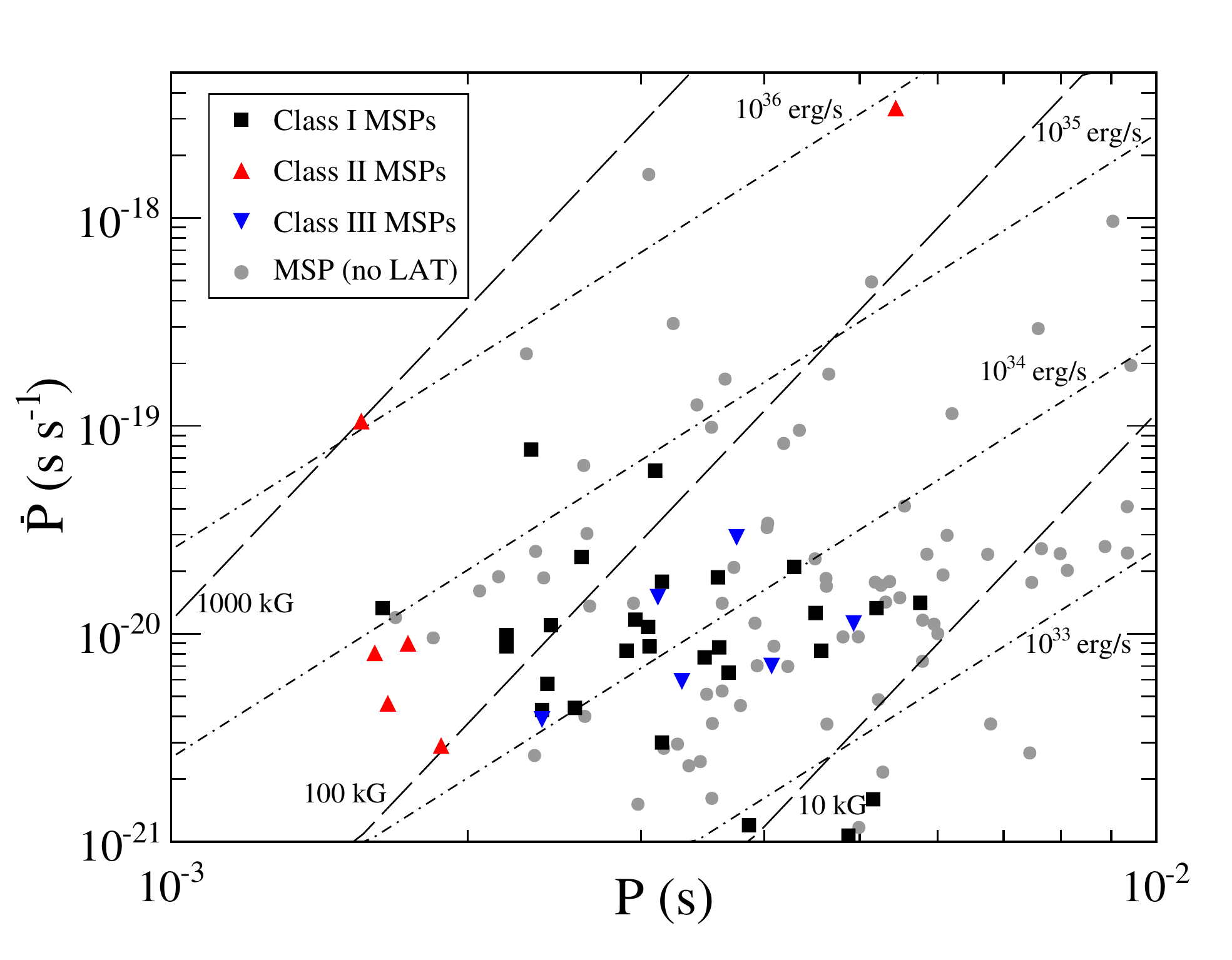}
\end{center}
\caption{\Pd\ versus \Per, LAT-detected MSPs are labeled by model class as indicated: Class I MSPs are black squares, Class II MSPs are light gray (red in the online version) upward-pointing triangles, and Class III MSPs are dark gray (blue in the online version) downward-pointing triangles.  Gray circles are known radio MSPs with no detection in 2PC.  Dashed lines show constant \Blc\ while dot-dashed lines show constant \Ed, both assuming $I=10^{45}$ g cm$^{2}$.  All but one Class II MSP has $P\ <$ 2 ms.\label{fig-PPdot}}
\end{figure}

Figure \ref{fig-PPdot} presents the LAT-detected MSPs, labeled by model class, on a \Per-\Pd\ diagram with lines of constant \Blc\ and \Ed\ as well as MSPs from the ATNF pulsar catalog\footnote{\url{http://www.atnf.csiro.au/research/pulsar/psrcat}} \citep{ATNFCat} with no pulsed gamma-ray detection in 2PC.

Table \ref{tbl-simpars} gives the simulation and fitting parameters for each MSP.  For Class I and II MSPs, we always use a hollow-cone radio model and only include a core beam if polarization data in the literature suggests there is evidence for such a component (noted in the table).  For Class II MSPs $\nu_{\rm sim}$ is not a parameter of the simulations and thus those entries are empty in Table \ref{tbl-simpars}.  We use events with reconstructed directions within 2\DG\ of each MSP with spectrally-derived probabilities of coming from the pulsar $\geq0.2$, except for PSRs J1939+2134 and J2047+1053 for which we impose a minimum probability of 0.01 and 0.1, respectively, in order to optimize the unweighted counts light curves of these two MSPs.  From these events we produced folded light curves with $n_{\gamma}$ fixed-width bins.

The light curves presented in 2PC use weighted counts \citep[where the weight of each photon is the probability that it originated from the pulsar based on a spectral and spatial model of the region,][]{Kerr11} while we use unweighted counts.  We made this choice as it allows us to use Poisson likelihood for the gamma-ray light curves.  Poisson likelihood more naturally matches sharp peaks, particularly for faint MSPs, than a $\chi^{2}$ statistic, which is necessary when using binned weighted-counts light curves.  While 2PC used the weighted H-test value to determine the number of bins in the gamma-ray light curves, we have adopted a different approach.  We only use numbers of bins which are an integer divisor of 180 (the number of bins in our simulations) and we choose the number for each MSP, largely by eye, to be as large as possible while preserving the significance of structures for the fitting procedure.  We have also decreased the binning of the radio profiles to be an integer divisor of 180 bins and to smooth some of the fine structure that our geometric models cannot reproduce.


\begin{deluxetable}{lcccccccc}
\tabletypesize{\scriptsize}
\tablecaption{Simulation and fitting parameters for each MSP\label{tbl-simpars}}
\tablewidth{0pt}

\tablehead{
\colhead{PSR} & \colhead{$P_{\rm sim}$} & \colhead{$\nu_{\rm sim}$} & \colhead{$n_{\gamma}$} & \colhead{$n_{\rm R}$} & \colhead{$b_{\gamma}$} & \colhead{$b_{\rm R}$} & \colhead{$\sigma_{\rm R}$} & \colhead{Model Class}\\
\colhead{}    & \colhead{(ms)} & \colhead{(MHz)}  & \colhead{(Bins)} & \colhead{(Bins)} & \colhead{(Counts/Bin)} & \colhead{(Arb. Units/Bin)} & \colhead{(Arb. Units)} & \colhead{}
}

\startdata
J0023+0923 & 3.5 & 2000 & 15 & 30 & 15.1060 & 0.8550 & 3.2930 & I\\
J0030+0451 & 4.5 & 1400 & 90 & 90 & 21.5035 & 0.1400 & 0.6534 & I\\
J0034$-$0534 & 1.5 & \nodata & 45 & 45 & 11.628 & 0.680 & 0.472 & II\\
J0101$-$6422 & 2.5 & 1400 & 30 & 60 & 9.1766 & 1.6400 & 13.4850 & I\\
J0102+4839 & 2.5 & 1400 & 30 & 30 & 11.3830 & 0.0060 & 0.0094 & I\\
J0218+4232 & 2.5 & 1400 & 30 & 60 & 79.1190 & 0.1972 & 0.1250 & I\\
J0340+4130 & 3.5 & 1400 & 60 & 60 & 7.5300 & 1.6790 & 2.8060 & III\\
J0437$-$4715\tablenotemark{a,c} & 5.5 & 1400 & 90 & 90 & 8.1020 & 0.0000 & 0.6743 & I\\
J0610$-$2100\tablenotemark{d} & 3.5 & 1400 & 20 & 20 & 8.5540 & 1.4460 & 2.1335 & I\\
J0613$-$0200\tablenotemark{a} & 3.5 & 1400 & 60 & 60 & 11.5090 & 0.2340 & 1.6630 & I\\
J0614$-$3329 & 3.5 & 1400 & 90 & 180 & 28.0340 & 3.0710 & 1.5770 & I\\
J0751+1807 & 3.5 & 1400 & 45 & 90 & 6.6950 & 0.6370 & 2.8380 & I\\
J1024$-$0719\tablenotemark{d} & 5.5 & 1400 & 15 & 15 & 4.5980 & 1.5440 & 6.2700 & I\\
J1124$-$3653 & 2.5 & 800 & 30 & 30 & 8.0760 & 0.0000 & 0.0007 & I\\
J1125$-$5825 & 3.5 & 1400 & 15 & 30 & 6.3028 & 0.0040 & 0.0122 & I\\
J1231$-$1411 & 3.5 & 1400 & 90 & 90 & 28.8400 & 0.3990 & 0.3540 & I\\
J1446$-$4701 & 2.5 & 1400 & 15 & 30 & 9.1440 & 0.0018 & 0.0025 & I\\
J1514$-$4946 & 3.5 & 1400 & 60 & 60 & 15.3750 & 0.4320 & 0.6260 & I\\
J1600$-$3053\tablenotemark{a,d} & 3.5 & 1400 & 30 & 30 & 6.1387 & 0.6525 & 6.4656 & I\\
J1614$-$2230 & 3.5 & 1400 & 60 & 60 & 8.9790 & 2.1540 & 10.0710 & I\\
J1658$-$5324\tablenotemark{e} & 2.5 & 1400 & 45 & 45 & 19.1750 & 1.3390 & 14.8830 & I\\
J1713+0747\tablenotemark{a} & 4.5 & 1400 & 15 & 30 & 14.0380 & 0.5160 & 8.9440 & I\\
J1741+1351\tablenotemark{b} & 3.5 & 1400 & 15 & 30 & 2.8650 & 0.0660 & 0.8280 & III\\
J1744$-$1134 & 4.5 & 1400 & 45 & 45 & 20.2670 & 0.6060 & 10.4710 & III\\
J1747$-$4036 & 1.5 & 1400 & 15 & 15 & 7.8620 & 3.3660 & 64.3190 & I\\
J1810+1744 & 1.5 & \nodata & 45 & 45 & 18.413 & 0.525 & 0.708 & II\\
J1823$-$3021A & 1.5 & \nodata & 30 & 30 & 5.948 & 1.924 & 1.457 & II\\
J1858$-$2216 & 2.5 & 800 & 30 & 60 & 4.9741 & 0.0026 & 0.0021 & III\\
J1902$-$5105 & 1.5 & \nodata & 45 & 45 & 16.729 & 1.878 & 21.175 & II\\
J1939+2134 & 1.5 & \nodata & 30 & 30 & 290.938 & 0.000 & 0.284 & II\\
J1959+2048 & 1.5 & \nodata & 45 & 45 & 9.521 & 0.691 & 0.765 & II\\
J2017+0603 & 2.5 & 1400 & 60 & 180 & 9.6560 & 0.2380 & 0.0926 & I\\
J2043+1711\tablenotemark{d} & 2.5 & 1400 & 60 & 60 & 12.1462 & 0.1135 & 0.0505 & I\\
J2047+1053\tablenotemark{e} & 4.5 & 800 & 20 & 60 & 11.0960 & 0.0031 & 0.0096 & I\\
J2051$-$0827 & 4.5 & 1400 & 15 & 15 & 8.1139 & 0.4159 & 1.0245 & I\\
J2124$-$3358 & 4.5 & 1400 & 60 & 60 & 14.3570 & 0.3410 & 1.8000 & III\\
J2214+3000 & 3.5 & 1400 & 45 & 90 & 20.3950 & 2.2670 & 1.0987 & III\\
J2215+5135 & 2.5 & 2000 & 15 & 30 & 17.9653 & 7.5109 & 5.0260 & I\\
J2241$-$5236 & 2.5 & 1400 & 45 & 90 & 19.8430 & 0.0360 & 0.4180 & I\\
J2302+4442 & 5.5 & 1400 & 90 & 90 & 10.4260 & 1.4130 & 2.2020 & I\\
\enddata

\tablenotetext{a}{Polarization suggests presence of a core beam.}
\tablenotetext{b}{Non-standard $\sigma_{\rm R}$ estimate, one-fourth the normal value used.}
\tablenotetext{c}{Non-standard $\sigma_{\rm R}$ estimate, one-third the normal value used.}
\tablenotetext{d}{Non-standard $\sigma_{\rm R}$ estimate, one-half the normal value used.}
\tablenotetext{e}{Non-standard $\sigma_{\rm R}$ estimate, twice the normal value used.}
\end{deluxetable}
\normalsize


The gamma-ray background levels were estimated as $b_{\gamma}=(N-S)/n_{\gamma}$, where $N$ is the total number of events passing the selection criteria for each MSP, $S$ is a proxy for the pulsar signal taken to be $S = \sum_{i}p_{i}$ with $p_{i}$ the probability that event $i$ is associated with the pulsar, and $n_{\gamma}$ is the number of bins in the light curve.  The radio background levels ($b_{\rm R}$) were estimated from fits of a constant value to the off-peak regions of the radio profiles using the estimated radio uncertainties ($\sigma_{\rm R}$, derived as described in Section \ref{lcfit} with exceptions noted in Table \ref{tbl-simpars}) divided by the number of radio bins ($n_{\rm R}$).

For the light curve fits, we matched the observed period to the closest simulated period ($P_{\rm sim}$) except for Class II MSPs for which we always use $P_{\rm sim} = 1.5$ ms.  For all but one of the Class II MSPs, this choice of $P_{\rm sim}$ matches the observed period well, the exception being PSR J1823$-$3021A with a period of 5.44 ms.  However, the period enters the simulations most strongly through the size of the polar cap, which means that using a shorter period will, at most, overestimate the size of any predicted off-pulse region.  For Class I and III MSPs we also matched the simulated radio frequency ($\nu_{\rm sim}$) to the observed frequency as closely as possible since this affects the assumed emission height and cone/core size.  For Class II MSPs we do not specify $\nu_{\rm sim}$ as we assume all frequencies come from an extended region in the magnetosphere.

\section{RESULTS}\label{res}
The best-fit parameters for Class I MSPs are given in Tables \ref{tbl-tpc} and \ref{tbl-og}; for Class II MSPs in Tables \ref{tbl-altpc}, \ref{tbl-alog} and \ref{tbl-lasg}; and for Class III MSPs in Table \ref{tbl-pspc}.  In each table we give the phase of the magnetic axis ($\Phi_{\mu}$) predicted from the models.  The observed and best-fit light curves for all 2PC MSPs are provided in Appendix \ref{app-fits} along with information about the initial radio and gamma-ray discoveries and a comparison of the best-fit geometries to other estimates in the literature, when they exist, for each MSP.

Sample confidence contours are shown in Appendix \ref{app-contours}.  All confidence contours are included in the auxiliary online material.  For contours that are not simply connected, the uncertainties on \al\ and \ze\ are estimated only from the region that contains the best-fit geometry.  As such, uncertainties for fits that are flagged as not simply connected should be used with care and, when possible, the confidence contours should be used directly.  For some of the brightest MSPs in our sample the resulting confidence contours, even after rescaling as discussed in Section \ref{lcfit}, suggest uncertainties on the order of a few degrees.  For these MSPs we remind readers of the systematic uncertainties discussed in Section \ref{systs}.

The light curve fitting is only done for $\alpha \in [1^{\circ},90^{\circ}]$ and $\zeta \in [0^{\circ},90^{\circ})$.  When using geometric models the light curves produced beyond \al=90\DG\ or \ze=90\DG\ are the same but shifted by 0.5 in phase and those with \al\ and \ze\ both $>90^{\circ}$ are exactly the same.  This arises due to the assumption that both poles have the same emission properties.  Thus, while the uncertainties we quote are constrained to the region with \al\ and \ze\ both $\leq90^{\circ}$, there are identical confidence regions when reflected across the 90\DG\ boundaries.

Sample skymaps of simulated gamma-ray and radio emission (constructed as detailed in Section \ref{sims}), corresponding to the best-fit models of the given MSP, are shown in Appendix \ref{app-skymaps}.  Emission maps for the best-fit models of all MSPs are included in the auxiliary online material.


\begin{deluxetable}{lccccccc}
\tabletypesize{\scriptsize}
\tablecaption{TPC model fit results for Class I MSPs\label{tbl-tpc}}
\tablewidth{0pt}

\tablehead{
\colhead{PSR} & \colhead{\lnl}  & \colhead{$n_{\rm dof}$} & \colhead{$\Phi_{\mu}$} & \colhead{\al} & \colhead{\ze} & \colhead{$w$} & \colhead{\fO}\\
\colhead{} & \colhead{} & \colhead{} & \colhead{} & \colhead{($\circ$)}  & \colhead{($\circ$)} & \colhead{(\% \Tpc)} & \colhead{}
}

\startdata
J0023+0923\tablenotemark{a} & 58.4 & 39 & 0.567 & 38$_{-17}^{+9}$ & 65$_{-5}^{+7}$ & 5.0$\pm$5.0 & 0.69$_{-0.14}^{+0.73}$\\
J0030+0451 & 581.1 & 174 & 0.044 & 74$\pm$2 & 55$_{-1}^{+3}$ & 2.5$\pm$2.5 & 1.13$_{-0.01}^{+0.07}$\\
J0101$-$6422\tablenotemark{a} & 131.7 & 84 & 0.333 & 30$_{-2}^{+11}$ & 84$_{-2}^{+5}$ & 0.0$_{-2.5}^{+7.5}$ & 0.74$_{-0.07}^{+0.33}$\\
J0102+4839\tablenotemark{a} & 102.0 & 54 & 0.100 & 43$_{-28}^{+35}$ & 70$_{-16}^{+11}$ & 0.0$_{-2.5}^{+7.5}$ & 0.46$_{-0.01}^{+0.74}$\\
J0218+4232 & 325.3 & 84 & 0.033 & 25$_{-1}^{+2}$ & 12$\pm$2 & 0.0$_{-2.5}^{+10.0}$ & 2.12$_{-0.08}^{+0.01}$\\
J0437$-$4715\tablenotemark{a} & 349.3 & 173 & 0.022 & 35$\pm$1 & 64$\pm$1 & 50.0$_{-5.0}^{+2.5}$ & 0.70$_{-0.01}^{+0.12}$\\
J0610$-$2100\tablenotemark{a} & 70.8 & 34 & 0.000 & 87$_{-17}^{+3}$ & 49$_{-21}^{+17}$ & 0.0$_{-2.5}^{+10.0}$ & 1.11$_{-0.64}^{+0.74}$\\
J0613$-$0200\tablenotemark{a} & 218.8 & 113 & 0.067 & 55$_{-3}^{+1}$ & 43$_{-2}^{+3}$ & 2.5$\pm$2.5 & 0.78$_{-0.07}^{+0.08}$\\
J0614$-$3329 & 1312.5 & 264 & 0.067 & 63$\pm$1 & 84$\pm$1 & 0.0$\pm$2.5 & 1.03$\pm$0.01\\
J0751+1807 & 206.3 & 129 & 0.100 & 21$_{-2}^{+10}$ & 69$\pm$2 & 5.0$_{-5.0}^{+2.5}$ & 0.60$_{-0.06}^{+0.27}$\\
J1024$-$0719\tablenotemark{a} & 50.8 & 24 & 0.033 & 66$_{-65}^{+18}$ & 73$_{-73}^{+11}$ & 0.0$_{-2.5}^{+10.0}$ & 0.83$_{-0.43}^{+2.17}$\\
J1124$-$3653\tablenotemark{a} & 102.9 & 54 & 0.067 & 13$_{-4}^{+48}$ & 69$_{-34}^{+3}$ & 2.5$_{-2.5}^{+7.5}$ & 0.55$_{-0.13}^{+2.45}$\\
J1125$-$5825\tablenotemark{a} & 40.9 & 39 & 0.033 & 29$_{-11}^{+29}$ & 71$_{-13}^{+10}$ & 10.0$_{-10.0}^{+2.5}$ & 0.60$_{-0.03}^{+0.97}$\\
J1231$-$1411\tablenotemark{a} & 835.7 & 174 & 0.022 & 26$_{-4}^{+3}$ & 69$\pm$1 & 0.0$\pm$2.5 & 0.50$_{-0.01}^{+0.03}$\\
J1446$-$4701\tablenotemark{a} & 55.6 & 39 & 0.033 & 17$_{-4}^{+57}$ & 68$_{-40}^{+5}$ & 2.5$_{-2.5}^{+7.5}$ & 0.57$_{-0.09}^{+0.61}$\\
J1514$-$4946 & 212.8 & 114 & 0.450 & 24$_{-2}^{+7}$ & 68$\pm$1 & 0.0$\pm$2.5 & 0.54$_{-0.05}^{+0.04}$\\
J1600$-$3053 & 84.3 & 53 & 0.033 & 61$_{-3}^{+10}$ & 37$_{-9}^{+5}$ & 2.5$_{-2.5}^{+5.0}$ & 0.72$_{-0.02}^{+0.57}$\\
J1614$-$2230 & 207.4 & 114 & 0.167 & 80$_{-20}^{+8}$ & 80$_{-4}^{+6}$ & 7.5$\pm$2.5 & 1.02$_{-0.09}^{+0.05}$\\
J1658$-$5324\tablenotemark{a} & 159.0 & 84 & 0.411 & 30$_{-16}^{+17}$ & 69$_{-13}^{+5}$ & 7.5$_{-7.5}^{+2.5}$ & 0.60$_{-0.08}^{+0.80}$\\
J1713+0747\tablenotemark{a} & 44.8 & 38 & 0.033 & 36$_{-4}^{+9}$ & 68$\pm$4 & 10.0$_{-10.0}^{+2.5}$ & 0.68$_{-0.04}^{+0.52}$\\
J1747$-$4036\tablenotemark{a} & 46.0 & 24 & 0.167 & 19$_{-18}^{+71}$ & 80$_{-80}^{+10}$ & 0.0$_{-2.5}^{+10.0}$ & 0.48$_{-0.26}^{+4.52}$\\
J2017+0603\tablenotemark{a} & 438.5 & 234 & 0.150 & 34$_{-2}^{+3}$ & 67$_{-1}^{+2}$ & 2.5$_{-2.5}^{+5.0}$ & 0.64$_{-0.02}^{+0.09}$\\
J2043+1711\tablenotemark{a} & 245.8 & 114 & 0.767 & 54$_{-13}^{+6}$ & 76$_{-3}^{+5}$ & 7.5$_{-2.5}^{+5.0}$ & 0.83$_{-0.13}^{+0.93}$\\
J2047+1053\tablenotemark{a} & 66.0 & 74 & 0.050 & 51$_{-4}^{+14}$ & 71$_{-2}^{+13}$ & 0.0$_{-2.5}^{+7.5}$ & 0.57$_{-0.05}^{+0.63}$\\
J2051$-$0827\tablenotemark{a} & 48.9 & 24 & 0.500 & 43$_{-22}^{+47}$ & 69$_{-42}^{+21}$ & 0.0$_{-2.5}^{+10.0}$ & 0.52$_{-0.05}^{+3.81}$\\
J2215+5135 & 50.8 & 39 & 0.900 & 18$_{-1}^{+20}$ & 71$_{-1}^{+4}$ & 0.0$_{-2.5}^{+7.5}$ & 0.48$_{-0.02}^{+0.22}$\\
J2241$-$5236\tablenotemark{a} & 233.0 & 129 & 0.056 & 20$\pm$1 & 76$\pm$1 & 2.5$\pm$2.5 & 0.63$_{-0.03}^{+0.02}$\\
J2302+4442 & 383.4 & 174 & 0.200 & 60$_{-2}^{+5}$ & 46$_{-5}^{+3}$ & 2.5$\pm$2.5 & 0.93$_{-0.11}^{+0.06}$\\
\enddata

\tablenotetext{a}{Confidence contours in \al\ and \ze\ are not simply connected or simple shapes, quoted uncertainties should be used with caution.}
\end{deluxetable}
\normalsize


\begin{deluxetable}{lcccccccc}
\tabletypesize{\scriptsize}
\tablecaption{OG model fit results for Class I MSPs\label{tbl-og}}
\tablewidth{0pt}

\tablehead{
\colhead{PSR} & \colhead{\lnl}  & \colhead{$n_{\rm dof}$} & \colhead{$\Phi_{\mu}$}  & \colhead{\al} & \colhead{\ze} & \colhead{$w_{\rm acc}$} & \colhead{$w_{\rm em}$} & \colhead{\fO}\\
\colhead{} & \colhead{} & \colhead{} & \colhead{} & \colhead{($\circ$)}  & \colhead{($\circ$)} & \colhead{(\% \Tpc)} & \colhead{(\% \Tpc)} & \colhead{}
}

\startdata
J0023+0923\tablenotemark{a} & 60.7 & 38 & 0.567 & 67$\pm$7 & 24$_{-1}^{+10}$ & 0.0$_{-2.5}^{+10.0}$ & 0.0$\pm$2.5 & 0.81$_{-0.53}^{+0.24}$\\
J0030+0451 & 506.9 & 173 & 0.011 & 88$_{-2}^{+1}$ & 68$\pm$1 & 2.5$\pm$2.5 & 0.0$\pm$2.5 & 0.98$_{-0.03}^{+0.01}$\\
J0101$-$6422\tablenotemark{a} & 135.7 & 83 & 0.833 & 90$\pm$21 & 35$_{-4}^{+47}$ & 5.0$\pm$5.0 & 2.5$\pm$2.5 & 0.89$_{-0.11}^{+0.08}$\\
J0102+4839\tablenotemark{a} & 97.8 & 53 & 0.133 & 60$_{-20}^{+30}$ & 76$_{-9}^{+6}$ & 0.0$_{-2.5}^{+5.0}$ & 0.0$\pm$2.5 & 0.73$_{-0.44}^{+0.39}$\\
J0218+4232\tablenotemark{a} & 2355.2 & 83 & 0.167 & 45$_{-35}^{+33}$ & 67$_{-47}^{+23}$ & 0.0$_{-2.5}^{+5.0}$ & 0.0$\pm$2.5 & 0.52$_{-0.47}^{+1.00}$\\
J0437$-$4715\tablenotemark{a} & 340.1 & 172 & 0.022 & 76$\pm$1 & 46$\pm$1 & 10.0$\pm$2.5 & 2.5$\pm$2.5 & 0.82$_{-0.17}^{+0.01}$\\
J0610$-$2100\tablenotemark{a} & 79.6 & 33 & 0.500 & 63$_{-40}^{+13}$ & 89$_{-21}^{+1}$ & 0.0$_{-2.5}^{+10.0}$ & 0.0$\pm$2.5 & 1.02$_{-0.74}^{+0.21}$\\
J0613$-$0200\tablenotemark{a} & 226.6 & 112 & 0.067 & 60$\pm$1 & 45$\pm$1 & 5.0$\pm$2.5 & 2.5$\pm$2.5 & 0.67$_{-0.34}^{+0.02}$\\
J0614$-$3329 & 1487.9 & 263 & 0.089 & 58$_{-2}^{+1}$ & 88$\pm$1 & 0.0$\pm$2.5 & 0.0$\pm$2.5 & 0.98$\pm$0.01\\
J0751+1807\tablenotemark{a} & 207.4 & 128 & 0.033 & 59$_{-2}^{+6}$ & 72$_{-1}^{+2}$ & 0.0$_{-2.5}^{+5.0}$ & 0.0$\pm$2.5 & 0.55$_{-0.22}^{+0.59}$\\
J1024$-$0719 & 48.2 & 23 & 0.033 & 66$_{-18}^{+23}$ & 73$_{-37}^{+11}$ & 0.0$_{-2.5}^{+10.0}$ & 0.0$_{-2.5}^{+5.0}$ & 0.64$_{-0.36}^{+0.60}$\\
J1124$-$3653\tablenotemark{a} & 108.5 & 53 & 0.067 & 17$_{-8}^{+35}$ & 73$_{-13}^{+5}$ & 5.0$\pm$5.0 & 2.5$\pm$2.5 & 0.34$_{-0.11}^{+1.66}$\\
J1125$-$5825\tablenotemark{a} & 40.2 & 38 & 0.033 & 84$_{-24}^{+5}$ & 60$_{-38}^{+4}$ & 2.5$_{-2.5}^{+7.5}$ & 0.0$\pm$2.5 & 1.17$_{-0.88}^{+0.02}$\\
J1231$-$1411\tablenotemark{a} & 696.8 & 173 & 0.056 & 88$\pm$1 & 67$\pm$1 & 2.5$\pm$2.5 & 0.0$\pm$2.5 & 0.99$_{-0.02}^{+0.12}$\\
J1446$-$4701\tablenotemark{a} & 51.7 & 38 & 0.033 & 80$_{-18}^{+9}$ & 25$_{-6}^{+17}$ & 5.0$\pm$5.0 & 0.0$\pm$2.5 & 0.76$_{-0.36}^{+0.29}$\\
J1514$-$4946\tablenotemark{a} & 221.6 & 113 & 0.450 & 25$_{-5}^{+7}$ & 68$\pm$1 & 0.0$_{-2.5}^{+10.0}$ & 0.0$\pm$2.5 & 0.35$_{-0.06}^{+0.50}$\\
J1600$-$3053 & 84.9 & 52 & 0.033 & 65$_{-5}^{+3}$ & 28$_{-1}^{+6}$ & 0.0$_{-2.5}^{+5.0}$ & 0.0$\pm$2.5 & 0.76$_{-0.12}^{+0.28}$\\
J1614$-$2230 & 204.1 & 113 & 0.150 & 64$_{-20}^{+8}$ & 88$_{-5}^{+2}$ & 2.5$_{-2.5}^{+5.0}$ & 0.0$\pm$2.5 & 0.88$_{-0.18}^{+0.06}$\\
J1658$-$5324 & 165.6 & 83 & 0.411 & 78$_{-14}^{+3}$ & 23$_{-5}^{+7}$ & 5.0$_{-5.0}^{+2.5}$ & 2.5$\pm$2.5 & 0.84$_{-0.19}^{+1.66}$\\
J1713+0747\tablenotemark{a} & 50.3 & 37 & 0.033 & 36$_{-4}^{+18}$ & 65$_{-2}^{+14}$ & 0.0$_{-2.5}^{+10.0}$ & 0.0$\pm$2.5 & 0.49$_{-0.21}^{+0.71}$\\
J1747$-$4036\tablenotemark{a} & 45.3 & 23 & 0.167 & 19$_{-13}^{+71}$ & 80$_{-68}^{+10}$ & 0.0$_{-2.5}^{+10.0}$ & 0.0$\pm$2.5 & 0.32$_{-0.21}^{+1.03}$\\
J2017+0603\tablenotemark{a} & 409.3 & 233 & 0.150 & 23$\pm$1 & 74$\pm$1 & 5.0$\pm$2.5 & 2.5$\pm$2.5 & 0.30$_{-0.01}^{+0.16}$\\
J2043+1711\tablenotemark{a} & 303.6 & 113 & 0.733 & 53$_{-26}^{+8}$ & 79$_{-4}^{+8}$ & 0.0$_{-2.5}^{+5.0}$ & 0.0$\pm$2.5 & 0.90$_{-0.48}^{+0.29}$\\
J2047+1053\tablenotemark{a} & 67.6 & 73 & 0.050 & 51$_{-5}^{+14}$ & 71$_{-1}^{+13}$ & 0.0$\pm$2.5 & 0.0$\pm$2.5 & 0.38$_{-0.05}^{+0.84}$\\
J2051$-$0827 & 44.4 & 23 & 0.500 & 68$_{-42}^{+22}$ & 42$_{-14}^{+48}$ & 0.0$_{-2.5}^{+10.0}$ & 0.0$\pm$2.5 & 1.08$_{-0.80}^{+0.16}$\\
J2215+5135 & 48.8 & 38 & 0.900 & 18$_{-1}^{+20}$ & 71$_{-1}^{+11}$ & 0.0$_{-2.5}^{+7.5}$ & 0.0$\pm$2.5 & 0.29$_{-0.01}^{+0.41}$\\
J2241$-$5236 & 287.7 & 128 & 0.056 & 19$\pm$1 & 75$\pm$1 & 0.0$\pm$2.5 & 0.0$\pm$2.5 & 0.44$\pm$0.01\\
J2302+4442\tablenotemark{a} & 390.2 & 173 & 0.189 & 65$_{-2}^{+3}$ & 37$_{-2}^{+3}$ & 0.0$_{-2.5}^{+5.0}$ & 0.0$\pm$2.5 & 0.98$_{-0.03}^{+0.16}$\\
\enddata

\tablenotetext{a}{Confidence contours in \al\ and \ze\ are not simply connected or simple shapes, quoted uncertainties should be used with caution.}
\end{deluxetable}
\normalsize


\begin{landscape}
\begin{deluxetable}{lccccccccccc}
\tabletypesize{\scriptsize}
\tablecaption{alTPC model fit results for Class II MSPs\label{tbl-altpc}}
\tablewidth{0pt}

\tablehead{
\colhead{PSR} & \colhead{\lnl}  & \colhead{$n_{\rm dof}$} & \colhead{$\Phi_{\mu}$} & \colhead{\al} & \colhead{\ze} & \colhead{$w_{\gamma}$} & \colhead{$w_{\rm R}$} &\colhead{$R_{\rm max}^{\gamma}$} & \colhead{$R_{\rm min}^{\rm R}$} & \colhead{$R_{\rm max}^{\rm R}$} & \colhead{\fO}\\
\colhead{} & \colhead{} & \colhead{} & \colhead{} & \colhead{($\circ$)}  & \colhead{($\circ$)} & \colhead{(\% \Tpc)} & \colhead{(\% \Tpc)} & \colhead{(\Rlc)} & \colhead{(\Rlc)} & \colhead{(\Rlc)}
}

\startdata
J0034$-$0534\tablenotemark{a} & 130.9 & 80 & 0.567 & 23$_{-13}^{+21}$ & 69$\pm$3 & 2.5$_{-2.5}^{+7.5}$ & 2.5$_{-2.5}^{+7.5}$ & 0.95$_{-0.10}^{+0.25}$ & 0.85$_{-0.71}^{+0.05}$ & 1.00$_{-0.30}^{+0.20}$ & 0.46$_{-0.06}^{+0.59}$\\
J1810+1744\tablenotemark{a} & 150.0 & 80 & 0.611 & 82$_{-10}^{+5}$ & 16$_{-10}^{+6}$ & 5.0$\pm$5.0 & 2.5$_{-2.5}^{+7.5}$ & 0.70$_{-0.05}^{+0.50}$ & 0.14$_{-0.05}^{+0.81}$ & 0.70$\pm$0.50 & 1.31$_{-0.56}^{+1.67}$\\
J1823$-$3021A\tablenotemark{a} & 83.4 & 50 & 0.633 & 46$_{-38}^{+20}$ & 52$_{-28}^{+17}$ & 2.5$_{-2.5}^{+7.5}$ & 0.0$_{-2.5}^{+10.0}$ & 0.85$_{-0.15}^{+0.35}$ & 0.65$_{-0.45}^{+0.30}$ & 0.70$\pm$0.50 & 0.94$_{-0.36}^{+0.95}$\\
J1902$-$5105\tablenotemark{a} & 164.2 & 80 & 0.189 & 10$_{-2}^{+5}$ & 73$_{-1}^{+3}$ & 0.0$_{-2.5}^{+10.0}$ & 0.0$_{-2.5}^{+10.0}$ & 0.95$\pm$0.25 & 0.80$_{-0.66}^{+0.10}$ & 0.95$_{-0.50}^{+0.05}$ & 0.49$_{-0.12}^{+1.71}$\\
J1939+2134\tablenotemark{a} & 137.3 & 50 & 0.433 & 88$_{-1}^{+2}$ & 88$_{-1}^{+2}$ & 0.0$_{-2.5}^{+10.0}$ & 0.0$\pm$2.5 & 0.85$_{-0.15}^{+0.35}$ & 0.75$_{-0.05}^{+0.15}$ & 0.95$\pm$0.10 & 1.09$_{-0.03}^{+0.05}$\\
J1959+2048\tablenotemark{a} & 130.4 & 80 & 0.522 & 56$_{-16}^{+11}$ & 85$_{-6}^{+3}$ & 0.0$_{-2.5}^{+5.0}$ & 0.0$_{-2.5}^{+10.0}$ & 0.95$_{-0.05}^{+0.25}$ & 0.14$_{-0.05}^{+0.81}$ & 0.95$_{-0.75}^{+0.25}$ & 0.89$_{-0.11}^{+1.25}$\\
\enddata

\tablenotetext{a}{Confidence contours in \al\ and \ze\ are not simply connected or simple shapes, quoted uncertainties should be used with caution.}
\end{deluxetable}
\normalsize


\begin{deluxetable}{lccccccccccccc}
\tabletypesize{\scriptsize}
\tablecaption{alOG model fit results for Class II MSPs\label{tbl-alog}}
\tablewidth{0pt}

\tablehead{
\colhead{PSR} & \colhead{\lnl}  & \colhead{$n_{\rm dof}$} & \colhead{$\Phi_{\mu}$} & \colhead{\al} & \colhead{\ze} & \colhead{$w_{\gamma,\rm acc}$} & \colhead{$w_{\gamma,\rm em}$} &  \colhead{$w_{\rm R,acc}$} &  \colhead{$w_{\rm R,em}$} &\colhead{$R_{\rm max}^{\gamma}$} & \colhead{$R_{\rm min}^{\rm R}$} & \colhead{$R_{\rm max}^{\rm R}$} & \colhead{\fO}\\
\colhead{} & \colhead{} & \colhead{} & \colhead{} & \colhead{($\circ$)}  & \colhead{($\circ$)} & \colhead{(\% \Tpc)} & \colhead{(\% \Tpc)} & \colhead{(\% \Tpc)} & \colhead{(\% \Tpc)} & \colhead{(\Rlc)} & \colhead{(\Rlc)} & \colhead{(\Rlc)}
}

\startdata
J0034$-$0534\tablenotemark{a} & 137.9 & 78 & 0.589 & 22$_{-17}^{+23}$ & 75$_{-5}^{+15}$ & 5.0$\pm$5.0 & 2.5$\pm$2.5 & 5.0$\pm$5.0 & 2.5$\pm$2.5 & 0.95$_{-0.15}^{+0.25}$ & 0.90$_{-0.76}^{+0.05}$ & 1.05$_{-0.35}^{+0.15}$ & 0.24$_{-0.06}^{+0.81}$\\
J1810+1744\tablenotemark{a} & 147.8 & 78 & 0.167 & 81$_{-10}^{+9}$ & 24$_{-14}^{+5}$ & 2.5$_{-2.5}^{+7.5}$ & 0.0$\pm$2.5 & 5.0$\pm$5.0 & 2.5$\pm$2.5 & 0.95$\pm$0.25 & 0.25$_{-0.11}^{+0.70}$ & 1.10$_{-0.45}^{+0.10}$ & 0.90$_{-0.72}^{+0.15}$\\
J1823$-$3021A & 81.2 & 48 & 0.467 & 42$_{-36}^{+48}$ & 78$_{-65}^{+12}$ & 5.0$\pm$5.0 & 2.5$\pm$2.5 & 5.0$\pm$5.0 & 2.5$\pm$2.5 & 1.10$_{-0.40}^{+0.10}$ & 0.95$_{-0.81}^{+0.05}$ & 1.00$_{-0.45}^{+0.20}$ & 0.45$_{-0.27}^{+2.23}$\\
J1902$-$5105\tablenotemark{a} & 159.2 & 78 & 0.189 & 10$\pm$2 & 73$\pm$1 & 0.0$_{-2.5}^{+7.5}$ & 0.0$\pm$2.5 & 0.0$_{-2.5}^{+7.5}$ & 0.0$\pm$2.5 & 0.95$_{-0.05}^{+0.10}$ & 0.80$_{-0.66}^{+0.15}$ & 0.95$_{-0.05}^{+0.10}$ & 0.30$_{-0.10}^{+0.22}$\\
J1939+2134\tablenotemark{a} & 135.7 & 48 & 0.433 & 72$\pm$1 & 85$\pm$1 & 0.0$_{-2.5}^{+10.0}$ & 0.0$\pm$2.5 & 2.5$_{-2.5}^{+7.5}$ & 0.0$\pm$2.5 & 0.80$_{-0.10}^{+0.40}$ & 0.90$_{-0.30}^{+0.05}$ & 1.05$\pm$0.15 & 1.01$_{-0.63}^{+0.19}$\\
J1959+2048 & 131.8 & 78 & 0.544 & 52$_{-42}^{+15}$ & 87$_{-13}^{+3}$ & 0.0$_{-2.5}^{+5.0}$ & 0.0$\pm$2.5 & 0.0$_{-2.5}^{+10.0}$ & 0.0$_{-2.5}^{+5.0}$ & 1.00$_{-0.10}^{+0.20}$ & 0.14$_{-0.05}^{+0.81}$ & 1.00$_{-0.40}^{+0.20}$ & 0.80$_{-0.59}^{+0.22}$\\
\enddata

\tablenotetext{a}{Confidence contours in \al\ and \ze\ are not simply connected or simple shapes, quoted uncertainties should be used with caution.}
\end{deluxetable}
\normalsize
\end{landscape}

\begin{deluxetable}{lccccccccccc}
\tabletypesize{\scriptsize}
\tablecaption{laSG model fit results for Class II MSPs\label{tbl-lasg}}
\tablewidth{0pt}

\tablehead{
\colhead{PSR} & \colhead{\lnl}  & \colhead{$n_{\rm dof}$} & \colhead{$\Phi_{\mu}$} & \colhead{\al} & \colhead{\ze} & \colhead{$w_{\gamma}$} & \colhead{$w_{\rm R}$} &\colhead{$\sigma_{\rm in}$} & \colhead{$\sigma_{\rm out}$} & \colhead{$s_{\rm f}$} & \colhead{\fO}\\
\colhead{} & \colhead{} & \colhead{} & \colhead{} & \colhead{($\circ$)}  & \colhead{($\circ$)} & \colhead{(\% \Tpc)} & \colhead{(\% \Tpc)} & \colhead{(\Rns)} & \colhead{(\Rns)} & \colhead{(\Rns)} & \colhead{}
}

\startdata
J0034$-$0534\tablenotemark{a} & 133.3 & 80 & 0.033 & 23$_{-18}^{+9}$ & 74$_{-34}^{+4}$ & 0.0$_{-2.5}^{+5.0}$ & 0.0$_{-2.5}^{+5.0}$ & 0.8$_{-0.5}^{+0.2}$ & 0.9$\pm$0.6 & 1.8$_{-0.6}^{+0.2}$ & 0.86$_{-0.54}^{+0.33}$\\
J1810+1744\tablenotemark{a} & 150.7 & 80 & 0.033 & 5$_{-3}^{+81}$ & 29$_{-27}^{+47}$ & 5.0$_{-5.0}^{+2.5}$ & 2.5$\pm$2.5 & 0.7$_{-0.6}^{+0.3}$ & 0.9$_{-0.6}^{+1.2}$ & 1.8$_{-0.6}^{+0.2}$ & 5.28$_{-5.02}^{+1.37}$\\
J1823$-$3021A\tablenotemark{a} & 87.7 & 50 & 0.567 & 78$_{-3}^{+10}$ & 45$_{-22}^{+6}$ & 5.0$_{-5.0}^{+2.5}$ & 0.0$_{-2.5}^{+5.0}$ & 0.1$_{-0.1}^{+0.9}$ & 0.3$_{-0.3}^{+1.8}$ & 1.4$_{-0.2}^{+0.6}$ & 0.60$_{-0.28}^{+2.33}$\\
J1902$-$5105\tablenotemark{a} & 166.1 & 80 & 0.167 & 56$\pm$8 & 16$_{-7}^{+8}$ & 0.0$_{-2.5}^{+5.0}$ & 5.0$_{-5.0}^{+2.5}$ & 0.3$_{-0.2}^{+0.3}$ & 2.1$_{-0.9}^{+0.3}$ & 1.6$\pm$0.4 & 0.92$_{-0.39}^{+0.62}$\\
J1939+2134\tablenotemark{a} & 154.3 & 50 & 0.833 & 33$\pm$12 & 41$_{-9}^{+6}$ & 0.0$_{-2.5}^{+5.0}$ & 2.5$\pm$2.5 & 0.2$_{-0.1}^{+0.3}$ & 0.3$_{-0.3}^{+0.6}$ & 1.4$_{-0.2}^{+0.6}$ & 1.51$_{-0.55}^{+0.40}$\\
J1959+2048 & 125.5 & 80 & 0.433 & 46$_{-38}^{+13}$ & 35$_{-27}^{+31}$ & 0.0$_{-2.5}^{+5.0}$ & 2.5$\pm$2.5 & 0.3$_{-0.2}^{+0.7}$ & 2.1$_{-1.8}^{+0.3}$ & 1.6$\pm$0.4 & 1.63$_{-1.17}^{+0.66}$\\
\enddata

\tablenotetext{a}{Confidence contours in \al\ and \ze\ are not simply connected or simple shapes, quoted uncertainties should be used with caution.}
\end{deluxetable}
\normalsize


\begin{deluxetable}{lccccccc}
\tabletypesize{\scriptsize}
\tablecaption{PSPC model fit results for Class III MSPs\label{tbl-pspc}}
\tablewidth{0pt}

\tablehead{
\colhead{PSR} & \colhead{\lnl}  & \colhead{$n_{\rm dof}$} & \colhead{$\Phi_{\mu}$} & \colhead{\al} & \colhead{\ze} & \colhead{\fO}\\
\colhead{} & \colhead{} & \colhead{} & \colhead{} & \colhead{($\circ$)}  & \colhead{($\circ$)} & \colhead{}
}

\startdata
J0340+4130\tablenotemark{a} & 197.4 & 115 & 0.050 & 43$_{-1}^{+6}$ & 73$\pm$7 & 1.25$_{-0.29}^{+0.20}$\\
J1741+1351\tablenotemark{a} & 33.4 & 40 & 0.300 & 46$_{-29}^{+11}$ & 80$_{-20}^{+7}$ & 1.06$_{-0.07}^{+0.74}$\\
J1744$-$1134\tablenotemark{a} & 174.8 & 85 & 0.033 & 51$_{-4}^{+23}$ & 85$_{-9}^{+3}$ & 1.04$_{-0.53}^{+1.00}$\\
J1858$-$2216 & 93.3 & 85 & 0.067 & 42$_{-5}^{+7}$ & 74$_{-4}^{+8}$ & 1.16$_{-0.20}^{+0.23}$\\
J2124$-$3358\tablenotemark{a} & 286.2 & 115 & 0.983 & 19$\pm$1 & 25$_{-2}^{+4}$ & 0.46$_{-0.02}^{+0.50}$\\
J2214+3000 & 207.2 & 130 & 0.056 & 59$_{-3}^{+1}$ & 86$\pm$2 & 1.05$\pm$0.02\\
\enddata

\tablenotetext{a}{Confidence contours in \al\ and \ze\ are not simply connected or simple shapes, quoted uncertainties should be used with caution.}
\end{deluxetable}
\normalsize
\clearpage


The gamma-ray and radio light curves of several 2PC MSPs have been modeled previously using similar models and methods.  \citet{Venter09} modeled the light curves of the first eight MSPs detected with the LAT, choosing the best geometries by eye with steps of 5\DG\ in both \al\ and \ze.  This approach put more weight on matching the gamma-ray light curves.  Our results agree well for PSRs J0030+0451, J0437$-$4715 (the TPC fit), and J1744$-$1134. For the other MSPs they fit, our likelihood method either prefers a different maximum (see the discussion of confidence contours with multiple maxima in Appendix \ref{app-contours}) or the best-fit is substantially different due to increased statistics and/or more emphasis on matching the radio light curve.

The light curves of PSR J0034$-$0534 have been modeled by \citet{J0034} using by-eye fits with alOG and alTPC models and by \citet{Venter12}.  The latter authors used a similar likelihood fitting technique for alOG and alTPC models and by-eye fits for the laSG model.  In both cases, the alOG and alTPC results agree with the best-fit parameters presented in Tables \ref{tbl-altpc} and \ref{tbl-alog}.  However, our laSG best-fit geometry does not agree with that reported by \citet{Venter12}, likely because their by-eye fits did not put emphasis on the low-level off-peak emission.

Both previous studies used the 320 MHz radio profile and not the 1400 MHz profile we use here.  At lower frequency, the peaks are at roughly equal heights whereas at 1400 MHz the peak near phase $\sim$0.8 is significantly lower than the other peak.  Compared to the best-fit parameters of \citet{Venter12}, we do find slightly higher values of $R_{\rm min}^{\rm R}$ and $R_{\rm max}^{\rm R}$, though the differences are not significant when considering the estimated uncertainties.  This is opposite of what one would expect from radius-to-frequency mapping studies \citep[e.g.,][]{Cordes78} and may point to additional differences between the radio emission observed in Class II MSPs and the single-altitude cone and/or core assumption.

\citet{Cognard11} fit the light curves of PSRs J2017+0603 and J2302+4442 with OG and TPC gamma-ray models, a hollow-cone beam, and a Markov chain Monte Carlo maximum likelihood procedure.  Our results agree well with theirs for PSR J2302+4442 but we find somewhat larger \al\ values for PSR J2017+0603.

Light curve fits for PSR J1823$-$3021A were reported by \citet{Freire11}, in the supplementary online material, using the alTPC and alOG models and a similar maximum likelihood technique.  These fits were done above 500 MeV due to issues with the background level estimation. Their best-fit values agree with ours within the quoted uncertainties, though our values of $R_{\rm min}^{\rm R}$ are noticeably greater than theirs, for both models.

\citet{Guillemot12} reported fits to the light curves of PSRs J1939+2134 (B1937+21) and J1959+2048 (B1957+20) using alOG and alTPC models and a similar maximum likelihood procedure.  \citet{Venter12} provided more details on these fits as well as by-eye fits using the laSG model.  For PSR J1939+2134 the fits agree within uncertainties, but our alTPC best-fit geometry agrees better with those from fitting radio polarization data \citep[also presented by][]{Guillemot12}.  For PSR J1959+2048 our alOG and alTPC fits agree with theirs, within the estimated uncertainties, while our best-fit laSG model has a slightly different geometry (though still within uncertainties).  Their fits used the 350 MHz radio profile and not the 1400 MHz profile.  The lower frequency profile only has two peaks, both of which are matched to a peak in the gamma-ray light curve.  The 1400 MHz profile has an additional peak that does not have a corresponding feature in the gamma-ray light curve.  This additional peak was not fit by our models but its position in phase may indicate that it is from either a core or conal component.

\citet{Kerr2012} modeled the gamma-ray and radio light curves of PSR J0101$-$6422 using TPC and OG gamma-ray models with a hollow-cone radio beam.  Our fits are compatible with theirs, but the models have difficulties producing both the gamma-ray and radio light curves.  This MSP may require a more complex radio model.

The gamma-ray and radio light curves of PSR J2043+1711 were fit by \citet{GuillemotJ2043} using TPC and OG gamma-ray models with a hollow-cone radio beam and the same maximum likelihood procedure.  Our best-fit results for this MSP are in agreement with theirs.

\citet{Du10} modeled only the gamma-ray light curves of PSRs J0030+0451, J0218+4232, and J0437$-$4715 using data from the first LAT pulsar catalog \citep{1PC} and an annular gap model, assuming that the magnetic axes of MSPs should have moved towards alignment with the spin axis.  For PSR J0030+0451, our fits do not agree well with theirs, finding larger \al\ values consistent with X-ray light curve modeling, and their model over predicts the observed level of off-peak emission.  For PSR J0218+4232, our OG fit is consistent with theirs; however, our TPC best-fit geometry has a substantially lower \ze\ value, necessary to match the radio profile.  For PSR J0437$-$4715, our TPC fit finds a similar \al\ with somewhat larger \ze.  Our OG fit finds a significantly larger \al\ and slightly lower \ze.

\citet{Du13} modeled the gamma-ray and radio light curves of PSRs J0034$-$0534, J0101$-$6422, and J0437$-$4715 using an annular gap model.  Their fits were done by eye but guided by the assumption that the inclination angles should be small \citep[consistent with the possible trend of decreasing \al\ with age for non-recycled pulsars, e.g.][]{Young2010} and by geometric estimates from other wavelengths when possible.  Their geometry for PSR J0034$-$0534 is in agreement with all of our fits, though their model does not produce any off-peak emission, which is interesting given that the annular gap emission geometry is different from our models.  For PSR J0101$-$6422 their models reproduce the data well and neither of our best-fit geometries agree with theirs.  Finally, our best-fit results do not agree with their quoted geometry for PSR J0437$-$4715 \citep[note that they fixed \ze\ \emph{a priori} to agree with the estimated orbital inclination angle,][]{vanStrat01,Hotan06}.  \citet{Bogdanov07} found it necessary to use a magnetic dipole that was offset from the center of the PC to fit the X-ray light curve of this MSP, a possibility we plan to explore in the future.  In comparing our fits to those of both \citet{Du10} and \citet{Du13}, it is important to note that their simulations used a static dipole magnetic field configuration.  It is unclear exactly how including the effects of rotation will change their best-fit geometries.

\citet{Petri11} modeled the observed gamma-ray and radio light curves of eight pulsars (three of which were MSPs) using a striped wind model for the HE emission and a polar cap beam for the radio profile.  For PSR J0030+0451 our fits agree well with those of \citet{Petri11} in that both \al\ and \ze\ should be $>50^{\circ}$ but we find $\alpha>\zeta$, the opposite of what he finds.  Additionally, we note that he had to increase the size of the polar cap emitting region to a significant portion of the entire neutron star surface to match the radio peaks and his predicted radio to gamma-ray phase lag is too large by 0.1 in phase.  For PSR J0218+4132 our OG fit results agree well with his geometry, though his model radio peaks are not broad enough and the double-peaked structure of the main component is not reproduced.  Our best-fit results for PSR J0437$-$4715 do not agree well with his model geometry (except for the \ze\ value of our OG fit) and while he does reproduce the basic light curve structures well the predicted phase lag is too large.

\citet{Espinoza13} identified PSRs J0218+4232, J2017+0603, J2124$-$3358, and J2302+4442 as W-type, suggesting that these were aligned rotators with \al\ close to 0\DG.  For PSRs J0218+4232, J2017+0603, and J2124$-$3358 the best-fit geometries have values of \al\ between 20\DG\ and 30\DG (excluding the OG fit for PSR J0218+4232 which does not reproduce the radio profile well), supporting the aligned rotator hypothesis. For PSR J2302+4232, both the TPC and OG models find \al\ near 60\DG.  While neither exactly reproduces the radio profile (see Appendix \ref{app-fits}) the gamma-ray light curve is reproduced well and \citet{thesis} showed that, for the same \al\ and \ze, increasing the altitude of the simulated radio emission leads to a better fit of the radio data.  PSR J1124$-$3653 would be identified as a W-type MSP and, for both the TPC and OG model fits, is found with a low \al, between 10\DG\ and 20\DG, which may support the separation of W-type and N-type MSPs based on the characteristics of their radio profiles.  However, there are several Class I MSPs with low values of \al, in some fits, that would likely be, or have been, classified as N-type, namely, PSRs J0751+1807, J1446$-$4701, J1514$-$4946, J2215+5135, and J2241$-$5236.  If our best-fit \al\ values are reliable, this brings into question the morphological distinction between the N-type and W-type MSPs of \citet{Espinoza13}.

\subsection{BEST-FIT PARAMETER DISTRIBUTIONS}\label{bestPars}
\begin{landscape}
\begin{figure}[h]
\begin{center}
\includegraphics[height=0.85\textheight]{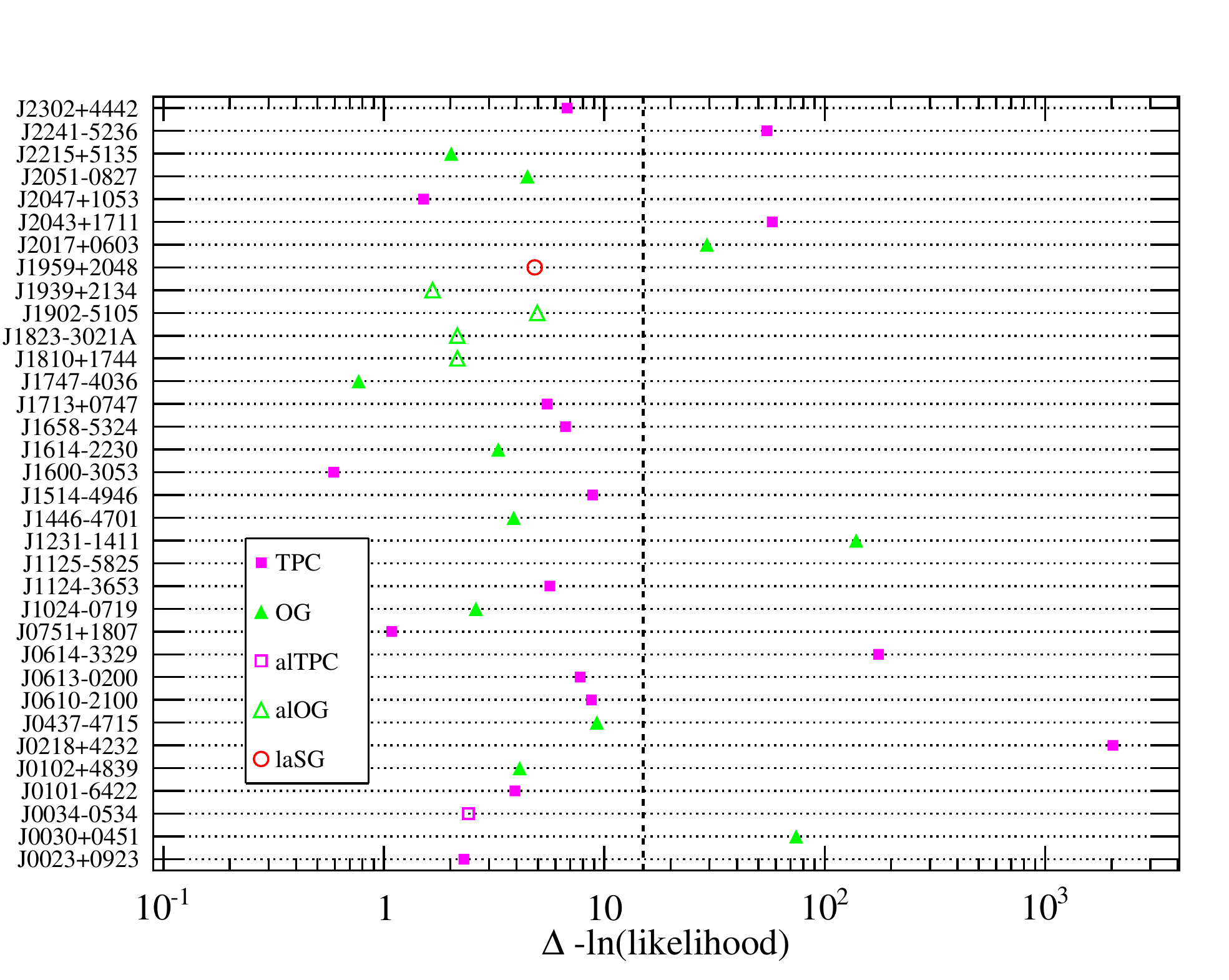}
\end{center}
\caption{Difference between the smallest \lnl\ value and the next best for all Class I and II MSPs.  The symbol for each MSP denotes which fit is best as indicated in the legend: TPC, light gray (pink in the online version) filled squares; OG, dark gray (green in the online version) filled upward-pointing triangles; alTPC, light gray (pink in the online version) open squares; alOG, dark gray (green in the online version) open upward-pointing triangles; and laSG, black (red in the online version) open circles.  The vertical dashed line indicates a \lnl\ difference of 15, our threshold for a significant preference of a particular model over the others.\label{fig-deltaLL}}
\end{figure}
\end{landscape}

Figure \ref{fig-deltaLL} presents the \lnl\ differences between the best and next-best models for all Class I and II MSPs.  We say that one model is significantly favored over the other for only seven MSPs, those with $\Delta$(\lnl)$\geq15$ indicated by the dashed vertical line.

\begin{figure}[h]
\begin{center}
\includegraphics[height=0.5\textheight]{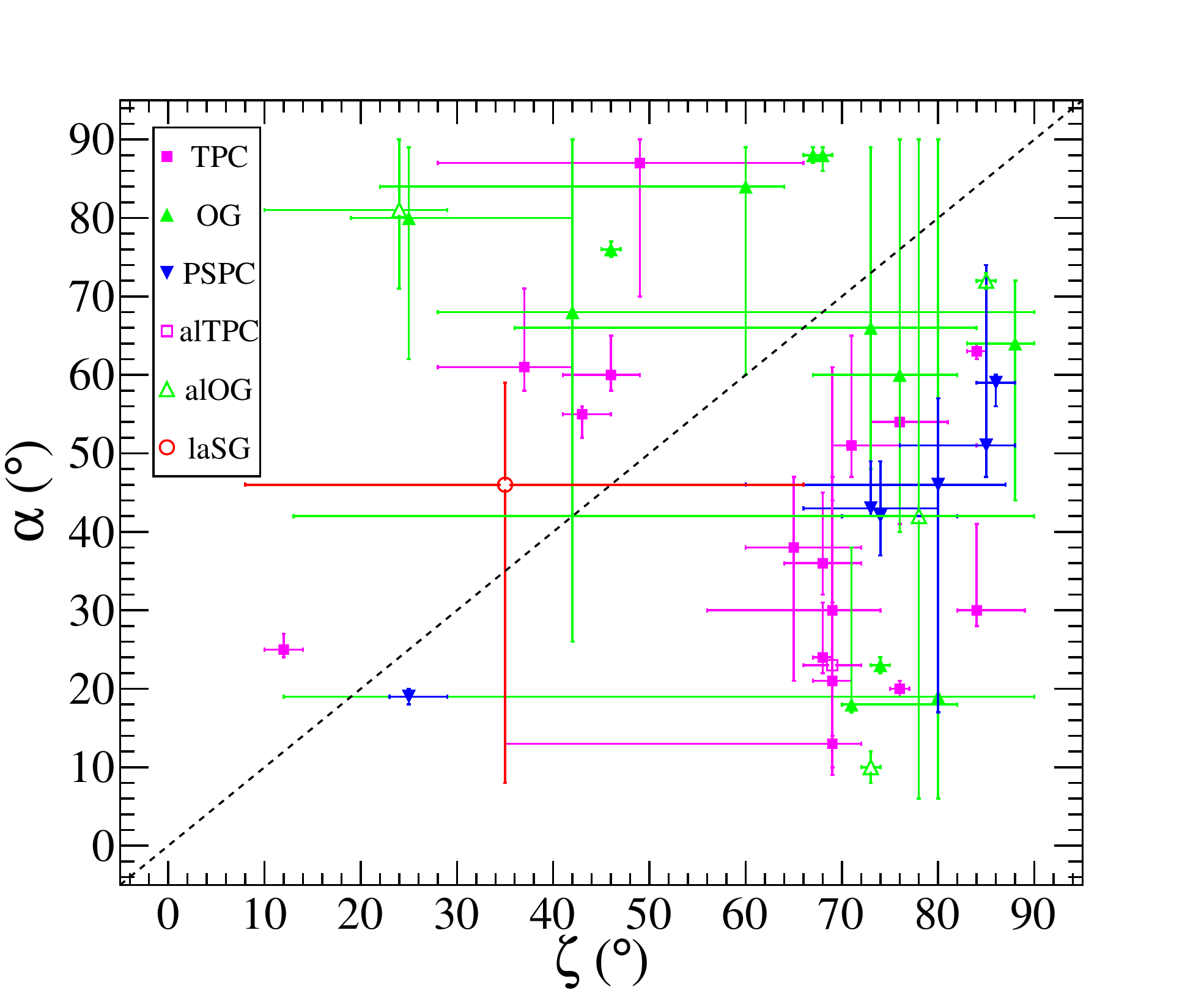}
\end{center}
\caption{Best-fit (\al,\ze) pairs for all 40 2PC MSPs.  The markers indicate which model is best as follows: TPC, light gray (pink in the online version) filled squares; OG, dark gray (green in the online version) filled upward-pointing triangles; PSPC, black (blue in the online version) filled downward-pointing triangles; alTPC, light gray (pink in the online version) open squares; alOG, dark gray (green in the online version) open upward-pointing triangles; and laSG, black (red in the online version) open circles.  Only the geometry for the fit with the smallest \lnl\ is shown for each MSP.    The dashed diagonal line indicates the magnetic axis with \al=\ze.\label{fig-bestGeoms}}
\end{figure}

Figure \ref{fig-bestGeoms} presents the best-fit (\al,\ze) pairs for each MSP, with only the values corresponding to the fit with the smallest \lnl\ shown for each MSP, even if the log-likelihood difference is $<15$.  There is a clear preference for \ze\ near 90\DG.  This is expected if pulsar spin axes are distributed randomly with respect to the Earth line of sight and for outer-magnetospheric gamma-ray models where the brightest emission is predicted near the spin equator.  However, the best-fit \al\ values seem to be uniformly distributed between 0\DG\ and 90\DG, which may have implications for MSP evolution after the recycling phase (see Section \ref{evolve}).

This contrasts with best-fit $\alpha > 40^\circ$ values found for non-recycled gamma-ray pulsars by \citet{Pierbattista13}.  While the \al\ values are spread out in the best-fit TPC models, the best-fit OG models generally seem to require $\alpha>60^\circ$.  This reflects the fact that emission in the OG models fills a relatively smaller part of (\al,\ze) phase space, so that pulsars are only visible over a restricted range of \al.  In the case of the PSPC models, most of the best-fit \al\ values lie in the range from $40^\circ$ to $60^\circ$. This gives the optimal level of off-peak emission and radio peak multiplicity.

\begin{figure}[h]
\begin{center}
\includegraphics[height=0.5\textheight]{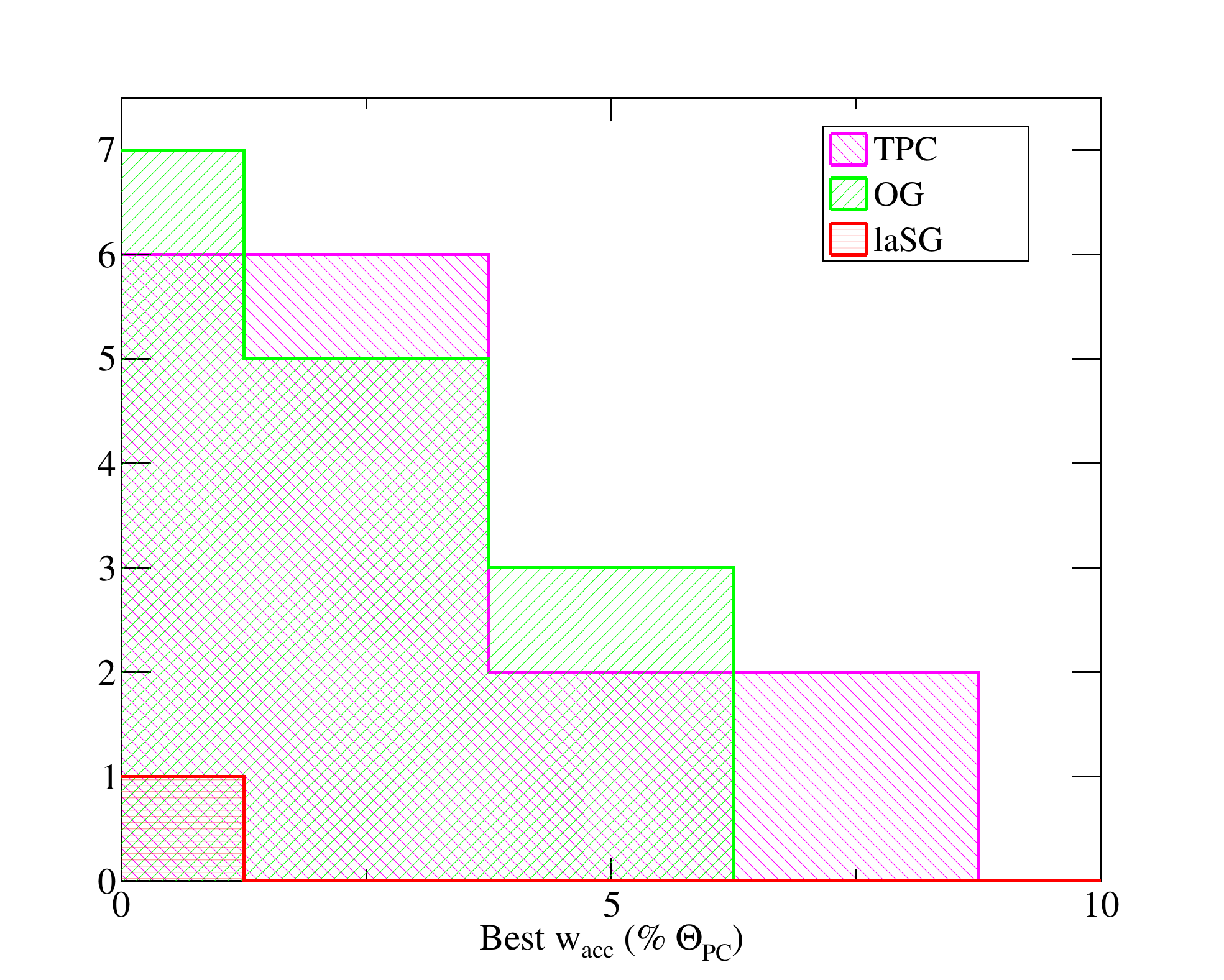}
\end{center}
\caption{Best-fit accelerating gap widths for Class I and II MSPs.  Only the value for the fit with the smallest \lnl\ is shown for each MSP.  TPC and alTPC fit values are shown in the light gray filled histogram (pink in the online version, filled with downward diagonal lines from left to right).  OG and alOG fit values are shown in the dark gray (green in the online version) histogram filled with upward diagonal lines from left to right.  Values from laSG fits are shown in the black (red in the online version) histogram filled with horizontal lines.\label{fig-bestWidths}}
\end{figure}

Figure \ref{fig-bestWidths} presents the best-fit accelerating gap widths for Class I and II MSPs, only the values corresponding to the fit with the smallest \lnl\ are shown.  Values from TPC and alTPC fits are grouped in the same histogram, as are values from OG and alOG fits.  The uncertainties in the Tables suggest that, in general, we cannot constrain the gap width parameters well, but it is clear that the light curve shapes prefer smaller widths.  Gaps with width 0 (i.e., emission only along one field line for each $l_{\rm ovc}$) are unphysical and suggest that the actual best-fit value is smaller than our resolution of 2.5\% \Tpc.

\begin{landscape}
\begin{figure}[h]
\begin{center}
\hspace*{-25pt}\includegraphics[height=0.9\textheight]{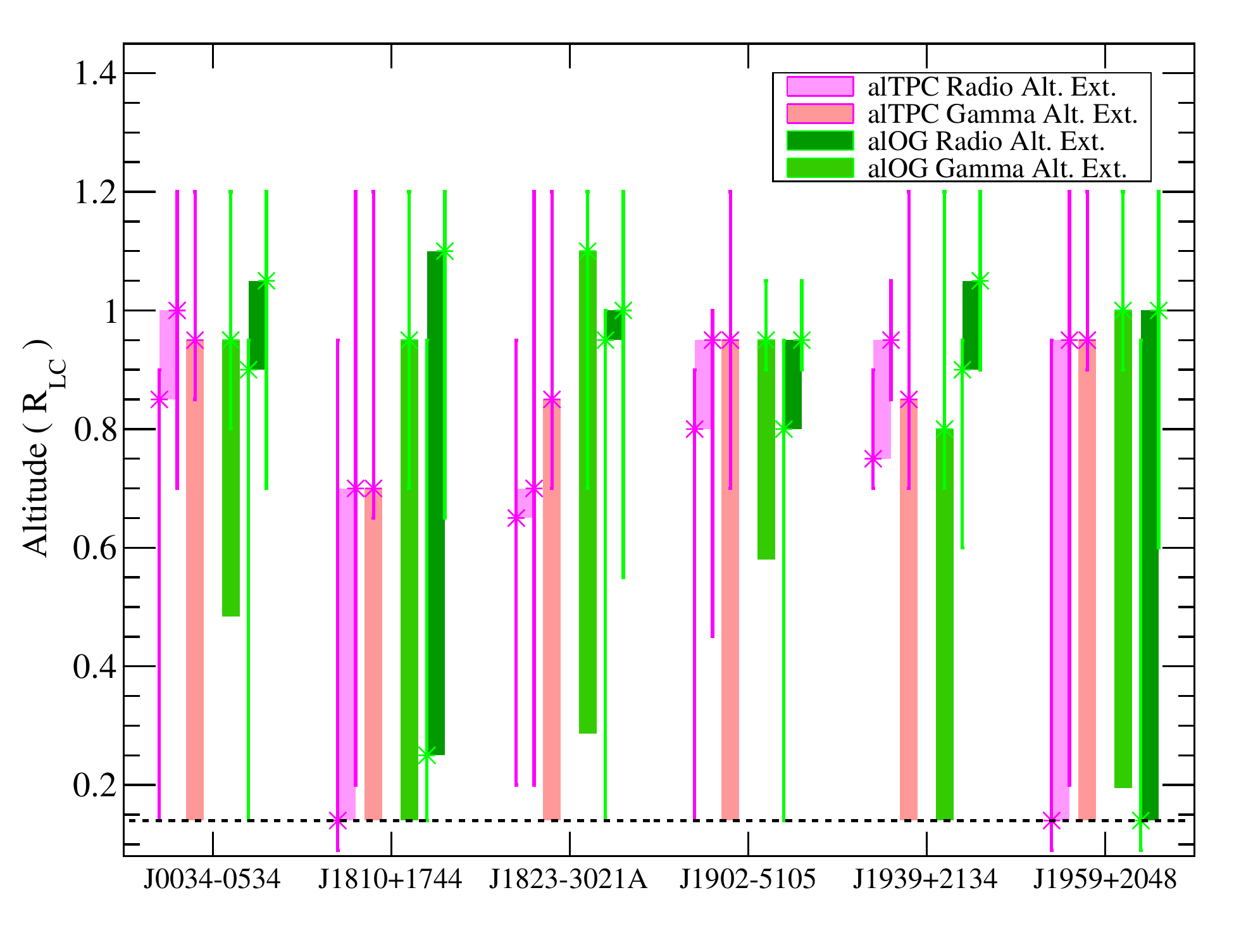}
\end{center}
\caption{Best-fit radio and gamma-ray emission altitudes from alTPC and alOG fits of all Class II MSPs.  The filled areas show the extent of the emission regions.  The asterisk points show the best-fit values and corresponding uncertainties.  The horizontal dashed line indicates the neutron star surface at 0.14 \Rlc\ for a 1.5 ms spin period.  The minimum altitudes of the gamma-ray emission regions for alOG models are set to the approximate minimum altitude of the NCS for the best-fit \al. \label{fig-altRanges}}
\end{figure}
\end{landscape}

Figure \ref{fig-altRanges} plots the best-fit emission altitudes for alTPC and alOG fits of all Class II MSPs, regardless of which fit is preferred by the likelihood in order to better demonstrate the behavior for both models.  The minimum radio emission altitudes for alOG fits in Figure \ref{fig-altRanges} are plotted as either the best-fit value or the average altitude of the NCS for the best-fit \al, whichever is larger.

The radio and gamma-ray emission regions typically have significant overlap, with the radio emission often concentrated towards the upper edge of the gamma-ray emission region.  These altitudes are significantly greater than typically assumed for other radio emission models (e.g., the cone and core beams we use for Class I and III MSPs, see Equation \ref{eqn-rkg}).  With the current statistics and use of the 95\% confidence level uncertainties we cannot constrain these altitudes well.

\section{DISCUSSION AND CONCLUSIONS}\label{disc}
The gamma-ray MSPs in 2PC (representing approximately one-third of the known gamma-ray pulsar population) have a greater variety of light curve morphology than the young gamma-ray pulsars.  While most MSP light curves are similar to those of young pulsars, showing the correlation of increasing radio lag with decreasing gamma-ray peak separation \citep{1PC,2PC}, a much larger number show phase alignment of radio and gamma-ray peaks and a new phenomenon, gamma-ray peaks leading the radio peaks, not observed in the young pulsar population.  We find that the MSP light curves resembling those of young gamma-ray pulsars are indeed well fit by the narrow gap models that best describe young pulsars \citep{RW10,Pierbattista13} and by standard radio core and cone geometry \citep{Rankin83,Arzoumanian02}.  

We have successfully fit the light curves showing aligned gamma-ray and radio peaks with the same models but assuming that the radio peaks are also caustics, and fitting for minimum and maximum radius of emission.  This implies that the radio emission may occur at high enough altitude, and may be extended enough along the last open field lines, that the emission is also strongly distorted into caustic patterns.  \citet{Manchester2005} and \citet{Ravi2010} have suggested that similar models are needed for the radio emission of all gamma-ray pulsars.  One prediction of this model, that the caustic peaks should have low linear polarization, seems to be borne out by radio observations of these pulsars showing that most in fact have little or no polarization \citep[also noted by][]{Espinoza13}.  We have fit the MSP light curves in which the gamma-ray peaks lead the radio peak(s) with PSPC models where gamma-ray emission occurs over the whole open-field volume and we are viewing the high-altitude portion that is highly aberrated to earlier phase.

The best fits provide estimates of \al\ and \ze, as well as emission gap width, for each model.  Additionally, the best fits for the altitude-limited models provide minimum and maximum radii for the radio and maximum radius for the gamma-ray emission.  We find best-fit \al\ values are nearly evenly distributed over a large range between 10\DG\ and 90\DG.  On the other hand, our best-fit \ze\ values are concentrated mostly at high values between 70\DG\ and 90\DG, due to the caustics being brighter at large \ze\ and the greater probability of viewing at large angles.

The large \al\ range partly reflects the wide radio beam size and smaller magnetospheres of MSPs, allowing both the radio and gamma-ray beams to be visible over a larger region of phase space than for longer-period pulsars.  Thus a larger range of impact parameter $\beta$ is allowed, restricting the number of radio-quiet MSPs to very few.  Indeed, there are no radio-quiet MSPs in 2PC, while half of the young gamma-ray pulsar population is radio-quiet.  This larger range of phase space accessible to observations may also partly explain the increased diversity in light curve profiles we observe for MSPs.  On the other hand, the large \al\ range for MSPs could be a result of their spin-up evolution (see Section \ref{evolve}).  Their relatively weak magnetic fields compared to young gamma-ray pulsars require some suppression, possibly by flux burying \citep{Romani90} or crustal plate tectonics \citep{Ruderman91}, during their accretion and spin-up phase.  Such processes may be accompanied by changes in \al.

We find that the classic radio core/cone geometry, with aberration and retardation included, is able to reasonably fit the properties of most MSP light curves, such as peak widths, number, and gamma/radio lag.
However, there is indication for some sources that the radio emission altitude we have assumed is too low, possibly suggesting that standard radius-to-frequency mapping is not applicable to MSPs \citep[suggested by][]{Kramer99}.  In some cases where the radio light curves have multiple peaks, such as PSRs J0613$-$0200, J1658$-$5324, and J2302+4442; both the width and radio lags are too small in our models.  It may be interesting to explore models with higher-altitude radio cones and with partially filled cones \citep{LM88}, which \citet{Kramer99} suggested could explain the apparently abnormal profile development with frequency of radio MSPs.  Additionally, turning on different sets of field lines in a partially-filled cone will lead to different predicted phase lags.

Our finding that the majority of MSP light curves are best fit by the TPC or OG models with narrow emission gaps was not expected prior to the launch of \fermi.  The formation of narrow accelerator gaps in these models requires sufficient multiplicity of electron-positron pairs to screen the parallel electric field throughout the rest of the open-field region outside the gaps.  The standard versions of these models assume dipole magnetic fields and the death lines for pair cascades lie above the \Pd\ of most MSPs \citep{Harding05,Zhang04}, suggesting they should be pair-starved.  We find that only a small fraction of MSP light curves are best fit with PSPC models, indicating that the pair death lines predicted by the standard models are too high.  One idea for lowering the pair death lines of MSPs is to invoke non-dipolar fields near the surface \citep{HM11a,HM11b,Takata12} that can produce off-set dipoles or polar caps, and larger parallel electric fields and radii of curvature.  Additional pair-production mechanisms than were traditionally assumed have also been explored, such as magnetic pair production at low altitudes in OG models \citep{Takata10} and photon-photon pair production near the surface in polar cap models \citep{ZQ98,HMZ02}.

\subsection{EFFICIENCIES}\label{eff}
\begin{figure}[h]
\begin{center}
\includegraphics[height=0.5\textheight]{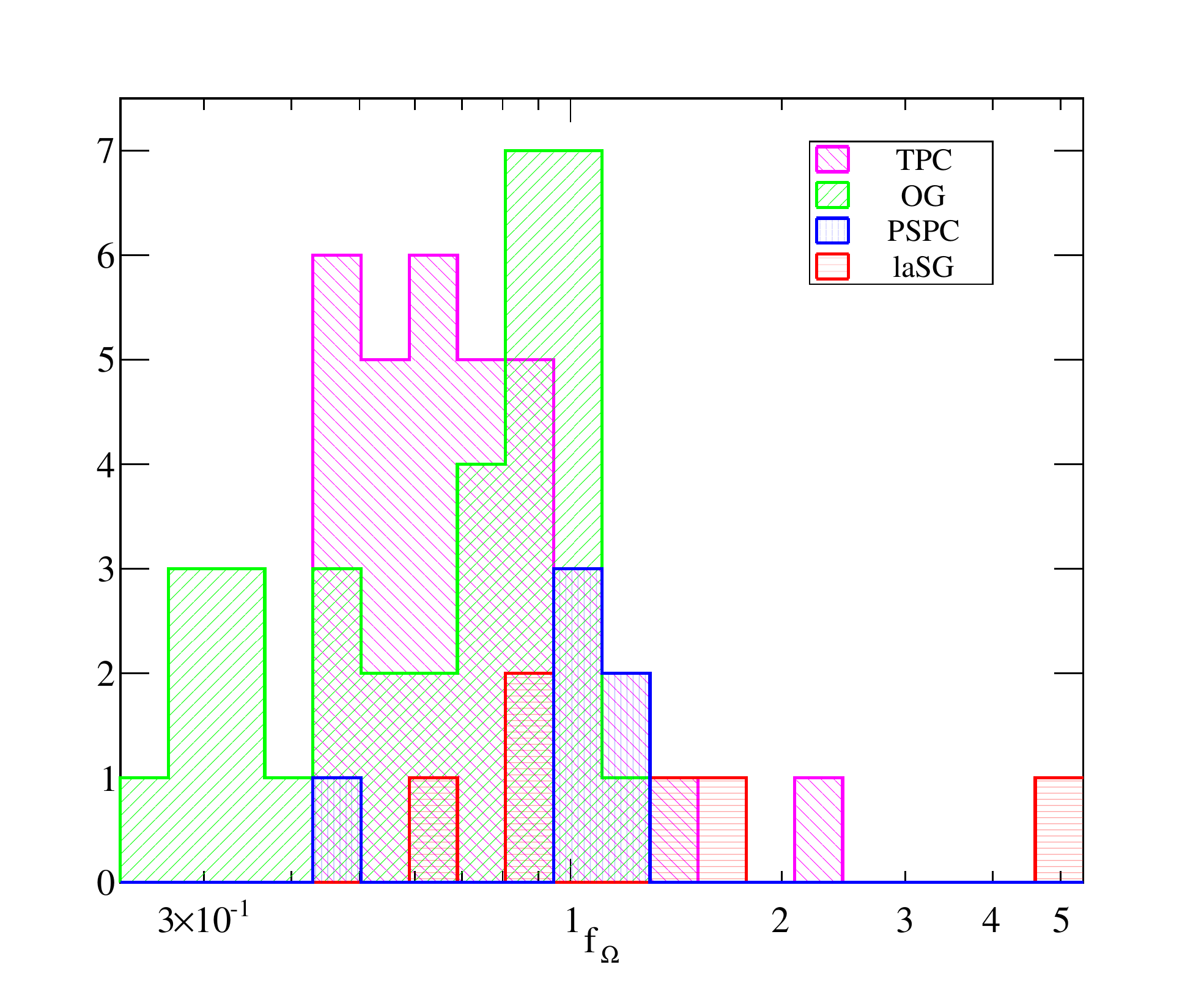}
\end{center}
\caption{Histogram of \fO\ values, regardless of which fit is preferred by the likelihood, for the TPC and alTPC models (light gray filled histogram, pink with downward diagonal lines from left to right in the online version), OG and alOG models (dark gray histogram, green in the online version, with upward diagonal lines from left to right), PSPC model (histogram filled with vertical lines, blue in the online version), and laSG model (histogram filled with horizontal lines, red in the online version).\label{fig-bestfOs}}
\end{figure}

In Tables \ref{tbl-tpc} through \ref{tbl-pspc}, for each MSP and model, are estimates of the beaming correction factor \fO\ \citep[Equation 4 of][reproduced in Equation \ref{eqn-fo}]{Watters09}, which accounts for the fact that MSP emission is not isotropic.  The \fO\ values for each fit are shown in Figure \ref{fig-bestfOs}, here we do not show only the best fits in order to better demonstrate the range of \fO\ values corresponding to realistic light curve shapes for each model.  We estimate \fO\ by summing the simulated emission ($F_{\rm sim}$) for all viewing angles at the best-fit \al\ and dividing by twice the sum of the simulated emission for the best-fit \al\ and \ze.

\begin{equation}\label{eqn-fo}
f_{\Omega}(\alpha,\zeta)\ =\ \frac{\iint F_{\rm sim}(\alpha,\zeta^{\prime},\phi)\sin(\zeta^{\prime})d\zeta^{\prime} d\phi}{2\int F_{\rm sim}(\alpha,\zeta,\phi)d\phi}
\end{equation}

Uncertainties for \fO\ are estimated from the range of values found when calculating \fO\ at each (\al,\ze) pair within the 95\% confidence-level contours.  The \fO\ values for gamma-ray pulsars are typically close to and/or consistent with 1 \citep[e.g.,][]{Watters09,Venter09,WR11} motivating the use of this value in 2PC when reporting $L_{\gamma}$.  There are cases where this assumption leads to efficiencies greater than 100\% (see column 10 of Table \ref{tbl-vitals}).  While these anomalous efficiencies could be due to overestimated distances or underestimated moments of inertia, viewing geometries resulting in \fO$<1$ could also alleviate the problem to some extent.

We find that the fits give $f_\Omega < 1$ in most cases, indicating that we are sampling emission that is above the average over $4\pi$.  For MSPs best fit with PSPC models, $f_\Omega >1$ in most cases since we are missing the brightest part of the emission concentrated at low altitudes near the polar caps.

\begin{figure}[h]
\begin{center}
\includegraphics[height=0.5\textheight]{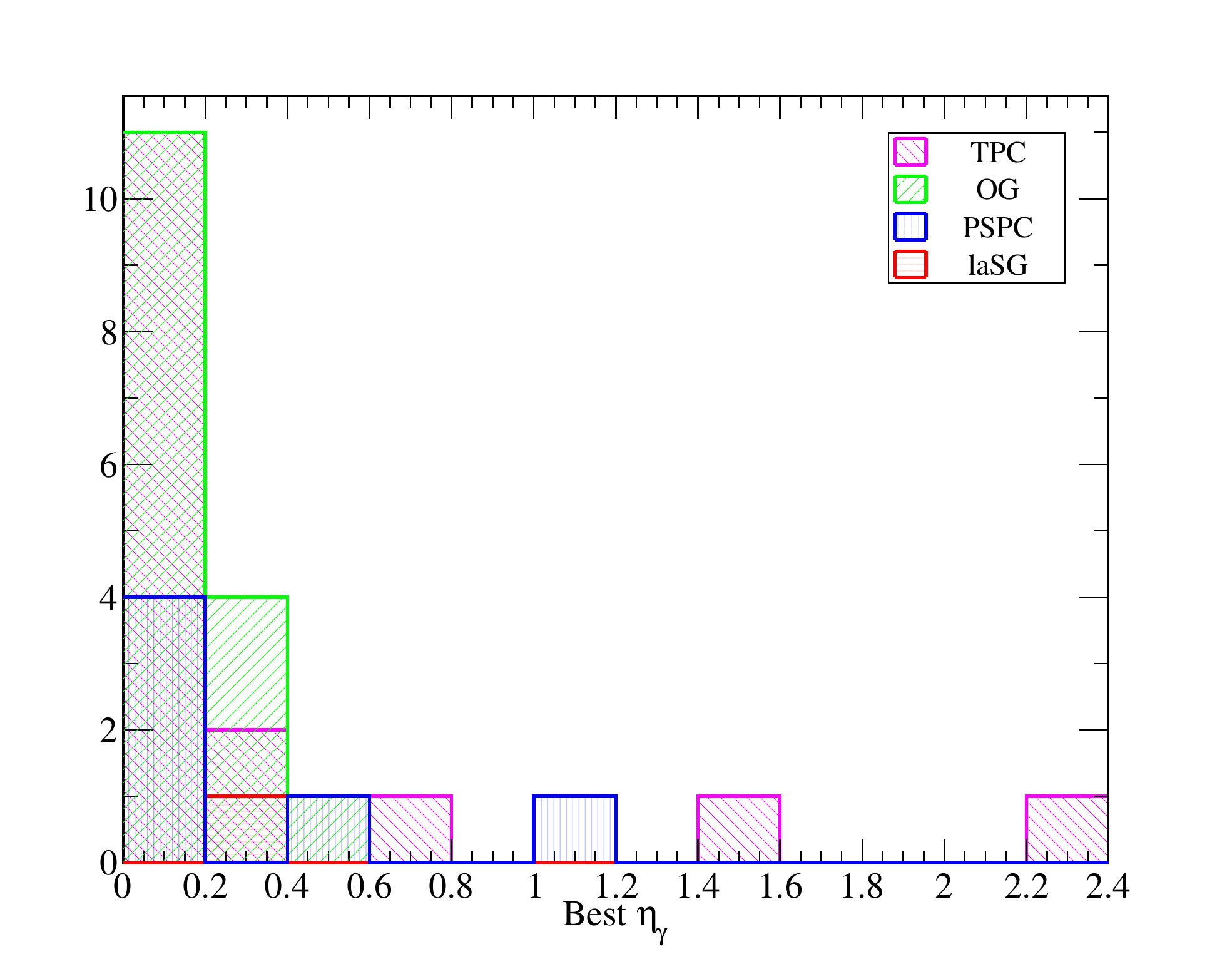}
\end{center}
\caption{Efficiencies corrected for best-fit \fO\ values, excluding the high efficiency outlier PSR J0610$-$2100 for visualization purposes.  Only the value for the fit with the smallest \lnl\ is shown for each MSP.  TPC and alTPC fit values are shown in the light gray filled histogram (pink in the online version) filled with downward diagonal lines from left to right.  OG and alOG fit values are shown in the dark gray (green in the online version) histogram filled with upward diagonal lines from left to right.  Values from laSG fits are shown in the black (red in the online version) histogram filled with horizontal lines from from the PSPC fits in the black (blue in the online version) histogram filled with vertical lines.\label{fig-bestEffs}}
\end{figure}

Figure \ref{fig-bestEffs} presents the distribution of $\eta_{\gamma}$ for each MSP, excluding PSR J0610$-$2100 with an suspiciously high $\eta_{\gamma}\approx13$, corrected with the \fO\ values for the fit with the smallest \lnl.  These efficiencies cover a large range from $\sim1$\% to $\sim1200$\% with the majority clustering near 10\%.

There are three MSPs in 2PC that have greater than 100\% efficiency when assuming \fO=1, PSRs J0610$-$2100, J0614$-$3329, and J2302+4442.  Assuming that the distances from 2PC are not severely overestimated, for these MSPs we would need \fO$<$ 0.08, 0.47, and 0.62, respectively, to move the efficiency less than 100\%.  All of the fits for these MSPs have \fO$\approx$1 with estimated uncertainties that do not include the necessary \fO\ values.  Thus, we must conclude that, if our models are correct, the high efficiencies of these MSPs are not strongly skewed by beaming effects.

A number of MSPs now have measured masses\footnote{A compilation of neutron star mass measurements is maintained by J.~M.~Lattimer and A.~W.~Steiner at\\ \url{http://www.stellarcollapse.org/nsmasses} \citep{Lattimer12}} significantly greater than the nominal $1.4 M_\odot$ \citep[e.g.,][]{Demorest10,Masses,Romani12,Antoniadis13}.
These larger masses could lead to larger moments of inertia (depending on the equation of state), increased \Ed\ values, and lower efficiencies.

\begin{figure}[h]
\begin{center}
\includegraphics[height=0.5\textheight]{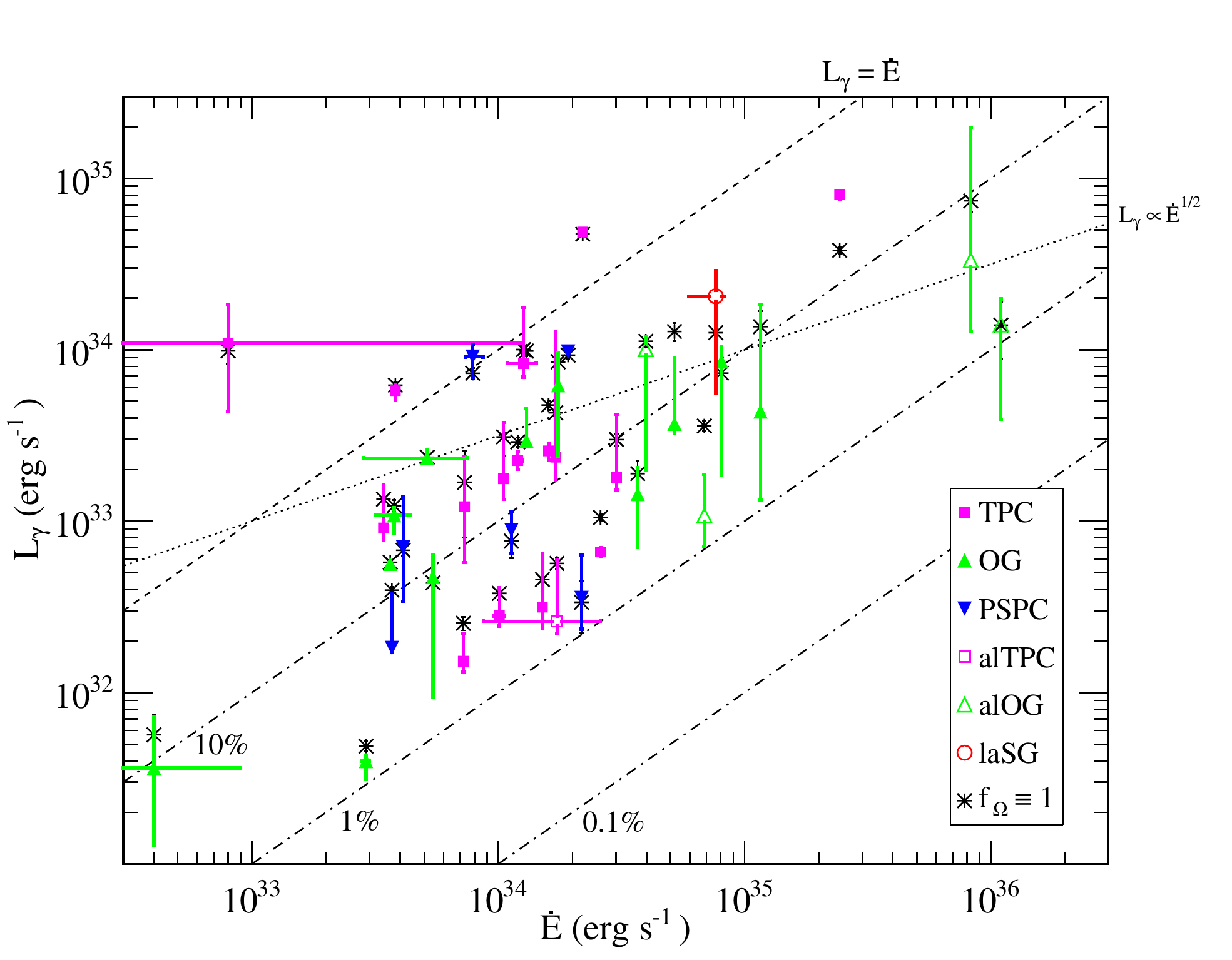}
\end{center}
\caption{$L_{\gamma}$ versus \Ed\ for the 2PC MSPs with \fO\ estimated from the light curve fit with the smallest \lnl\ value.  The 2PC values assuming $f_{\Omega}\equiv1$ are shown as asterisks.  The \Ed\ values have been derived assuming Shklovskii and Galactic acceleration corrections, as in 2PC, sometimes leading to visible horizontal error bars.  The marker styles are as indicated in Figure \ref{fig-bestGeoms}.  The dashed line shows 100\% efficiency with the dash-dotted lines showing 10\%, 1\%, and 0.1\% efficiency, as labeled.  The dotted line shows $L_{\gamma}\propto\sqrt{\dot{E}}$, with arbitrary normalization.\label{fig-LgammaEdot}}
\end{figure}

Figure \ref{fig-LgammaEdot} plots $L_{\gamma}$ versus \Ed\ as is done in 2PC but using our \fO\ estimates from the light curve fitting rather than assuming \fO=1 as was done in 2PC (asterisk points).  The luminosities follow a roughly linear trend with \Ed. There also seems to be a weak separation of TPC-preferred vs.~OG-preferred light curve fits with respect to \Ed, with TPC-preferred for lower values of \Ed.

\subsection{POST-RECYCLING EVOLUTION}\label{evolve}
\begin{figure}[h]
\begin{center}
\includegraphics[height=0.5\textheight]{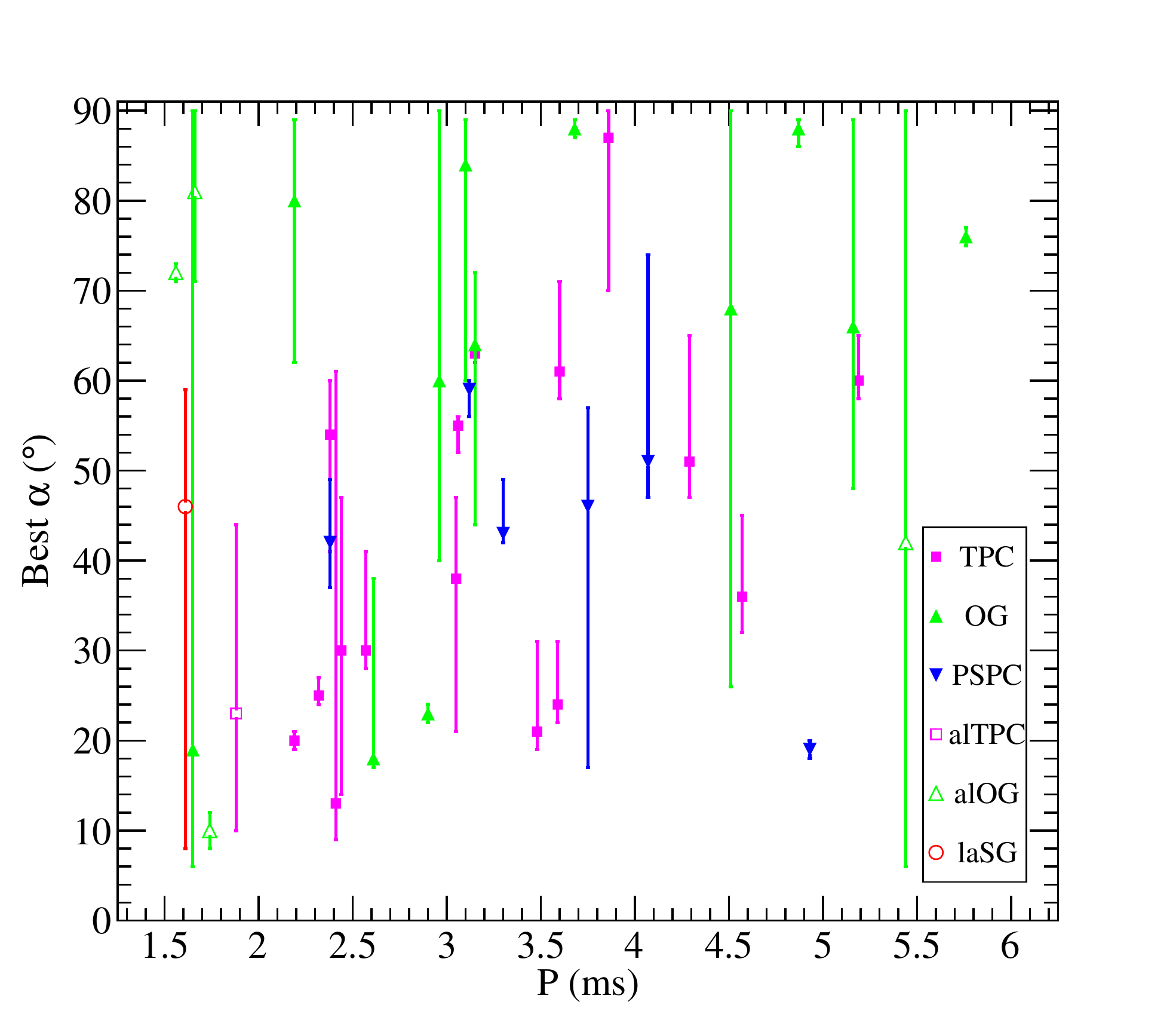}
\end{center}
\caption{Best-fit \al\ versus $P$.  Only the value for the fit with the smallest \lnl\ is shown for each MSP.  The marker styles are as indicated in Figure \ref{fig-bestGeoms}.\label{fig-AlphaPeriod}}
\end{figure}

Figure \ref{fig-AlphaPeriod} plots the best-fit \al\ values against the measured \Per.  The Class I MSPs (filled squares and upward-pointing triangles) display the most significant trend in \al\ vs.~\Per, with a Pearson correlation coefficient of 0.51 and chance probability (two-sided p-value) of 5.6$\times10^{-3}$.  The Pearson correlation coefficient tests the linear relationship between two sets of quantities, a positive (negative) coefficient indicates a positive (negative) slope and absolute values of the coefficient closer to zero indicate weaker correlation.  The chance probability for this trend increases to 1.6$\times10^{-2}$ when including the Class III MSPs and 6.4$\times10^{-2}$ when considering all MSPs.  It should be noted that the Pearson correlation coefficient does not incorporate the uncertainties on the values.  In some cases, the uncertainties in \al\ are large, so these trends are suggestive at best.

A study of isolated, non-recycled pulsars by \citet{Young2010} suggested that as a pulsar spins down the spin and magnetic axes should move towards alignment (i.e., \al\ approaches 0\DG).  This is in contrast to the study of \citet{BN07}, which suggested that the torque exerted by the surface current on the neutron star will tend to move \al\ towards 90\DG\ where the energy loss is minimal.  \citet{Ruderman91} suggested that spinning up a neutron star with a strong magnetic field ($B_{\rm surf}\gtrsim10^{10}$ G) via accretion should also move these two axes toward alignment.  Thus, one might expect a correlation of increasing magnetic alignment with decreasing MSP period, for which we see evidence.  However, the radio light curve and polarization data of PSR J1939+2134, with \Per\ = 1.56 ms, strongly prefer an orthogonal rotator \citep{Guillemot12}.  This may pose a challenge to models that suggest accretion should lead to nearly aligned systems or that this MSP is not a recycled pulsar \citep[as has also been suggested by][]{Ruderman91}.

\citet{CRZ98} noted an abundance of MSPs with both orthogonal and aligned geometries compared to the non-recycled population, based largely on considerations of radio profiles more than on polarimetry or other geometrical constraints, and explained this as a natural consequence of the interactions between the superfluid neutrons and superconducting protons inside a neutron star.  These authors outlined two specific magnetic field geometries that lead to either an aligned or orthogonal rotator, and suggested that pulsars with field configurations between the two cases would be somewhere between aligned and orthogonal.

Our distribution of best-fit \al\ values (Figure \ref{fig-bestGeoms}) would seem to be in line with the predictions of \citet{CRZ98}; however, under their model only the field configuration which led to an orthogonal rotator would be able to produce MSPs with the shortest periods ($\sim1.5$ ms) and we find four MSPs with $P<2$ ms and best-fit $\alpha<50^{\circ}$ (though it is not clear if predicted polarization properties of the laSG model make it viable for PSR J1959+2048 and the error on the value for PSR J1747$-$4036 is large).  Additionally, it is unclear if the sunspot-like geometry they assume is compatible with the VRD used in our simulations, even at a significant fraction of \Rlc.

\citet{Ruderman91} noted that a neutron star with a weak magnetic field spinning down should move towards an orthogonal alignment \citep[in agreement with][though their findings did not depend on the magnetic field strength]{BN07}, which might explain the possible trend of increasing \al\ with increasing \Per\ hinted at by the Pearson coefficient.

\begin{figure}[h]
\begin{center}
\includegraphics[height=0.5\textheight]{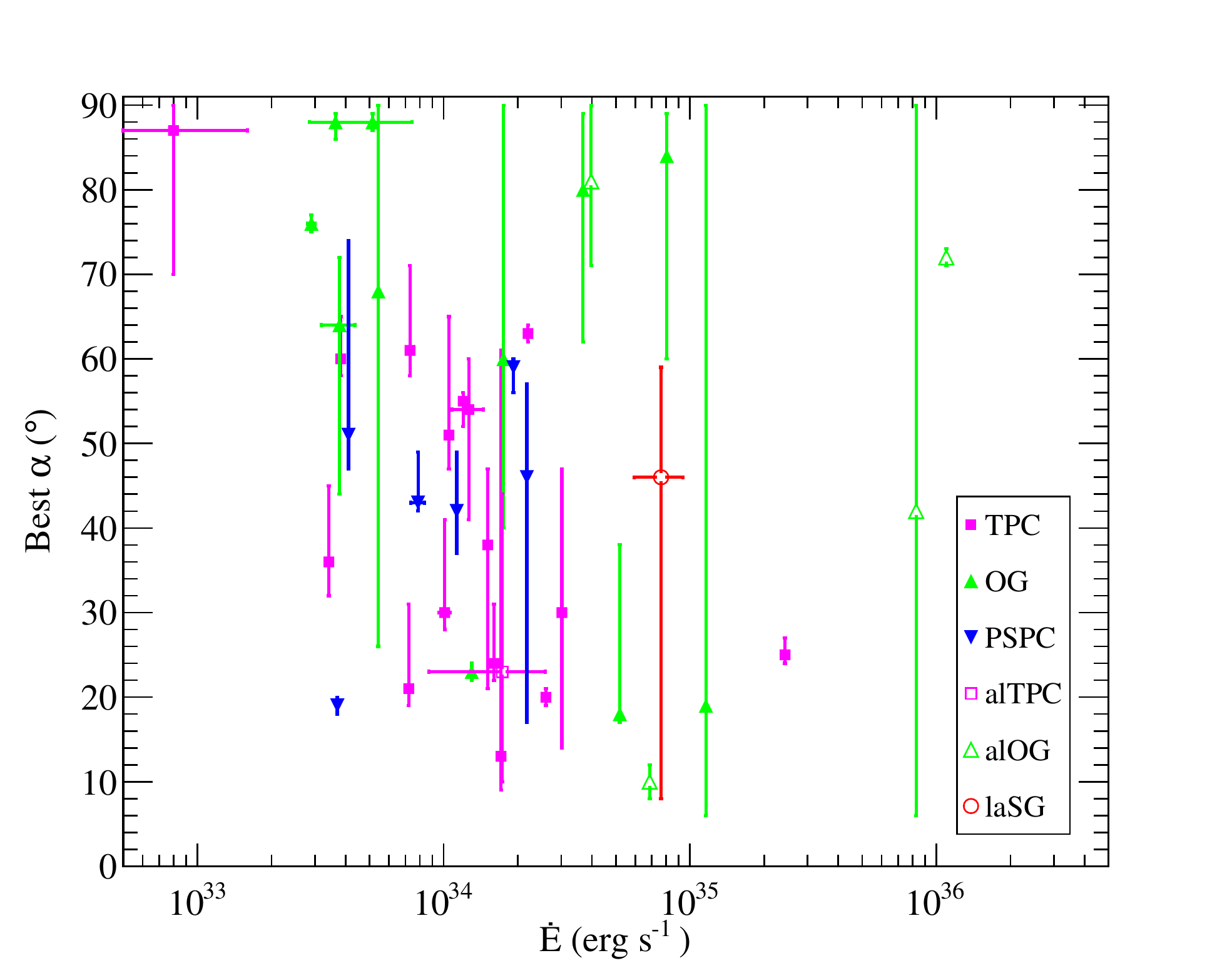}
\end{center}
\caption{Best-fit \al\ versus \Ed.  Only the value for the fit with the smallest \lnl\ is shown for each MSP.  The marker styles are as indicated in Figure \ref{fig-bestGeoms}.\label{fig-AlphaEdot}}
\end{figure}

Figure \ref{fig-AlphaEdot} plots the best-fit \al\ values against the derived \Ed.  The Class I MSPs display the most significant trend in \al\ vs.~$\log_{10}(\dot{E})$, with a Pearson correlation coefficient of -0.47 (indicating anti-correlation) and chance probability of 1.3$\times10^{-2}$.  No correlation is found when considering MSPs best fit by the OG and alOG models alone, but when considering only MSPs best fit by the TPC and alTPC models the chance probability only increases to 1.5$\times10^{-2}$.  When considering Class I and III MSPs the chance probability increases to 1.9$\times10^{-2}$ and when considering all MSPs the chance probability is 1.3$\times10^{-1}$.  Similar to the trends in \al\ vs.~\Per, the uncertainties in \al\ are large in some cases and these trends are suggestive at best.

From the Pearson coefficients, there is no clear trend in \al\ as a function of \Ed\ for the OG or PSPC fits in Figure \ref{fig-AlphaEdot}. The OG model seems to prefer larger values for \al, and this is to be expected, since there is no visible emission from this model when \al\ is too low (reflecting the fact that there is no emission generated below the NCS). For the TPC model, however, the Pearson coefficient hints at a linear trend with negative slope between \al\ and \Ed.  Increasing \Ed\ generally points to younger age, and if the trend is confirmed, to a smaller \al\ in this case. Given that values of \ze\ near 90\DG\ are usually preferred, this points to larger impact angles for the younger, faster-spinning MSPs in our sample.  This corresponds to the fact that the TPC emission is indeed visible for larger values of $\beta$ compared to the OG model, since it includes low-altitude emission not present in the OG geometry. Furthermore, since all MSPs in this plot are radio-loud, we expect that the radio beams of MSPs with high \Ed\ will generally be at higher altitudes, and therefore wider, so that one may probe smaller \al.

On the other hand, there may be hints of an underlying evolution of \al\ with \Ed, e.g., movement toward the equator with age.  
Is the implied, inverse trend due to unfavorable geometry, selection effects, or a systematic evolution of \al\ toward larger values?  
A population synthesis approach would be necessary to disentangle the effects of visibility and obliquity evolution. We can mention in this context, though, that the relatively large radio beams of MSPs compared to those of the younger pulsars give us the opportunity to probe any supposed evolution of \al\ better, though in a different magnetic field strength regime, given the larger parameter space over which we see both radio and gamma-ray emission. This is because the values of \al\ and \ze\ are much less constrained when only fitting the gamma-ray profiles of radio-quiet pulsars.

\subsection{MSPS WITH SIGNIFICANT OFF-PEAK EMISSION}\label{offpeak}
An important discriminator between different emission models is the level of emission predicted outside of the main gamma-ray peaks.  In particular, models that restrict emission to be above the NCS need large $|\beta|$ to produce significant off-peak emission.  One particular example is PSR J0613$-$0200 for which \citet{thesis} found an OG model fit with $|\beta|\sim30^{\circ}$.  However, that analysis only used a hollow-cone radio model whereas we have included a core component for this MSP, supported by polarization measurements, and requiring the core to be seen excludes this geometry.  As a consequence, our best-fit OG geometry does not predict any off-peak emission for PSR J0613$-$0200 and misses the possible low-level peak immediately preceding the main gamma-ray peak.

In 2PC the off-peak interval was defined and analyzed for each pulsar with the goal of assessing the probable nature of any significant emission (e.g., from a pulsar wind or from the magnetosphere).  No 2PC MSPs were found to have off-peak emission consistent with an origin in the pulsar wind but three were found to have off-peak emission that was consistent with a magnetospheric origin (i.e., point-like with a significantly cutoff spectrum).  Of these three MSPs, two are Class III (PSRs J0340+4130 and J2124$-$3358) and one is Class I (PSR J2302+4442).  The TPC model is preferred for PSR J2302+442, though not significantly.  The PSPC fits for PSRs J0340+4130 and J2124$-$3358 do predict significant off-peak emission, consistent with the 2PC analysis.

There are 11 Class I and 4 Class II MSPs for which the best-fit geometries are the same to within a few degrees for the standard and altitude-limited TPC and OG models, neglecting laSG fits.  Of these Class I MSPs the likelihood prefers the TPC model for eight and the OG for three, but the preference is only significant ($\geq 15$) for three MSPs (all preferring the TPC model).  Of these Class II MSPs the likelihood prefers the alTPC model for two and the alOG model for two, but never significantly.  While some of these MSPs are faint, resulting in small log-likelihood difference and no clear reason for the preference, in many cases the preferred model is obvious from the suggestion, or not, of emission above the background level in the off-peak region.  Only two of these MSPs, PSRs J0034$-$0534 and J2043+1711, are found to have significant off-peak emission in 2PC. For both MSPs the origin of the emission is unclear, while neither show any evidence for extension the significance of a cutoff in the off-peak spectrum is just below the threshold for classification as possibly magnetospheric.

Although we have fit Class I and II MSPs with narrow gap models that are geometric versions of SG (TPC) and OG acceleration and radiation models, we find that neither provides the best-fit solution for all MSPs.  In fact, the best-fit solutions divide roughly equally between the two models.  The OG models best fit the MSP light curves that have no off-peak and high bridge emission, such as PSR J0030+0451.  The TPC models best fit the light curves having some suggestion of off-peak emission, such as PSR J0034$-$0534.  Higher off-peak emission may be correlated with larger $\beta$ values that are allowed for both young radio-quiet pulsars and MSPs.  In SG models, off-peak emission comes from regions below the NCS on leading-edge field lines.  
We can speculate that an emission geometry that would fit all MSP light curves might be an SG model in which the emission below the NCS either decreases toward the stellar surface and/or is dependent on magnetic azimuth.  Such a hybrid model might also explain the Class I and II MSPs where the standard and altitude-limited TPC and OG models find similar geometries and the preferred fit is chosen largely based on the apparent presence, or not, of off-peak emission.

\subsection{NATURE OF THE CLASS II MSPS}\label{class2}
It is still unclear what parameters give rise to the Class II MSPs.  In particular, while the gamma-ray light curves look qualitatively similar to those of Class I MSPs why do the gamma-ray and radio components appear at nearly the same phase?  This question has been explored by several authors \citep{thesis,Venter12,Espinoza13} but no definitive conclusions have been reached.

\citet{AKH05} predicted that aligned radio and HE profiles could occur for pulsars with $P < 50$ ms.  With the exception of PSR J1823$-$3021A, all of the Class II MSPs have $P<2$ ms; however, PSR J1747$-$4036 has $P = 1.65$ ms but is Class I.  Aside from PSRs J1939+2134 and J1810+1744, the radio profiles of Class II MSPs show essentially no linear polarization, in line with the prediction that caustics have low linear polarization \citep{Dyks04a}.  \citet{Kramer99} analyzed the degree of linear polarization of PSR J1939+2134 as a function of frequency and found evidence for depolarization, best fit by a power law with index $-0.76\pm0.07$, steeper than any other MSP in their study.  For PSR J1810+1744 the situation is unclear as no polarimetric observations have yet been reported.

\begin{figure}[h]
\begin{center}
\includegraphics[height=0.48\textheight]{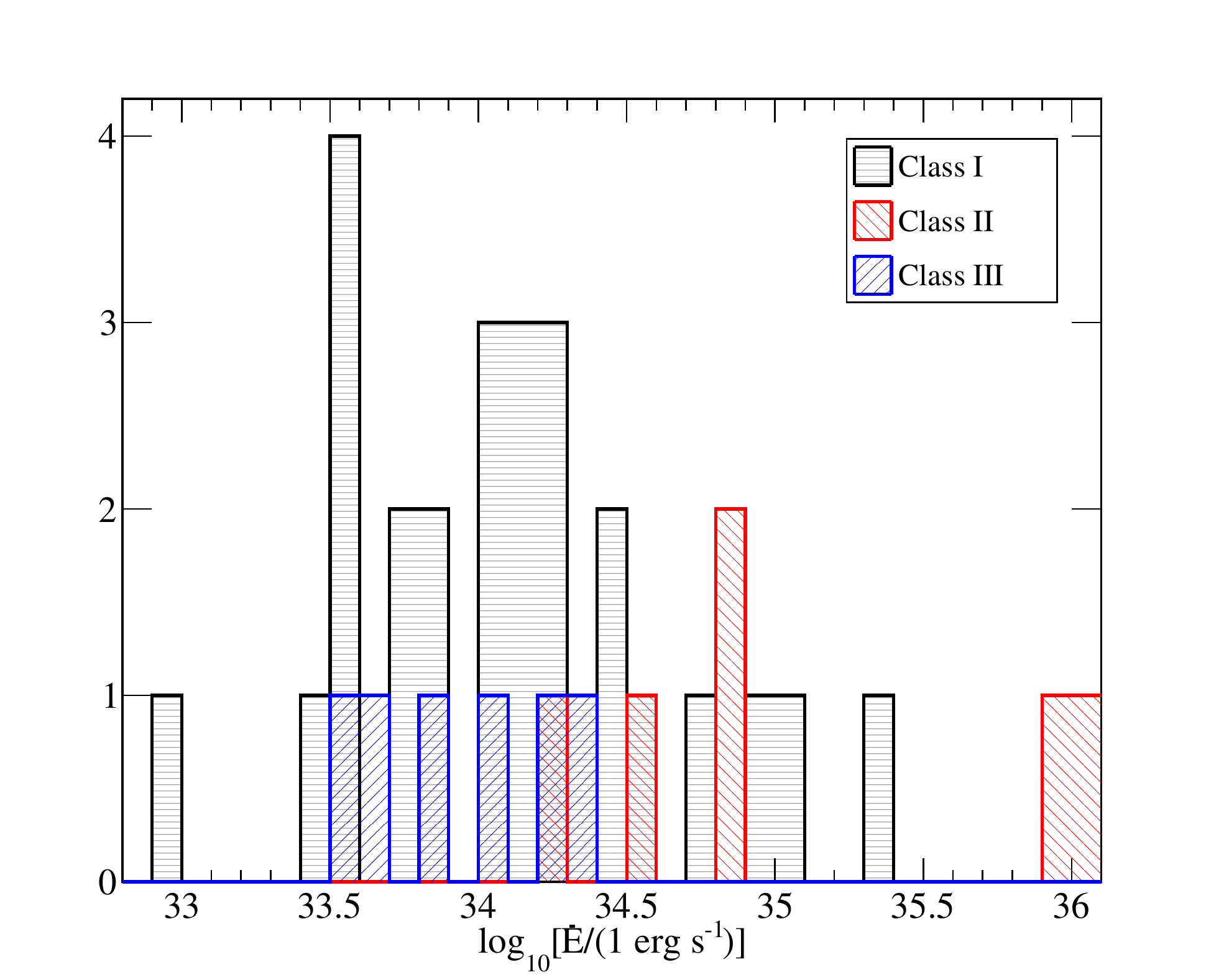}
\includegraphics[height=0.48\textheight]{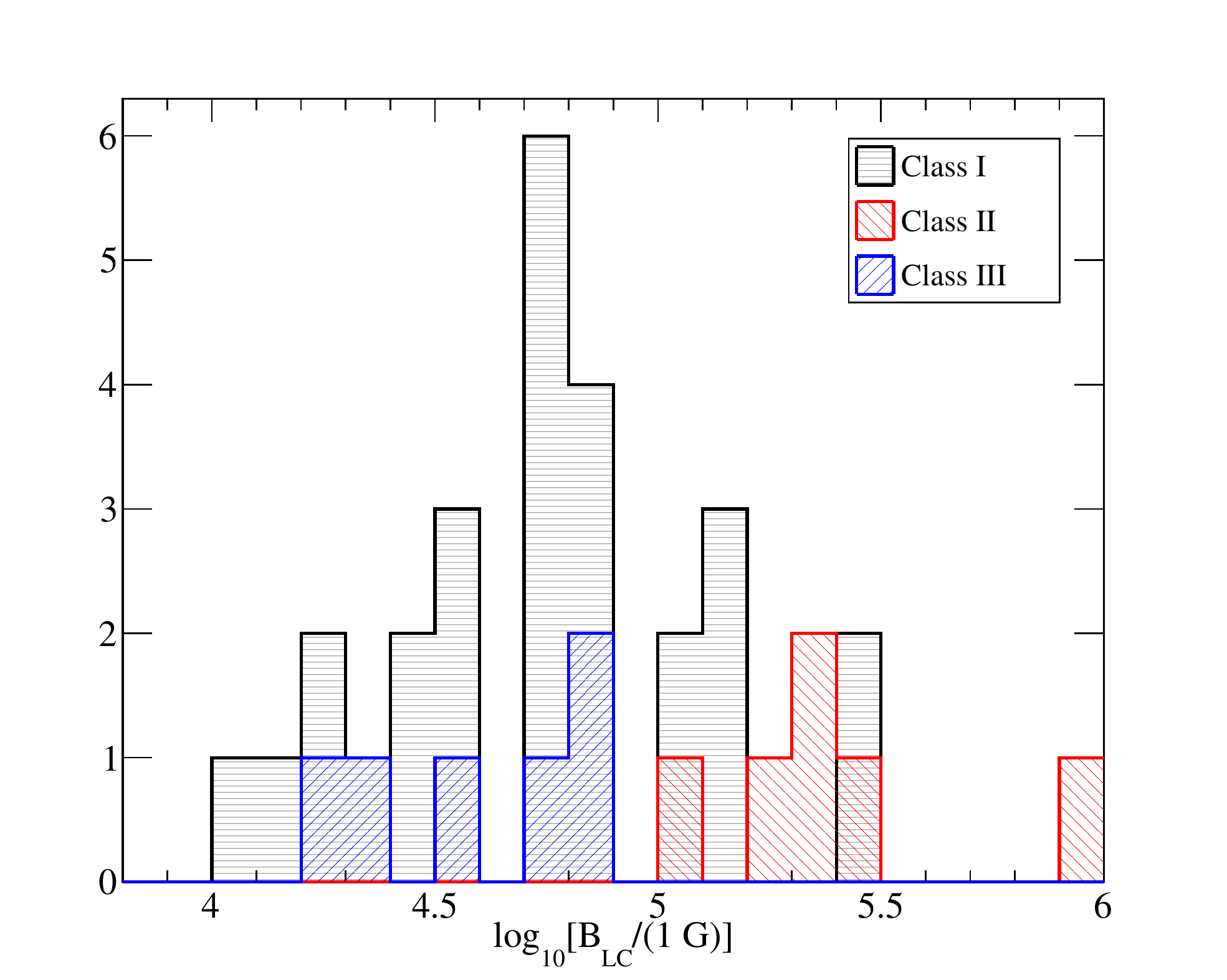}
\end{center}
\caption{Distributions of \Ed\ (\emph{top}) and \Blc\ (\emph{bottom}) by model Class.  The black histograms filled with horizontal lines are for the Class I MSPs.  The light gray (red in the online version) histograms filled with downward diagonal lines from left to right are for the Class II MSPs.  The dark gray filled histograms (blue in the online version, filled with upward diagonal lines from left to right) are for the Class III MSPs.\label{fig-classEdBlc}}
\end{figure}

Figure \ref{fig-classEdBlc} presents the \Ed\ (top) and \Blc\ (bottom) values for the 2PC MSPs separated by model class.  Similar to the findings of \citet{Espinoza13} but with more MSPs, a Kolmogorov-Smirnov test comparing the Class II MSPs (their A-type) to the rest of the LAT-detected MSPs in our sample returns a probability of 0.004 that the \Blc\ values are drawn from the same distribution and a probability of 0.02 for the \Ed\ values.  These probabilities are slightly larger than those reported by \citet{Espinoza13} but still suggest that the Class II MSPs stand out most strongly when comparing \Blc.

\citet{Espinoza13} compared the radio spectral indices of their A-type MSPs to those of the N-type and W-type MSPs in their sample.  A Kolmogorov-Smirnov test indicated a 0.6\% probability that the two sets belonged to the same distribution.  They also compared the \Blc\ values of their A-type MSPs with the rest of the gamma-ray MSP population using a Kolmogorov-Smirnov test and found a 0.02\% probability that they belong to the same distribution.  \citet{Espinoza13} also pointed out that three of the A-type MSPs are known to emit giant radio pulses, as is PSR J0218+4232, a W-type MSP.  Significant gamma-ray pulsations have recently been detected from PSR B1821$-$24 in the globular cluster M28 \citep{wuM28A,JohnsonB1821}, which is also known to emit giant pulses.  The first gamma-ray, and X-ray, light curve peaks occur at the same phase from which the giant pulses have been detected but the identification of this MSP as A-type is unclear.  Giant pulses have also been detected from the Crab pulsar \citep{CrabGPs}, the only non-recycled pulsar with the main radio and gamma-ray peaks occurring at the same phase.

It is not clear what physical characteristics divide MSPs into the three model classes we have identified.  Class I and II do not show a preference for any particular viewing geometry, although Class III MSPs all have $\zeta > \alpha$.  Since very few Class I and II MSPs were expected originally, perhaps class membership is governed by other characteristics that determine whether narrow gaps can form in these MSPs, such as surface magnetic field structure.

The laSG model is usually not preferred.  It requires small gap widths, with the gamma-ray widths generally smaller than the radio widths.  The preferred positions of the peak emission seem to be about 2 \Rns\ with the best-fit \al\ and \ze\ generally exceeding 40\DG\ and 70\DG, respectively.

The large number of gamma-ray MSPs discovered by the \fermi\ LAT has raised some intriguing questions not only about MSP emission and evolution, but also about pulsar emission physics in general.  Continued study of these sources is sure to yield more surprises and lead to more answers.

\clearpage

\acknowledgments

\begin{center}
\emph{ACKNOWLEDGMENTS}
\end{center}

The \fermi\ LAT Collaboration acknowledges generous ongoing support from a number of agencies and institutes that have supported both the development and the operation of the LAT as well as scientific data analysis.  These include the National Aeronautics and Space Administration and the Department of Energy in the United States, the Commissariat \`a l'Energie Atomique and the Centre National de la Recherche Scientifique / Institut National de Physique Nucl\'eaire et de Physique des Particules in France, the Agenzia Spaziale Italiana and the Istituto Nazionale di Fisica Nucleare in Italy, the Ministry of Education, Culture, Sports, Science and Technology (MEXT), High Energy Accelerator Research Organization (KEK) and Japan Aerospace Exploration Agency (JAXA) in Japan, and the K.~A.~Wallenberg Foundation, the Swedish Research Council and the Swedish National Space Board in Sweden.

Additional support for science analysis during the operations phase is gratefully acknowledged from the Istituto Nazionale di Astrofisica in Italy and the Centre National d'\'Etudes Spatiales in France.

Our research used the radio profiles published in \citet{2PC} and we gratefully acknowledge the efforts of radio astronomers in support of \fermi\ LAT pulsar science.  The Parkes radio telescope is part of the Australia Telescope which is funded by the Commonwealth Government for operation as a National Facility managed by CSIRO.  The Robert C. Byrd Green Bank Telescope (GBT) is operated by the National Radio Astronomy Observatory, a facility of the National Science Foundation operated under cooperative agreement by Associated Universities, Inc.  The Arecibo Observatory is operated by SRI International under a cooperative agreement with the National Science Foundation (AST-1100968), and in alliance with Ana G. Méndez-Universidad Metropolitana, and the Universities Space Research Association.  The Nan\c{c}ay Radio Observatory is operated by the Paris Observatory, associated with the French Centre National de la Recherche Scientifique (CNRS).  The Westerbork Synthesis Radio Telescope is operated by Netherlands Foundation for Radio Astronomy, ASTRON.

Portions of this research performed at the Naval Research Laboratory are sponsored by NASA DPR S-15633-Y.  C.~V.~is supported by the South African National Research Foundation.  A.~K.~H.~acknowledges support from NASA Astrophysics Theory grants and \fermi\ Guest Investigator grants 10-Fermi10-C4-0054 and 09-FERMI309-0076.

\bibliographystyle{apj}
\bibliography{modelPSRcat}

\appendix
\begin{center}
\textbf{APPENDIX}
\end{center}
\renewcommand{\thefigure}{\Alph{section}.\arabic{figure}}
\makeatletter
\@addtoreset{figure}{section}
\makeatother
\section{ALL LIGHT CURVE FITS}\label{app-fits}
\begin{figure}
\begin{center}
\includegraphics[height=0.6\textheight]{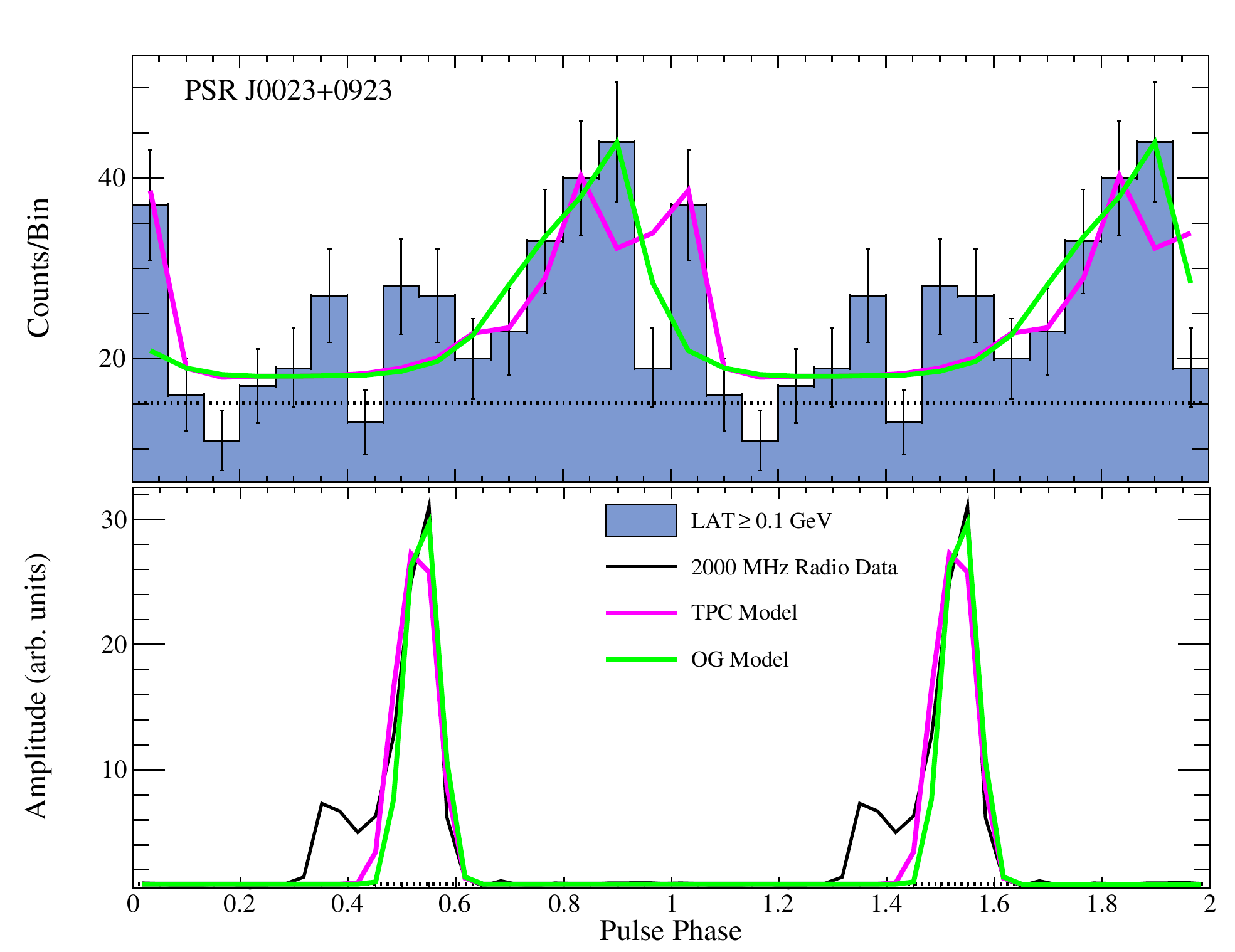}
\end{center}
\caption{Best-fit gamma-ray (\emph{top}) and radio (\emph{bottom}) light curves for PSR J0023+0923.  The dashed line (solid green in the online version) is the OG model, best-fit parameters in Table \ref{tbl-og}.  The solid gray line (pink in the online version) is the TPC model, best-fit parameters are given in Table \ref{tbl-tpc}.  In both panels, the dotted horizontal lines are the estimated background levels given in Table \ref{tbl-simpars}.\label{app-J0023LCs}}
\end{figure}

Figure \ref{app-J0023LCs} presents the observed and best-fit light curves for PSR J0023+0923.  This is a 3.05 ms pulsar in a binary system with a 0.14 day orbital period discovered in radio observations of unassociated LAT sources with pulsar-like characteristics (Hessels et al.~in preparation).  This MSP was first announced as a gamma-ray pulsar in 2PC.  Both the OG and TPC model fit the gamma-ray light curve well, though neither best-fit geometry matches the smaller radio peak near phase 0.4.  \citet{Breton13} modeled the optical light curve of this MSP and found $i\ =\ 58^{\circ}\pm14^{\circ}$ and were able to exclude $i\lesssim40^{\circ}$.  This does agree with the value of \ze\ for our TPC fit, within uncertainties, and it better matches the lack of observed radio eclipses.  Only the TPC confidence contours can accomodate $\zeta\ =\ 58^{\circ}$, with a slightly lower value of \al\ that that in Table \ref{tbl-tpc}.

\begin{figure}
\begin{center}
\includegraphics[height=0.6\textheight]{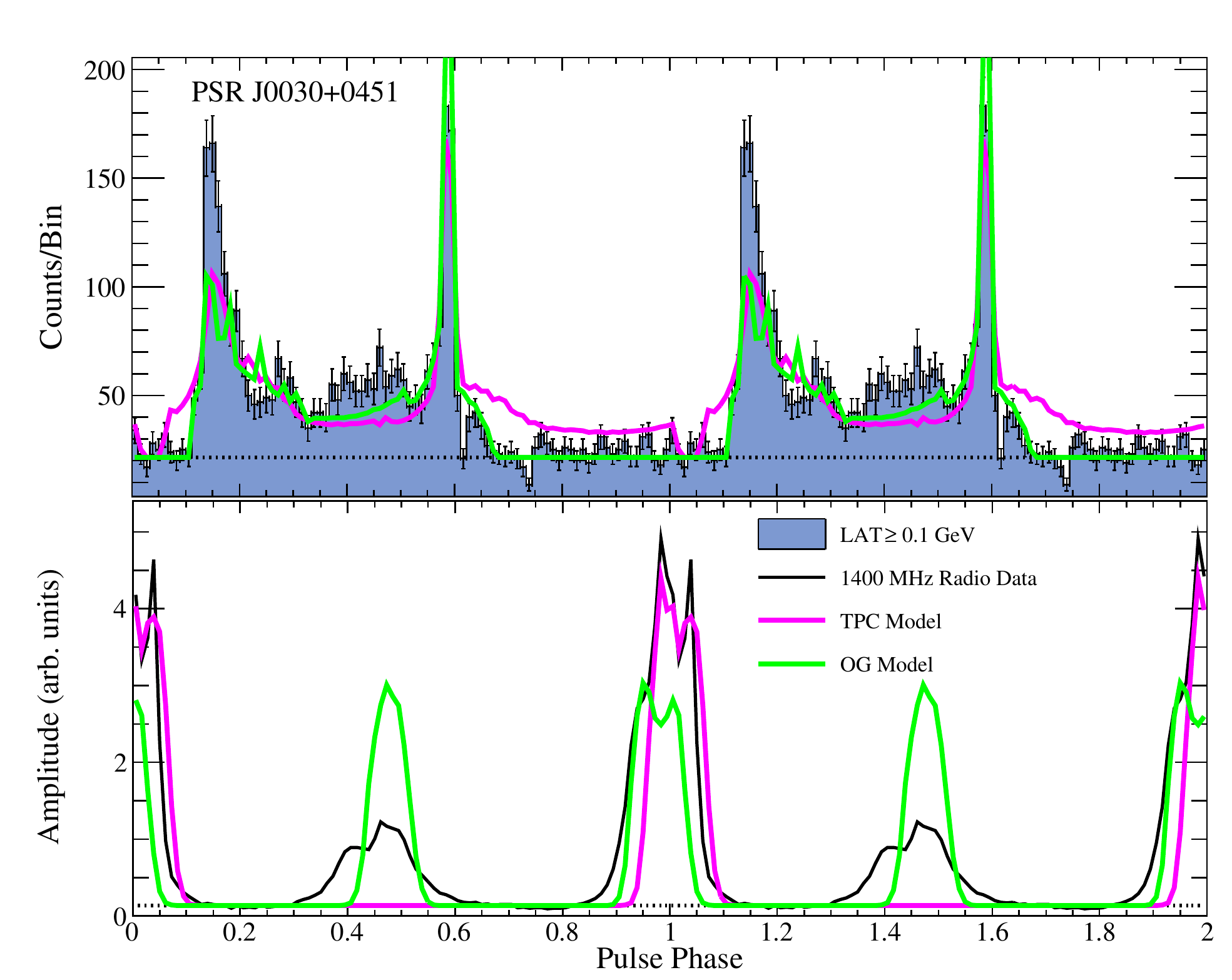}
\end{center}
\caption{Best-fit gamma-ray (\emph{top}) and radio (\emph{bottom}) light curves for PSR J0030+0451.  The dashed line (solid green in the online version) is the OG model, best-fit parameters in Table \ref{tbl-og}.  The solid gray line (pink in the online version) is the TPC model, best-fit parameters are given in Table \ref{tbl-tpc}.  In both panels, the dotted horizontal lines are the estimated background levels given in Table \ref{tbl-simpars}.\label{app-J0030LCs}}
\end{figure}

Figure \ref{app-J0030LCs} presents the observed and best-fit light curves for PSR J0030+0451.  This is an isolated pulsar with a 4.87 ms spin period first discovered by \citet{Lommen00}.  Gamma-ray pulsations from this MSP were first reported by \citet{AbdoJ0030}.  \citet{Lommen00} also attempted to fit the polarimetric data for this MSP but could not uniquely constrain the viewing geometry.  They did suggest that the pulsar was either nearly aligned or nearly orthogonal, and \citet{Bogdanov08} ruled out the nearly aligned geometry by modeling the thermal X-ray pulsations.  Fits with the TPC and OG model both find values of \al\ near 90\DG, but the OG fit is significantly preferred by the likelihood as the TPC model over predicts the level of off-peak emission and only the OG model produces a radio interpulse.

\begin{figure}
\begin{center}
\includegraphics[height=0.6\textheight]{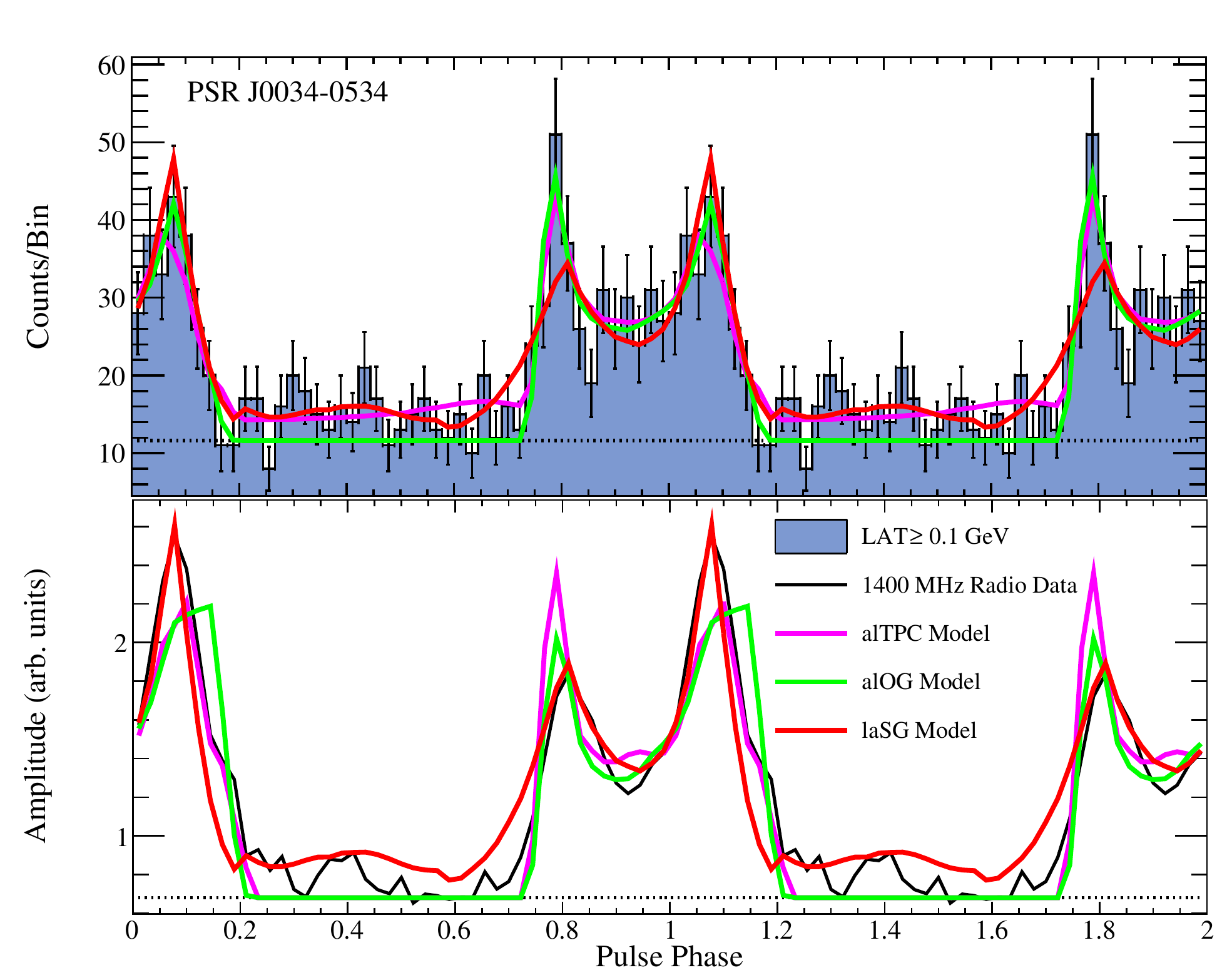}
\end{center}
\caption{Best-fit gamma-ray (\emph{top}) and radio (\emph{bottom}) light curves for PSR J0034$-$0534.  The dashed line (solid green in the online version) is the alOG model, best-fit parameters in Table \ref{tbl-alog}.  The solid gray line (pink in the online version) is the alTPC model, best-fit parameters are given in Table \ref{tbl-altpc}.  The dash-dot, light gray line (solid red in the online version) is the laSG model, best-fit parameters are given in Table \ref{tbl-lasg}.  In both panels, the dotted horizontal lines are the estimated background levels given in Table \ref{tbl-simpars}.\label{app-J0034LCs}}
\end{figure}

Figure \ref{app-J0034LCs} presents the observed and best-fit light curves of PSR J0034$-$0534.  This is a 1.88 ms pulsar in a binary system with a 1.6 day orbital period discovered by \citet{Bailes94}.  Gamma-ray pulsations from this MSP were first reported by \citet{J0034}.  The radio emission from this MSP shows very low levels of linear polarization, consistent with being unpolarized \citep{Stairs99}.  The alTPC, alOG, and laSG models all fit the observed light curves well, but only the laSG and alTPC models predict significant levels of off-peak emission.  The laSG model does the best at matching the relative radio peak heights but underestimates the gamma-ray peak at phase 0.8.

\begin{figure}
\begin{center}
\includegraphics[height=0.6\textheight]{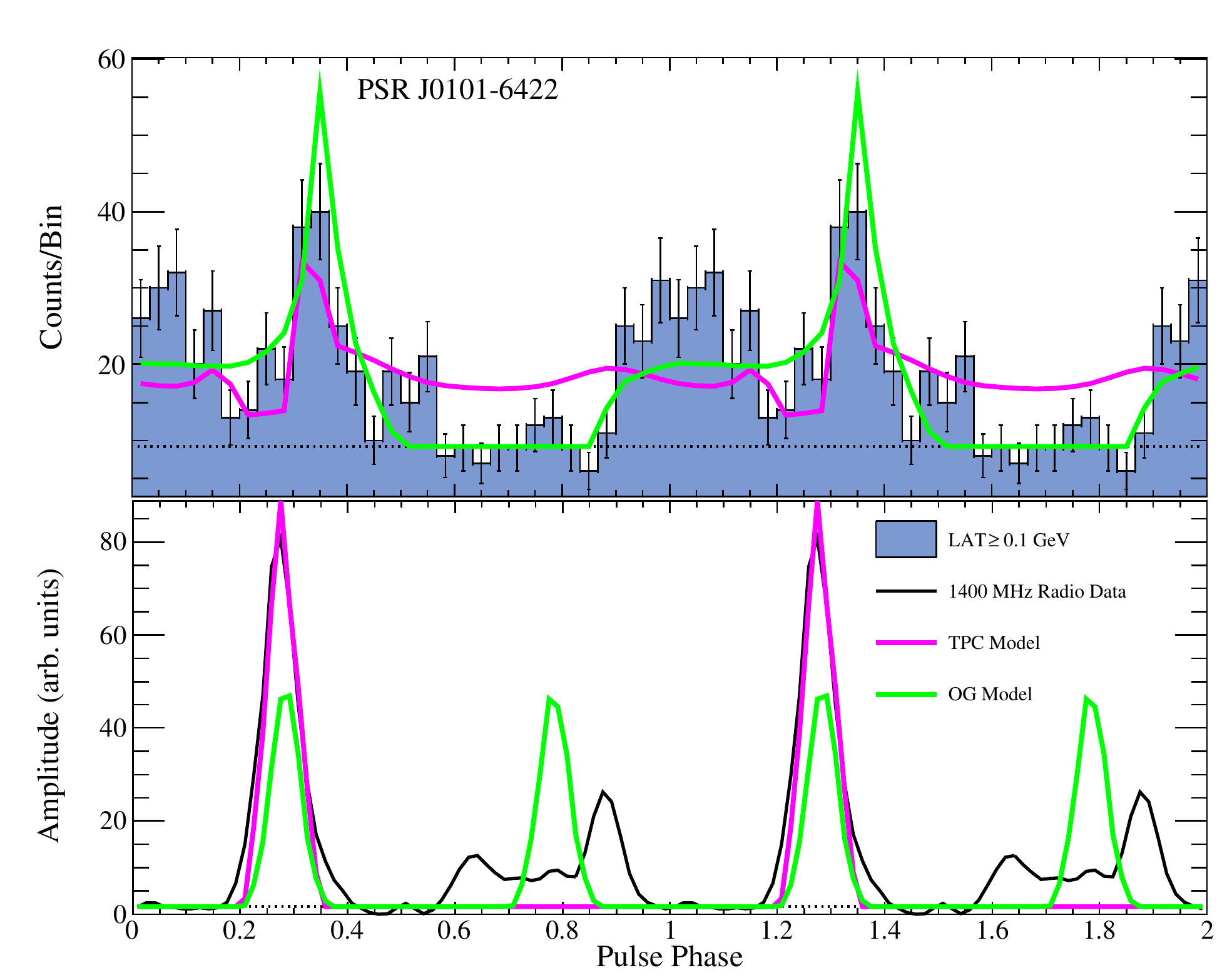}
\end{center}
\caption{Best-fit gamma-ray (\emph{top}) and radio (\emph{bottom}) light curves for PSR J0101$-$6422.  The dashed line (solid green in the online version) is the OG model, best-fit parameters in Table \ref{tbl-og}.  The solid gray line (pink in the online version) is the TPC model, best-fit parameters are given in Table \ref{tbl-tpc}.  In both panels, the dotted horizontal lines are the estimated background levels given in Table \ref{tbl-simpars}.\label{app-J0101LCs}}
\end{figure}

Figure \ref{app-J0101LCs} presents the observed and best-fit light curves of PSR J0101$-$6422.  This is a 2.57 ms pulsar in a binary system with a 1.8 day orbital period discovered by \citet{Kerr2012}, who also announced this as a gamma-ray MSP, in radio observations of unassociated LAT sources with pulsar-like characteristics.  Neither the TPC or OG model is able to simultaneously match the gamma-ray and radio light curves. The OG model fits the gamma-rays reasonably well and is the only model to predict more than one radio peak.

\begin{figure}
\begin{center}
\includegraphics[height=0.6\textheight]{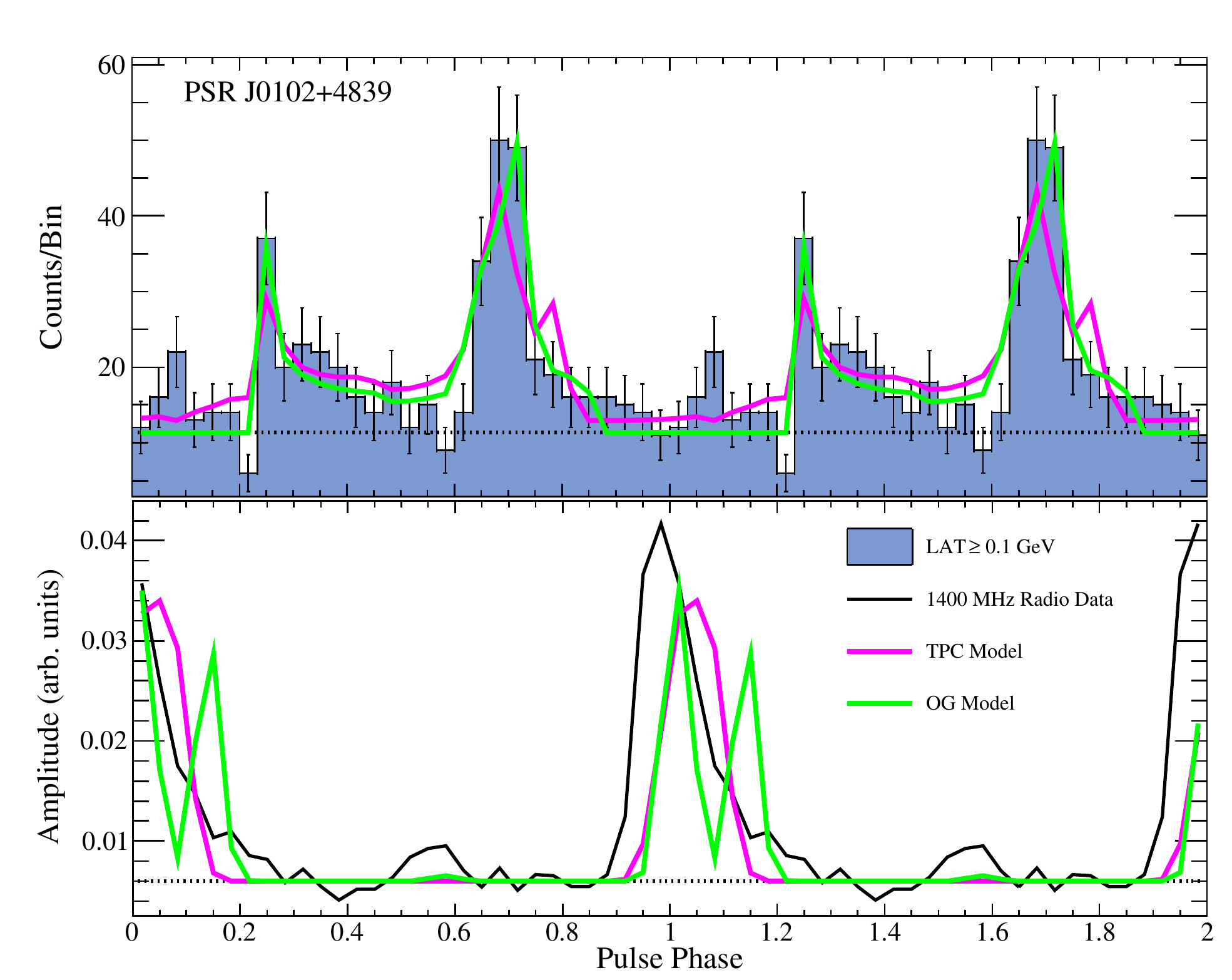}
\end{center}
\caption{Best-fit gamma-ray (\emph{top}) and radio (\emph{bottom}) light curves for PSR J0102+4839.  The dashed line (solid green in the online version) is the OG model, best-fit parameters in Table \ref{tbl-og}.  The solid gray line (pink in the online version) is the TPC model, best-fit parameters are given in Table \ref{tbl-tpc}.  In both panels, the dotted horizontal lines are the estimated background levels given in Table \ref{tbl-simpars}.\label{app-J0102LCs}}
\end{figure}

Figure \ref{app-J0102LCs} presents the observed and best-fit light curves of PSR J0102+4839.  This is a 2.96 ms pulsar in a binary system with a 1.7 day orbital period discovered in radio observations of unassociated LAT sources with pulsar-like characteristics (Bengale et al.~in preparation).  This MSP was first announced as a gamma-ray pulsar in 2PC.  Both the OG and TPC model are able to reproduce the observed gamma-ray light curve well, but both also predict a radio lag that is too small.  The best-fit geometry corresponding to the TPC fit does match the observed radio profile better.

\begin{figure}
\begin{center}
\includegraphics[height=0.6\textheight]{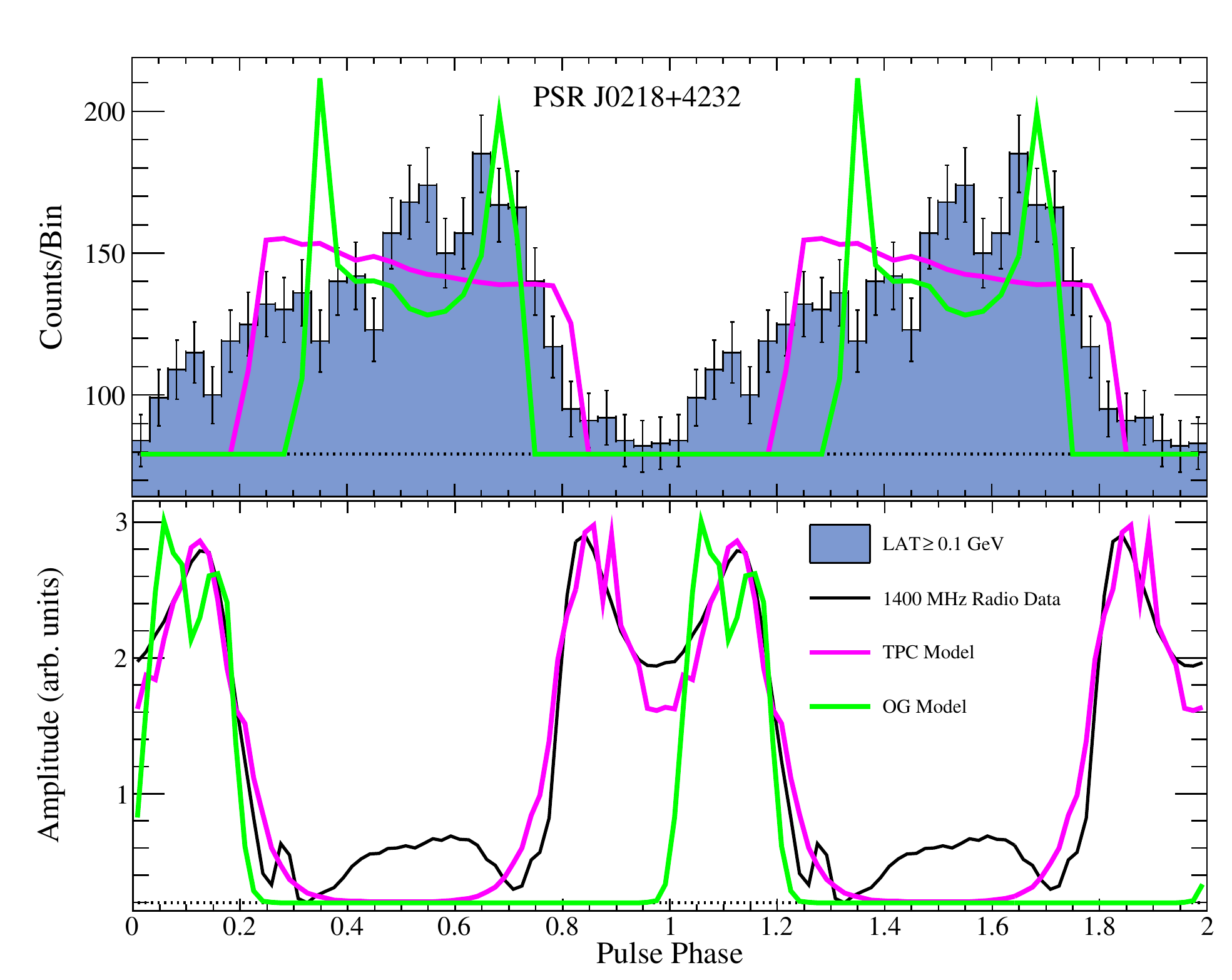}
\end{center}
\caption{Best-fit gamma-ray (\emph{top}) and radio (\emph{bottom}) light curves for PSR J0218+4232.  The dashed line (solid green in the online version) is the OG model, best-fit parameters in Table \ref{tbl-og}.  The solid gray line (pink in the online version) is the TPC model, best-fit parameters are given in Table \ref{tbl-tpc}.  In both panels, the dotted horizontal lines are the estimated background levels given in Table \ref{tbl-simpars}.\label{app-J0218LCs}}
\end{figure}

Figure \ref{app-J0218LCs} presents the observed and best-fit light curves of PSR J0218+4232.  This is a 2.32 ms pulsar in a binary system with a 2.0 day orbital period discovered by \citet{Navarro95}.  Low-significance pulsations ($\sim3\sigma$) using \emph{EGRET} data were reported by \citet{Kuiper00}, with a firm pulsed detection using LAT data first reported by \citet{MSPpop}.  \citet{Stairs99} reported polarization measurements for this MSP at 410 and 610 MHz and found $\alpha = 8^{\circ}\pm11^{\circ}(15^{\circ})$ for the 410 (610) MHz data, with unconstrained $\beta$, from RVM fits.  Neither the TPC or OG model fits exactly match the gamma-ray light curve of this MSP, but only the TPC model fit matches the radio reasonably well.  The best-fit \al\ from the TPC model agrees with the findings of \citet{Stairs99} within statistical and systematic uncertainties, while that from the OG model does not, largely due to the fact that only accepting emission from above the NCS tends to exclude low-\al\ solutions.

\begin{figure}
\begin{center}
\includegraphics[height=0.6\textheight]{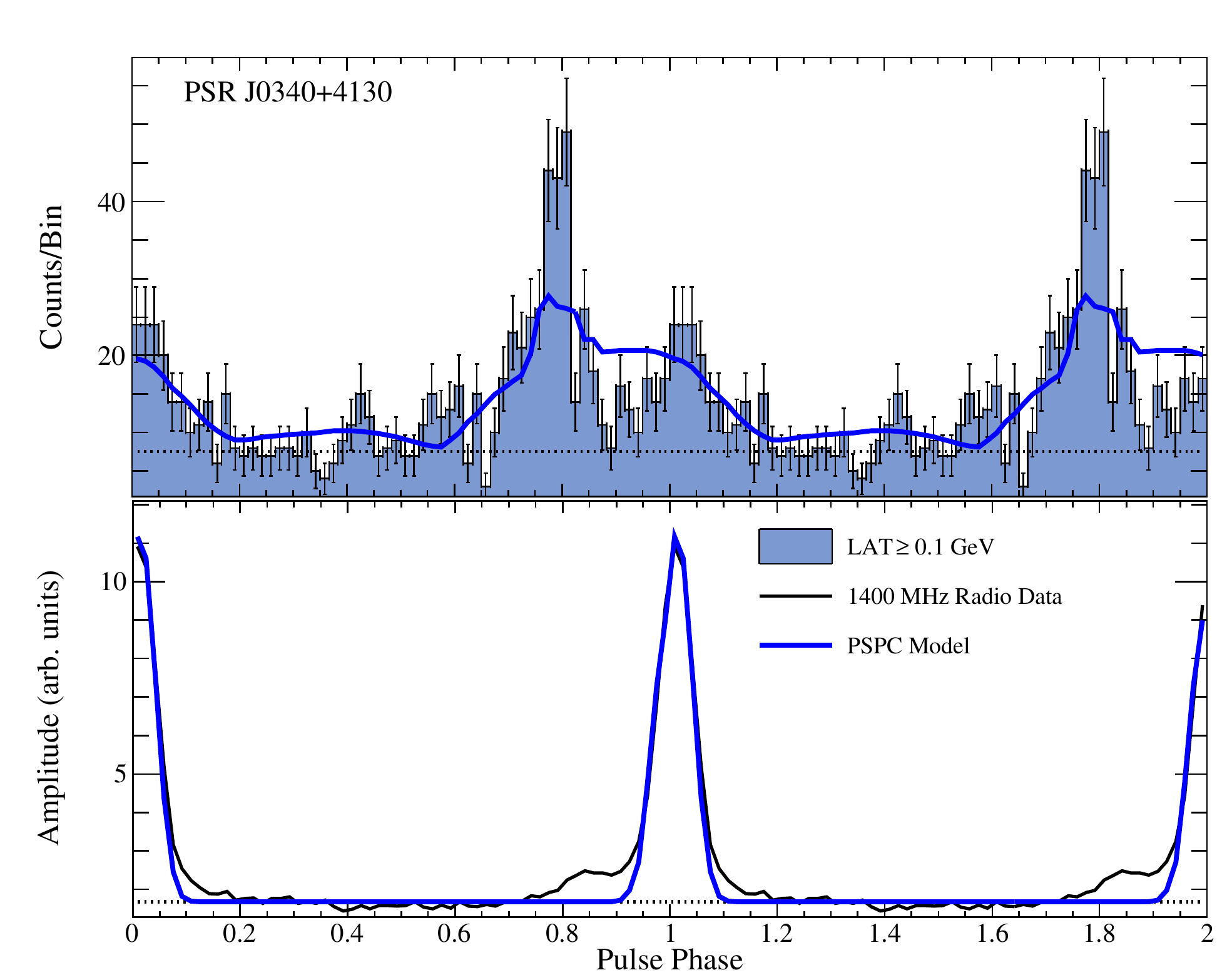}
\end{center}
\caption{Best-fit gamma-ray (\emph{top}) and radio (\emph{bottom}) light curves for PSR J0340+4130.  The solid gray line (blue in the online version) is the PSPC model, best-fit parameters are given in Table \ref{tbl-pspc}.  In both panels, the dotted horizontal lines are the estimated background levels given in Table \ref{tbl-simpars}.\label{app-J0340LCs}}
\end{figure}

Figure \ref{app-J0340LCs} presents the observed and best-fit light curves of PSR J0340+4130.  This is an isolated 3.30 ms pulsar discovered in radio observations of unassociated LAT sources with pulsar-like characteristics (Bengale et al.~in preparation).  This MSP was first announced as a gamma-ray pulsar in 2PC.  The PSPC model is able to reproduce both the radio and gamma-ray light curves of this MSP reasonably well.  The gamma-ray fit does not find two well-separated peaks, nor is the main peak sharp enough.  

\begin{figure}
\begin{center}
\includegraphics[height=0.6\textheight]{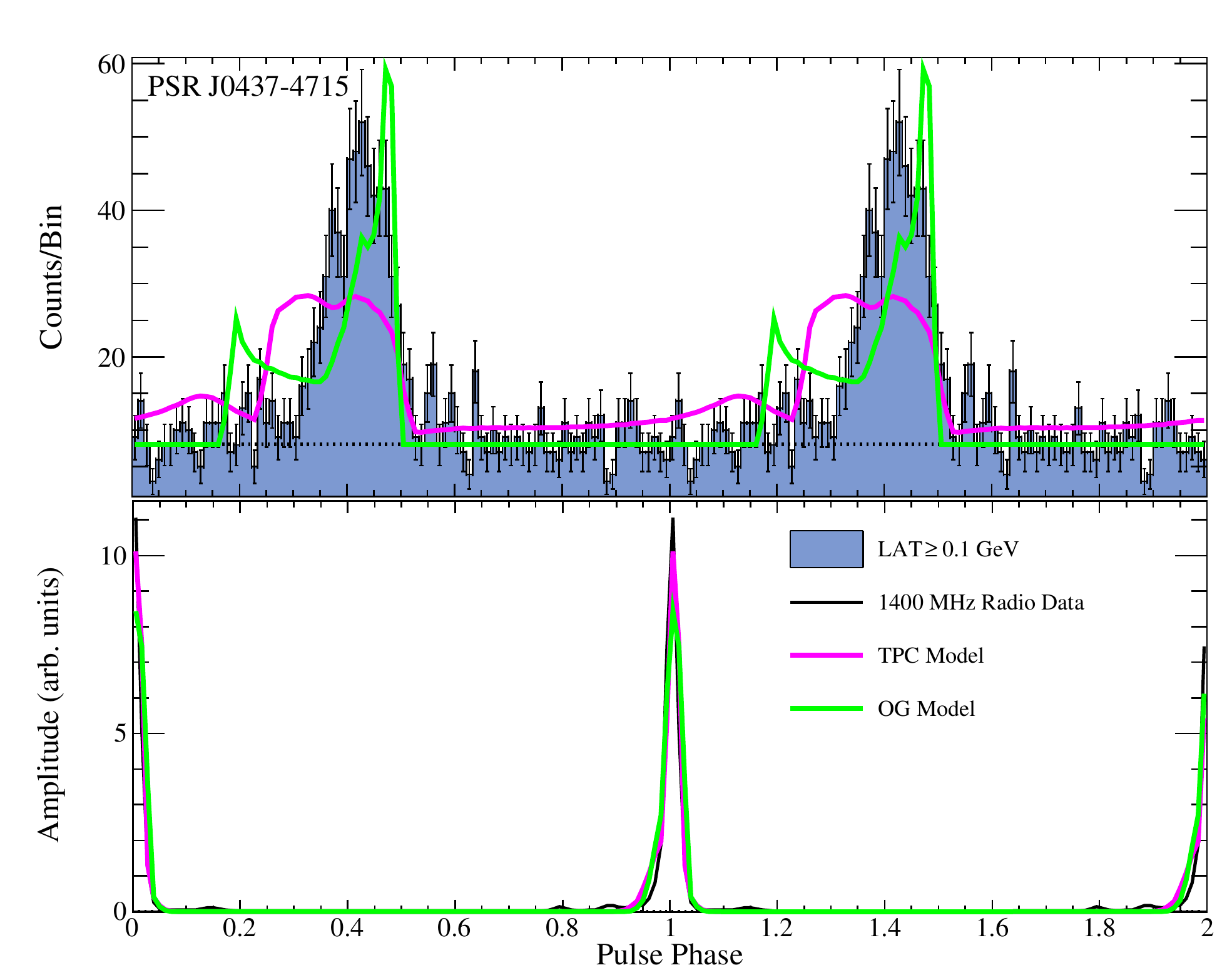}
\end{center}
\caption{Best-fit gamma-ray (\emph{top}) and radio (\emph{bottom}) light curves for PSR J0437$-$4715.  The dashed line (solid green in the online version) is the OG model, best-fit parameters in Table \ref{tbl-og}.  The solid gray line (pink in the online version) is the TPC model, best-fit parameters are given in Table \ref{tbl-tpc}.  In both panels, the dotted horizontal lines are the estimated background levels given in Table \ref{tbl-simpars}.\label{app-J0437LCs}}
\end{figure}

Figure \ref{app-J0437LCs} presents the observed and best-fit light curves of PSR J0437$-$4715.  This is a 5.76 ms pulsar in a binary system with a 5.7 day orbital period first discovered by \citet{Johnston93}.  Gamma-ray pulsations were first reported by \citet{MSPpop}.  \citet{MJ95} and \citet{Navarro97} both presented polarimetric observations of this MSP, including circular polarization with sense reversal near the profile center indicative of a core component.  Thus, in our fits we used both a hollow-cone and core component and require that the best-fit \ze\ intersects the core component.  \citet{MJ95} attempted to constrain the viewing geometry of PSR J0437$-$4715 using RVM fits but were unsuccessful.  They did present RVM curves for $\alpha = 145^{\circ}$ and $\zeta = 140^{\circ}$ which seemed to represent the data well (these values would correspond to $\alpha = 35^{\circ}$ and $\zeta = 40^{\circ}$ in our fits).  \citet{GK97} applied the relativistic RVM of \citet{Blaskiewicz91} to the same data and found $\alpha = 20^{\circ}$ and $\beta = -4^{\circ}$.  These authors first found geometries which reproduced the basic profile characteristics and then calculated the predicted polarization properties and compared to the observations, so the values they report are not really fits and do not come with any uncertainty estimates.  \citet{Bogdanov07} modeled the thermal X-ray light curves of this MSP and found an offset-dipole geometry was necessary to match the data.  They assumed that the spin and orbit axis were aligned and thus used \ze=42\DG, from the timing analyses of \citet{vanStrat01} and \citet{Hotan06} who found orbital inclination angles $i = 42\fdg75\pm0\fdg09$ and $42\fdg5\pm0\fdg2$, respectively.  With the offset dipole and this assumed \ze\ \citet{Bogdanov07} found acceptable fits to the observed X-ray light curves.  This value of \ze\ would argue against the geometry found by \citet{GK97}.  The predicted radio light curves from both the TPC and OG fits match the observed profile well (with this binning many of the lower-level features of the profile are smoothed out), but neither exactly matches the gamma-ray light curve.  The OG model best-fit finds \ze=46\DG, in better agreement with the orbital inclination measurements than the value of 64\DG\ found by the TPC fit.  In the future, fits to the radio and gamma-ray light curve of this MSP with an offset-dipole geometry will be useful to more directly compare with the results of \citet{Bogdanov07}.

\begin{figure}
\begin{center}
\includegraphics[height=0.6\textheight]{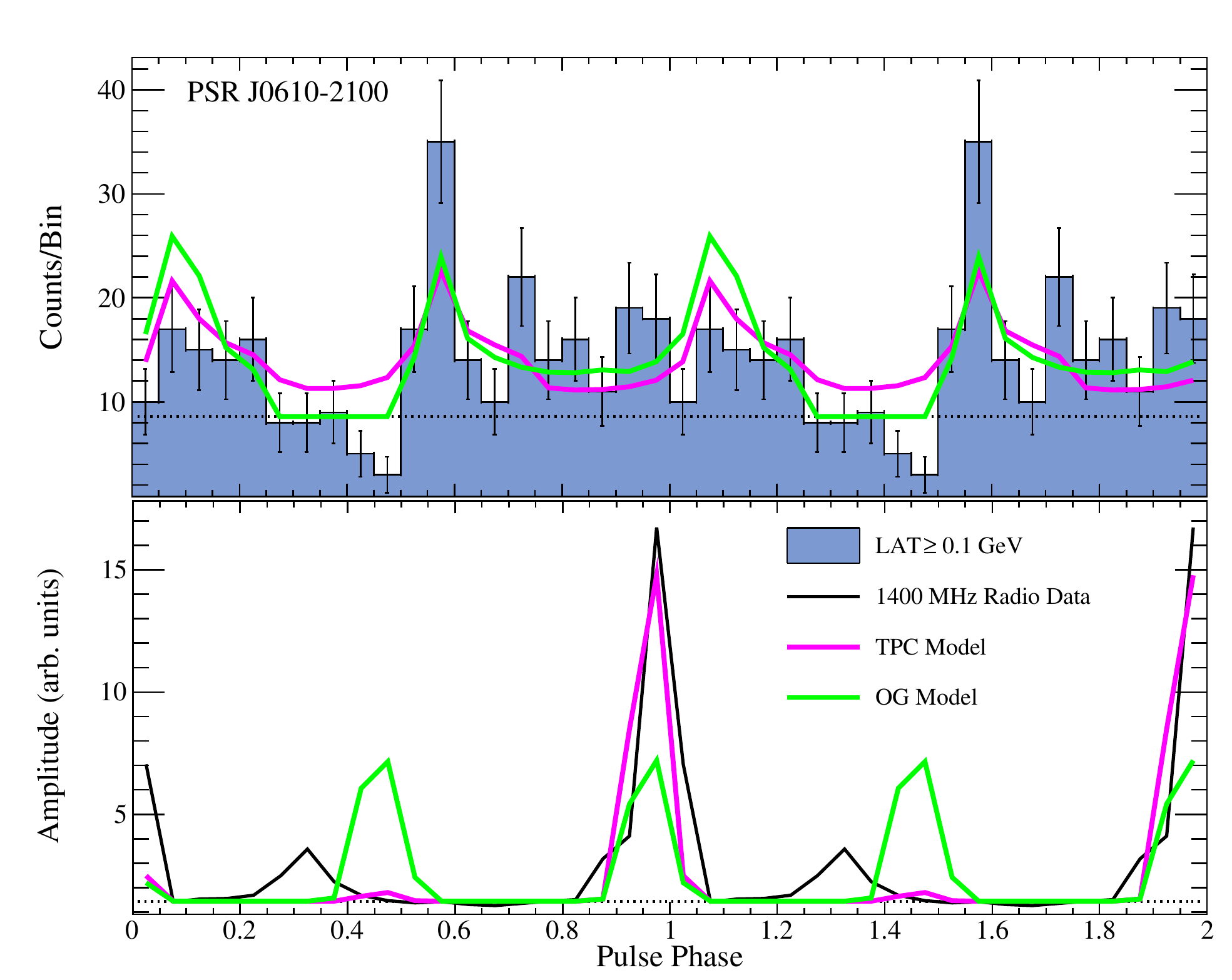}
\end{center}
\caption{Best-fit gamma-ray (\emph{top}) and radio (\emph{bottom}) light curves for PSR J0610$-$2100.  The dashed line (solid green in the online version) is the OG model, best-fit parameters in Table \ref{tbl-og}.  The solid gray line (pink in the online version) is the TPC model, best-fit parameters are given in Table \ref{tbl-tpc}.  In both panels, the dotted horizontal lines are the estimated background levels given in Table \ref{tbl-simpars}.\label{app-J0610LCs}}
\end{figure}

Figure \ref{app-J0610LCs} presents the observed and best-fit light curves of PSR J0610$-$2100.  This is a 3.86 ms pulsar in a binary system with a 0.29 day orbital period discovered by \citet{Burgay06}.  Gamma-ray pulsations from this MSP were first reported by \citet{Espinoza13}.  Both the OG and TPC models are able to reproduce the gamma-ray light curve of this MSP reasonably well and both predict that the light curve should consist of 2 peaks, though more statistics are necessary to confirm the existence of a second peak.  The predicted radio light curve from the TPC fit matches the main radio peak better, but both models predict two radio peaks, with different relative heights, though the second peak is not at the correct phase in either model.

\begin{figure}
\begin{center}
\includegraphics[height=0.6\textheight]{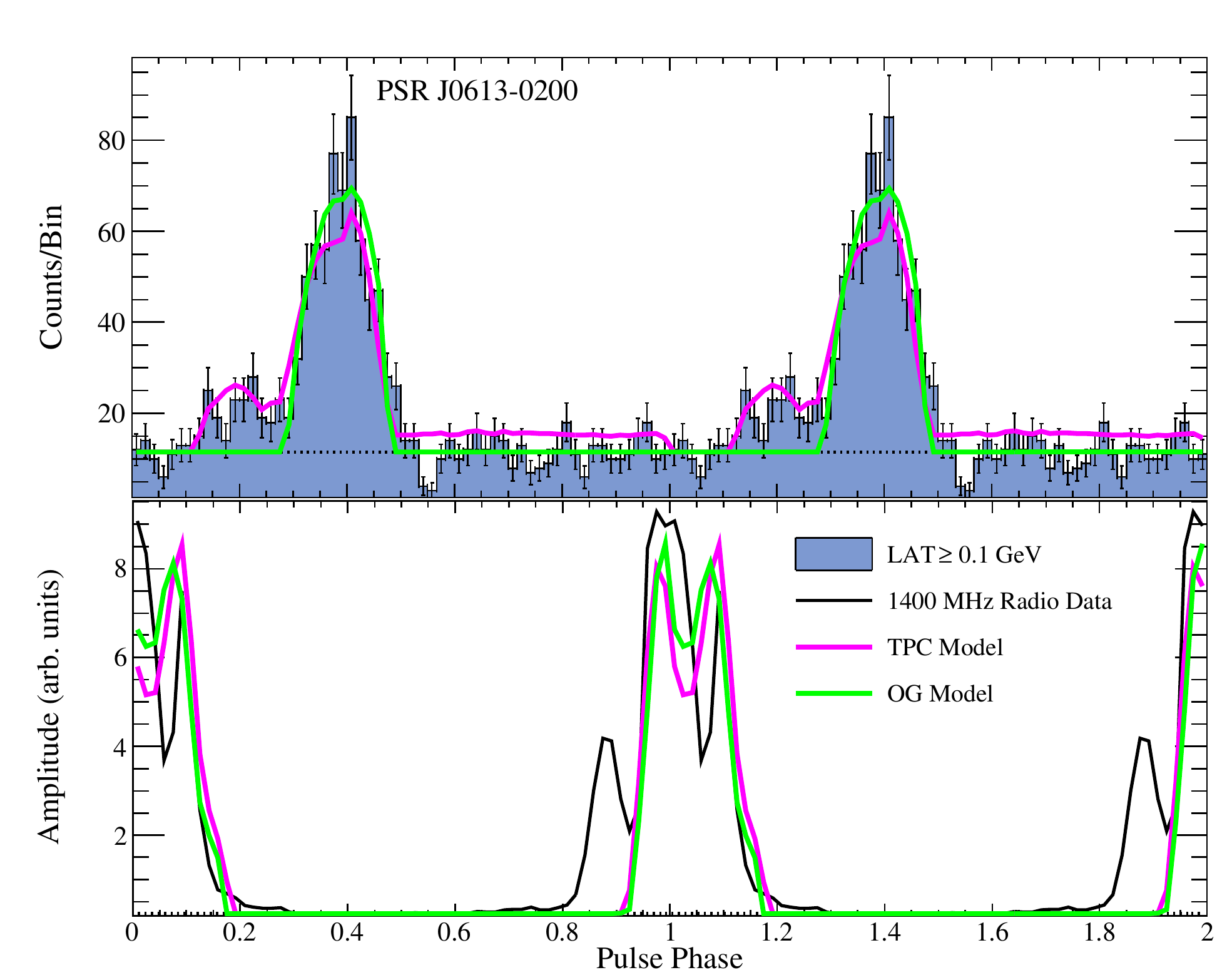}
\end{center}
\caption{Best-fit gamma-ray (\emph{top}) and radio (\emph{bottom}) light curves for PSR J0613$-$0200.  The dashed line (solid green in the online version) is the OG model, best-fit parameters in Table \ref{tbl-og}.  The solid gray line (pink in the online version) is the TPC model, best-fit parameters are given in Table \ref{tbl-tpc}.  In both panels, the dotted horizontal lines are the estimated background levels given in Table \ref{tbl-simpars}.\label{app-J0613LCs}}
\end{figure}

Figure \ref{app-J0613LCs} presents the observed and best-fit light curves of PSR J0613$-$0200.  This is a 3.06 ms pulsar in a binary system with a 1.2 day orbital period discovered by \citet{Lorimer95}.  Gamma-ray pulsations from this MSP were first reported by \citet{MSPpop}.  Polarimetric observations for this MSP have been reported by \citet{Xil98}, \citet{Stairs99}, \citet{Ord2004}, \citet{Manchester04}, and \citet{Yan11} with all but \citet{Manchester04} reporting circular polarization with sense reversal through the main pulse component.  As such, we use a hollow-cone and core radio beam in our fits and require that the best-fit \ze\ intersects the core component.  Both the OG and TPC model reproduce the gamma-ray light curve well, but only the TPC model matches the possible precursor to the main peak at phase 0.2.  Neither best-fit geometry leads to a wide enough radio pulse or to the correct peak multiplicity.

\begin{figure}
\begin{center}
\includegraphics[height=0.6\textheight]{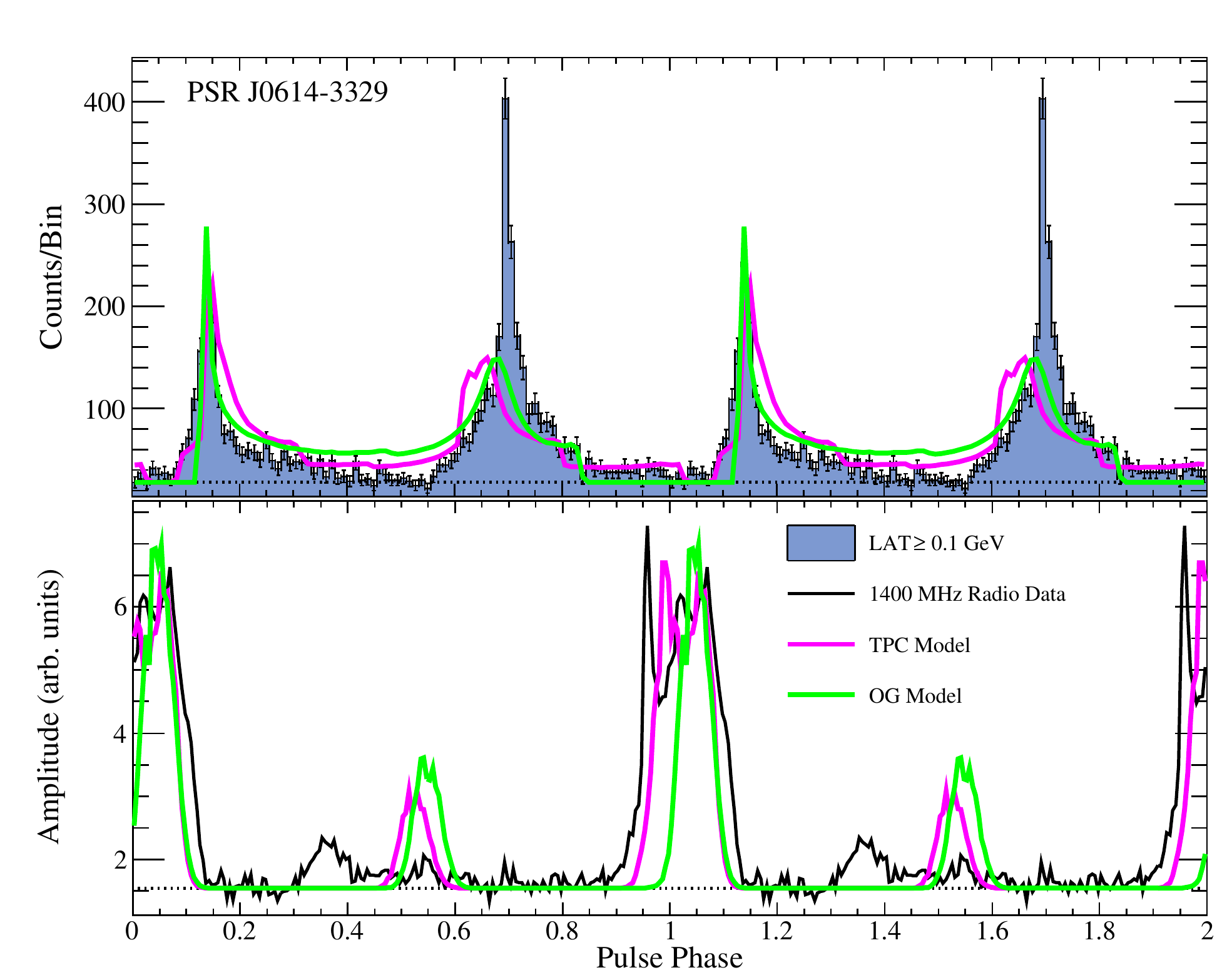}
\end{center}
\caption{Best-fit gamma-ray (\emph{top}) and radio (\emph{bottom}) light curves for PSR J0614$-$3329.  The dashed line (solid green in the online version) is the OG model, best-fit parameters in Table \ref{tbl-og}.  The solid gray line (pink in the online version) is the TPC model, best-fit parameters are given in Table \ref{tbl-tpc}.  In both panels, the dotted horizontal lines are the estimated background levels given in Table \ref{tbl-simpars}.\label{app-J0614LCs}}
\end{figure}

Figure \ref{app-J0614LCs} presents the observed and best-fit light curves of PSR J0614$-$3329.  This is a 3.15 ms pulsar in a binary system with a 54 day orbital period discovered by \citet{Ransom11}, who also reported the first detection of gamma-ray pulsations from this MSP, in radio observations of LAT unassociated sources with pulsar-like characteristics.  While both the OG and TPC model predict a two-peaked gamma-ray light curve, neither quite matches the observed peak separation nor the peak-height ratio.  Both best-fit geometries do predict two radio peaks but the smaller peak is not at the correct phase and the main peak is not wide enough, though the TPC fit does come close.

\begin{figure}
\begin{center}
\includegraphics[height=0.6\textheight]{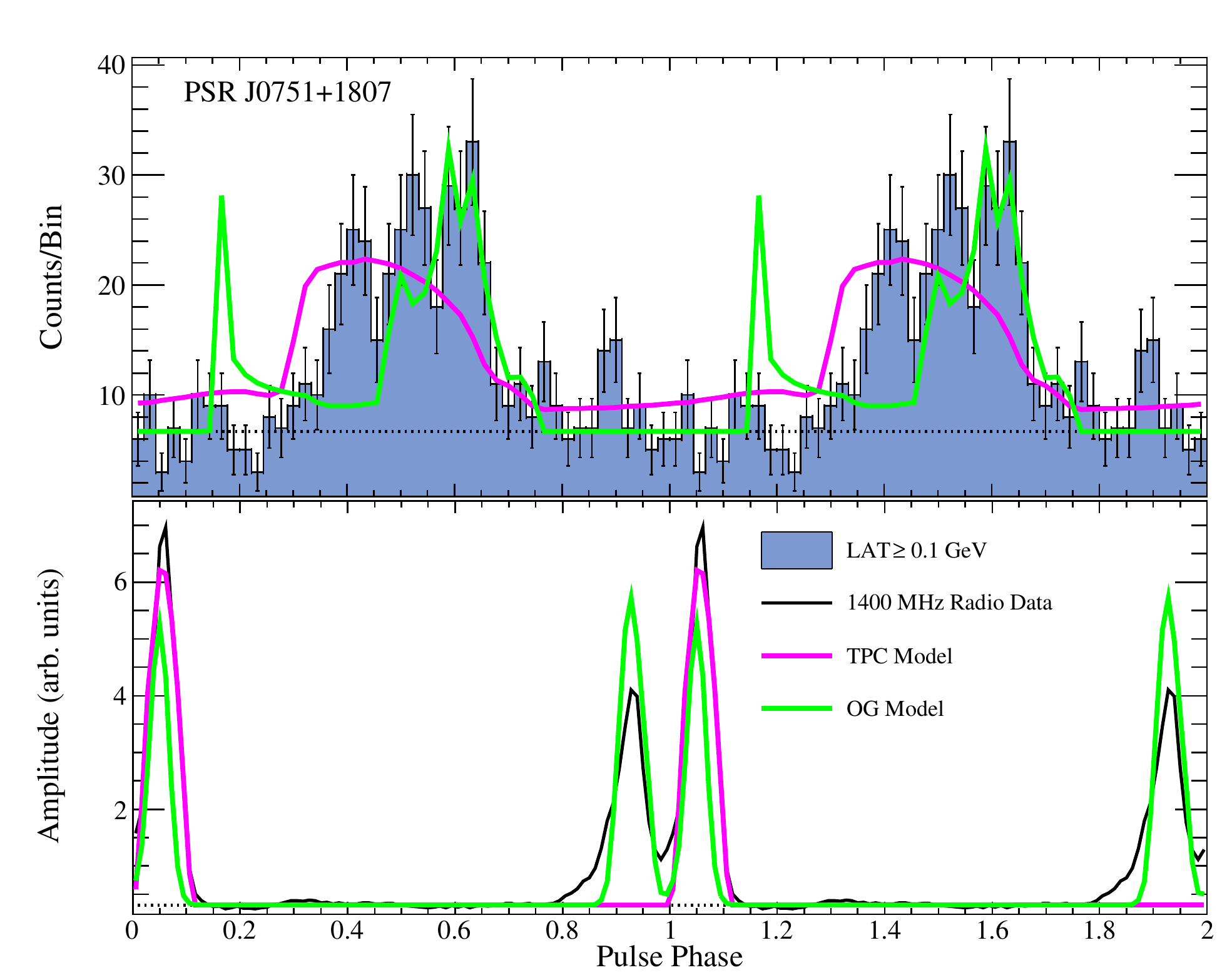}
\end{center}
\caption{Best-fit gamma-ray (\emph{top}) and radio (\emph{bottom}) light curves for PSR J0751+1807.  The dashed line (solid green in the online version) is the OG model, best-fit parameters in Table \ref{tbl-og}.  The solid gray line (pink in the online version) is the TPC model, best-fit parameters are given in Table \ref{tbl-tpc}.  In both panels, the dotted horizontal lines are the estimated background levels given in Table \ref{tbl-simpars}.\label{app-J0751LCs}}
\end{figure}

Figure \ref{app-J0751LCs} presents the observed and best-fit light curves for PSR J0751+1807.  This is a 3.48 ms pulsar in a binary system with a 0.26 day orbital period discovered by \citet{Lundgren95}.  Gamma-ray pulsations from this MSP were first reported by \citet{MSPpop}.  \citet{Xil98} reported polarimetric observations of this MSP but make no predictions on the viewing geometry.  Neither the TPC nor OG model matches the three peaks evident in the gamma-ray light curve and only the best-fit geometry corresponding to the OG fit predicts two closely-spaced radio peaks.

\begin{figure}
\begin{center}
\includegraphics[height=0.6\textheight]{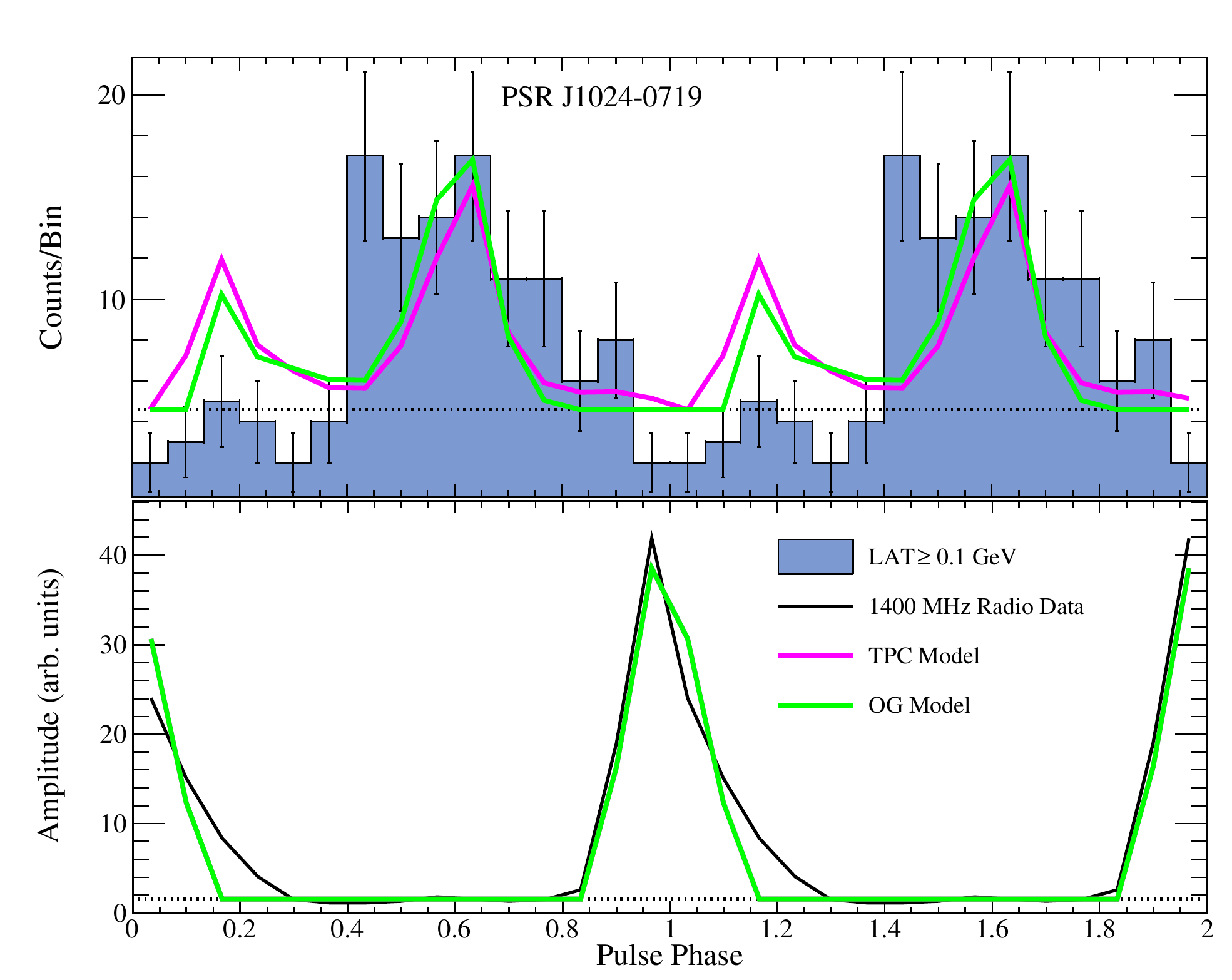}
\end{center}
\caption{Best-fit gamma-ray (\emph{top}) and radio (\emph{bottom}) light curves for PSR J1024$-$0719.  The dashed line (solid green in the online version) is the OG model, best-fit parameters in Table \ref{tbl-og}.  The solid gray line (pink in the online version) is the TPC model, best-fit parameters are given in Table \ref{tbl-tpc}.  The best-fit geometries are the same resulting in the same radio, model light curve resulting in the OG fit being drawn on top of the TPC fit.  In both panels, the dotted horizontal lines are the estimated background levels given in Table \ref{tbl-simpars}.\label{app-J1024LCs}}
\end{figure}

Figure \ref{app-J1024LCs} presents the observed and best-fit light curves of PSR J1024$-$0719.  This is an isolated 5.16 ms pulsar discovered by \citet{Bailes97}.  Gamma-ray pulsations from this MSP were first reported by \citet{Espinoza13}.  \citet{Yan11} reported polarimetric observations of this MSP, the position-angle swing of the linear polarization is flat, suggesting our line of sight clips the edge of a cone.  They also noted weak circular polarization with sense-reversal through the main component.  The circular polarization appears to have substructure and at least three overlapping components, so identification of a core beam is uncertain.   Since \citet{Yan11} do not specifically call this out as evidence for a core component, the results we report in Section \ref{res} only use a hollow-cone beam for the radio model.  However, when we include a core beam and require that the best-fit \ze\ intersects this component, the likelihood finds the same best-fit geometry.  The best-fit model does predict that the core component is much weaker than the cone.  Neither the OG nor the TPC model reproduces the observed gamma-ray light curve well with both predicting two peaks.  However, this MSP is relatively faint in gamma rays and the estimated background level may be too high.  Future modeling with more statistics will be interesting.

\begin{figure}
\begin{center}
\includegraphics[height=0.6\textheight]{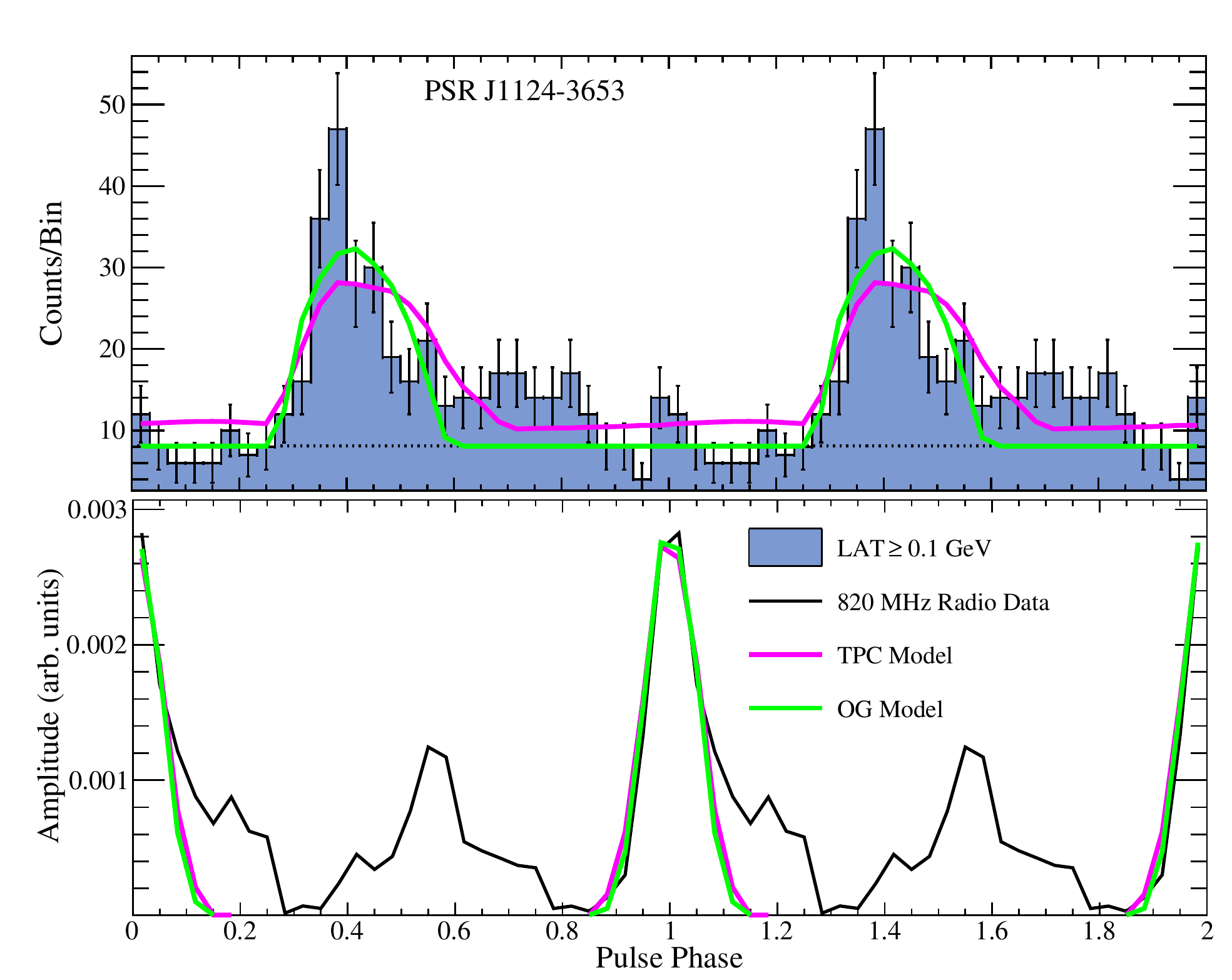}
\end{center}
\caption{Best-fit gamma-ray (\emph{top}) and radio (\emph{bottom}) light curves for PSR J1124$-$3653.  The dashed line (solid green in the online version) is the OG model, best-fit parameters in Table \ref{tbl-og}.  The solid gray line (pink in the online version) is the TPC model, best-fit parameters are given in Table \ref{tbl-tpc}.  The best-fit geometries are the same resulting in the same radio, model light curve resulting in the OG fit being drawn on top of the TPC fit.  In both panels, the dotted horizontal lines are the estimated background levels given in Table \ref{tbl-simpars}.\label{app-J1124LCs}}
\end{figure}

Figure \ref{app-J1124LCs} presents the observed and best-fit light curves of PSR J1124$-$3653.  This is a 2.41 ms pulsar in a binary system with a 0.23 day orbital period discovered in radio observations of unassociated LAT sources with pulsar-like characteristics (Bengale et al.~in preparation).  This MSP was first announced as a gamma-ray pulsar in 2PC.  Both the TPC and OG models reproduce the gamma-ray light curve well but only match the main radio peak. 

\begin{figure}
\begin{center}
\includegraphics[height=0.6\textheight]{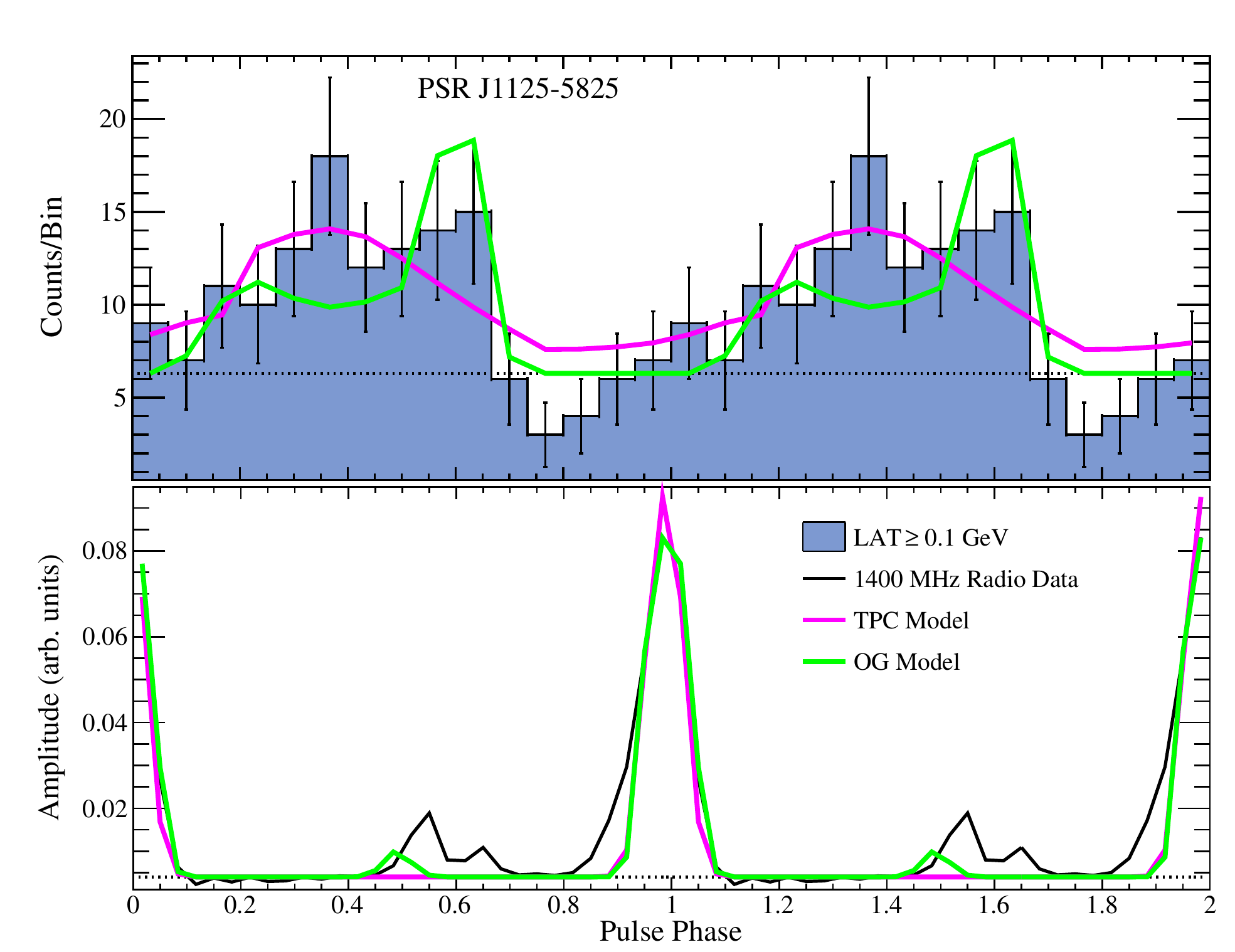}
\end{center}
\caption{Best-fit gamma-ray (\emph{top}) and radio (\emph{bottom}) light curves for PSR J1125$-$5825.  The dashed line (solid green in the online version) is the OG model, best-fit parameters in Table \ref{tbl-og}.  The solid gray line (pink in the online version) is the TPC model, best-fit parameters are given in Table \ref{tbl-tpc}.  In both panels, the dotted horizontal lines are the estimated background levels given in Table \ref{tbl-simpars}.\label{app-J1125LCs}}
\end{figure}

Figure \ref{app-J1125LCs} presents the observed and best-fit light curves of PSR J1125$-$5825.  This is a 3.10 ms pulsar in a binary system with a 76 day orbital period discovered by \citet{Bates11}.  \citet{Keith12} reported a detection of pulsed gamma rays from this MSP with a significance of 4.9$\sigma$.  They also reported polarimetric observations of this MSP and used an RVM fit of only the main pulse component to predict \al=128\DG and $\beta=-21^{\circ}$ with no quoted uncertainties.  Both the TPC and OG model can qualitatively reproduce the observed gamma-ray light curve, though more statistics would lead to a more well-defined profile.  Both models match the main radio peak but only the OG model fit geometry predicts more than 1 peak, though the second peak is not at the correct phase.  The viewing geometry found by \citet{Keith12} would correspond to \al=52\DG\ and \ze=73\DG\ in our fits.  Both the TPC and OG models find values of \ze\ which are compatible with their estimate, but only the TPC \al\ value is consistent with 52\DG\ within the quoted uncertainties.

\begin{figure}
\begin{center}
\includegraphics[height=0.6\textheight]{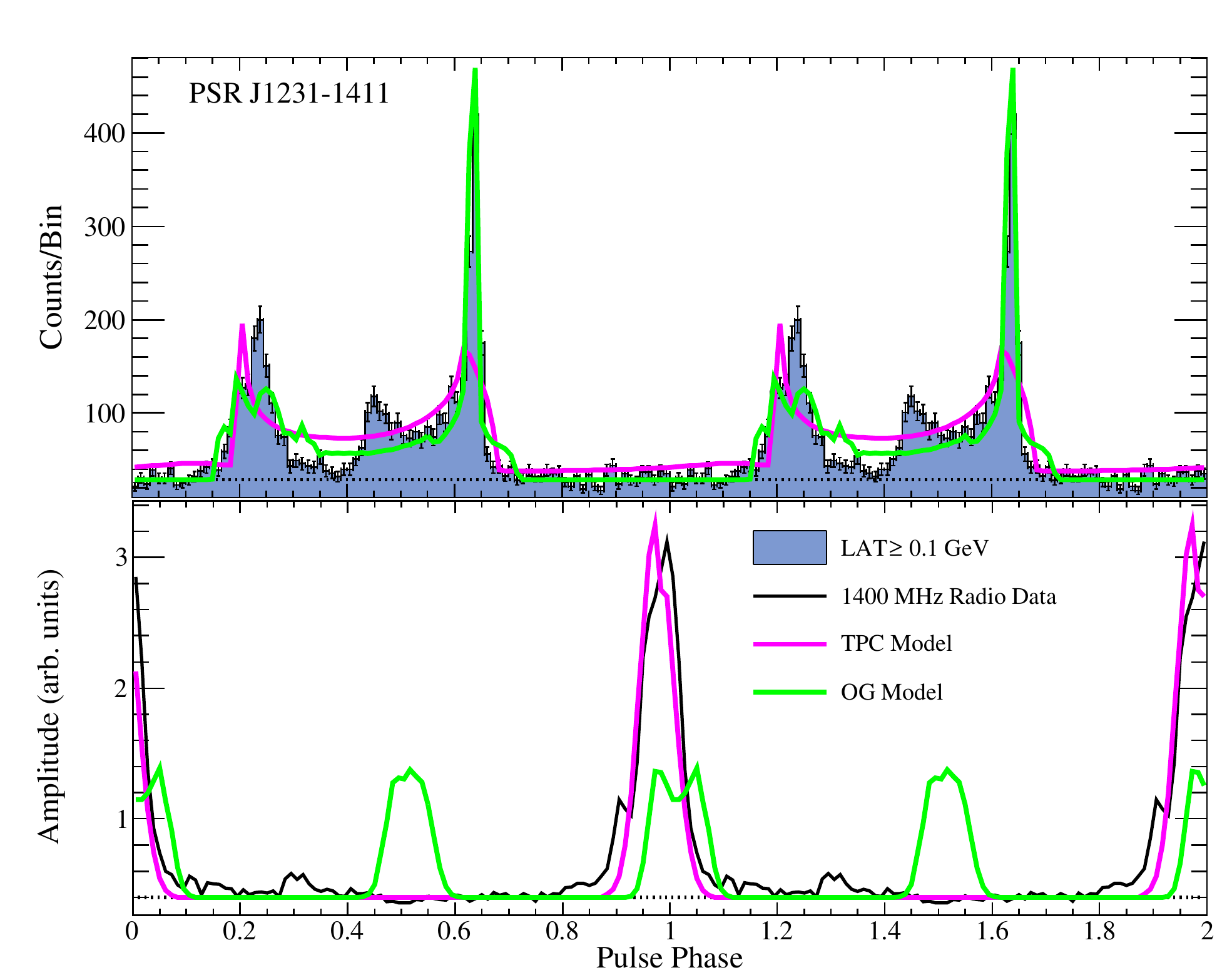}
\end{center}
\caption{Best-fit gamma-ray (\emph{top}) and radio (\emph{bottom}) light curves for PSR J1231$-$1411.  The dashed line (solid green in the online version) is the OG model, best-fit parameters in Table \ref{tbl-og}.  The solid gray line (pink in the online version) is the TPC model, best-fit parameters are given in Table \ref{tbl-tpc}.  In both panels, the dotted horizontal lines are the estimated background levels given in Table \ref{tbl-simpars}.\label{app-J1231LCs}}
\end{figure}

Figure \ref{app-J1231LCs} presents the observed and best-fit light curves of PSR J1231$-$1411.  This is a 3.68 ms pulsar in a binary system with a 1.9 day orbital period discovered by \citet{Ransom11}, who also reported the first detection of gamma-ray pulsations from this MSP, in radio observations of unassociated LAT sources with pulsar-like characteristics.  Both the TPC and OG models can reproduce approximately correct phase separation for the gamma-ray peaks, with the OG model matching the shapes better.  The best-fit geometry corresponding to the TPC fit matches the radio profile better than the OG model fit, which predicts a nearly orthogonal geometry leading to two radio peaks, one from each pole.

\begin{figure}
\begin{center}
\includegraphics[height=0.6\textheight]{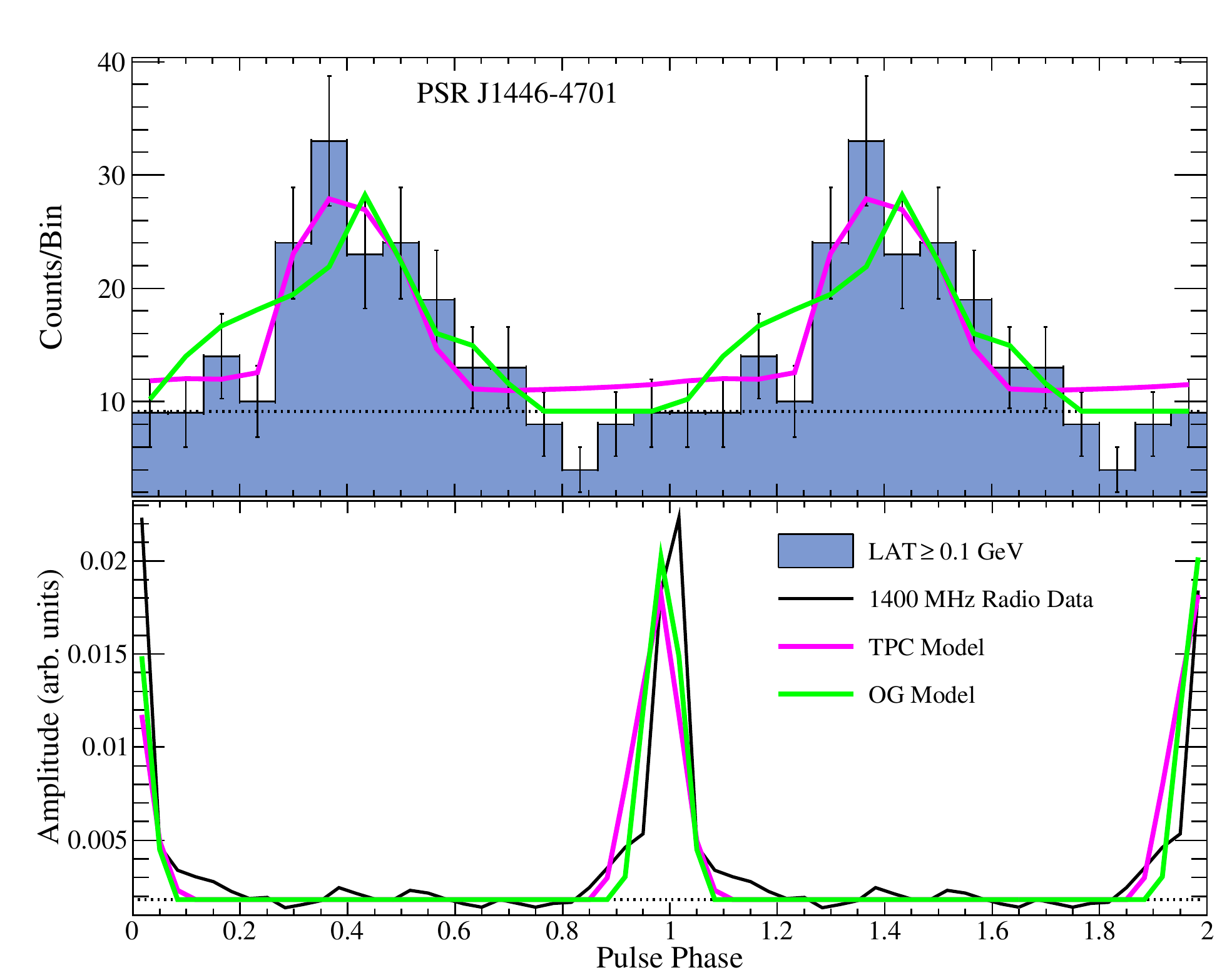}
\end{center}
\caption{Best-fit gamma-ray (\emph{top}) and radio (\emph{bottom}) light curves for PSR J1446$-$4701.  The dashed line (solid green in the online version) is the OG model, best-fit parameters in Table \ref{tbl-og}.  The solid gray line (pink in the online version) is the TPC model, best-fit parameters are given in Table \ref{tbl-tpc}.  In both panels, the dotted horizontal lines are the estimated background levels given in Table \ref{tbl-simpars}.\label{app-J1446LCs}}
\end{figure}

Figure \ref{app-J1446LCs} presents the observed and best-fit light curves for PSR J1446$-$4701.  This is a 2.19 ms pulsar in a binary system with a 0.28 day orbital period discovered by \citet{Keith12}, who also reported the first detection of gamma-ray pulsations from this MSP.  Both the TPC and OG model reproduce the gamma-ray light curve of PSR J1446$-$4701 well, with the TPC model matching the position of the peak slightly better.  Both models also match the radio profile well, though with slightly larger predicted phase lag than what is observed.

\begin{figure}
\begin{center}
\includegraphics[height=0.6\textheight]{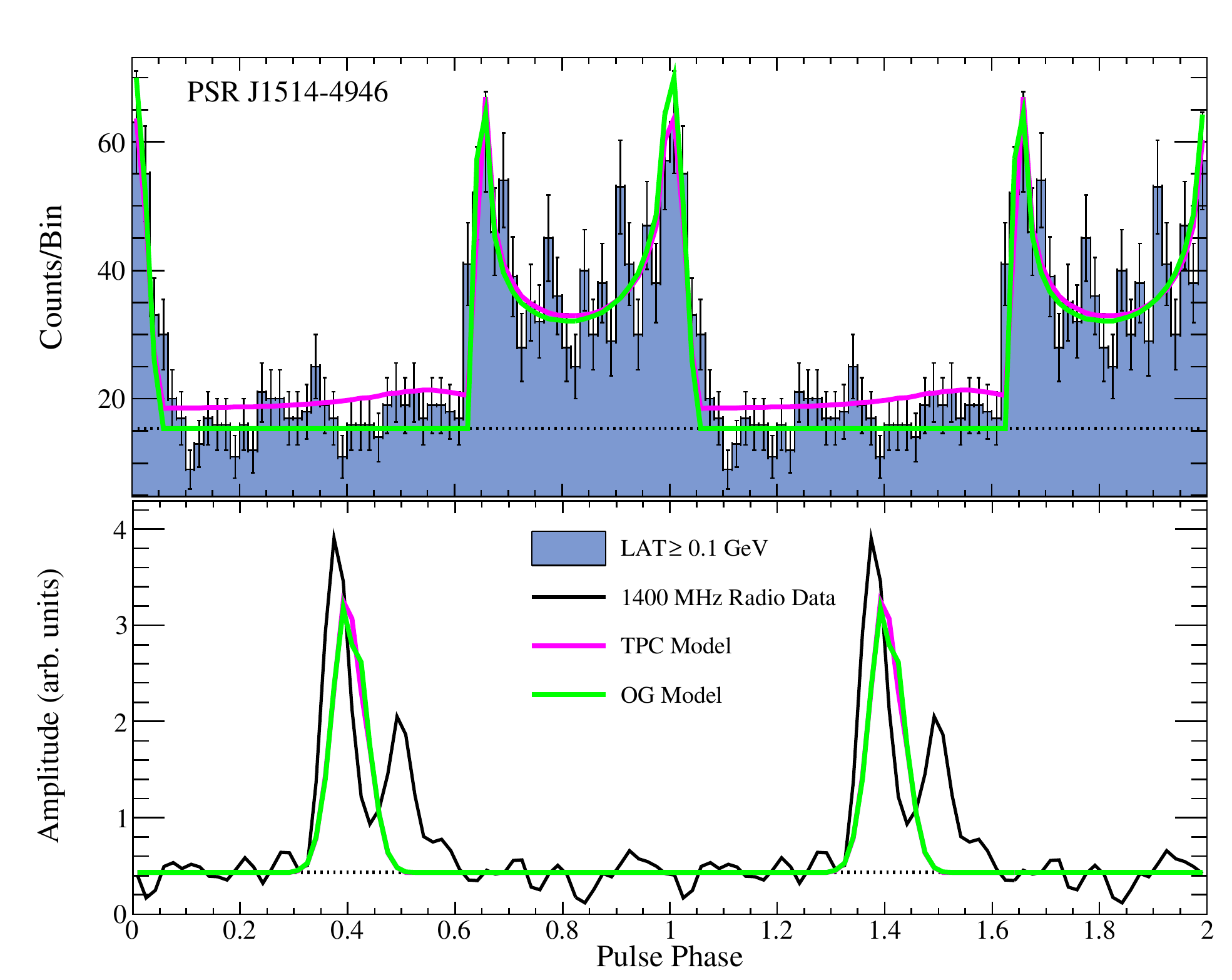}
\end{center}
\caption{Best-fit gamma-ray (\emph{top}) and radio (\emph{bottom}) light curves for PSR J1514$-$4946.  The dashed line (solid green in the online version) is the OG model, best-fit parameters in Table \ref{tbl-og}.  The solid gray line (pink in the online version) is the TPC model, best-fit parameters are given in Table \ref{tbl-tpc}.  In both panels, the dotted horizontal lines are the estimated background levels given in Table \ref{tbl-simpars}.\label{app-J1514LCs}}
\end{figure}

Figure \ref{app-J1514LCs} presents the observed and best-fit light curves of PSR J1514$-$4946.  This is a 3.58 ms pulsar in a binary system with a 1.9 day orbital period discovered in a radio observations of unassociated LAT sources with pulsar-like characteristics and first announced by \citet{Kerr2012} (with more details in Camilo et al.~in preparation).  This MSP was first announced as a gamma-ray pulsar in 2PC.  Both the TPC and OG models match the gamma-ray light curves of this MSP well, but neither model is able to reproduce the two closely-spaced radio peaks but the qualitative properties of the predicted radio emission agree well with observations.

\begin{figure}
\begin{center}
\includegraphics[height=0.6\textheight]{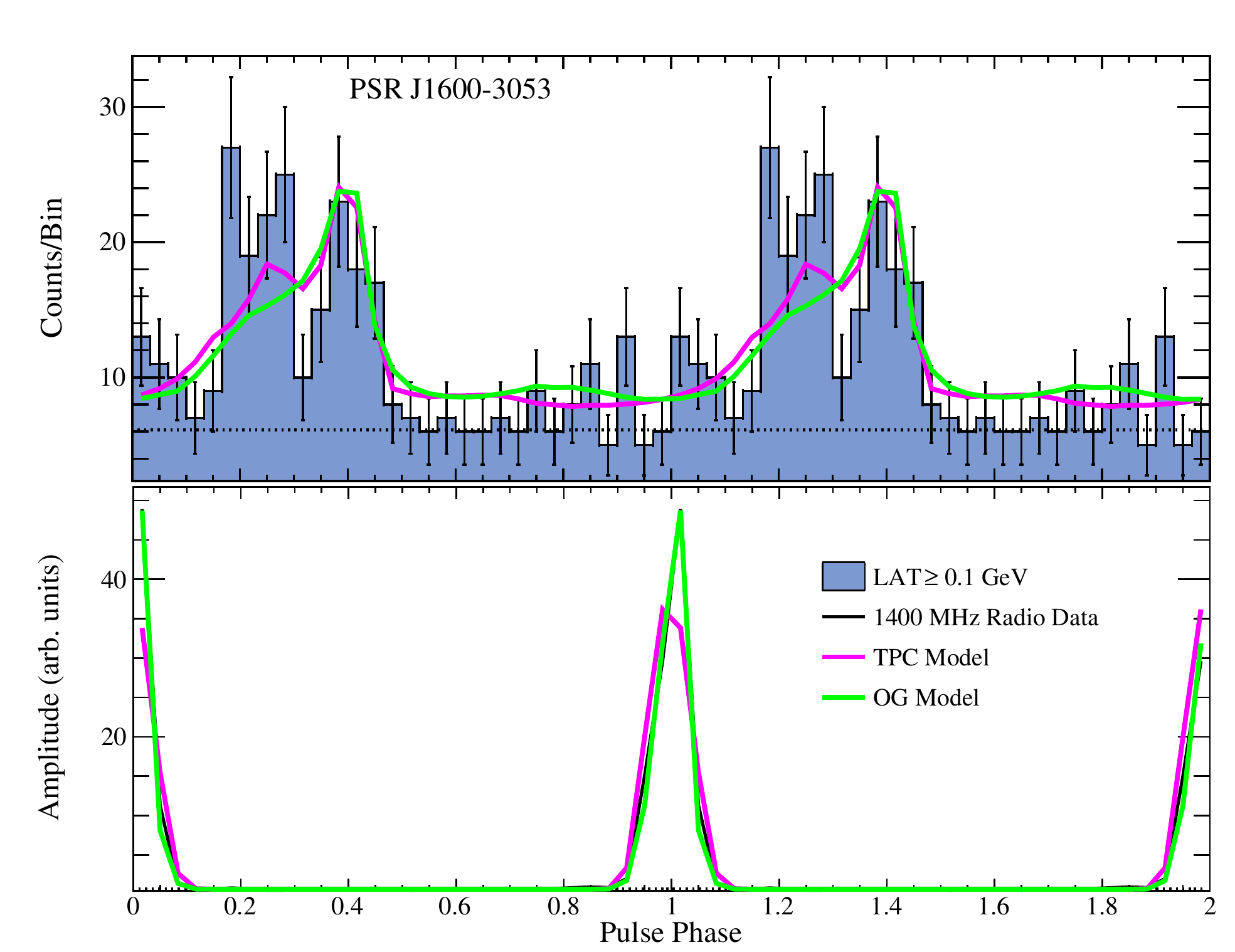}
\end{center}
\caption{Best-fit gamma-ray (\emph{top}) and radio (\emph{bottom}) light curves for PSR J1600$-$3053.  The dashed line (solid green in the online version) is the OG model, best-fit parameters in Table \ref{tbl-og}.  The solid gray line (pink in the online version) is the TPC model, best-fit parameters are given in Table \ref{tbl-tpc}.  In both panels, the dotted horizontal lines are the estimated background levels given in Table \ref{tbl-simpars}.\label{app-J1600LCs}}
\end{figure}

Figure \ref{app-J1600LCs} presents the observed and best-fit light curves of PSR J1600$-$3053.  This is a 3.60 ms pulsar in a binary system with a 14 day orbital period discovered by \citet{Jacoby07}.  \citet{Ord2004} and \citet{Yan11} reported polarimetric observations for this MSP and noted sense reversal in the circular polarization.  As such, we have used both a hollow-cone beam and a core beam when modeling the light curves of this pulsar and required the best-fit \ze\ to intersects the core component.  The best-fit geometries corresponding to fits with the TPC and OG models both reproduce the observed radio profile well.  However, neither fit is able to properly match the first gamma-ray peak.  This MSP is relatively faint in gamma rays and more statistics may lead to a sharper profile and drive the likelihood to a geometry which better matches the data.

\begin{figure}
\begin{center}
\includegraphics[height=0.6\textheight]{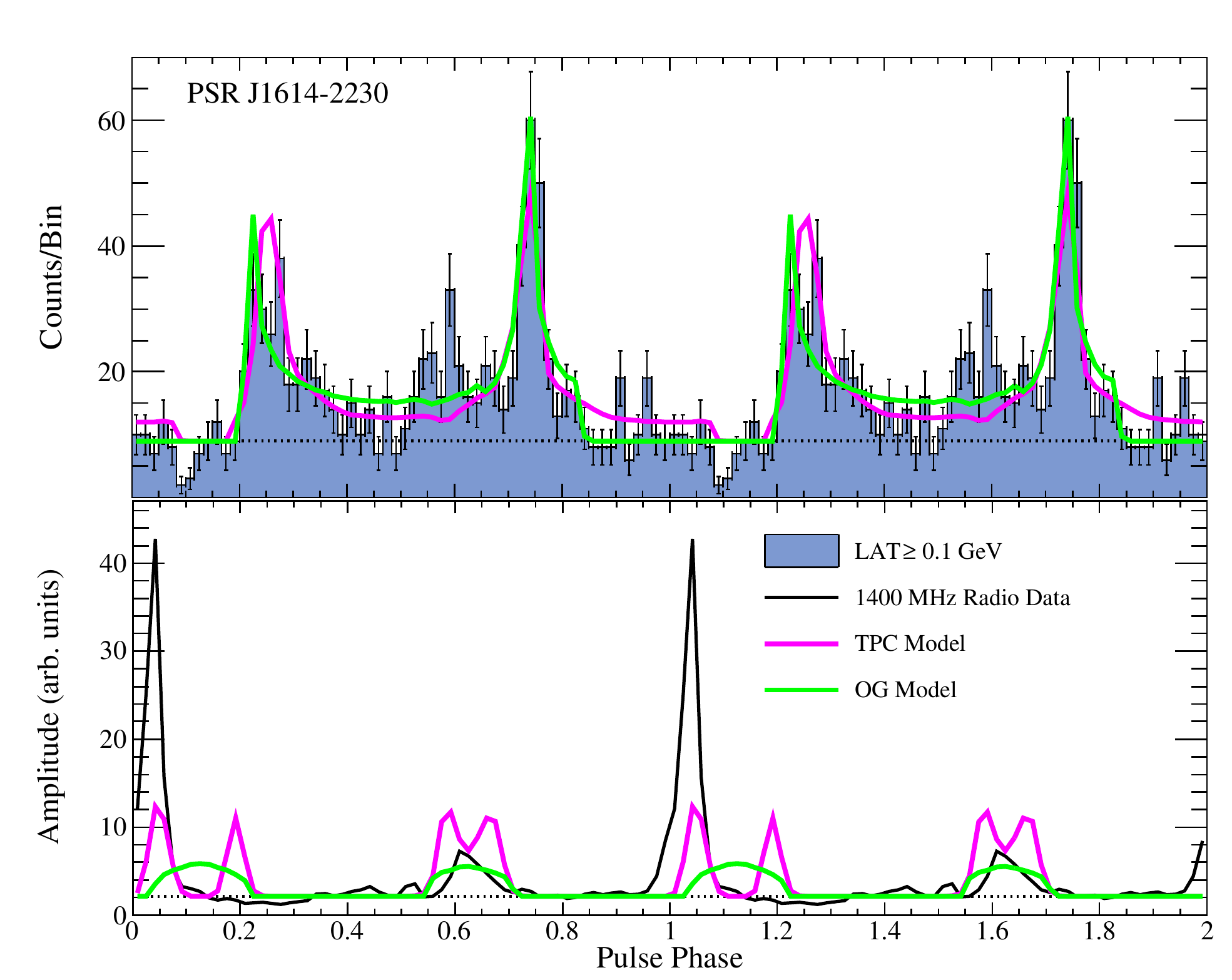}
\end{center}
\caption{Best-fit gamma-ray (\emph{top}) and radio (\emph{bottom}) light curves for PSR J1614$-$2230.  The dashed line (solid green in the online version) is the OG model, best-fit parameters in Table \ref{tbl-og}.  The solid gray line (pink in the online version) is the TPC model, best-fit parameters are given in Table \ref{tbl-tpc}.  In both panels, the dotted horizontal lines are the estimated background levels given in Table \ref{tbl-simpars}.\label{app-J1614LCs}}
\end{figure}

Figure \ref{app-J1614LCs} presents the observed and best-fit light curves of PSR J1614$-$2230.  This is a 3.15 ms pulsar in a binary system with a 8.7 day orbital period discovered by \citet{Crawford06} in radio observations of unassociated \emph{EGRET} sources.  Gamma-ray pulsations from this MSP were first reported by \citet{MSPpop}.  \citet{Demorest10} used the Shapiro delay to measure the mass of this pulsar to be $(1.97\pm0.04)\ M_{\odot}$ and the inclination angle with respect to the orbit axis to be $i = 88\fdg17\pm0\fdg02$.  If we assume that the spin and orbit axes are aligned, this can be taken as a constraint on \ze, but we only compare our best-fit values to this and do not use it to limit the likelihood analysis.  Both the OG and TPC model match the observed gamma-ray light curve well, but neither exactly matches the radio profile.  While both best-fit geometries predict radio emission from two poles, the relative peak heights and phases are wrong.  Both models find values of \ze\ near 90\DG\ with the OG model agreeing best with $\zeta = (88_{-5}^{+2})^{\circ}$.

\begin{figure}
\begin{center}
\includegraphics[height=0.6\textheight]{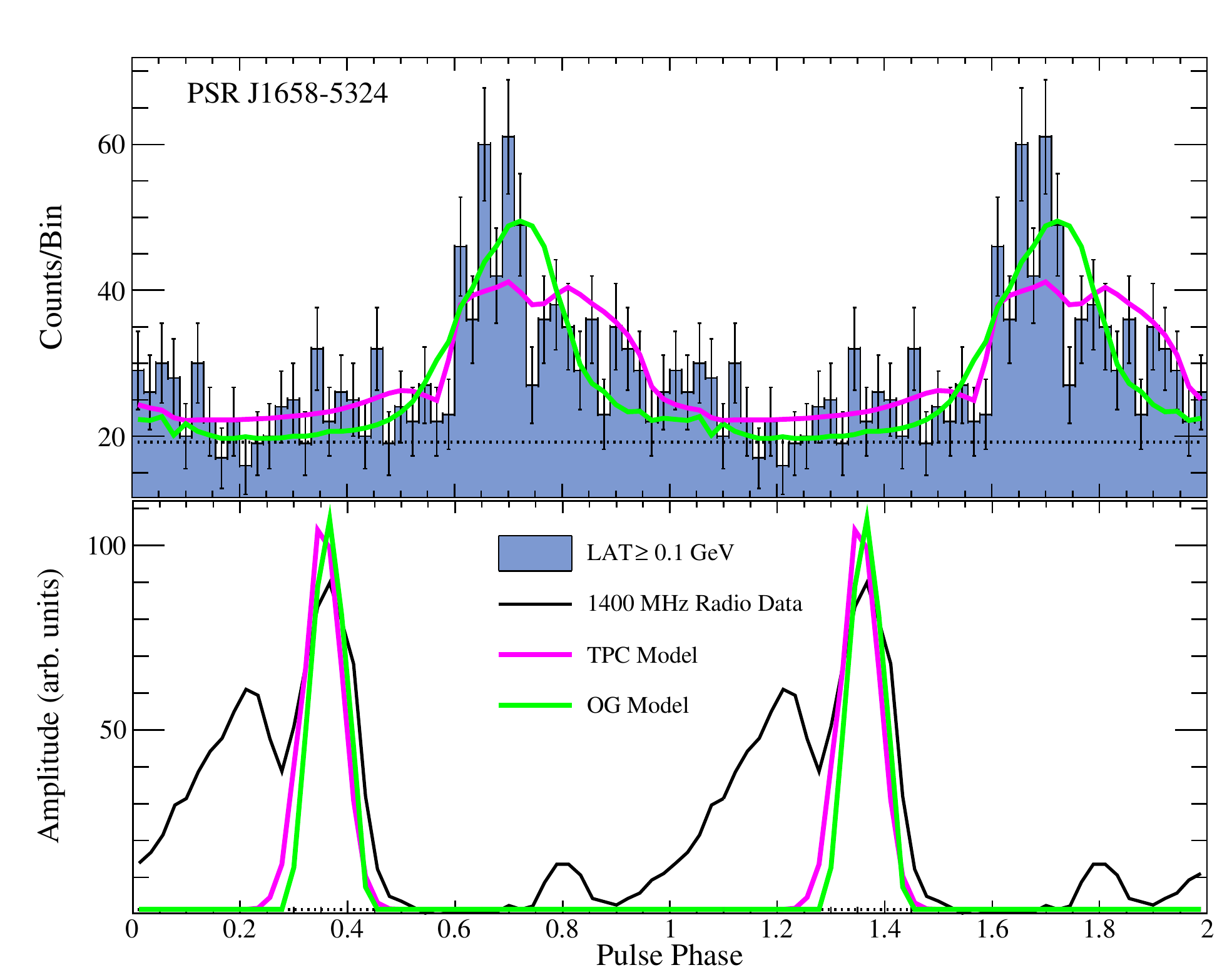}
\end{center}
\caption{Best-fit gamma-ray (\emph{top}) and radio (\emph{bottom}) light curves for PSR J1658$-$5324.  The dashed line (solid green in the online version) is the OG model, best-fit parameters in Table \ref{tbl-og}.  The solid gray line (pink in the online version) is the TPC model, best-fit parameters are given in Table \ref{tbl-tpc}.  In both panels, the dotted horizontal lines are the estimated background levels given in Table \ref{tbl-simpars}.\label{app-J1658LCs}}
\end{figure}

Figure \ref{app-J1658LCs} presents the observed and best-fit light curves of PSR J1658$-$5324.  This is an isolated 2.44 ms pulsar discovered in radio observations of unassociated LAT sources with pulsar like characteristics, first announced by \citet{Kerr2012} (with more details in Camilo et al.~in preparation).  This MSP was first announced as a gamma-ray pulsar in 2PC.  Both the OG and TPC model match the main gamma-ray peak of PSR J1658$-$5324 well, with the TPC model predicting two very closely-spaced peaks.  The best-fit geometries for both fits are able to match the highest radio peak but they miss the broad peak preceding this and the small peak near phase 0.8.

\begin{figure}
\begin{center}
\includegraphics[height=0.6\textheight]{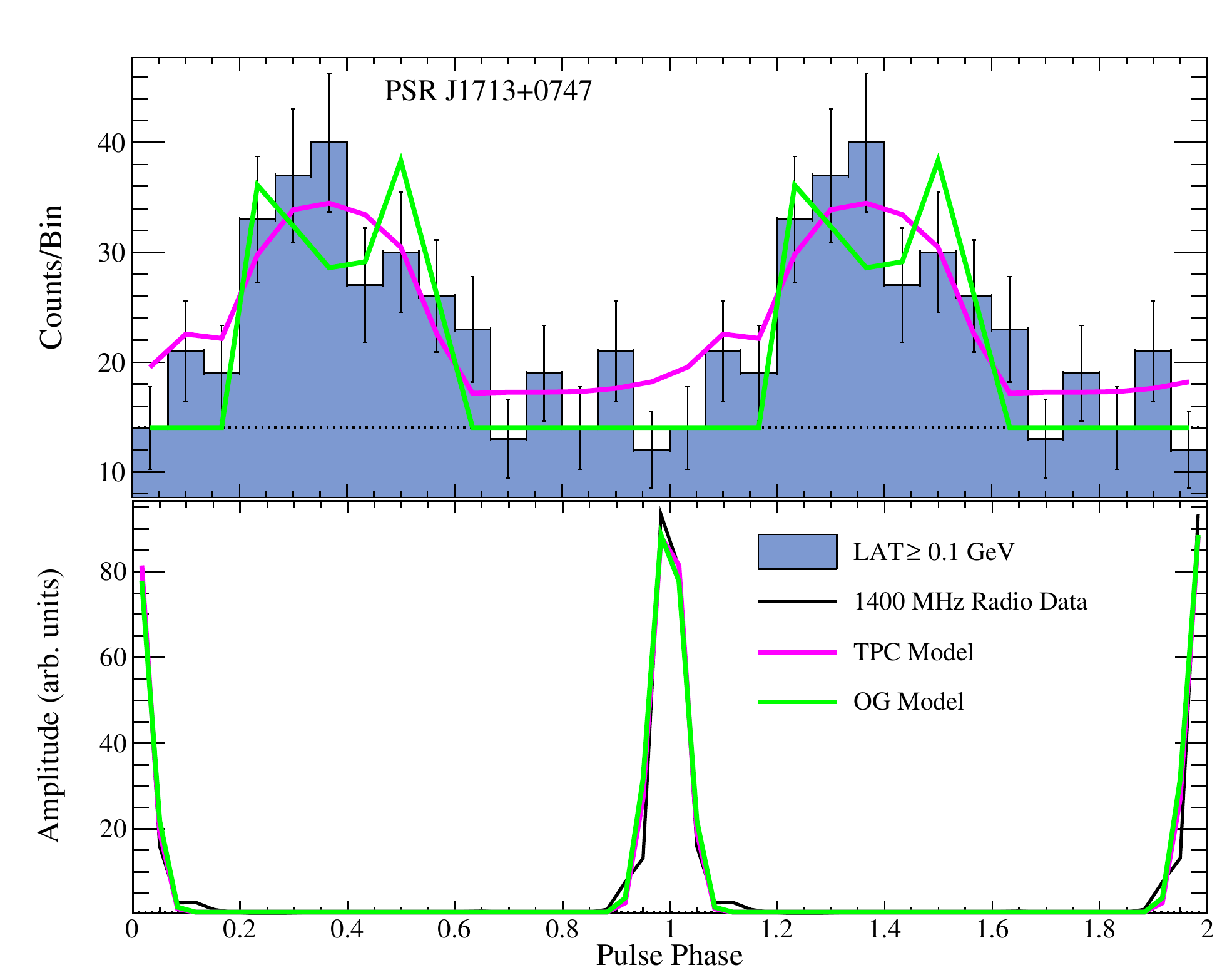}
\end{center}
\caption{Best-fit gamma-ray (\emph{top}) and radio (\emph{bottom}) light curves for PSR J1713+0747.  The dashed line (solid green in the online version) is the OG model, best-fit parameters in Table \ref{tbl-og}.  The solid gray line (pink in the online version) is the TPC model, best-fit parameters are given in Table \ref{tbl-tpc}.  In both panels, the dotted horizontal lines are the estimated background levels given in Table \ref{tbl-simpars}.\label{app-J1713LCs}}
\end{figure}

Figure \ref{app-J1713LCs} presents the observed and best-fit light curves of PSR J1713+0747.  This is a 4.57 ms pulsar in a binary system with a 68 day orbital period discovered by \citet{Foster93}.  Gamma-ray pulsations from this MSP were first reported by \citet{Espinoza13}.  Several authors have reported polarimetric observations of PSR J1713+0747 \citep[e.g.,][]{Xil98,Stairs99,Ord2004,Yan11} and all measure some degree of circular polarization, with sense reversal in some cases, but none have attempted RVM fits for this MSP.  We use both a hollow-cone and core beam to model the radio light curve of this MSP and require that the best-fit \ze\ intersects the core beam.  This pulsar is one of the best-timed and several authors have used Shapiro delay measurements to constrain the orbital inclination angle.  \citet{Camilo94} were the first to report such measurements for this MSP and they estimated $i=70^{\circ}$, though the fits were not well constrained.  \citet{Spalaver05} and \citet{Hotan06} reported more precise timing measurements and found $i = 71\fdg9\pm1\fdg8$ and $(76.6_{-2.0}^{+1.5})^{\circ}$, respectively.  As with PSR J1614$-$2230, these values can be used as constraints on \ze, assuming alignment of the spin and orbit axes, but we only compare our best-fit values to these and do not include them as limits in the likelihood fit.  Both the TPC and OG model are able to reproduce the observed gamma-ray light curve of this MSP well.  The TPC model matches the position of the main peak well with the OG model predicting two closely-spaced peaks, but more statistics will be necessary to determine if this MSP has one or two gamma-ray peaks.  The best-fit geometries for both fits reproduce the observed radio profile well.  Our TPC fit finds \ze=$68^{\circ}\pm4^{\circ}$ and our OG fit finds \ze$=(65_{-4}^{+18})^{\circ}$.  Both fits agree well, within uncertainties, with the constraints from timing.

\begin{figure}
\begin{center}
\includegraphics[height=0.6\textheight]{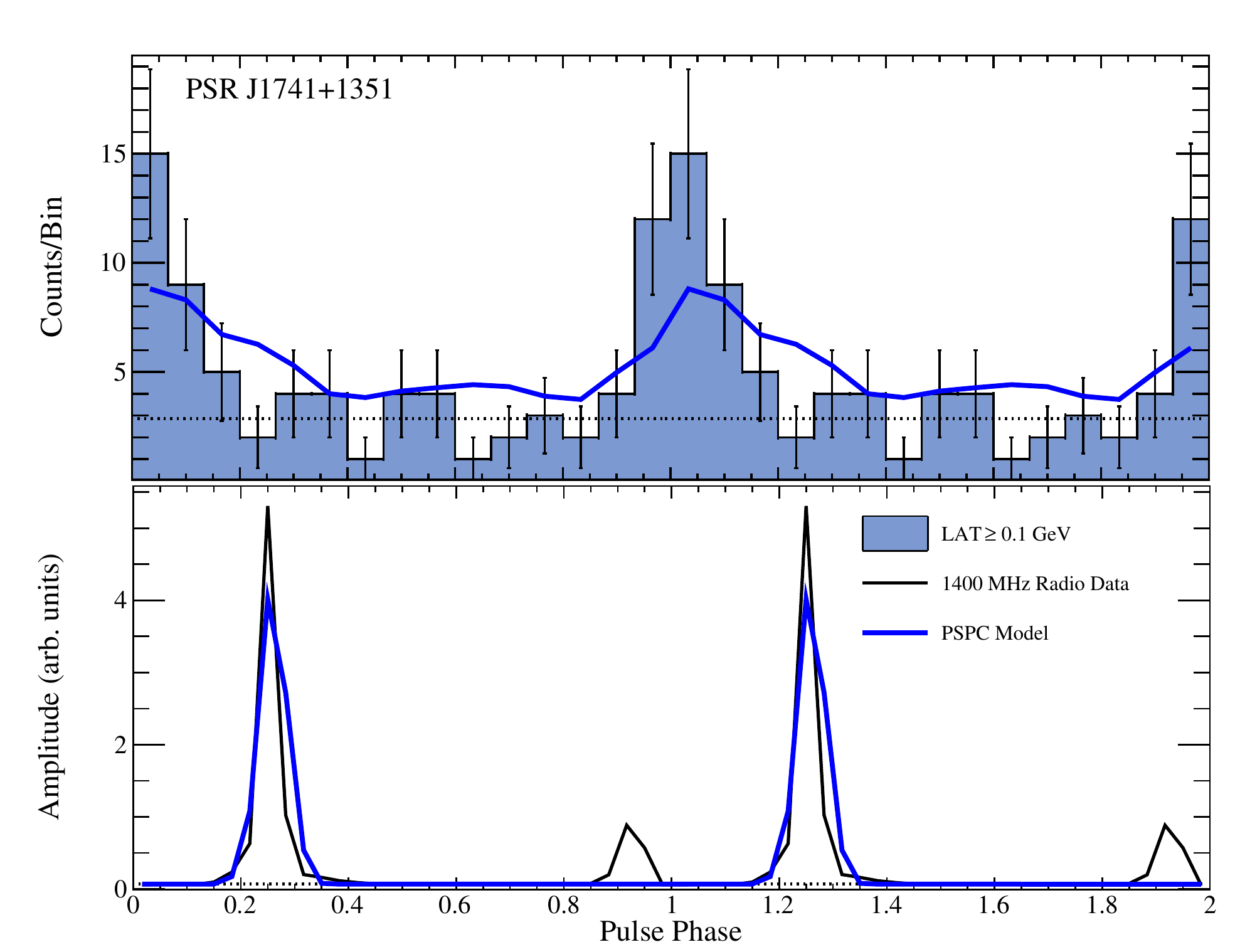}
\end{center}
\caption{Best-fit gamma-ray (\emph{top}) and radio (\emph{bottom}) light curves for PSR J1741+1351.  The solid gray line (blue in the online version) is the PSPC model, best-fit parameters are given in Table \ref{tbl-pspc}.  In both panels, the dotted horizontal lines are the estimated background levels given in Table \ref{tbl-simpars}.\label{app-J1741LCs}}
\end{figure}

Figure \ref{app-J1741LCs} presents the observed and best-fit light curves of PSR J1741+1351.  This is a 3.75 ms pulsar in a binary system with a 16 day orbital period discovered by \citet{Jacoby07}.  Gamma-ray pulsations from this MSP were first reported by \citet{Espinoza13}.  The PSPC model is able to reproduce the single gamma-ray peak of this MSP well, though the predicted peak is not sharp enough and it is unclear if the predicted off-peak emission is seen with the LAT.  The best-fit geometry matches the main radio peak well but does not reproduce the smaller peak at phase 0.9.

\begin{figure}
\begin{center}
\includegraphics[height=0.6\textheight]{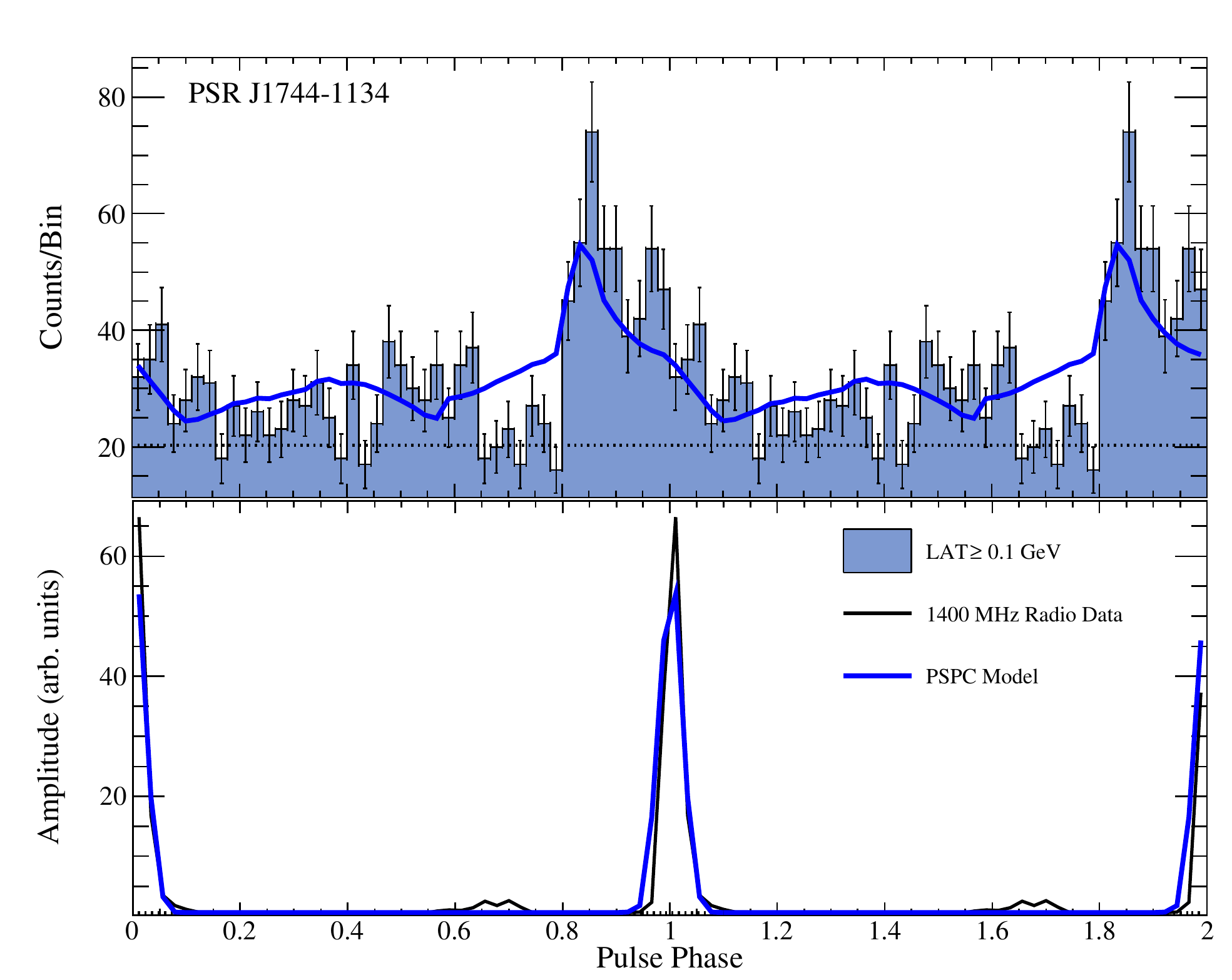}
\end{center}
\caption{Best-fit gamma-ray (\emph{top}) and radio (\emph{bottom}) light curves for PSR J1744$-$1134.  The solid gray line (blue in the online version) is the PSPC model, best-fit parameters are given in Table \ref{tbl-pspc}.  In both panels, the dotted horizontal lines are the estimated background levels given in Table \ref{tbl-simpars}.\label{app-J1744LCs}}
\end{figure}

Figure \ref{app-J1744LCs} presents the observed and best-fit light curves of PSR J1744$-$1134.  This is an isolated 4.07 ms pulsar discovered by \citet{Bailes97}.  Gamma-ray pulsations from this MSP were first reported by \citet{MSPpop}.  Polarimetric observations of this pulsar all present flat position-angle swings favoring a cone beam \citep{Xil98,Stairs99,Ord2004,Yan11}.  Additionally, \citet{Yan11} suggest that $\beta>0$ based on the assumption that the position-angle swing in the main component is a continuation of that in the precursor.  The PSPC model is able to reproduce the observed gamma-ray light curve well, and the best-fit geometry matches the main radio peak but not the low-level precursor.

\begin{figure}
\begin{center}
\includegraphics[height=0.6\textheight]{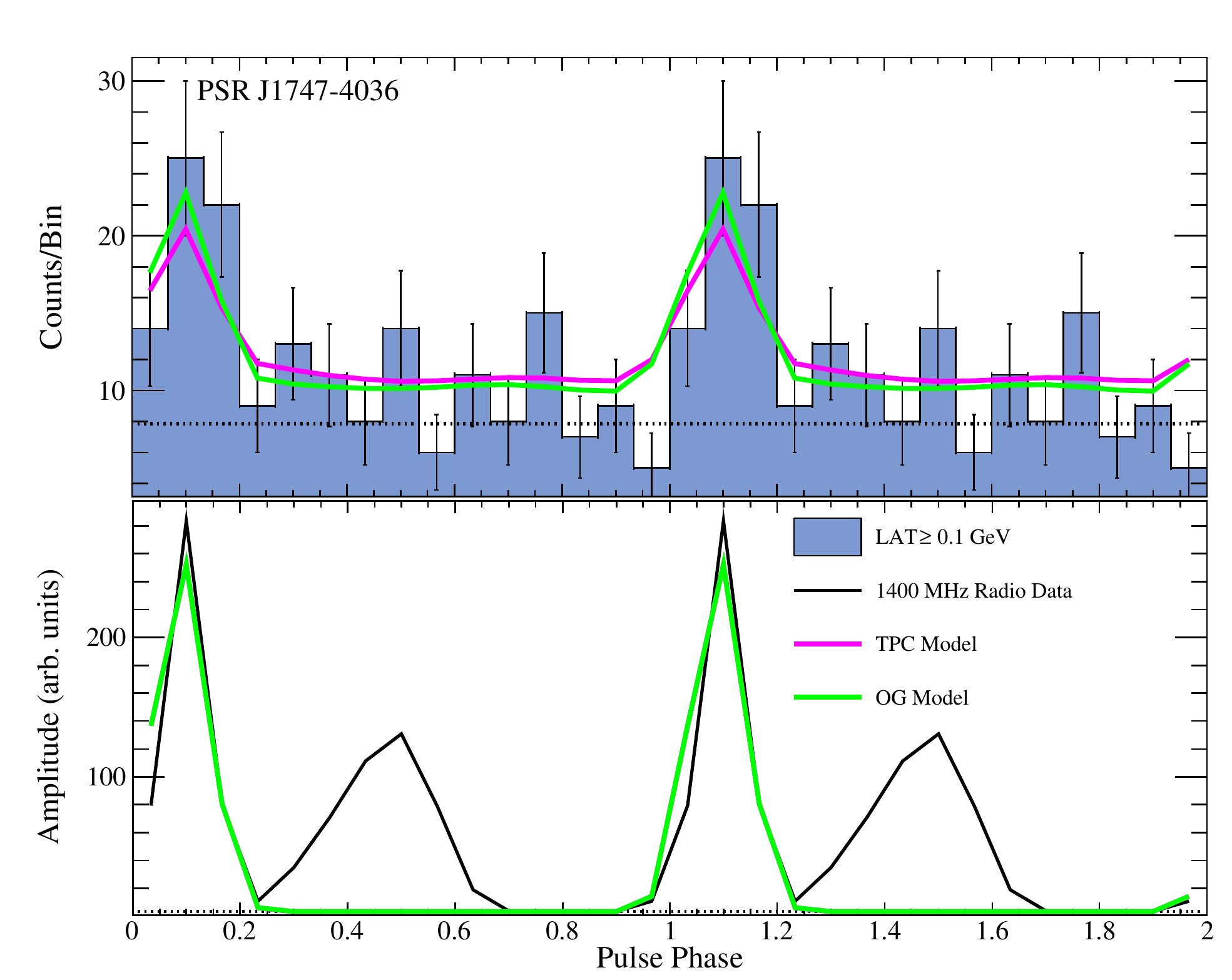}
\end{center}
\caption{Best-fit gamma-ray (\emph{top}) and radio (\emph{bottom}) light curves for PSR J1747$-$4036.  The dashed line (solid green in the online version) is the OG model, best-fit parameters in Table \ref{tbl-og}.  The solid gray line (pink in the online version) is the TPC model, best-fit parameters are given in Table \ref{tbl-tpc}.  The best-fit geometries are the same resulting in the same radio, model light curve resulting in the OG fit being drawn on top of the TPC fit.  In both panels, the dotted horizontal lines are the estimated background levels given in Table \ref{tbl-simpars}.\label{app-J1747LCs}}
\end{figure}

Figure \ref{app-J1747LCs} presents the observed and best-fit light curves of PSR J1747$-$4036.  This is an isolated 1.65 ms discovered in radio observations of unassociated LAT sources with pulsar-like spectra, first announced by \citet{Kerr2012} (with more details presented in Camilo et al.~in preparation).  This MSP was first announced as a gamma-ray pulsar in 2PC.  The gamma-ray peak and main radio peak are nearly aligned in phase but there is no gamma-ray component matching the second radio peak and the polarization properties of this MSP (see Camilo et al.~in preparation) are not in line with most other Class II MSPs.  Both the OG and TPC models reproduce the gamma-ray light curve well and match the main radio peak, but neither can reproduce the second radio peak.

\begin{figure}
\begin{center}
\includegraphics[height=0.6\textheight]{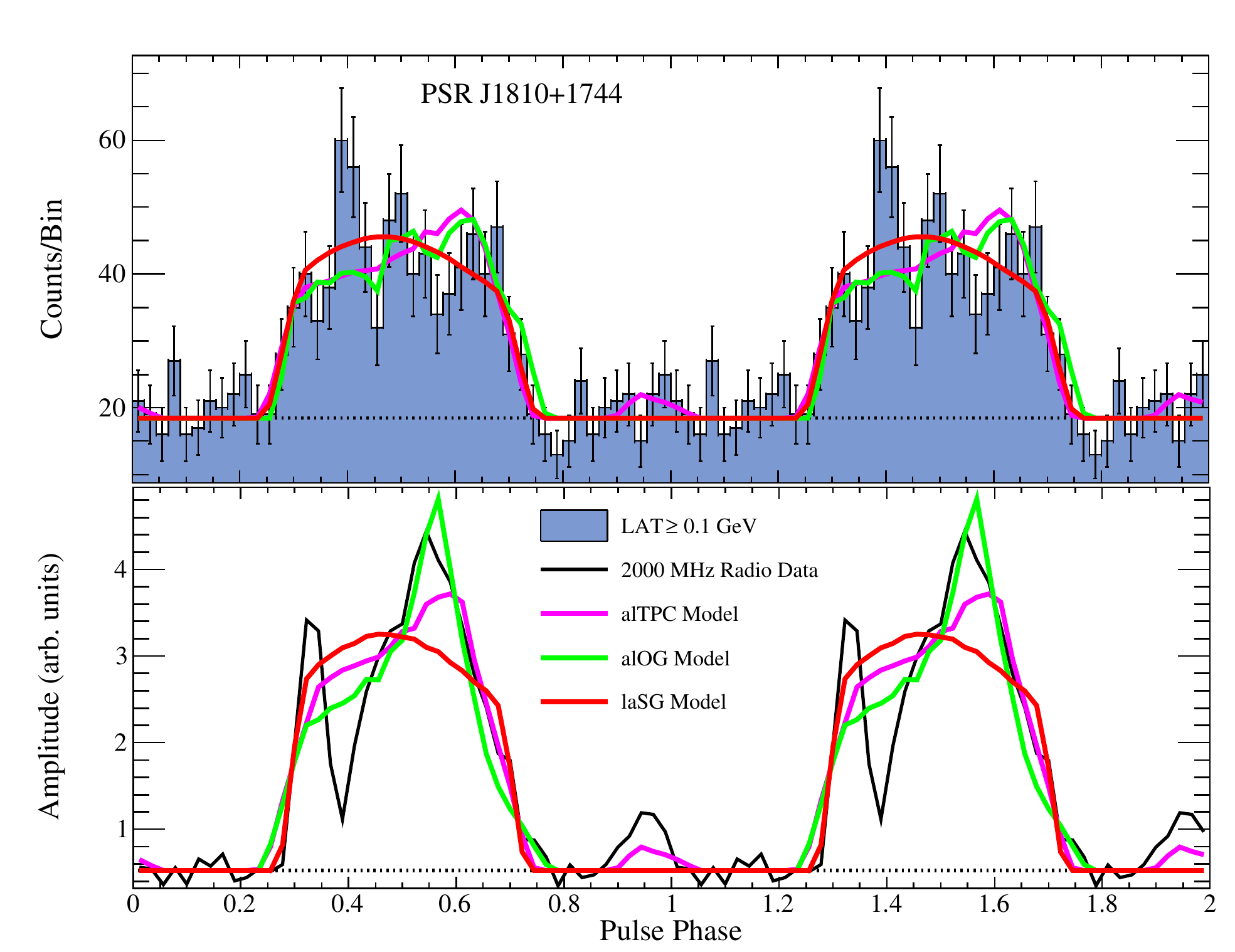}
\end{center}
\caption{Best-fit gamma-ray (\emph{top}) and radio (\emph{bottom}) light curves for PSR J1810+1744.  The dashed line (solid green in the online version) is the alOG model, best-fit parameters in Table \ref{tbl-alog}.  The solid gray line (pink in the online version) is the alTPC model, best-fit parameters are given in Table \ref{tbl-altpc}.  The dash-dot, light gray line (solid red in the online version) is the laSG model, best-fit parameters are given in Table \ref{tbl-lasg}.  In both panels, the dotted horizontal lines are the estimated background levels given in Table \ref{tbl-simpars}.\label{app-J1810LCs}}
\end{figure}

Figure \ref{app-J1810LCs} presents the observed and best-fit light curves of PSR J1810+1744.  This is a 1.66 ms pulsar in a binary system with a 0.15 day orbital period discovered in radio observations of unassociated LAT sources with pulsar-like characteristics (Hessels et al.~in preparation).  This MSP was first announced as a gamma-ray pulsar in 2PC.  The broad gamma-ray and radio peaks occur at the same phase, but the shapes do not match exactly and it is unclear if the smaller radio peak at phase 0.9 has a matching feature in gamma rays.  The alTPC, alOG, and laSG models all match the qualitative features of the gamma-ray and radio light curves, with the alOG model coming the closest to reproducing the three gamma-ray peaks in the observed light curve and only the alTPC model matching the smaller radio peak.  The optical light curve of this MSP has been modeled by \citet{Breton13} and \citet{SH14}.  The first authors found $i\ =\ 48^{\circ}\pm7^{\circ}$, though they note that their fit was poor and unable to match all of the data points.  The latter authors used two different models, finding $i\ =\ 56.75^{\circ}\pm2.25^{\circ}$ and $54.75^{\circ}\pm2.75$, compatible with the results of \citet{Breton13}; however, they were unable to produce fits that led to a realistic neutron star mass or X-ray efficiency.  This does not agree well with any of our fit values of \ze\ for this MSP and only the confidence contours for the laSG fit allow for solutions with $\zeta\sim55^{\circ}$.

\begin{figure}
\begin{center}
\includegraphics[height=0.6\textheight]{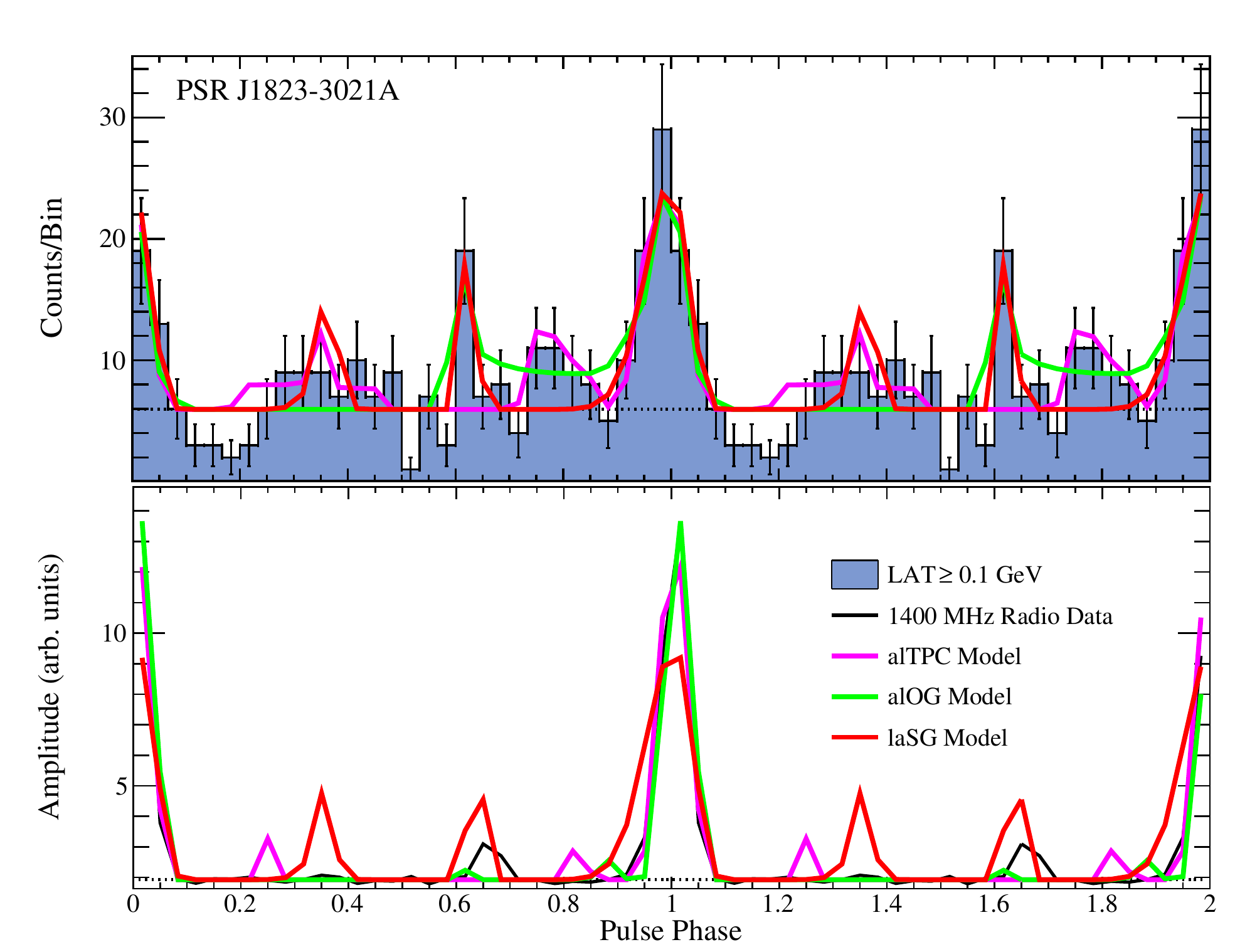}
\end{center}
\caption{Best-fit gamma-ray (\emph{top}) and radio (\emph{bottom}) light curves for PSR J1823$-$3021A.  The dashed line (solid green in the online version) is the alOG model, best-fit parameters in Table \ref{tbl-alog}.  The solid gray line (pink in the online version) is the alTPC model, best-fit parameters are given in Table \ref{tbl-altpc}.  The dash-dot, light gray line (solid red in the online version) is the laSG model, best-fit parameters are given in Table \ref{tbl-lasg}.  In both panels, the dotted horizontal lines are the estimated background levels given in Table \ref{tbl-simpars}.\label{app-J1823LCs}}
\end{figure}

Figure \ref{app-J1823LCs} presents the observed and best-fit light curves of PSR J1823$-$3021A.  This is an isolated 5.44 ms pulsar discovered in the globular cluster NGC 6624 by \citet{Biggs94}.  Gamma-ray pulsations from this MSP were first reported by \citet{Freire11} who suggested, based on the inferred gamma-ray efficiency, that the \Pd\ (and thus \Ed, see Table \ref{tbl-vitals}) of this MSP was largely intrinsic and, therefore, this is an extremely-luminous gamma-ray pulsar.  \citet{Stairs99} reported polarimetric observations of this MSP, which show no position-angle swing across the pulse and, in fact, are consistent with a mean polarization of 0\%.  The estimated background level in the top panel of Figure \ref{app-J1823LCs} may be overestimated, making it unclear how large the off-peak interval is and how many components the gamma-ray light curve has.  The alTPC, alOG, and laSG models all match the main gamma-ray peak well but predict different peak multiplicities.  The predicted radio profiles suffer from the same issue.

\begin{figure}
\begin{center}
\includegraphics[height=0.6\textheight]{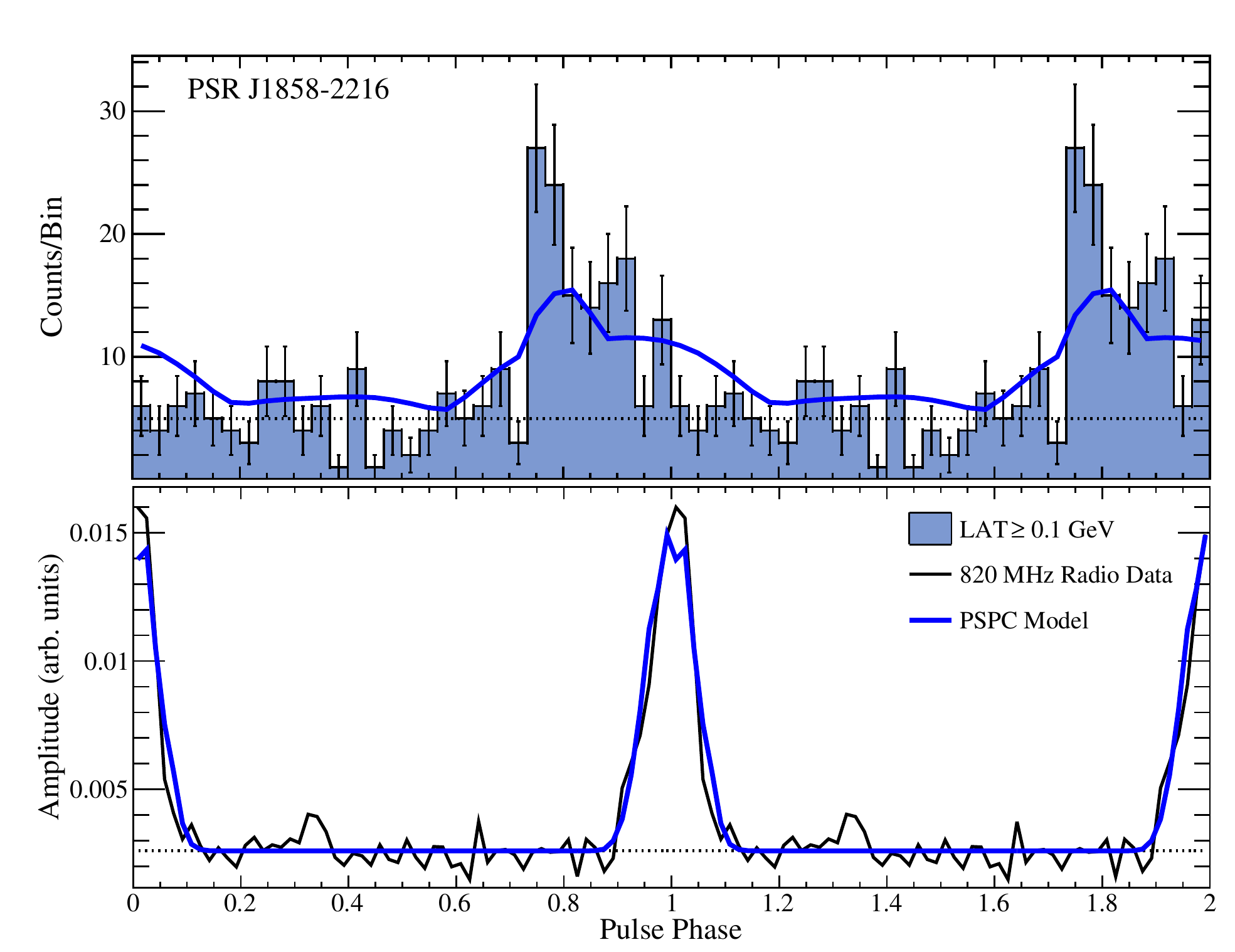}
\end{center}
\caption{Best-fit gamma-ray (\emph{top}) and radio (\emph{bottom}) light curves for PSR J1858$-$2216.  The solid gray line (blue in the online version) is the PSPC model, best-fit parameters are given in Table \ref{tbl-pspc}.  In both panels, the dotted horizontal lines are the estimated background levels given in Table \ref{tbl-simpars}.\label{app-J1858LCs}}
\end{figure}

Figure \ref{app-J1858LCs} presents the observed and best-fit light curves of PSR J1858$-$2216.  This is 2.38 ms pulsar in a binary system with a 46 day orbital period discovered in radio observations of unassociated LAT sources with pulsar-like characteristics (Sanpa-Arsa et al.~in preparation).  This MSP was first announced as a gamma-ray pulsar in 2PC.  The PSPC model reproduces the gamma-ray peak well, though the predicted peak is not sharp enough and the predicted level of off-peak emission may not be seen with LAT.  The best-fit geometry matches the main radio peak well but does not match the possible low-level feature between phases of 0.2 and 0.4, though it is unclear, from our background level estimate, if this feature is real.

\begin{figure}
\begin{center}
\includegraphics[height=0.6\textheight]{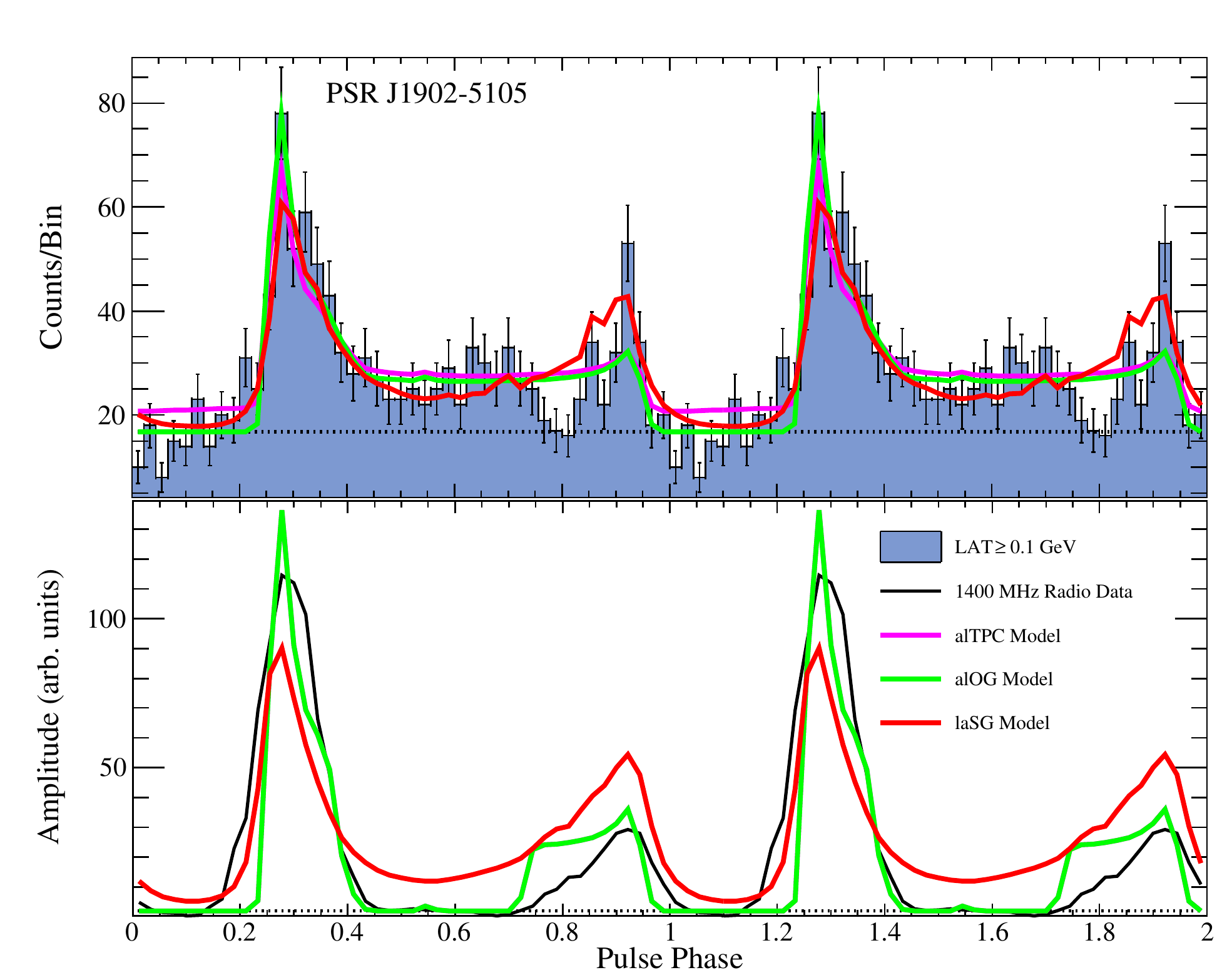}
\end{center}
\caption{Best-fit gamma-ray (\emph{top}) and radio (\emph{bottom}) light curves for PSR J1902$-$5105.  The dashed line (solid green in the online version) is the alOG model, best-fit parameters in Table \ref{tbl-alog}.  The solid gray line (pink in the online version) is the alTPC model, best-fit parameters are given in Table \ref{tbl-altpc}.  The dash-dot, light gray line (solid red in the online version) is the laSG model, best-fit parameters are given in Table \ref{tbl-lasg}.  In the radio panel, the altitude-limited models predict the same radio light curve resulting in the alOG fit being drawn on top of the alTPC fit.  In both panels, the dotted horizontal lines are the estimated background levels given in Table \ref{tbl-simpars}.\label{app-J1902LCs}}
\end{figure}

Figure \ref{app-J1902LCs} presents the observed and best-fit light curves of PSR J1902$-$5105.  This is a 1.74 ms pulsar in a binary system with a 2.0 day orbital period discovered in radio observations of unassociated LAT sources with pulsar-like characteristics, first announced by \citet{Kerr2012} (with more details in Camilo et al.~in preparation).  This MSP was first announced as a gamma-ray pulsar in 2PC.  The radio emission from this pulsar shows little to no polarization (Camilo et al.~in preparation).  The alOG, alTPC, and laSG models all match the observed gamma-ray and radio light curves of this MSP well.  None of the models match the sharpness of the gamma-ray peak near phase 0.9 properly, but this may be due to the likelihood putting some emphasis at matching the apparent bridge emission.  None of the models exactly match the shapes of the radio peaks but the alTPC and alOG models do better at predicting the proper peak-height ratio.

\begin{figure}
\begin{center}
\includegraphics[height=0.6\textheight]{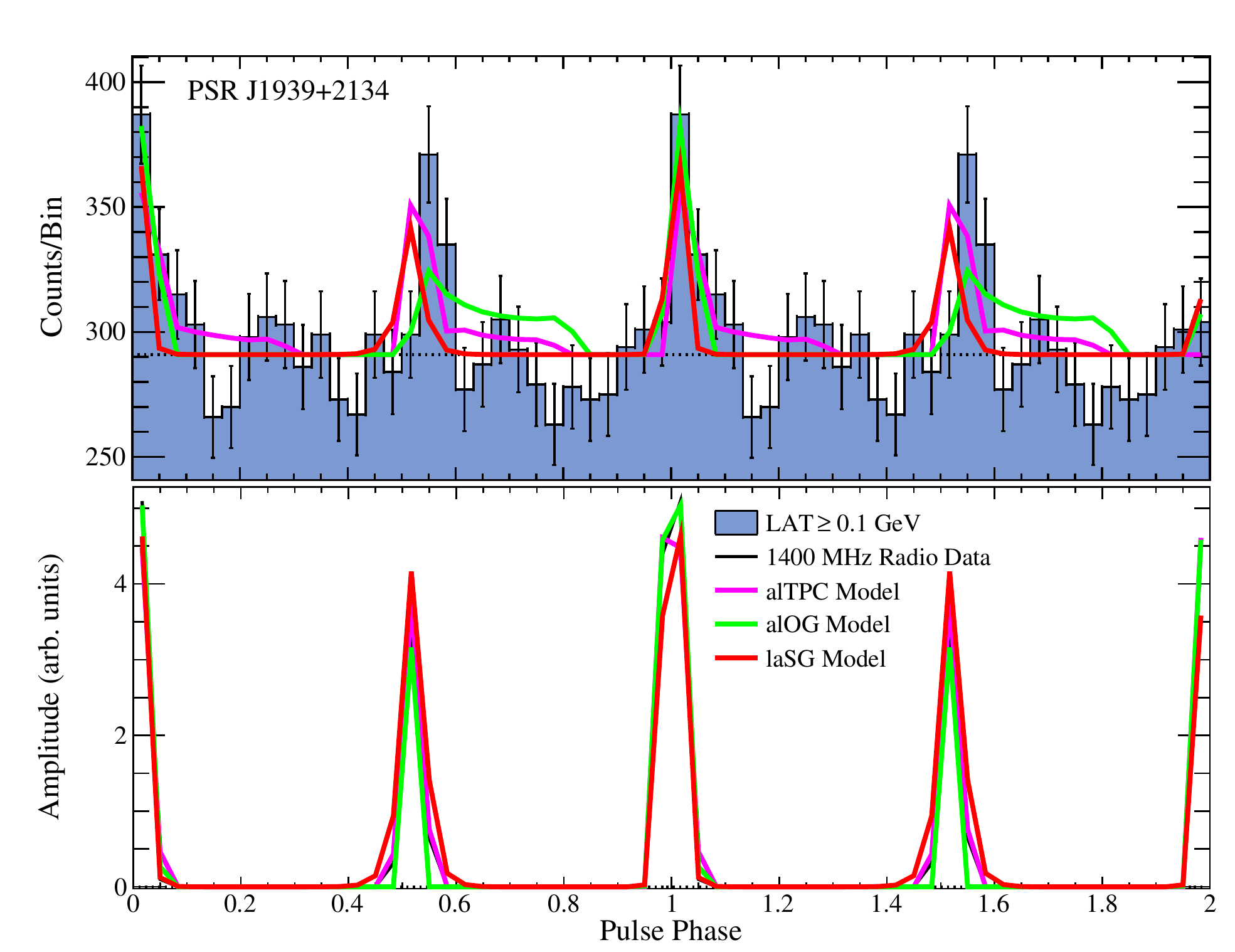}
\end{center}
\caption{Best-fit gamma-ray (\emph{top}) and radio (\emph{bottom}) light curves for PSR J1939+2134.  The dashed line (solid green in the online version) is the alOG model, best-fit parameters in Table \ref{tbl-alog}.  The solid gray line (pink in the online version) is the alTPC model, best-fit parameters are given in Table \ref{tbl-altpc}.  The dash-dot, light gray line (solid red in the online version) is the laSG model, best-fit parameters are given in Table \ref{tbl-lasg}.  In both panels, the dotted horizontal lines are the estimated background levels given in Table \ref{tbl-simpars}.\label{app-J1939LCs}}
\end{figure}

Figure \ref{app-J1939LCs} presents the observed and best-fit light curves of PSR J1939+2134 \citep[B1937+21, the first MSP ever discovered,][]{Backer82}.  This is an isolated 1.56 ms pulsar.  Gamma-ray pulsations from this MSP were first reported by \citet{Guillemot12} who also predicted the viewing geometry from RVM fits to be \al=89\DG\ and $\beta=-3^{\circ}$ with a radio emission altitude of 0.65 \Rlc.  Of the Class II MSPs for which polarimetric observations have been reported, this is the only one with significant levels of polarized emission.  Several authors have reported polarimetric observations for this MSP \citep{Ashworth83,Stinebring83,SC83,TS90,Xil98,Stairs99,Ord2004,Yan11} and all agree on the basic properties of the emission.  The near 180\DG\ phase separation of the peaks suggests an orthogonal rotator viewed near the spin equator (i.e., both \al\ and \ze\ near 90\DG), \citet{Stinebring83} noted that the polarization is consistent with this geometry if the depolarization near the inner peak edges was a sign of orthogonal mode switches.  This depolarization for an orthogonal rotator  would also be expected from outer-magnetospheric emission models \citep{Dyks04a}.  \citet{TS90} noted that the narrowness of the peaks is at odds with the typical interpretation of a core beam for an orthogonal rotator.  A caustic origin of the emission does naturally explain the narrow peaks, but so would partially-filled cone beams \citep{Kramer99}.  The alOG, alTPC, and laSG models all predict the observed gamma-ray and radio light curves well, though none exactly matches the phase of the gamma-ray peak near phase 0.5.  Only the best-fit geometry corresponding to the alTPC fit agrees well with the RVM fits of \citet{Guillemot12}, though the confidence contours for the alOG fit are not simply connected and do allow a fit consistent with their values within the 95\% confidence-level region.

\begin{figure}
\begin{center}
\includegraphics[height=0.6\textheight]{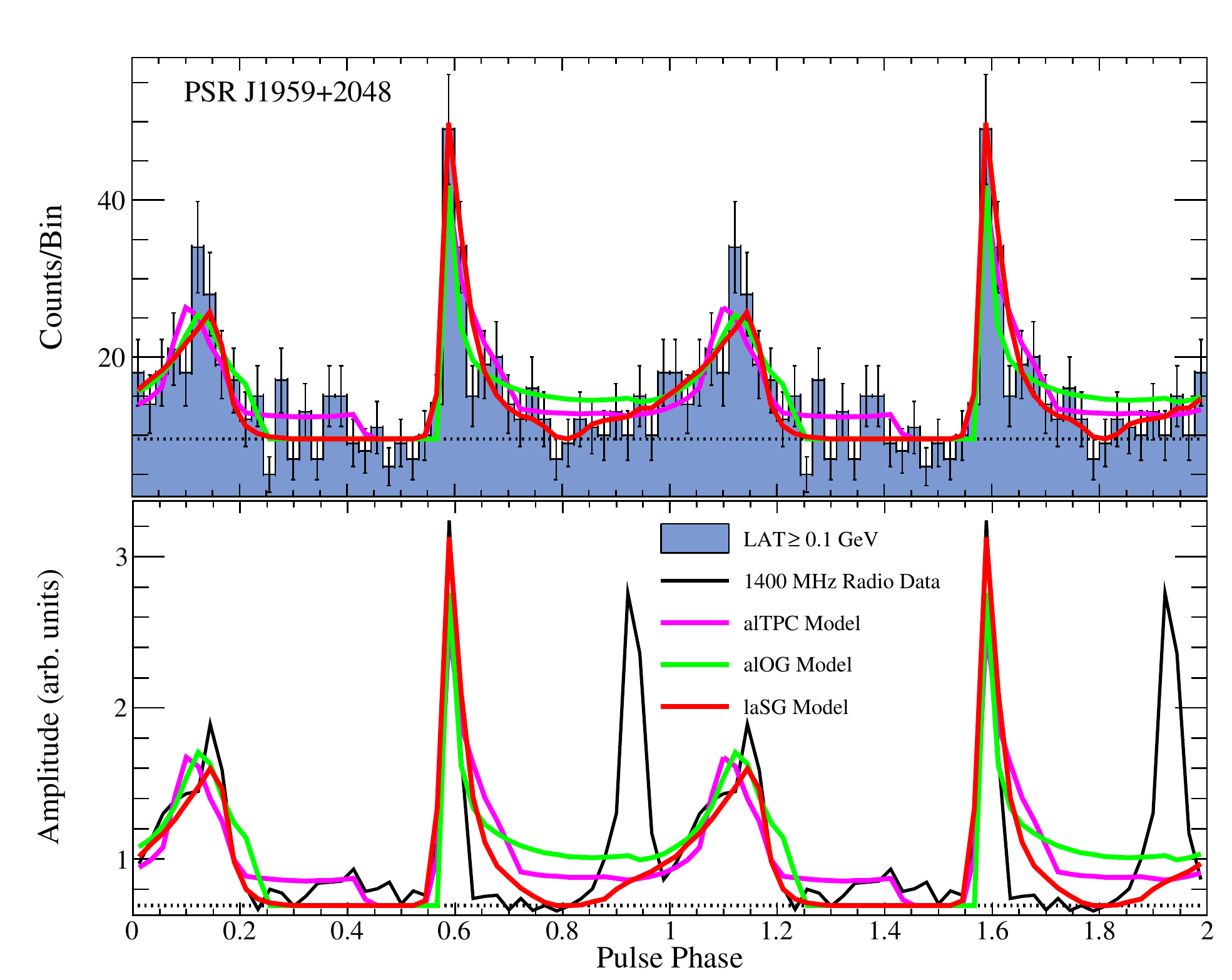}
\end{center}
\caption{Best-fit gamma-ray (\emph{top}) and radio (\emph{bottom}) light curves for PSR J1959+2048.  The dashed line (solid green in the online version) is the alOG model, best-fit parameters in Table \ref{tbl-alog}.  The solid gray line (pink in the online version) is the alTPC model, best-fit parameters are given in Table \ref{tbl-altpc}.  The dash-dot, light gray line (solid red in the online version) is the laSG model, best-fit parameters are given in Table \ref{tbl-lasg}.  In both panels, the dotted horizontal lines are the estimated background levels given in Table \ref{tbl-simpars}.\label{app-J1959LCs}}
\end{figure}

Figure \ref{app-J1959LCs} presents the observed and best-fit light curves of PSR J1959+2048 \citep[B1957+20, the first ``black-widow'' pulsar ever discovered,][]{Fruchter88}.  This is a 1.61 ms pulsar in a binary system with a 0.38 day orbital period and a very low-mass ($0.022\ M_{\odot}$) companion.  The companion is thought to have reached such a low mass from ablation by the pulsar wind, thus the ``black widow'' moniker.  Gamma-ray pulsations were first reported from PSR J1959+2048 by \citet{Guillemot12}.  \citet{TS90} reported polarimetric observations of this MSP, finding very-low levels of linear polarization ($<2$\% of the total intensity).  \citet{Reynolds07} used optical observations of the companion of PSR J1959+2048 to constrain the inclination of the orbit to be $i=65^{\circ}\pm2^{\circ}$, assuming a pulsar mass between 1.3 and 1.9 $M_{\odot}$.  The alOG, alTPC, and laSG models are all able to reproduce the observed gamma-ray light curves well and match the first and third radio peaks, but none of the models can match the second radio peak which does not have a corresponding feature in the gamma-ray light curve.  The second radio peak, near phase 0.9, is not present at lower frequencies.  None of the models find best-fit \ze\ values that agree well with the orbital inclination of \citet{Reynolds07}, though the laSG value is consistent within our estimated uncertainties.

\begin{figure}
\begin{center}
\includegraphics[height=0.6\textheight]{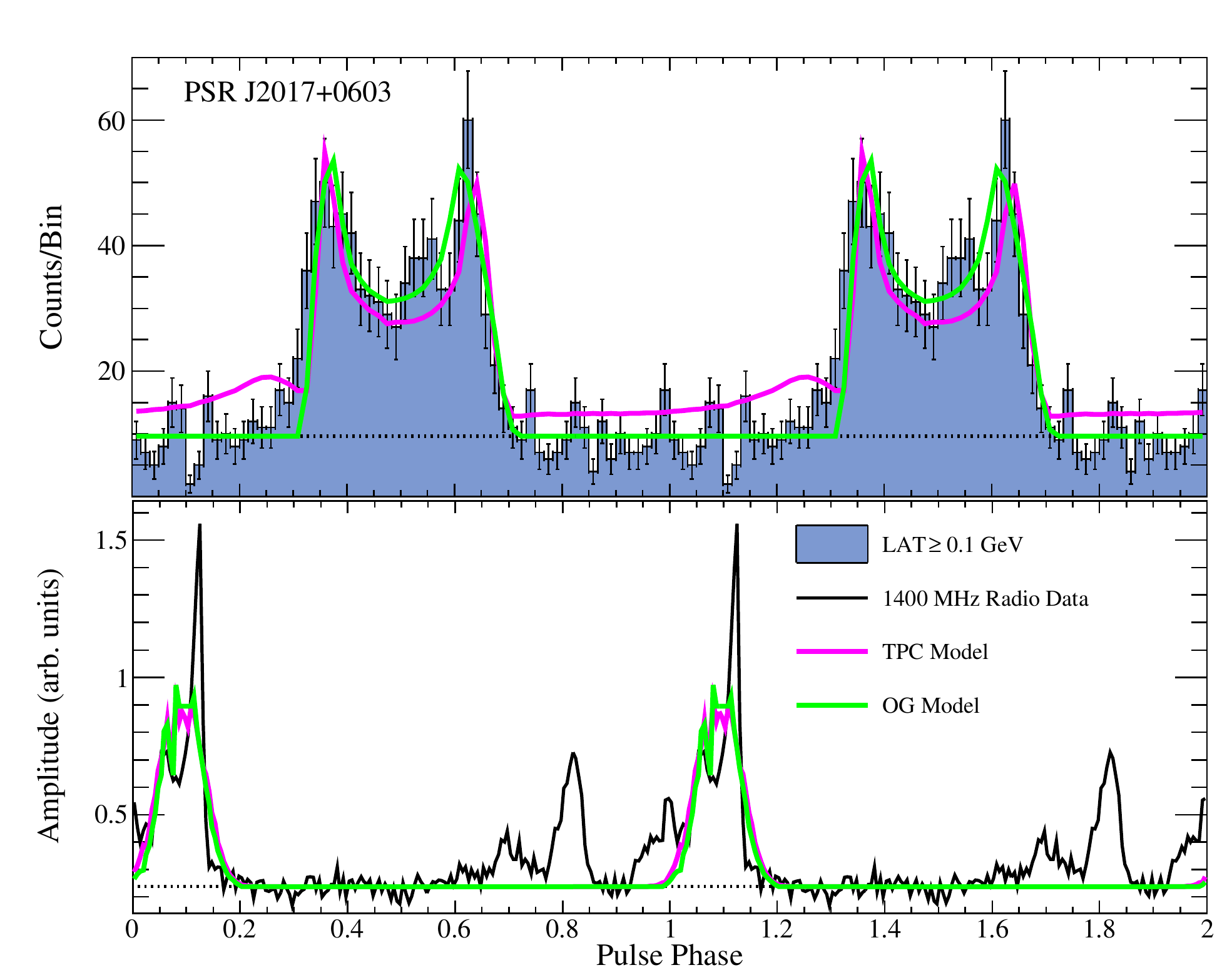}
\end{center}
\caption{Best-fit gamma-ray (\emph{top}) and radio (\emph{bottom}) light curves for PSR J2017+0603.  The dashed line (solid green in the online version) is the OG model, best-fit parameters in Table \ref{tbl-og}.  The solid gray line (pink in the online version) is the TPC model, best-fit parameters are given in Table \ref{tbl-tpc}.  In both panels, the dotted horizontal lines are the estimated background levels given in Table \ref{tbl-simpars}.\label{app-J2017LCs}}
\end{figure}

Figure \ref{app-J2017LCs} presents the observed and best-fit light curves for PSR J2017+0603.  This is 2.90 ms pulsar in a binary system with a 2.2 day orbital period discovered by \citet{Cognard11}, who also reported the first detection of gamma-ray pulsations from this MSP, in radio observations of unassociated LAT sources with pulsar-like characteristics.  The OG and TPC models both reproduce the gamma-ray peaks well, though the TPC model over predicts the off-peak emission and predicts a precursor at phase 0.3 which is not present in the data.  Neither model results in a best-fit geometry which reproduces all of the radio peaks, it is clear that a more complex radio emission model is needed to explain the observed light curve.

\begin{figure}
\begin{center}
\includegraphics[height=0.6\textheight]{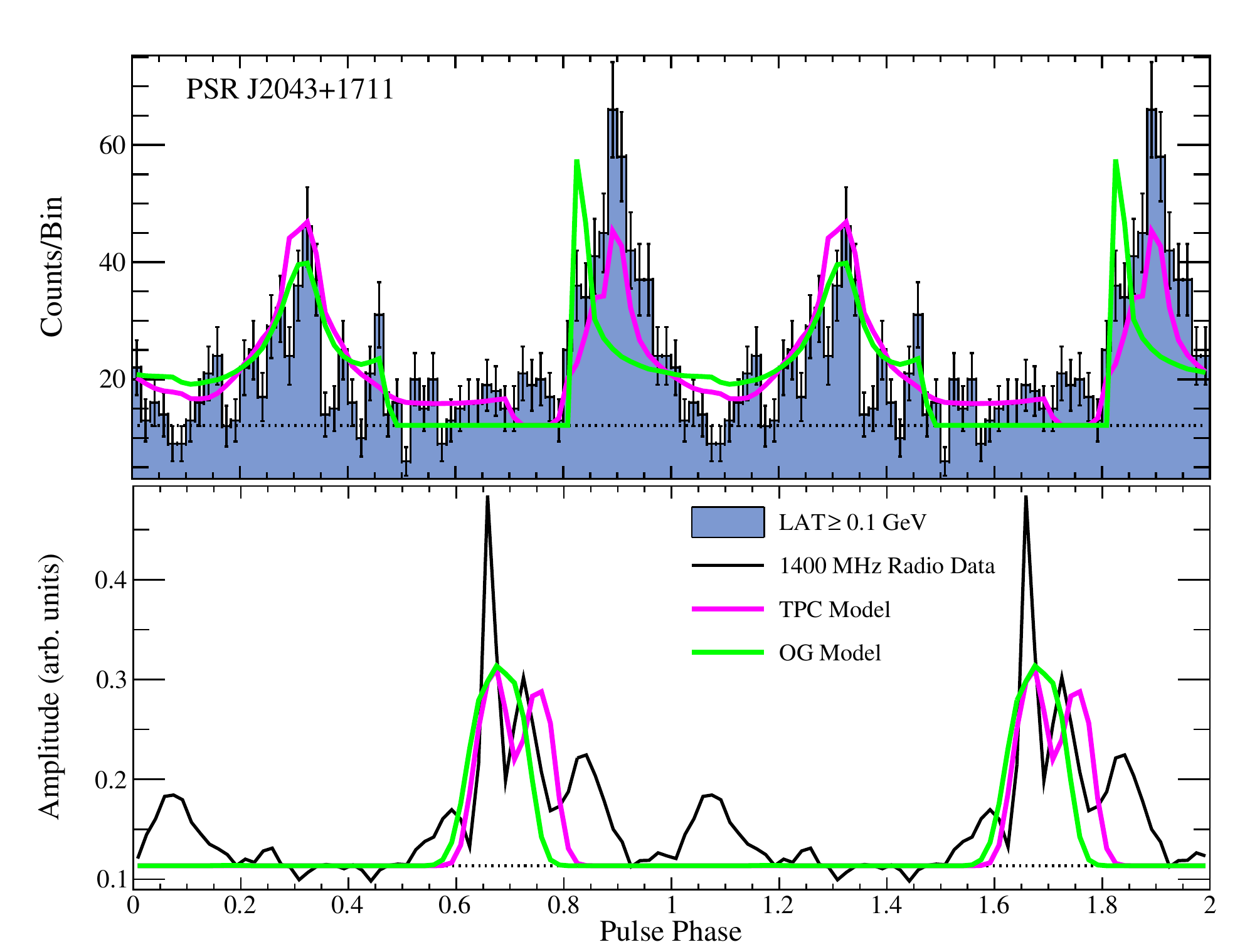}
\end{center}
\caption{Best-fit gamma-ray (\emph{top}) and radio (\emph{bottom}) light curves for PSR J2043+1711.  The dashed line (solid green in the online version) is the OG model, best-fit parameters in Table \ref{tbl-og}.  The solid gray line (pink in the online version) is the TPC model, best-fit parameters are given in Table \ref{tbl-tpc}.  In both panels, the dotted horizontal lines are the estimated background levels given in Table \ref{tbl-simpars}.\label{app-J2043LCs}}
\end{figure}

Figure \ref{app-J2043LCs} presents the observed and best-fit light curves for PSR J2043+1711.  This is a 2.38 ms pulsar in a binary system with a 1.5 day orbital period discovered by \citet{GuillemotJ2043}, who also reported the first detection of gamma-ray pulsations from this MSP, in radio observations of unassociated LAT sources with pulsar-like characteristics.  The OG and TPC model both reproduce the observed gamma-ray light curve well, though the TPC model matches the peak separation and off-peak emission level better.  The best-fit geometries using both models predict radio profiles that agree qualitatively with the observations, but neither can match all of the observed components.

\begin{figure}
\begin{center}
\includegraphics[height=0.6\textheight]{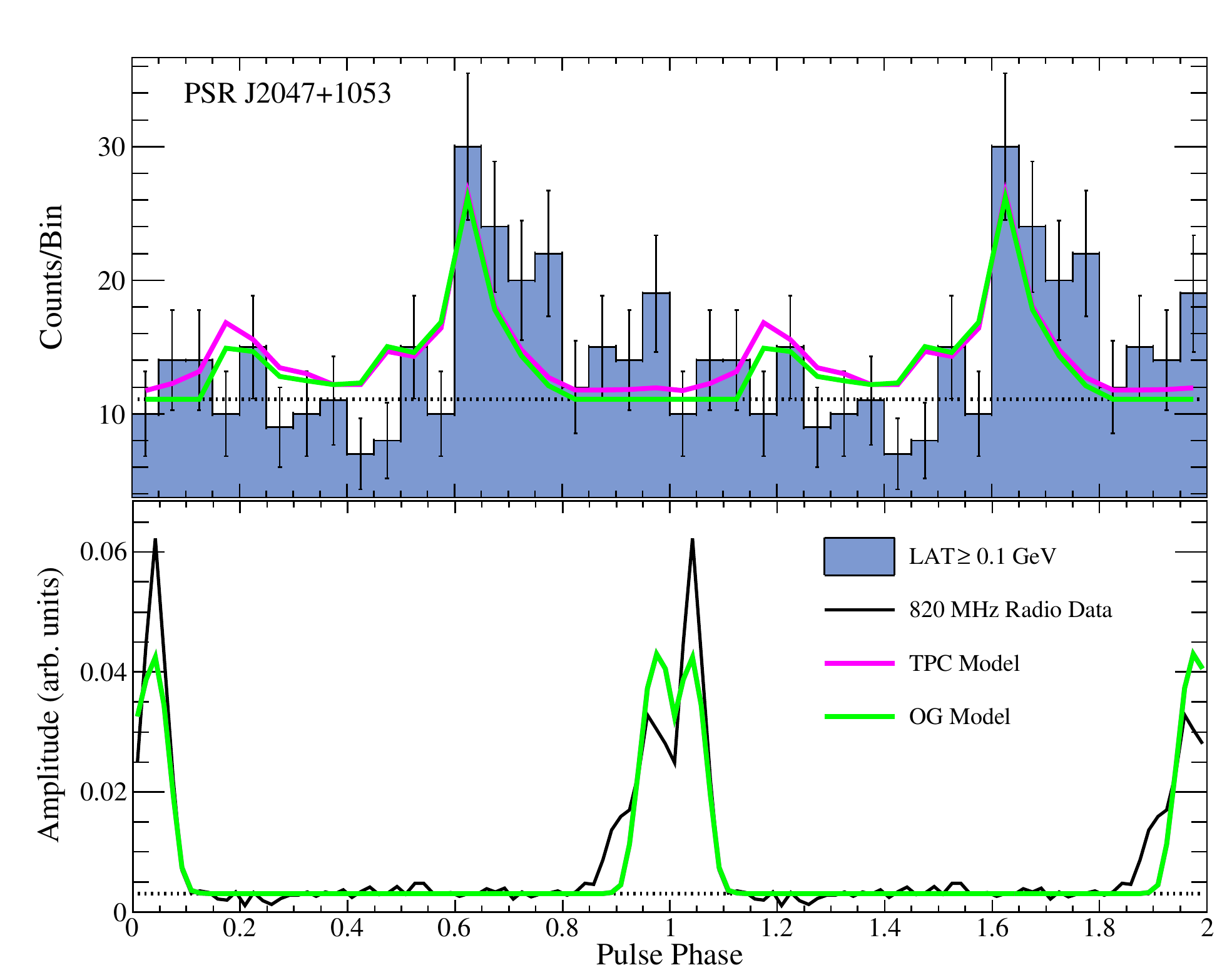}
\end{center}
\caption{Best-fit gamma-ray (\emph{top}) and radio (\emph{bottom}) light curves for PSR J2047+1053.  The dashed line (solid green in the online version) is the OG model, best-fit parameters in Table \ref{tbl-og}.  The solid gray line (pink in the online version) is the TPC model, best-fit parameters are given in Table \ref{tbl-tpc}.  The best-fit geometries are the same resulting in the same radio, model light curve resulting in the OG fit being drawn on top of the TPC fit.  In both panels, the dotted horizontal lines are the estimated background levels given in Table \ref{tbl-simpars}.\label{app-J2047LCs}}
\end{figure}

Figure \ref{app-J2047LCs} presents the observed and best-fit light curves of PSR J2047+1053. This is a 4.29 ms pulsar in a binary system with a 0.12 day orbital period discovered in radio observations of unassociated LAT sources with pulsar-like characteristics (Sanpa-Arsa et al.~in preparation).  This MSP was first announced as a gamma-ray pulsar in 2PC.  Both the TPC and OG model match the phase of the gamma-ray peak well but neither exactly matches the shape.  Both models also seem to predict the off-peak interval to be from phase 0.8 to 1.1, which does not agree well with the data.  The best-fit geometries of both fits predict two closely-spaced radio peaks, though neither exactly matches the shape and peak-height ratio.

\begin{figure}
\begin{center}
\includegraphics[height=0.6\textheight]{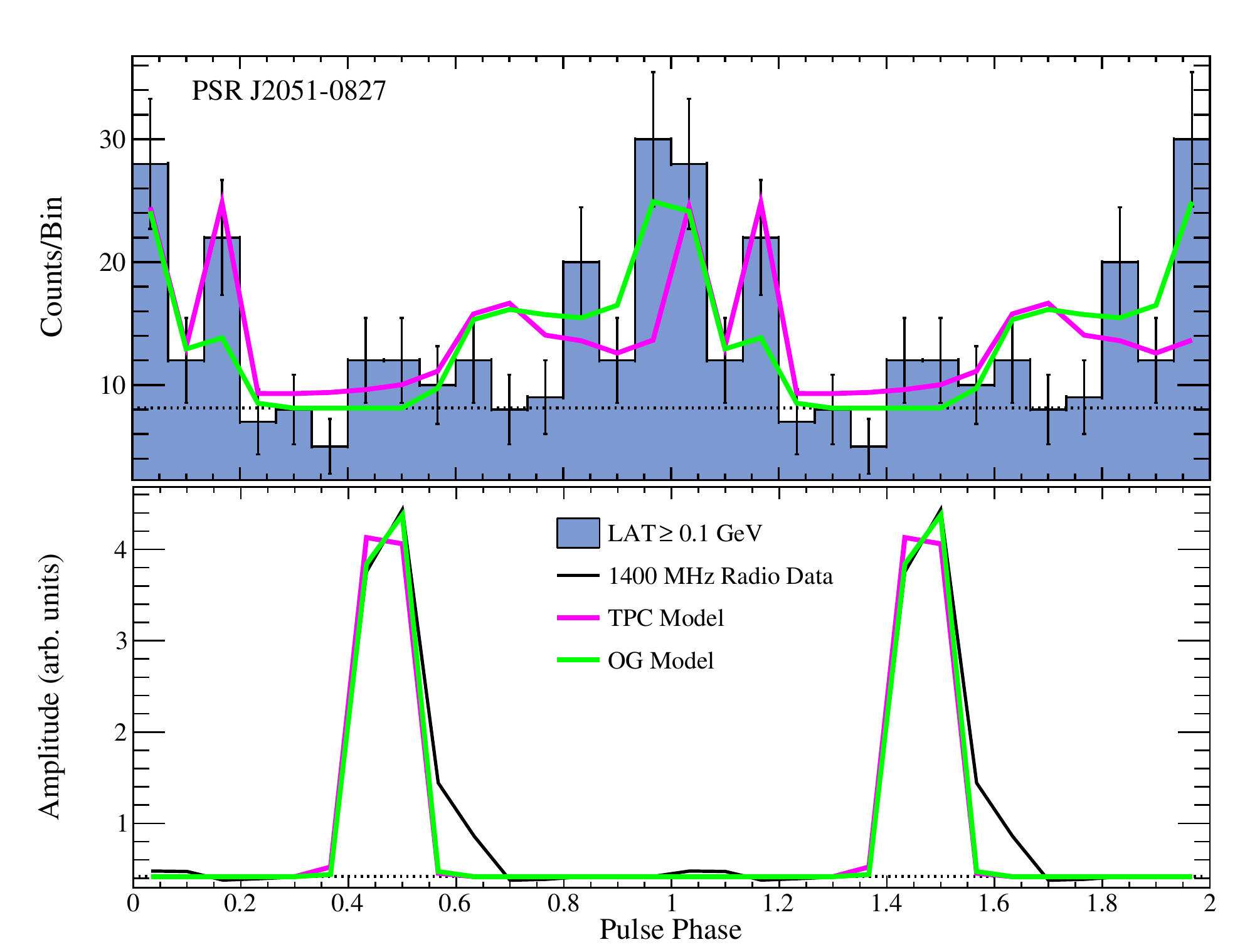}
\end{center}
\caption{Best-fit gamma-ray (\emph{top}) and radio (\emph{bottom}) light curves for PSR J2051$-$0827.  The dashed line (solid green in the online version) is the OG model, best-fit parameters in Table \ref{tbl-og}.  The solid gray line (pink in the online version) is the TPC model, best-fit parameters are given in Table \ref{tbl-tpc}.  In both panels, the dotted horizontal lines are the estimated background levels given in Table \ref{tbl-simpars}.\label{app-J2051LCs}}
\end{figure}

Figure \ref{app-J2051LCs} presents the observed and best-fit light curves of PSR J2051$-$0827.  This is a 4.51 ms pulsar in a binary system with a 0.10 day orbital period discovered by \citet{Stappers93} with eclipses seen for approximately 10\% of the orbital period at 436 MHz.  The observation of eclipses suggests that the orbital inclination angle is $\gtrsim60^{\circ}$.  Gamma-ray pulsations were first reported from this MSP by \citet{WuJ2051} and later by \citet{Espinoza13}.  Polarimetric observations of PSR J2051$-$0827 have been reported by several authors \citep{Xil98,Stairs99,Ord2004} who all agree that there is some level of linear and circular polarization but do not attempt to constrain the viewing geometry.  The OG and TPC model are both unable to reproduce the gamma-ray light curve well, though with current statistics it is unclear how many peaks are in the light curve of this MSP.  Fits with both models do reproduce the observed radio profile well.  Only the TPC model finds a best-fit $\zeta>60^{\circ}$, consistent with the observation of radio eclipses.

\begin{figure}
\begin{center}
\includegraphics[height=0.6\textheight]{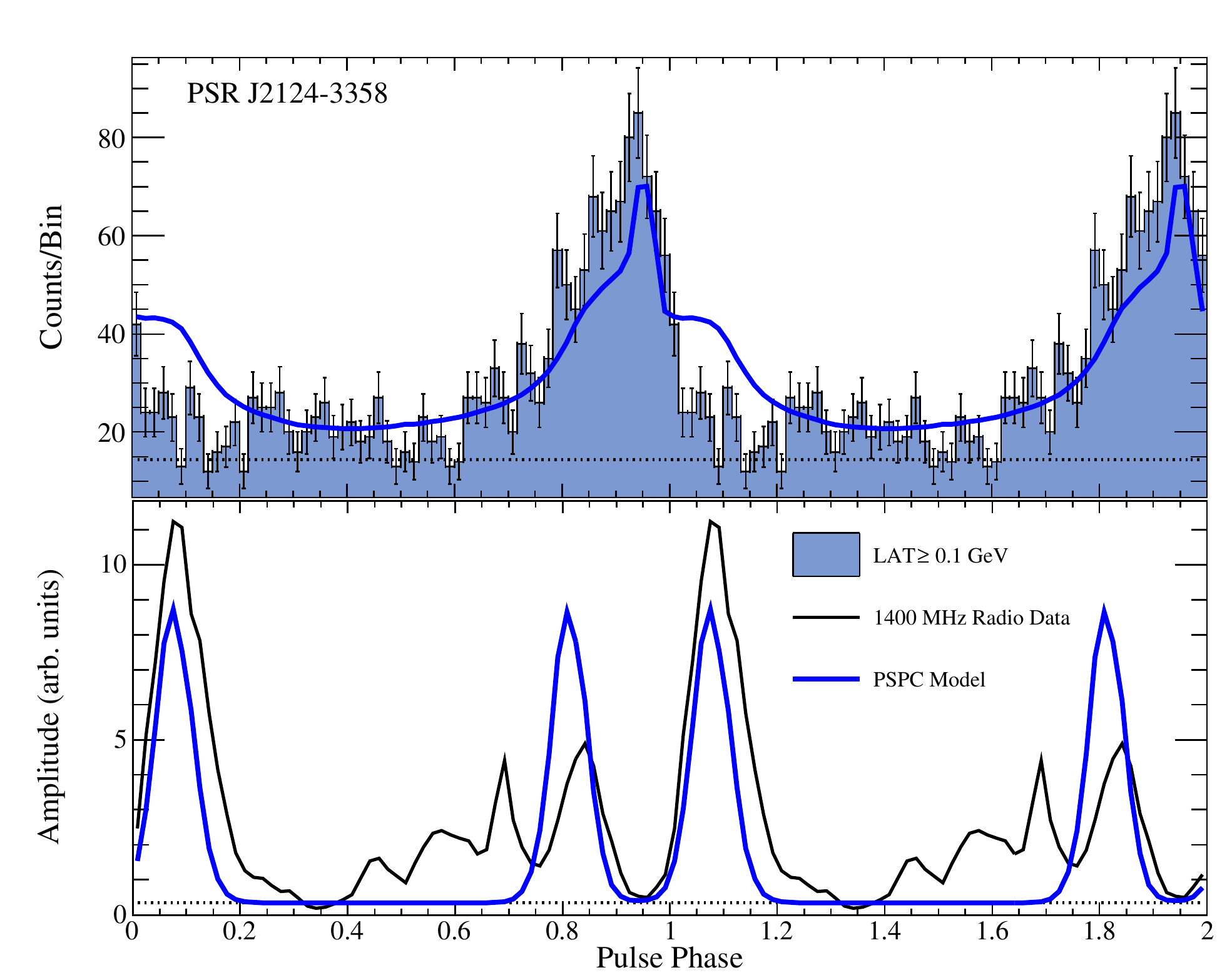}
\end{center}
\caption{Best-fit gamma-ray (\emph{top}) and radio (\emph{bottom}) light curves for PSR J2124$-$3358.  The solid gray line (blue in the online version) is the PSPC model, best-fit parameters are given in Table \ref{tbl-pspc}.  In both panels, the dotted horizontal lines are the estimated background levels given in Table \ref{tbl-simpars}.\label{app-J2124LCs}}
\end{figure}

Figure \ref{app-J2124LCs} presents the observed and best-fit light curves of PSR J2124$-$3358.  This is an isolated 4.93 ms pulsar discovered by \citet{Bailes97}.  Gamma-ray pulsations from this MSP were first reported by \citet{MSPpop}.  \citet{Ord2004}, \citet{Manchester04}, and \citet{Yan11} have reported polarimetric observations for this pulsar.  \citet{Manchester04} attempted RVM fits to the position-angle swing and found \al=$48^{\circ}\pm3^{\circ}$ and \ze=$67^{\circ}\pm5^{\circ}$ but noted a large degree of covariance in the parameters, suggesting that a more realistic range of values would be $\alpha\in[20^{\circ},60^{\circ}]$ and $\zeta\in[27^{\circ},80^{\circ}]$.  \citet{Yan11} argued that if the position angles of linearly-polarized emission for the peak between $-0.3$ and $-0.2$ in phase (corresponding to the peak near phase 0.85 in Figure \ref{app-J2124LCs}) were increased by 90\DG, the position-angle swing would match RVM predictions for $\alpha\sim25^{\circ}$ and $\beta>0$.  \citet{Bogdanov08} modeled the thermal X-ray light curve of this MSP but the fits were unconstrained in \al\ and \ze.  The PSPC model is able to reproduce the observed gamma-ray light curve of PSR J2124$-$3358 well, though the trailing shoulder on the main peak is too high in the model.  However, our hollow-cone beam model is unable to reproduce all the features observed in the radio profile, though we do match the peaks at phases 0.10 and 0.85 well.  Our best-fit geometry does predict $\beta>0$ with \al\ consistent with the prediction of \citet{Yan11} when systematic uncertainties are considered.

\begin{figure}
\begin{center}
\includegraphics[height=0.6\textheight]{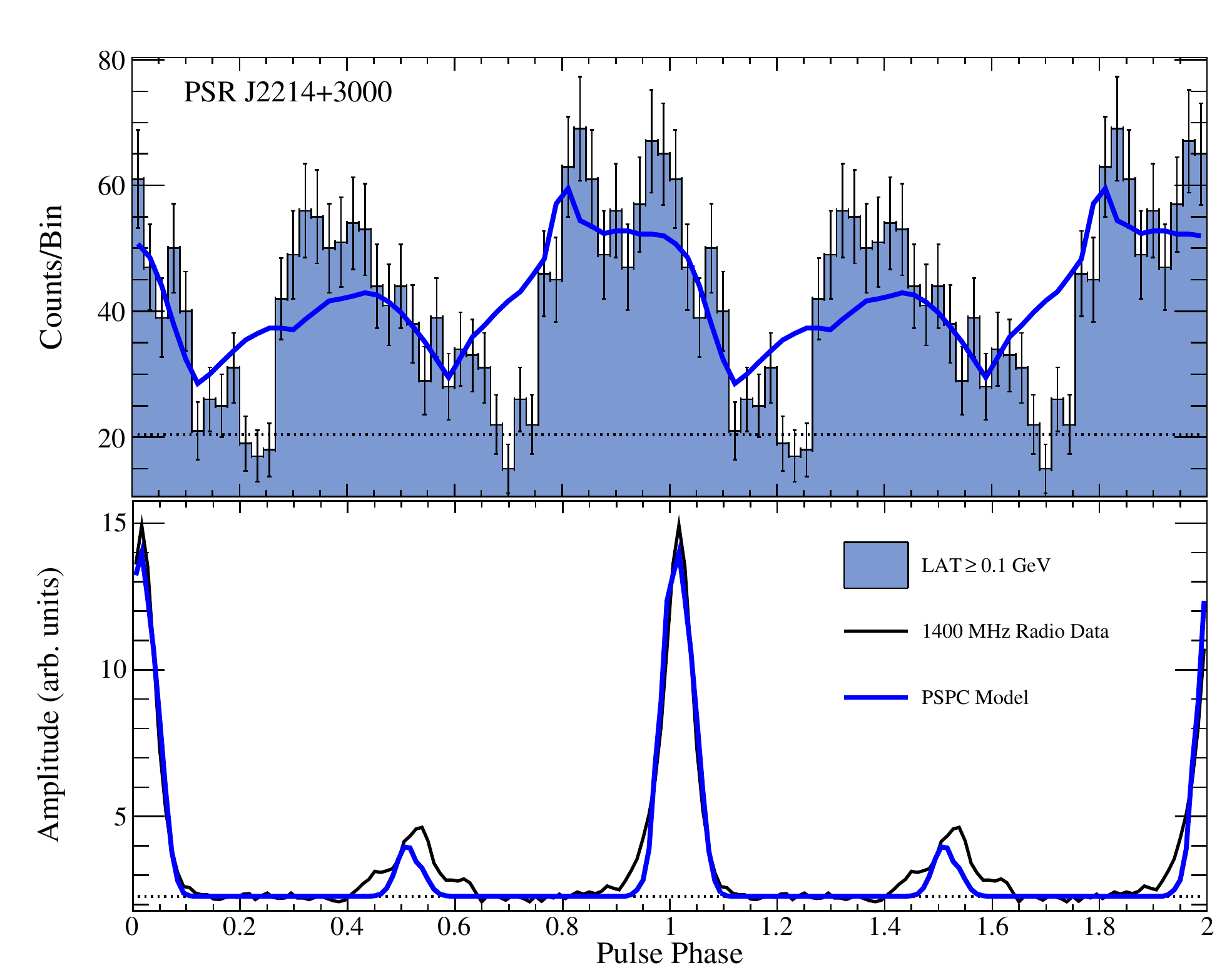}
\end{center}
\caption{Best-fit gamma-ray (\emph{top}) and radio (\emph{bottom}) light curves for PSR J2214+3000.  The solid gray line (blue in the online version) is the PSPC model, best-fit parameters are given in Table \ref{tbl-pspc}.  In both panels, the dotted horizontal lines are the estimated background levels given in Table \ref{tbl-simpars}.\label{app-J2214LCs}}
\end{figure}

Figure \ref{app-J2214LCs} presents the observed and best-fit light curves of PSR J2214+3000.  This is a 3.12 ms pulsar in a binary system with a 0.40 day orbital period discovered by \citet{Ransom11}, who also reported the first detection of gamma-ray pulsations from this MSP, in radio observations of unassociated LAT sources with pulsar-like characteristics.  PSR J2214+3000 is a ``black-widow'' pulsar but no radio eclipses have been reported, this suggests that our line of sight is significantly above the orbital plane.  The PSPC model is able to reproduce the two broad peaks observed in the gamma-ray light curve well, though the shapes are not exactly correct.  The best-fit geometry from this fit matches the observed radio profile well (both the locations in phase and peak-height ratio).  This geometry does predict \ze=$86^{\circ}\pm2^{\circ}$ which is at odds with the lack of radio eclipses in a system which is thought to have ablated away a significant fraction of the mass of its binary companion.

\begin{figure}
\begin{center}
\includegraphics[height=0.6\textheight]{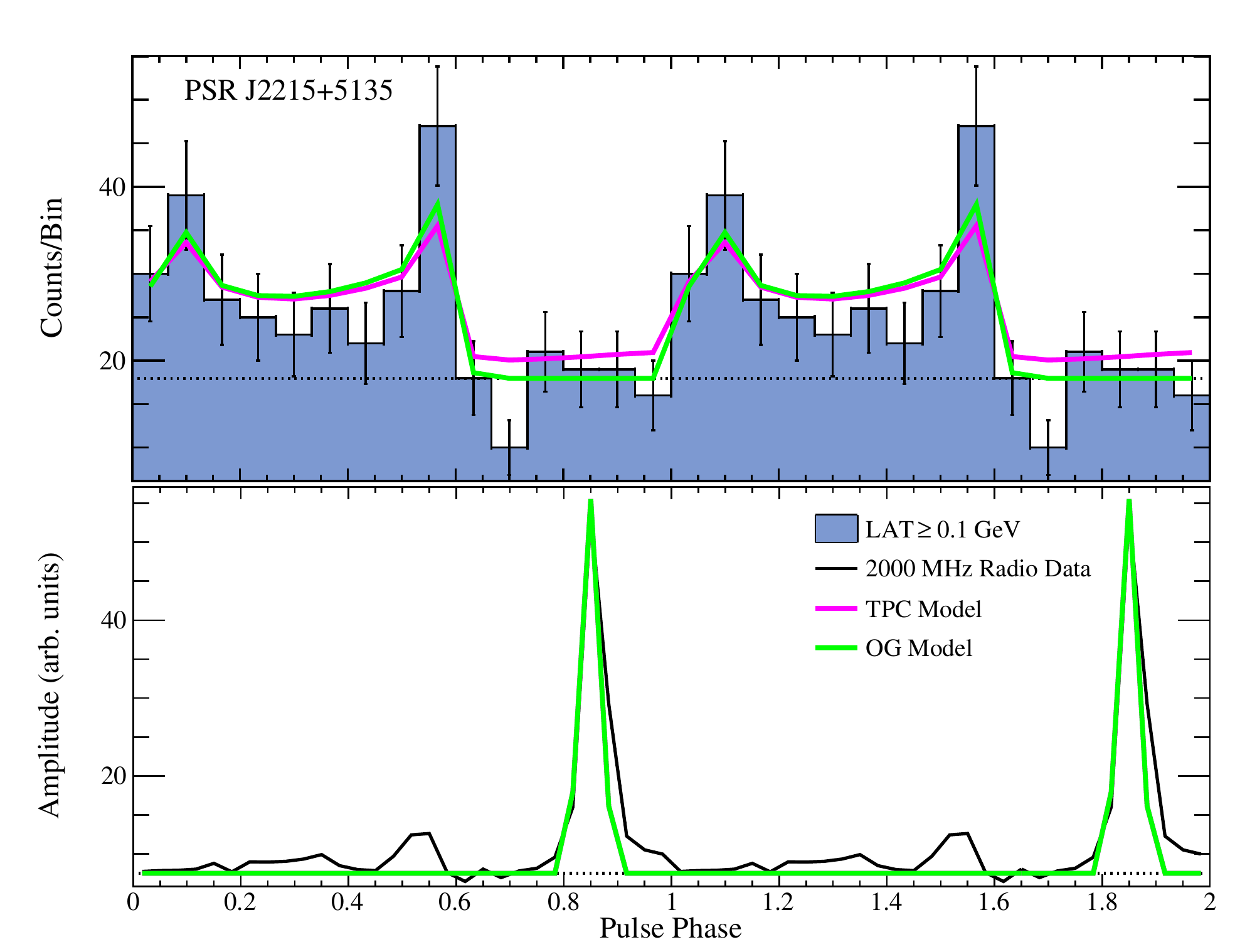}
\end{center}
\caption{Best-fit gamma-ray (\emph{top}) and radio (\emph{bottom}) light curves for PSR J2215+5135.  The dashed line (solid green in the online version) is the OG model, best-fit parameters in Table \ref{tbl-og}.  The solid gray line (pink in the online version) is the TPC model, best-fit parameters are given in Table \ref{tbl-tpc}.  The best-fit geometries are the same resulting in the same radio, model light curve resulting in the OG fit being drawn on top of the TPC fit.  In both panels, the dotted horizontal lines are the estimated background levels given in Table \ref{tbl-simpars}.\label{app-J2215LCs}}
\end{figure}

Figure \ref{app-J2215LCs} presents the observed and best-fit light curves of PSR J2215+5135.  This is a 2.61 ms pulsar in a binary system with a 0.17 day orbital period discovered in radio observations of unassociated LAT sources with pulsar-like characteristics (Hessels et al.~in preparation).  This MSP was first announced as a gamma-ray pulsar in 2PC.  Both the TPC and OG models reproduce the observed gamma-ray light curve well, and both match the main radio peak but miss the smaller peak near phase 0.5.  The optical light curve of this MSP has been modeled by \citet{Breton13} and \citet{SH14}.  The first authors found $i\ =\ 66^{\circ}\pm16^{\circ}$.  The latter authors used two different models, finding $i\ =\ (51.7^{+2.3}_{-1.5})^{\circ}$ and $(51.6^{+2.7}_{-2.1})^{\circ}$, significantly different from the results of \citet{Breton13}.  Both the TPC and OG best-fit values of \ze\ agree well with the results of \citet{Breton13}.  The OG confidence contours can accomodate values of \ze\ compatible with the best-fit $i$ values of \citet{SH14}, with \al\ near 90$^{\circ}$, at the 99\% confidence level, but the TPC contours can not.

\begin{figure}
\begin{center}
\includegraphics[height=0.6\textheight]{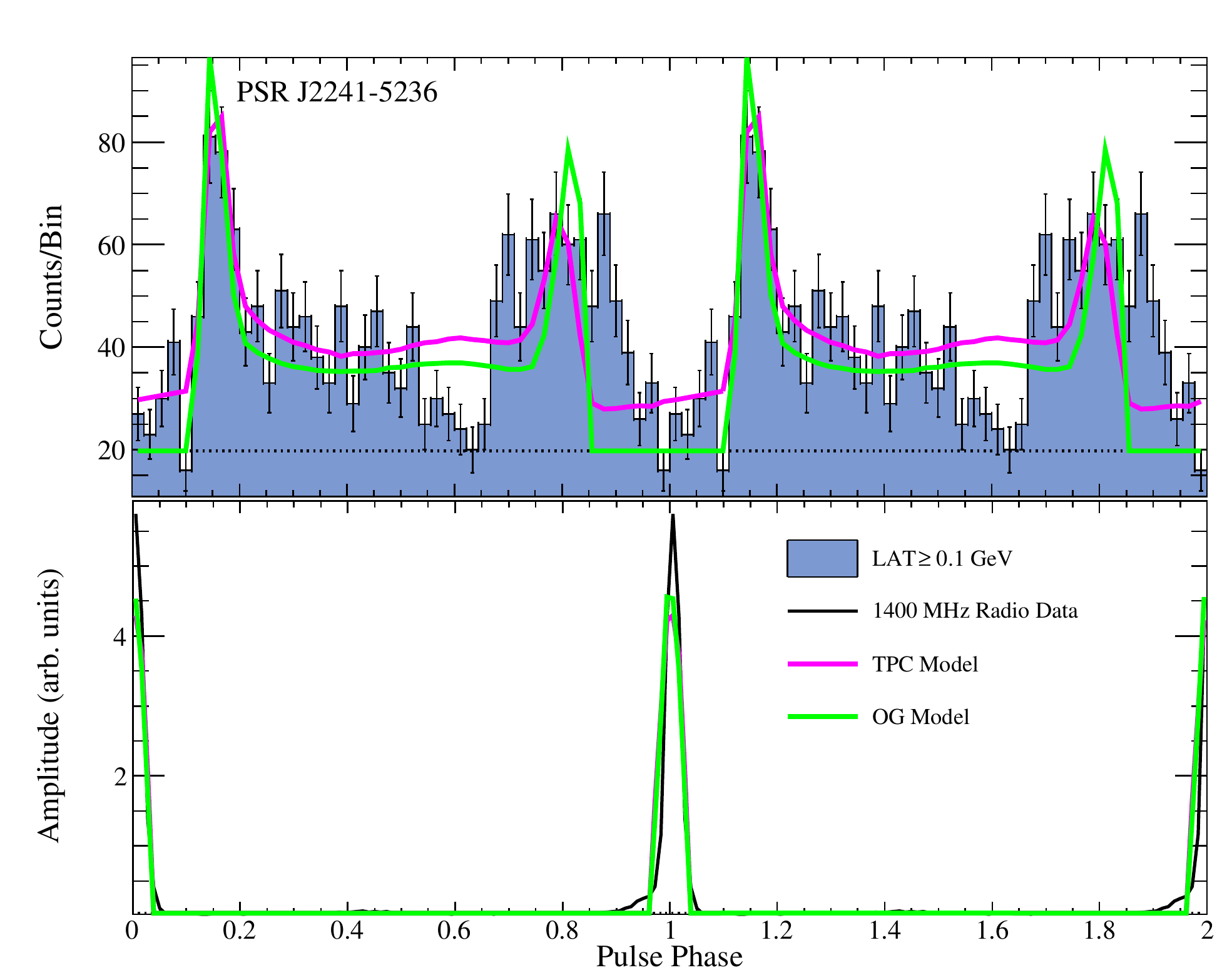}
\end{center}
\caption{Best-fit gamma-ray (\emph{top}) and radio (\emph{bottom}) light curves for PSR J2241$-$5236.  The dashed line (solid green in the online version) is the OG model, best-fit parameters in Table \ref{tbl-og}.  The solid gray line (pink in the online version) is the TPC model, best-fit parameters are given in Table \ref{tbl-tpc}.  The best-fit geometries are nearly the same resulting in similar radio, model light curve resulting in the OG fit being drawn on top of the TPC fit.  In both panels, the dotted horizontal lines are the estimated background levels given in Table \ref{tbl-simpars}.\label{app-J2241LCs}}
\end{figure}

Figure \ref{app-J2241LCs} presents the observed and best-fit light curves of PSR J2241$-$5236.  This is a 2.19 ms pulsar in a binary system with a 0.15 day orbital period discovered by \citet{Keith11}, who also reported the first detection of gamma-ray pulsations from this MSP, in radio observations of unassociated LAT sources with pulsar-like characteristics.  These authors also reported polarimetric observations for this pulsar but did not attempt to constrain the viewing geometry.  The OG and TPC model both predict two-peaked gamma-ray light curves with the correct separation, but neither model properly matches the broadness of the peak near phase 0.8.  Both models match the radio profile well.

\begin{figure}
\begin{center}
\includegraphics[height=0.6\textheight]{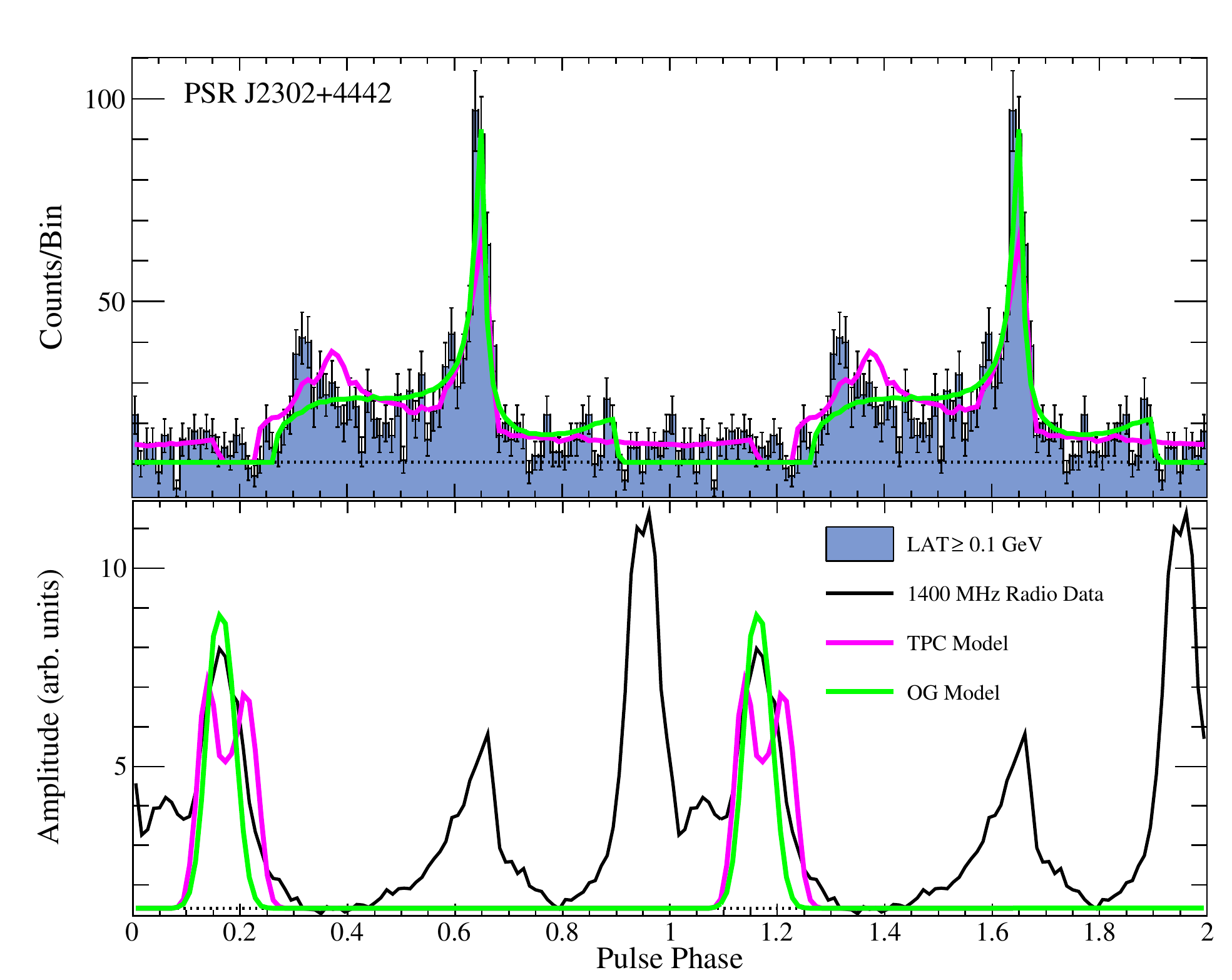}
\end{center}
\caption{Best-fit gamma-ray (\emph{top}) and radio (\emph{bottom}) light curves for PSR J2302+4442.  The dashed line (solid green in the online version) is the OG model, best-fit parameters in Table \ref{tbl-og}.  The solid gray line (pink in the online version) is the TPC model, best-fit parameters are given in Table \ref{tbl-tpc}.  In both panels, the dotted horizontal lines are the estimated background levels given in Table \ref{tbl-simpars}.\label{app-J2302LCs}}
\end{figure}

Figure \ref{app-J2302LCs} presents the observed and best-fit light curves of PSR J2302+4442.  This is a 5.19 ms pulsar in a binary system with a 51 day orbital period discovered by \citet{Cognard11}, who also reported the first detection of gamma-ray pulsations from this MSP, in radio observations of unassociated LAT sources with pulsar-like characteristics.  Both the TPC and OG model match the second gamma-ray peak well but only the TPC model predicts the first peak, though not at the correct phase.  The OG model also has difficulty matching the off-peak emission apparent in the top panel of Figure \ref{app-J2302LCs}.  Neither model is able to match all of the observed radio components.  \citet{thesis} also fit the light curves of this MSP with a similar likelihood technique and found results compatible with ours.  They noted that for the geometry corresponding to the TPC fit, substantially increasing the emission altitude of the radio cone would make the cone broader and more closely match the observed profile while leaving the predicted gamma-ray light curve unchanged.  This would not explain the radio peak near phase 0.65.

\clearpage

\section{SELECT CONFIDENCE CONTOURS}\label{app-contours}
This appendix presents several example confidence contours for each model class.  For a given MSP, if more than one fit exists the range of the x axes (\ze) are matched but the y axes (\al) are optimized for each fit.  The confidence contours show the best-fit positions from the Tables in Section \ref{res} as stars with 68\% (light gray, red in the online version), 95\% (dark gray, yellow in the online version), and 99\% (black, green in the online version) confidence contours.

\subsection{CONTOURS FOR CLASS I MSPS}\label{app-cls1cont}

\begin{figure}
\begin{center}
\includegraphics[height=0.6\textheight]{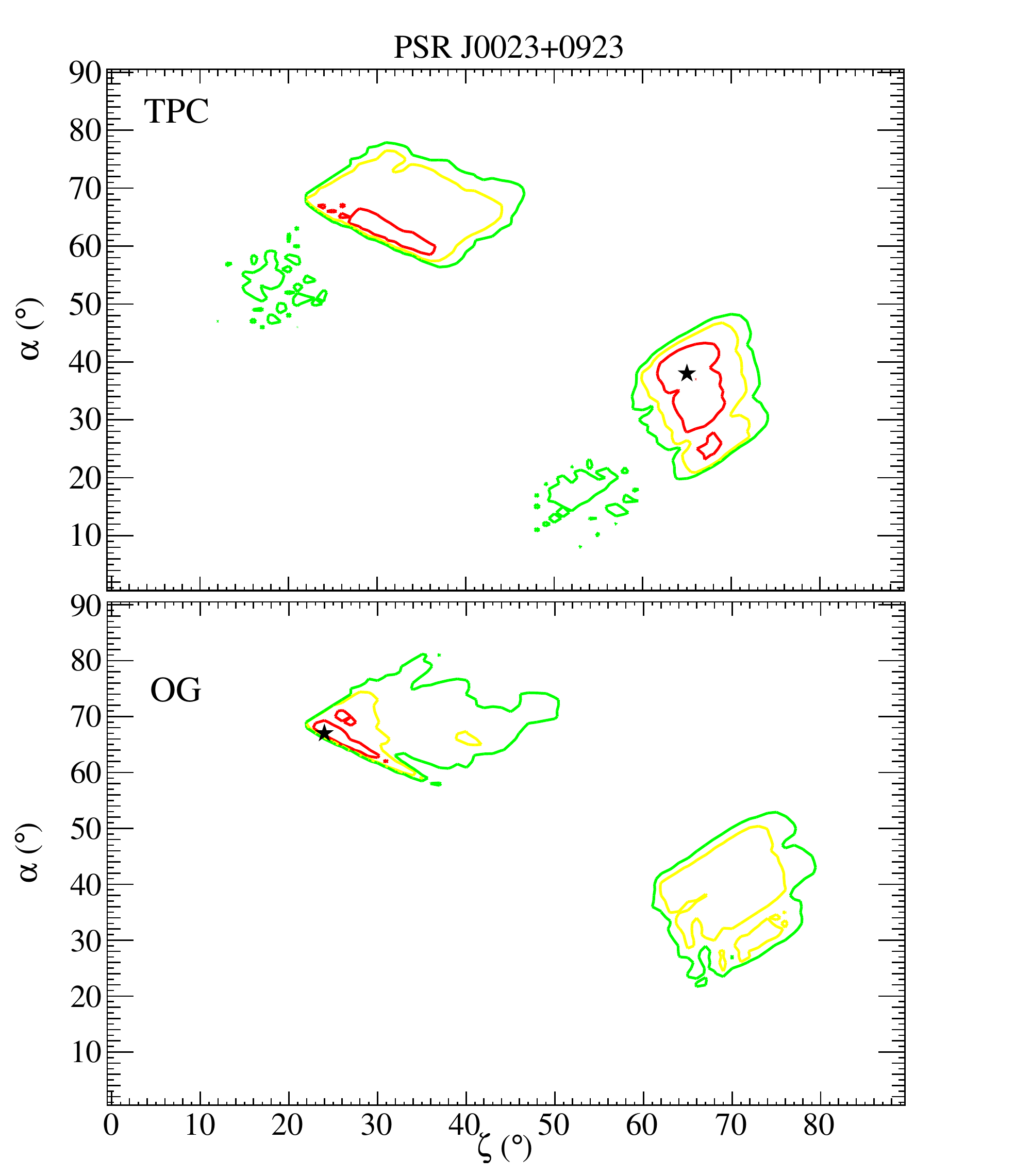}
\end{center}
\caption{Confidence contours in the \al-\ze\ plane for PSR J0023+0923 fit with the TPC \emph{(top)} and OG \emph{(bottom)} models.  The star in each panel marks the best-fit geometry given in Tables \ref{tbl-tpc} and \ref{tbl-og}.  The light gray (red in the online version) contour is 68\% confidence, dark gray (yellow in the online version) is 95\%, and black (green in the online version) is 99\%.\label{app-J0023cont}}
\end{figure}

Figures \ref{app-J0023cont} and \ref{app-J2047cont} show the confidence contours for PSR J0023+0923 and J2047+1053, respectively, using the TPC and OG models.  These are examples of confidence contours that are not simply connected, with two maxima that are almost reflections across the \al=\ze\ line.  For MSPs with this type of confidence contour, changing the radio uncertainty by a factor of 2 can lead to the best-fit jumping from one maxima to the other.  This leads to systematic uncertainties which seem large (changes on the order of 20\DG\ to 30\DG) but that stay within the confidence contours if, not the uncertainties quoted in Section \ref{res}.

The confidence contours in Figure \ref{app-J2047cont} are very elliptical and tilted with respect to the \al\ and \ze\ axes.  Thus, in addition to not being simply connected, the estimated uncertainties in Tables \ref{tbl-tpc} and \ref{tbl-og} overestimate the true size of the contour containing the best-fit.

\begin{figure}
\begin{center}
\includegraphics[height=0.6\textheight]{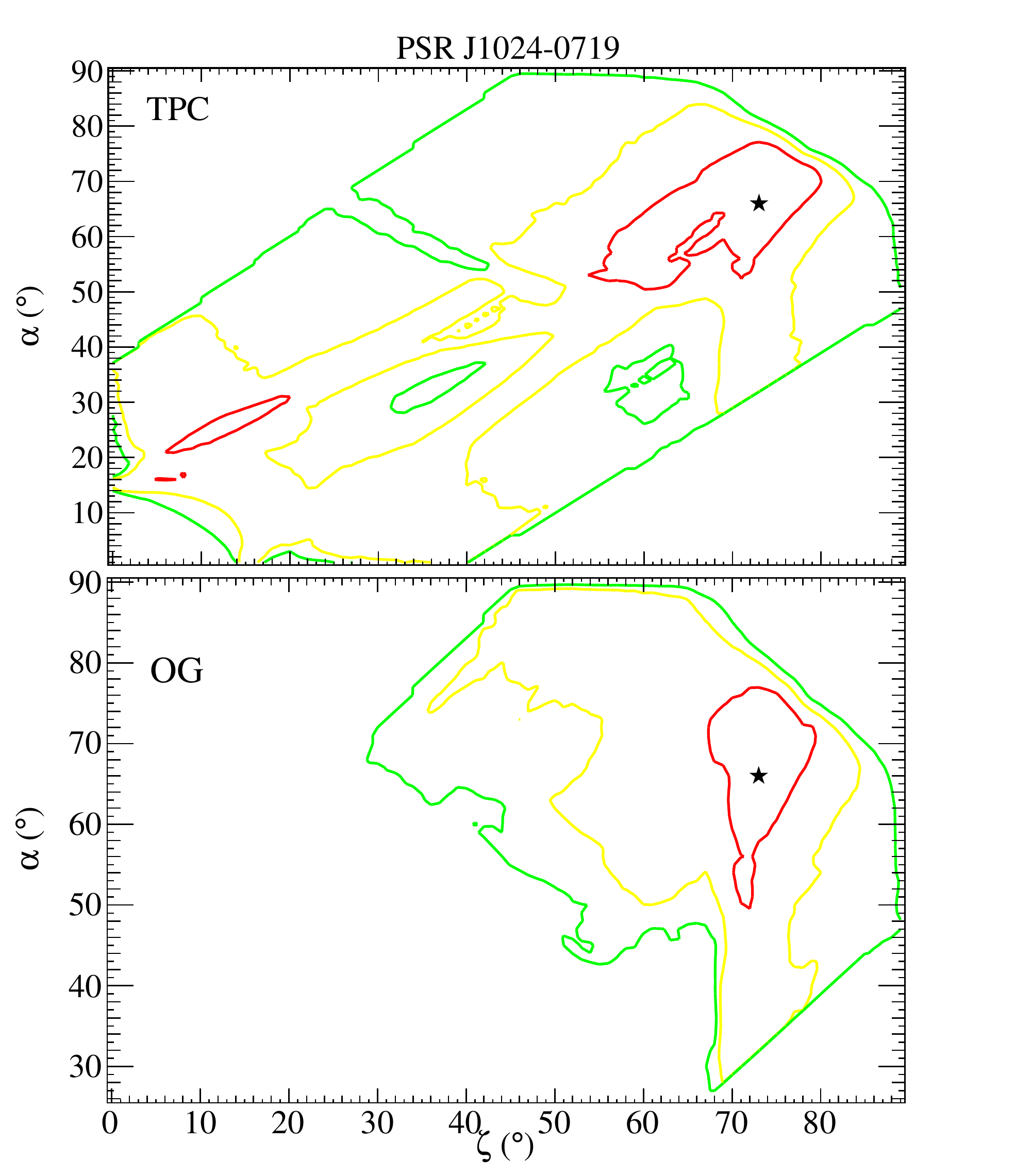}
\end{center}
\caption{Confidence contours in the \al-\ze\ plane for PSR J1024$-$0719 fit with the TPC \emph{(top)} and OG \emph{(bottom)} models.  The star in each panel marks the best-fit geometry given in Tables \ref{tbl-tpc} and \ref{tbl-og}.  The light gray (red in the online version) contour is 68\% confidence, dark gray (yellow in the online version) is 95\%, and black (green in the online version) is 99\%.\label{app-J1024cont}}
\end{figure}

Figure \ref{app-J1024cont} presents confidence contours for PSR J1024$-$0719.  These are examples of confidence contours which cover nearly the entire allowed parameter space (the corners with large \al\ with small \ze\ and small \al\ with large \ze\ are excluded because of the requirement that the line of sight intersects the radio cone and the small \al\ with small \ze\ region is excluded for the OG model due to only having emission above the NCS).  The geometry for this pulsar is not well constrained due to it's faint gamma-ray emission.  More statistics will improve our constraints.  Radio polarization measurements and RVM fits would also constrain the allowed parameter space.

\begin{figure}
\begin{center}
\includegraphics[height=0.6\textheight]{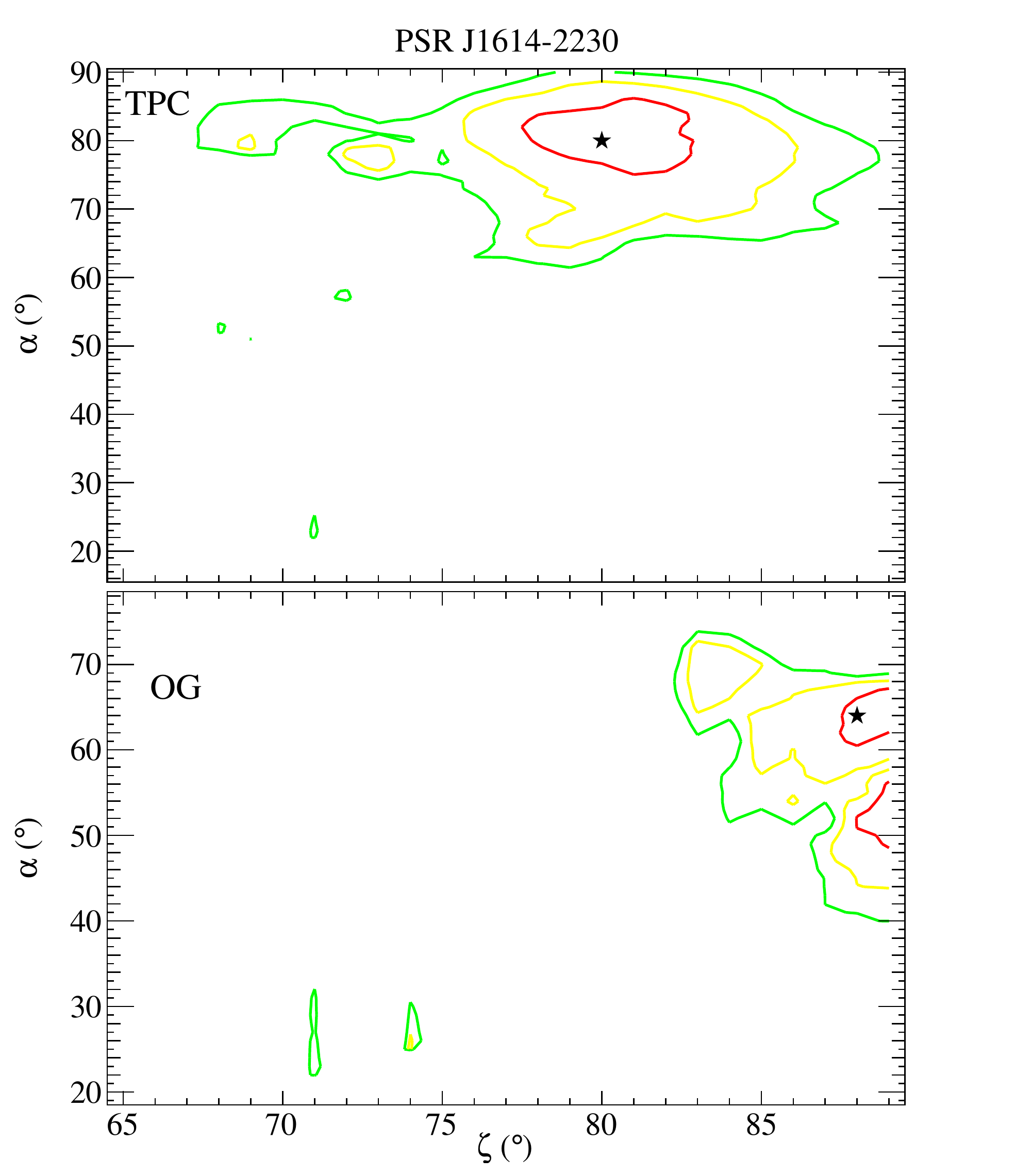}
\end{center}
\caption{Confidence contours in the \al-\ze\ plane for PSR J1614$-$2230 fit with the TPC \emph{(top)} and OG \emph{(bottom)} models.  The star in each panel marks the best-fit geometry given in Tables \ref{tbl-tpc} and \ref{tbl-og}.  The light gray (red in the online version) contour is 68\% confidence, dark gray (yellow in the online version) is 95\%, and black (green in the online version) is 99\%.\label{app-J1614cont}}
\end{figure}

\begin{figure}
\begin{center}
\includegraphics[height=0.6\textheight]{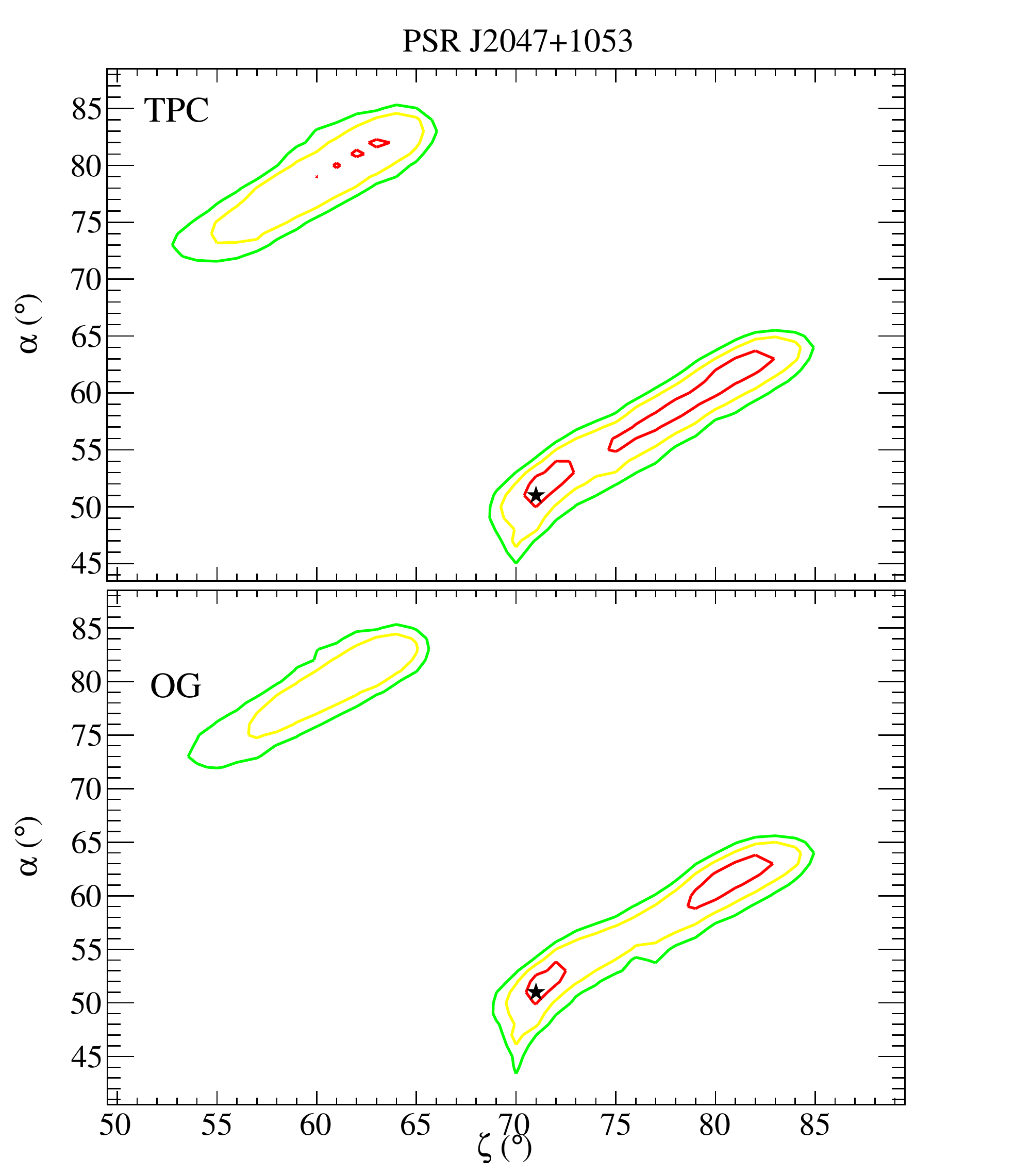}
\end{center}
\caption{Confidence contours in the \al-\ze\ plane for PSR J2047+1053 fit with the TPC \emph{(top)} and OG \emph{(bottom)} models.  The star in each panel marks the best-fit geometry given in Tables \ref{tbl-tpc} and \ref{tbl-og}.  The light gray (red in the online version) contour is 68\% confidence, dark gray (yellow in the online version) is 95\%, and black (green in the online version) is 99\%.\label{app-J2047cont}}
\end{figure}

Figure \ref{app-J1614cont} presents the confidence contours for PSR J1614$-$2230.  These contours, while not simple shapes, are well described by the estimated uncertainties in Tables \ref{tbl-tpc} and \ref{tbl-og} with the 95\% confidence level regions mostly concentrated in one area.

\clearpage

\clearpage

\clearpage

\clearpage

\subsection{CONTOURS FOR CLASS II MSPS}\label{app-cls2cont}

\begin{figure}
\begin{center}
\includegraphics[height=0.75\textheight]{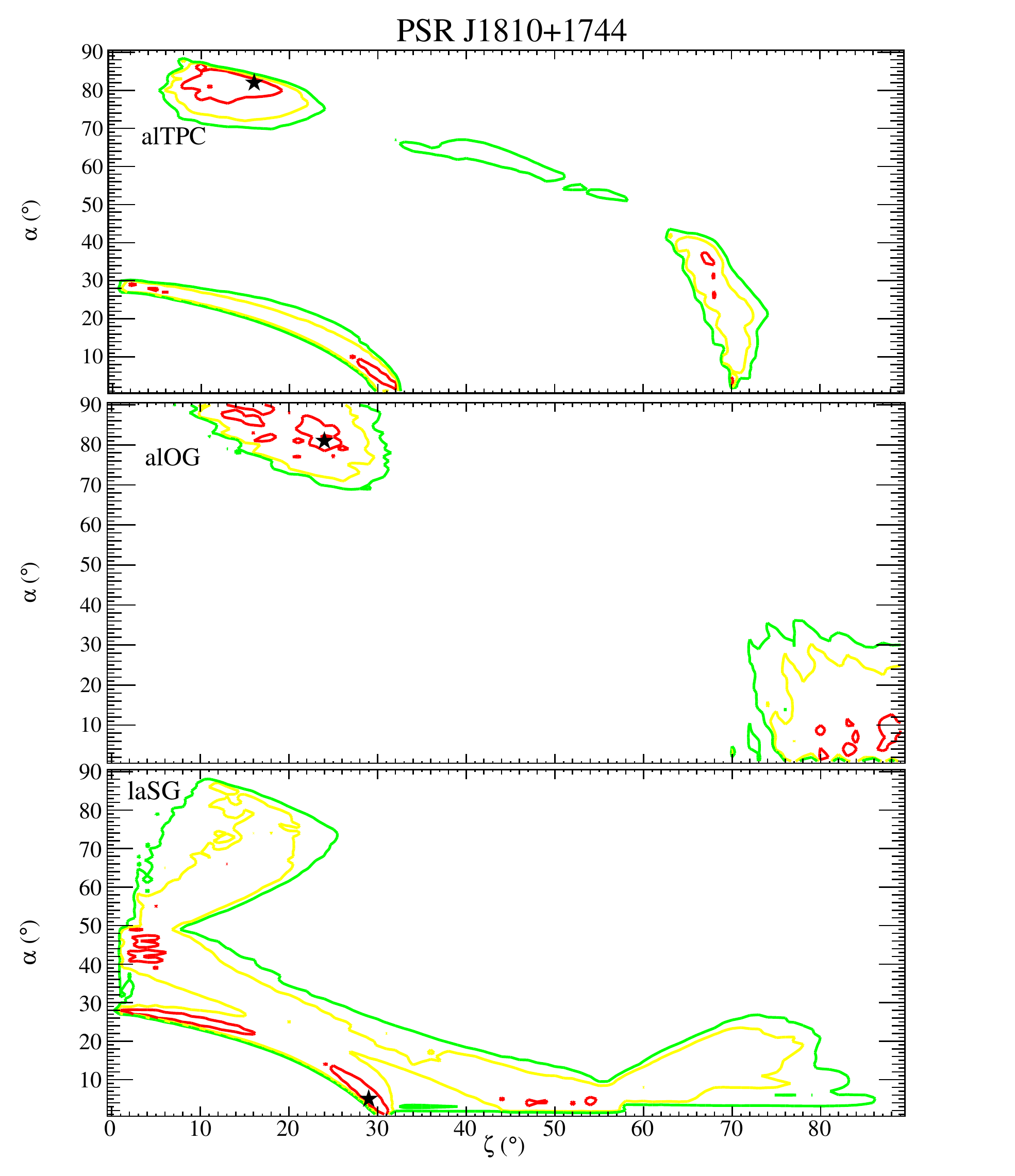}
\end{center}
\caption{Confidence contours in the \al-\ze\ plane for PSR J1810+1744 fit with the alTPC \emph{(top)}, alOG \emph{(middle)}, and laSG \emph{(bottom)} models.  The star in each panel marks the best-fit geometry given in Tables \ref{tbl-altpc}, \ref{tbl-alog}, and \ref{tbl-lasg}.  The light gray (red in the online version) contour is 68\% confidence, dark gray (yellow in the online version) is 95\%, and black (green in the online version) is 99\%.\label{app-J1810cont}}
\end{figure}

Figure \ref{app-J1810cont} presents the confidence contours for PSR J1810+1744 using the alTPC, alOG, and laSG models.  Similar to Figures \ref{app-J0023cont} and \ref{app-J2047cont}, the alTPC and alOG contours show maxima which are not simply connected, there are three regions with 95\% confidence level contours for the alTPC model.  The laSG contours are connected but the shape is not simple and not easily described by the estimated uncertainties in Table \ref{tbl-lasg}.

\begin{figure}
\begin{center}
\includegraphics[height=0.75\textheight]{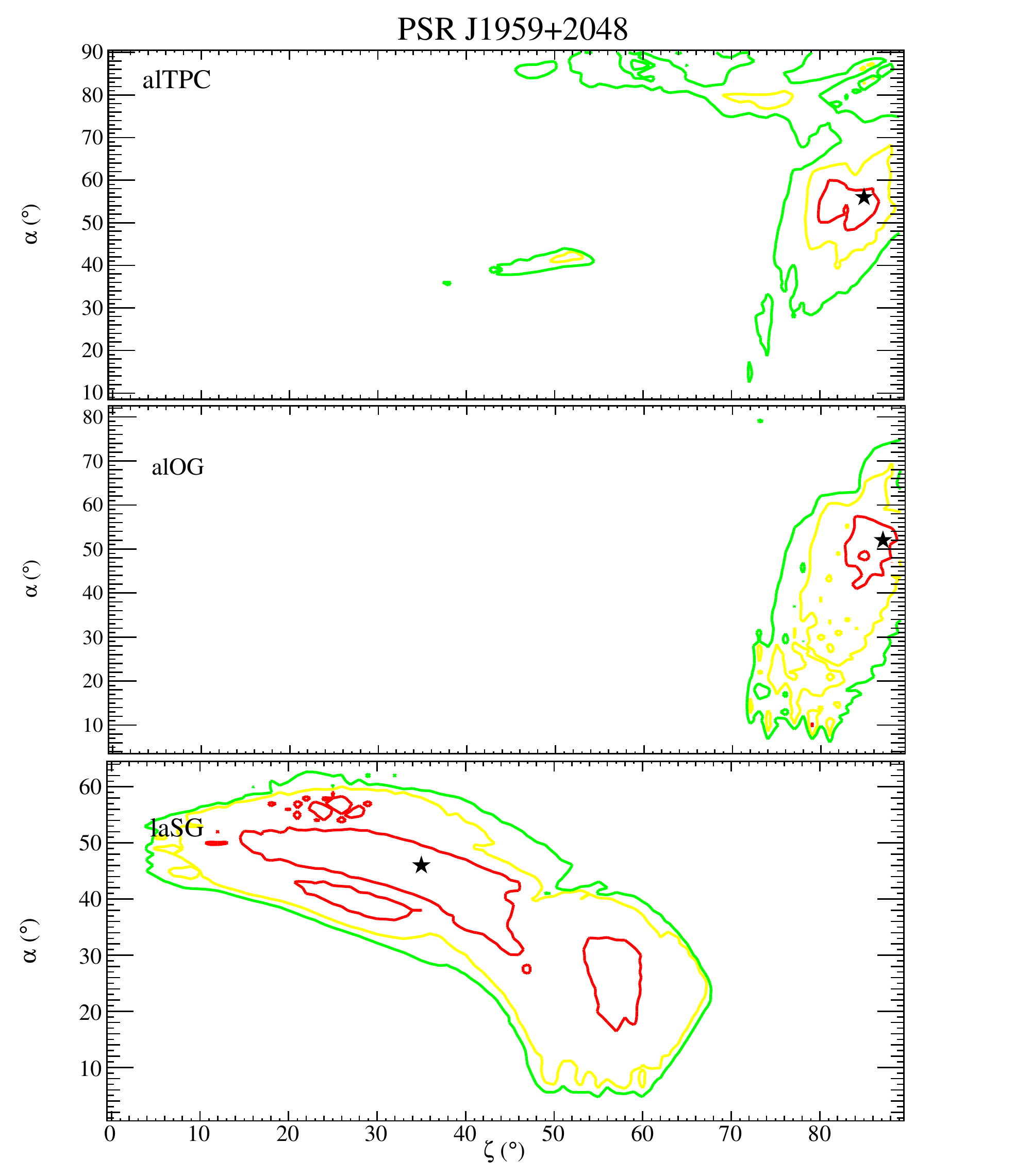}
\end{center}
\caption{Confidence contours in the \al-\ze\ plane for PSR J1959+2048 fit with the alTPC \emph{(top)}, alOG \emph{(middle)}, and laSG \emph{(bottom)} models.  The star in each panel marks the best-fit geometry given in Tables \ref{tbl-altpc}, \ref{tbl-alog}, and \ref{tbl-lasg}.  The light gray (red in the online version) contour is 68\% confidence, dark gray (yellow in the online version) is 95\%, and black (green in the online version) is 99\%.\label{app-J1959cont}}
\end{figure}

Figure \ref{app-J1959cont} presents the confidence contours for PSR J1959+2048.  These confidence contours are connected and generally simple shapes.  Both these contours and those for PSR J1810+1744 show a common case where the alTPC and alOG best-fit geometries are similar but the laSG geometry is quite different, preferring small values of both \al\ and \ze.

\clearpage

\clearpage

\subsection{CONTOURS FOR CLASS III MSPS}\label{app-cls3cont}

\begin{figure}
\begin{center}
\includegraphics[height=0.5\textheight]{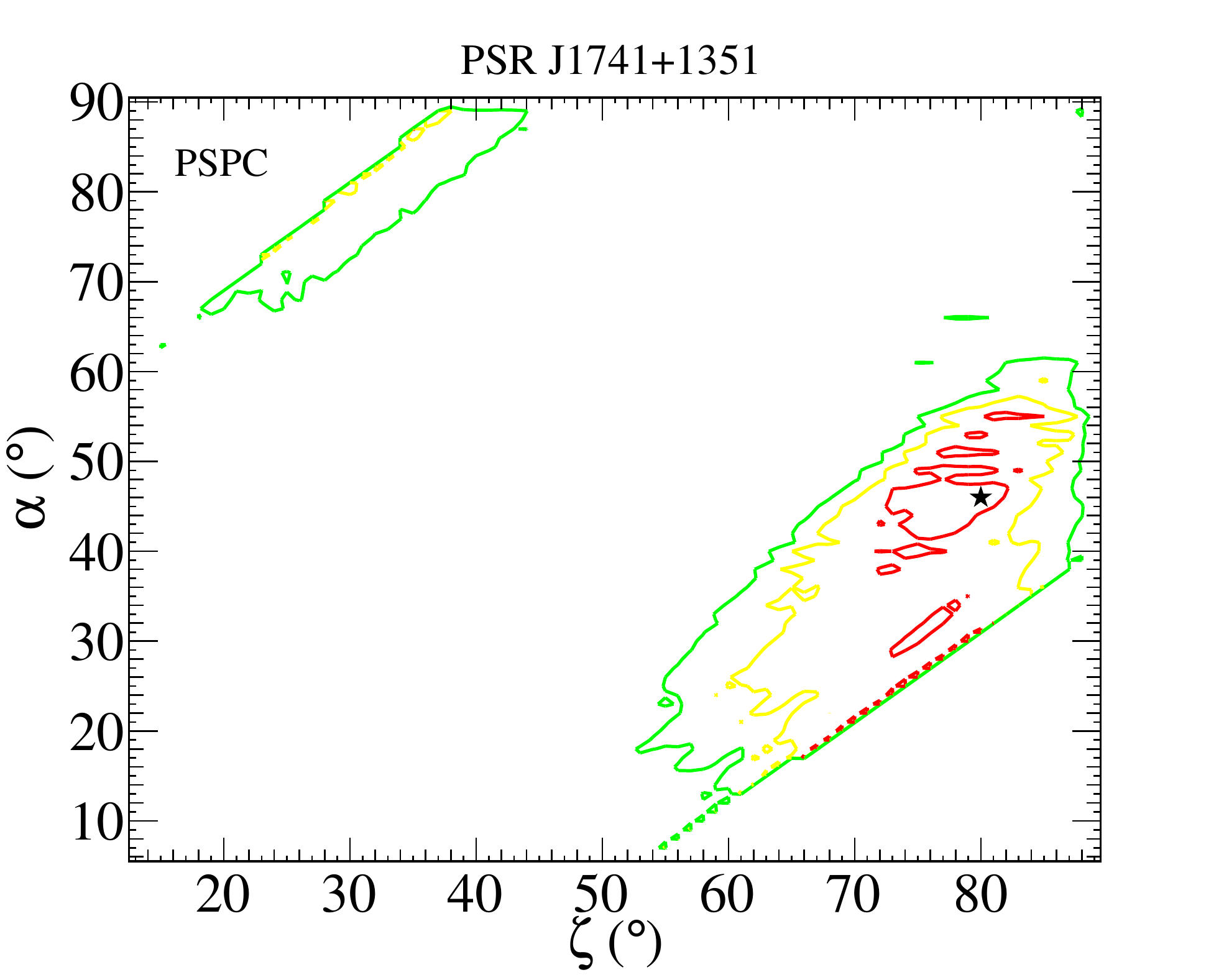}
\end{center}
\caption{Confidence contours in the \al-\ze\ plane for PSR J1741+1351 fit with the PSPC model.  The star marks the best-fit geometry given in Table \ref{tbl-pspc}.  The light gray (red in the online version) contour is 68\% confidence, dark gray (yellow in the online version) is 95\%, and black (green in the online version) is 99\%.\label{app-J1741cont}}
\end{figure}

Figure \ref{app-J1741cont} presents the confidence contours for PSR J1741+1351 fit with the PSPC model.  These contours show maxima which are not simply connected, similar to Figures \ref{app-J0023cont}, \ref{app-J2047cont}, and \ref{app-J1810cont} for Class I and II MSPs.  These maxima do not show the same level of symmetry as in the other cases in which the models have emission coming from narrow gaps.

\begin{figure}
\begin{center}
\includegraphics[height=0.5\textheight]{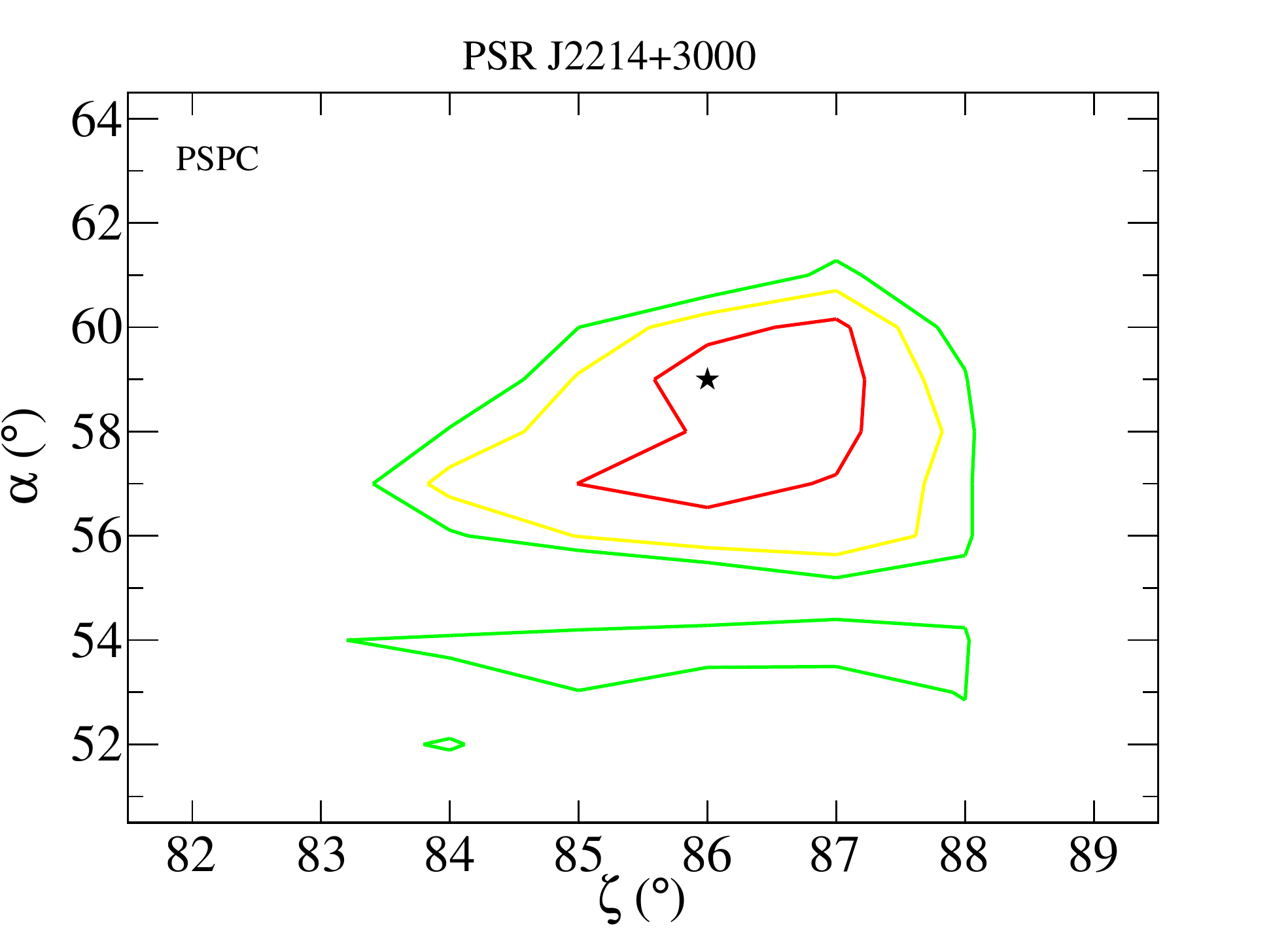}
\end{center}
\caption{Confidence contours in the \al-\ze\ plane for PSR J2214+3000 fit with the PSPC model.  The star marks the best-fit geometry given in Table \ref{tbl-pspc}.  The light gray (red in the online version) contour is 68\% confidence, dark gray (yellow in the online version) is 95\%, and black (green in the online version) is 99\%.\label{app-J2214cont}}
\end{figure}

Figure \ref{app-J2214cont} presents the confidence contours for PSR J2214+3000.  This is among the brighter MSPs in gamma rays leading to tighter constraints on the emission geometry.  This is an example where the uncertainties quoted in Table \ref{tbl-pspc} describe the confidence region well.

\clearpage

\clearpage

\section{SIMULATED EMISSION SKYMAPS}\label{app-skymaps}
This appendix presents several example simulated emission skymaps, corresponding to the best-fit model parameters, for each model class.

\subsection{EMISSION SKYMAPS FOR CLASS I MSPS}\label{app-cls1maps}

\begin{figure}
\hspace*{-25pt}\includegraphics[height=0.7\textheight]{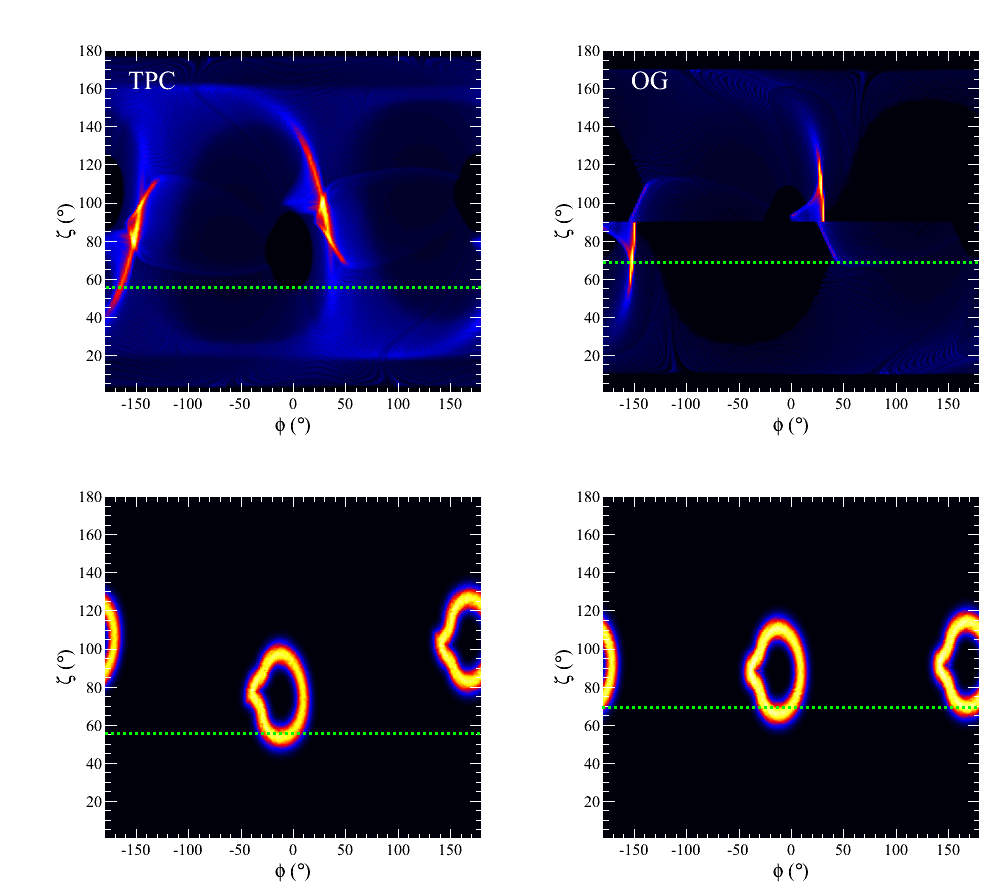}
\caption{Simulated gamma-ray (\emph{top}) and radio (\emph{bottom}) emission skymaps for PSR J0030+0451 fit with the TPC (\emph{left}, \al=74\DG) and OG (\emph{right}, \al=88\DG) models.  The white (green in the online version) dashed line in each panel denotes the best-fit \ze.  The color scale increases from black to dark gray to light gray to white (black to blue to red to yellow to white in the online version).\label{app-J0030map}}
\end{figure}

Figure \ref{app-J0030map} presents the simulated emission skymaps for PSR J0030+0451 corresponding to the best-fit TPC and OG models with a hollow-cone radio beam.  Both models have high values of \al\ and \ze.  Only in the OG model does the line of sight intersect both radio cones leading to an interpulse.  Even though emission is only collected from below the null charge surface, this geometry would allow for off-peak emission for \ze\ near 0\DG\ or 180\DG\ but the predicted radio profile would be much different, if the cone was intersected at all.

\begin{figure}
\hspace*{-25pt}\includegraphics[height=0.7\textheight]{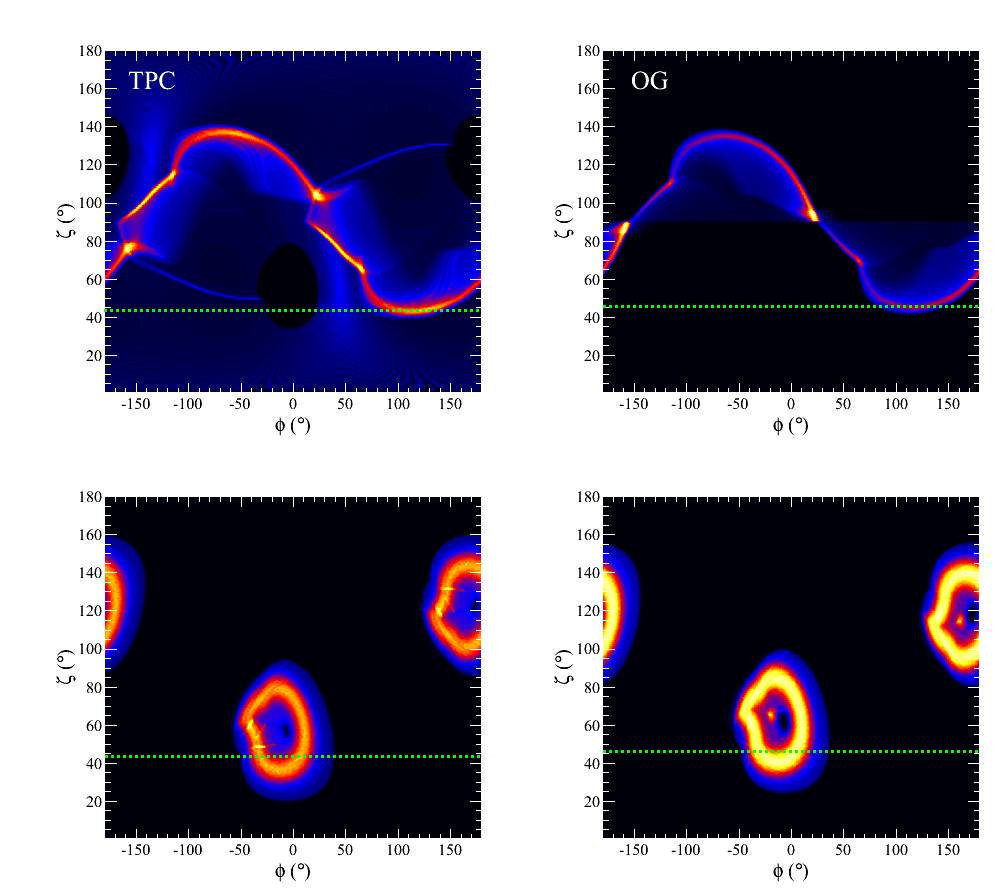}
\caption{Simulated gamma-ray (\emph{top}) and radio (\emph{bottom}) emission skymaps for PSR J0613$-$0200 fit with the TPC (\emph{left}, \al=55\DG) and OG (\emph{right}, \al=60\DG) models.  The white (green in the online version) dashed line in each panel denotes the best-fit \ze.  The color scale increases from black to dark gray to light gray to white (black to blue to red to yellow to white in the online version).\label{app-J0613map}}
\end{figure}

Figure \ref{app-J0613map} presents the simulated emission skymaps for PSR J0613$-$0200.  The radio model for this MSP included both a hollow-cone and a core beam, the latter being obvious in the lower level emission extending beyond the cone (compare to the lower panels of Figure \ref{app-J0030map}, for instance).  With more moderate values of \al, these skymaps demonstrate how no off-peak emission is predicted from the OG model.

\begin{figure}
\hspace*{-25pt}\includegraphics[height=0.7\textheight]{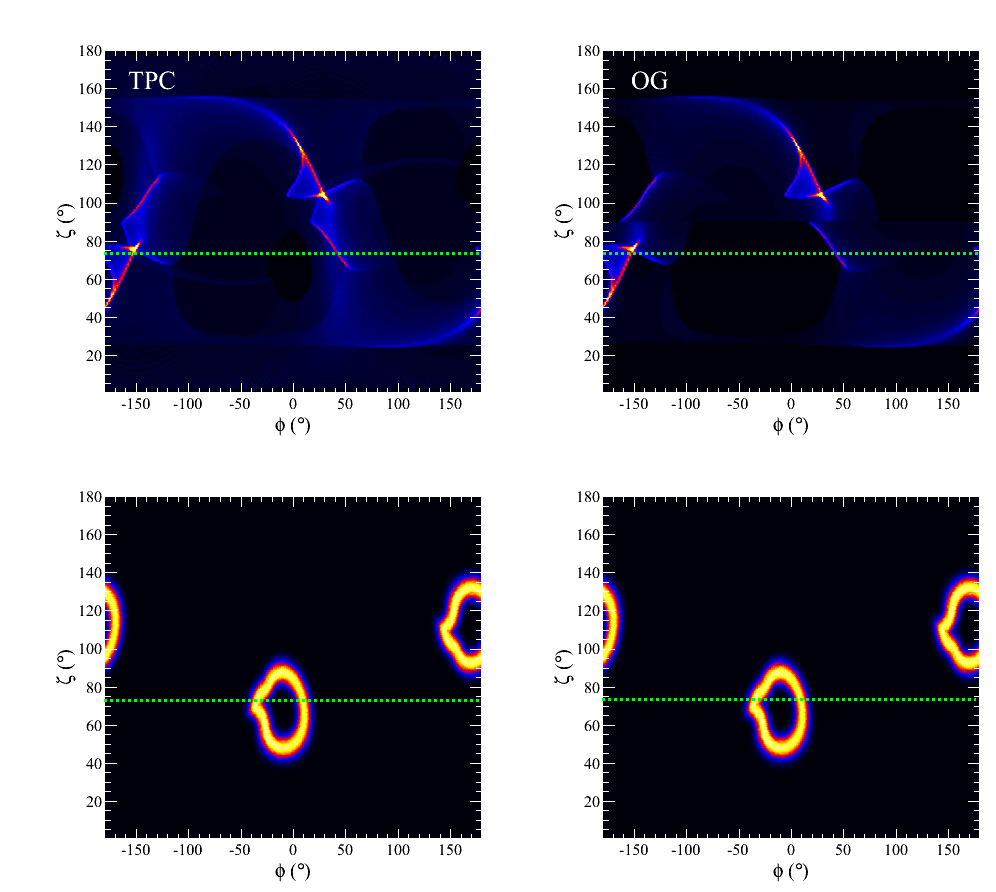}
\caption{Simulated gamma-ray (\emph{top}) and radio (\emph{bottom}) emission skymaps for PSR J1024$-$0719 fit with the TPC (\emph{left}, \al=66\DG) and OG (\emph{right}, \al=66\DG) models.  The white (green in the online version) dashed line in each panel denotes the best-fit \ze.  The color scale increases from black to dark gray to light gray to white (black to blue to red to yellow to white in the online version).\label{app-J1024map}}
\end{figure}

Figure \ref{app-J1024map} presents the simulated emission skymaps for PSR J1024$-$0719.  This is an example where the best-fit geometries for the TPC and OG models are the same.  From the lower panels, this geometry would predict two radio peaks spaced very close to each other, but when we reduce the number of bins they merge into one peak (see Figure \ref{app-J1024LCs}).  While there does appear to be structure in the radio profile shown in 2PC, it is clear that our best-fit \ze\ is not correct; however, there is structure in the observed profile which may be matched by a slightly higher \ze\ which can lead to multiple peaks where the emission level does not drop all the way to zero inbetween.

\begin{figure}
\hspace*{-25pt}\includegraphics[height=0.7\textheight]{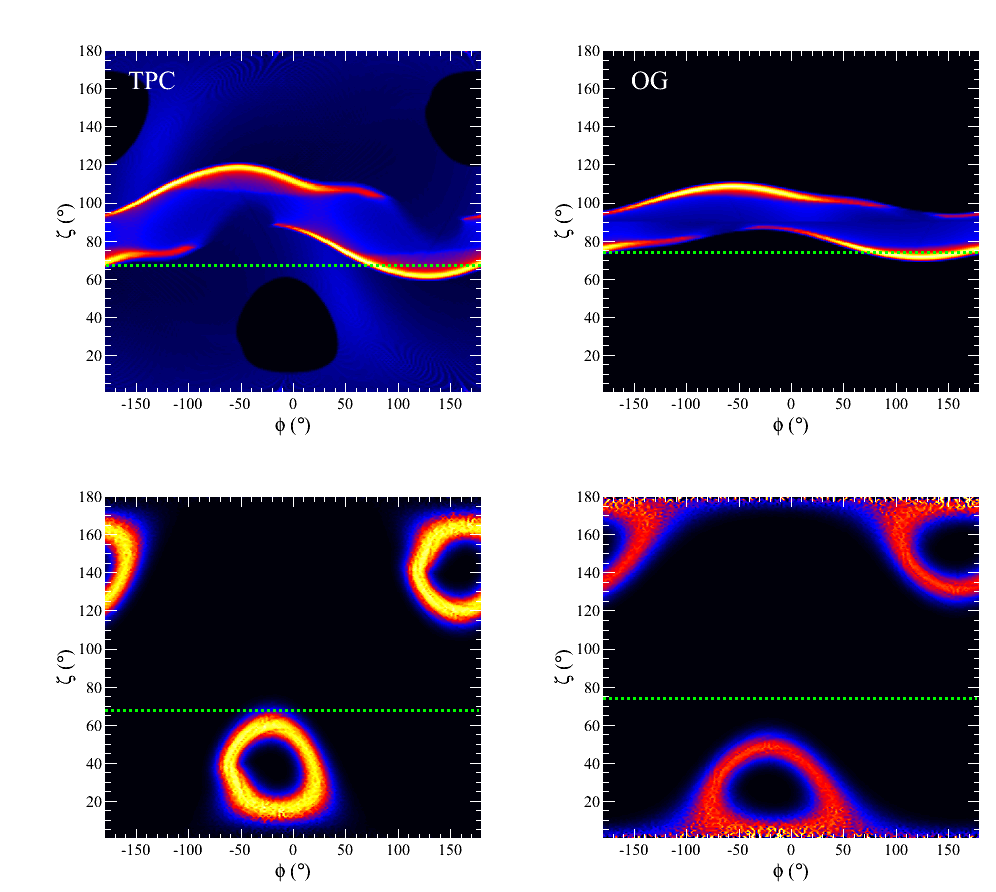}
\caption{Simulated gamma-ray (\emph{top}) and radio (\emph{bottom}) emission skymaps for PSR J2017+0603 fit with the TPC (\emph{left}, \al=34\DG) and OG (\emph{right}, \al=23\DG) models.  The white (green in the online version) dashed line in each panel denotes the best-fit \ze.  The color scale increases from black to dark gray to light gray to white (black to blue to red to yellow to white in the online version).\label{app-J2017map}}
\end{figure}

Figure \ref{app-J2017map} presents the simulated emission skymaps for PSR J2017+0603.  This is another case where the TPC and OG best-fit geometries are similar, with low \al\ and high \ze.  For the radio skymap corresponding to the OG fit, lower-right panel, the solid angle correction to the bins near \al=0\DG\ leads to very bright pixels in this range making the rest of the cone appear artificially dimmer compared to the TPC case with slightly larger \al.

\clearpage

\clearpage

\clearpage

\clearpage

\subsection{EMISSION SKYMAPS FOR CLASS II MSPS}\label{app-cls2maps}

Figure \ref{app-J1823map} presents the simulated emission skymaps for PSR J1823$-$3021A corresponding to the best-fit alTPC, alOG, and laSG models.  The alTPC and alOG skymaps corresponding to the radio models, lower panels, are examples of emission occurring at high altitudes in the magnetosphere and only over a small extent.  The laSG skymaps show an example of the radio emission region having a smaller emitting gap width than the gamma-ray emission.  The laSG geometry appears to be a large hollow cone, similar to the radio model used for Class I and III MSPs but covering a greater portion of the skymap.

Figure \ref{app-J1959map} presents the simulated emission skymaps for PSR J1959+2048.  The alTPC and alOG skymaps show examples of the gamma-ray and radio emission regions having the same minimum and maximum altitudes.  The laSG skymaps for this MSP show instances where the emission pattern on the sky is more complex than a hollow cone, contrary to the laSG skymaps of PSR J1823$-$3021A.

\begin{landscape}
\begin{figure}
\centering
\hspace*{-45pt}\includegraphics[width=1.25\textwidth]{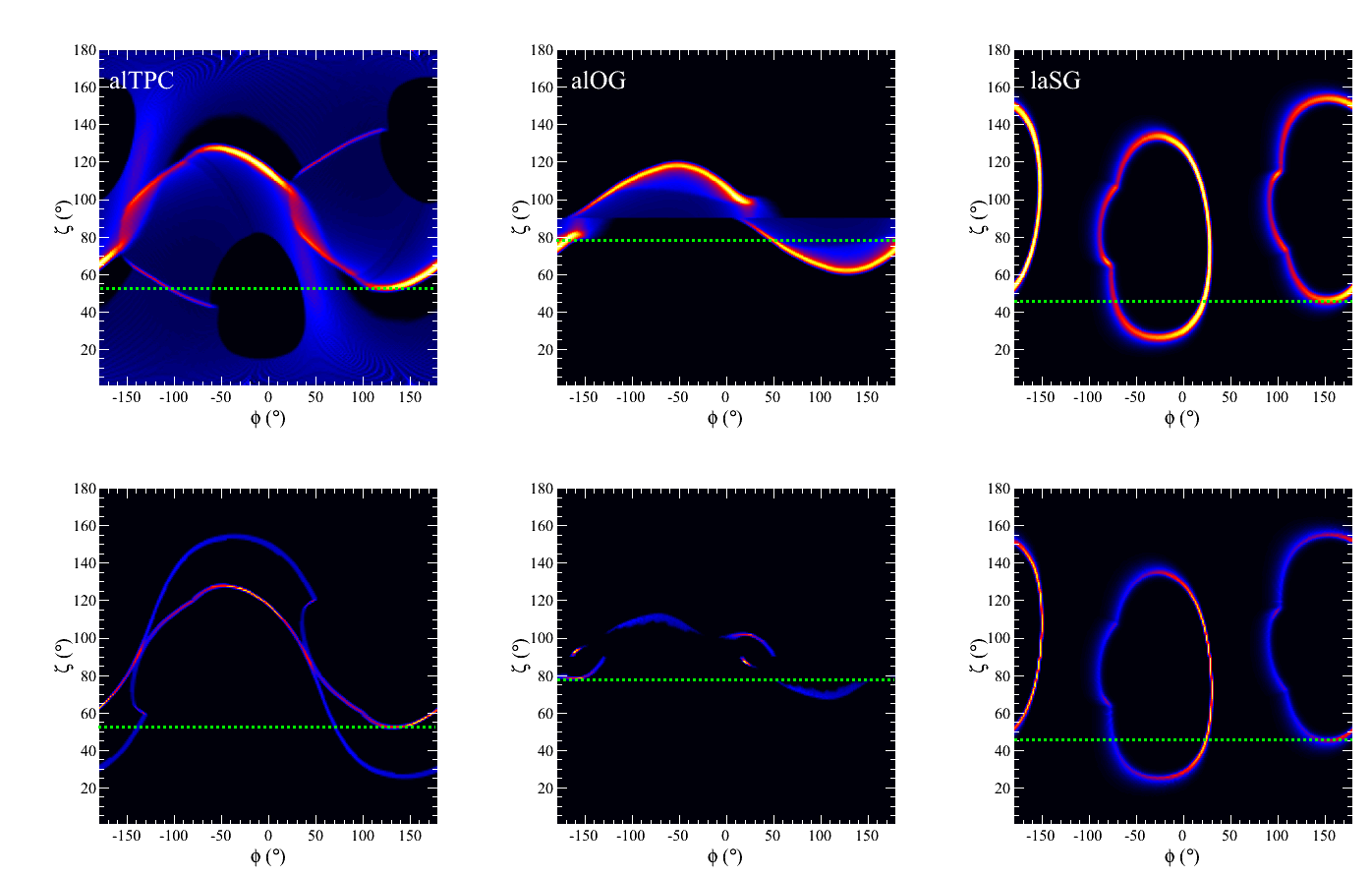}
\caption{Simulated gamma-ray (\emph{top}) and radio (\emph{bottom}) emission skymaps for PSR J1823$-$3021A fit with the alTPC (\emph{left}, \al=46\DG), alOG (\emph{middle}, \al=42\DG), and laSG (\emph{right}, \al=78\DG) models.  The white (green in the online version) dashed line in each panel denotes the best-fit \ze.  The color scale increases from black to dark gray to light gray to white (black to blue to red to yellow to white in the online version).\label{app-J1823map}}
\end{figure}
\clearpage

\begin{figure}
\centering
\hspace*{-45pt}\includegraphics[width=1.25\textwidth]{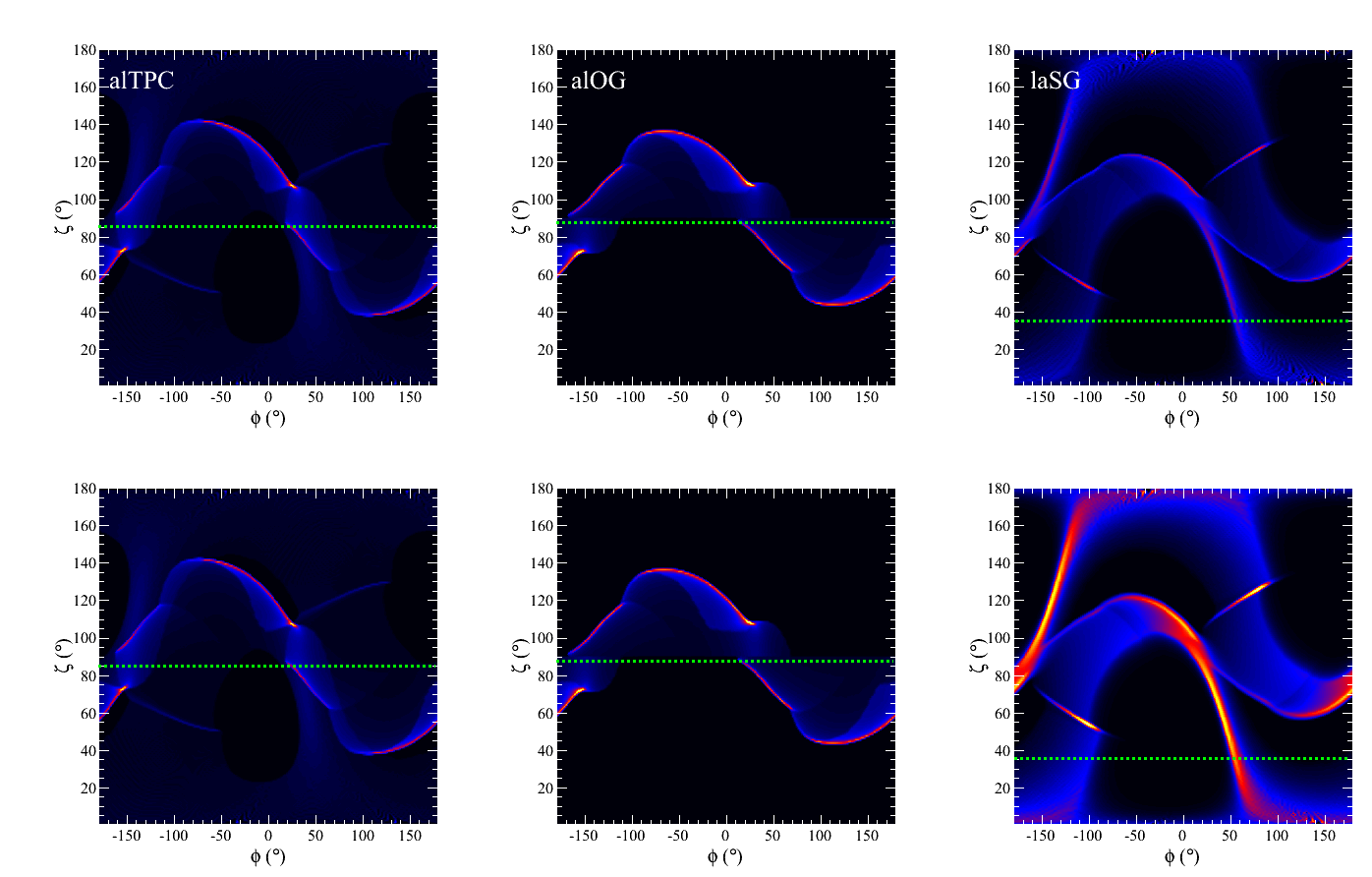}
\caption{Simulated gamma-ray (\emph{top}) and radio (\emph{bottom}) emission skymaps for PSR J1959+2048 fit with the alTPC (\emph{left}, \al=56\DG), alOG (\emph{middle}, \al=52\DG), and laSG (\emph{right}, \al=46\DG) models.  The white (green in the online version) dashed line in each panel denotes the best-fit \ze.  The color scale increases from black to dark gray to light gray to white (black to blue to red to yellow to white in the online version).\label{app-J1959map}}
\end{figure}
\clearpage
\end{landscape}

\subsection{EMISSION SKYMAPS FOR CLASS III MSPS}\label{app-cls3maps}

\begin{figure}
\begin{center}
\hspace*{-15pt}\includegraphics[height=0.7\textheight]{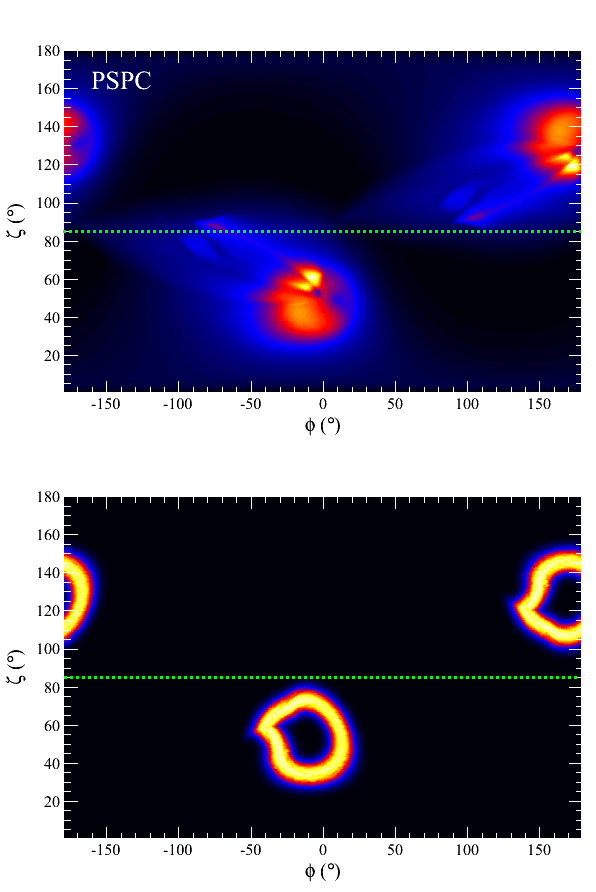}
\end{center}
\caption{Simulated gamma-ray (\emph{top}) and radio (\emph{bottom}) emission skymaps for PSR J1744$-$1134 fit with the PSPC model (\al=51\DG).  The white (green in the online version) dashed line in each panel denotes the best-fit \ze.  The color scale increases from black to dark gray to light gray to white (black to blue to red to yellow to white in the online version).\label{app-J1744map}}
\end{figure}

\begin{figure}
\begin{center}
\hspace*{-15pt}\includegraphics[height=0.7\textheight]{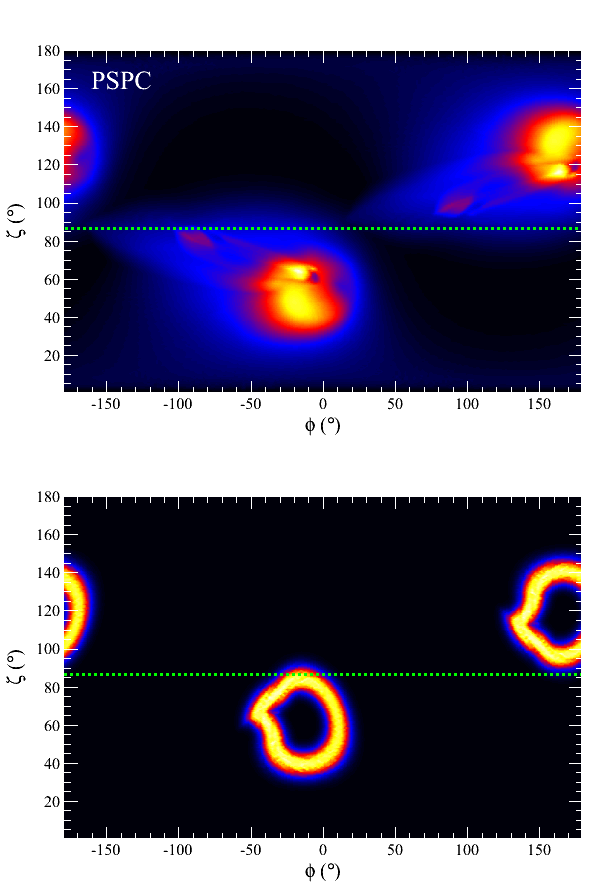}
\end{center}
\caption{Simulated gamma-ray (\emph{top}) and radio (\emph{bottom}) emission skymaps for PSR J2214+3000 fit with the PSPC model (\al=59\DG).  The white (green in the online version) dashed line in each panel denotes the best-fit \ze.  The color scale increases from black to dark gray to light gray to white (black to blue to red to yellow to white in the online version).\label{app-J2214map}}
\end{figure}

Figures \ref{app-J1744map} and \ref{app-J2214map} show the simulated emission skymaps for PSR J1744$-$1134 and J2214+3000 corresponding to the best-fit PSPC models.  Both gamma-ray emission maps, top panels, demonstrate how this emission model can lead to emission over most, if not all, of the pulse phase.  The radio emission skymap for PSR J1744$-$1134, bottom panel of Figure \ref{app-J1744map}, shows how just clipping the cone can lead to a sharp radio pulse (the dashed line appears to miss the cone but that is just an artifact of the color scale, the emission does extend slightly beyond the cone visible with this scaling due to the Gaussian profile).  The radio emission skymap for PSR J2214+3000, bottom panel of Figure \ref{app-J2214map} demonstrates how the line of sight can intersect both cones but still give radio peaks with very different intensities.

\clearpage

\clearpage

\end{document}